\documentclass[final,3p,times]{elsarticle}

\usepackage{amssymb}
\usepackage{color}

\biboptions{sort&compress}
\journal{Physics Reports}

\usepackage[utf8]{inputenc}
\usepackage[T1]{fontenc}                
\usepackage{epsfig,dcolumn}          
\usepackage{dcolumn}
\graphicspath{ {figures/} }
\usepackage{amsmath}
\usepackage{bbm}
\usepackage{amsfonts}
\usepackage{xspace}

\usepackage[final]{pdfpages}
\usepackage{lipsum}
\usepackage{enumitem} 
\usepackage{cancel}
\usepackage[usenames,dvipsnames,svgnames,table]{xcolor}
\usepackage{float}

\usepackage{multirow}

\usepackage[nottoc,notlot,notlof]{tocbibind}

\usepackage{bm} 
\usepackage{mfirstuc} 
\usepackage{soul,color}

\allowdisplaybreaks

\usepackage{hyperref}
\hypersetup{colorlinks = true,linktocpage=true,    allcolors=SeaGreen}

\usepackage[capitalise]{cleveref}

\newcommand{\be}{\begin{equation}}
\newcommand{\ee}{\end{equation}}

\newcommand{\degree}{{\rm o}}
\newcommand{\im}{\mbox{Im }}
\newcommand{\re}{\mbox{Re }}

\newcommand{\gev}{{\mbox{GeV}\,}}
\newcommand{\mev}{{\mbox{MeV}\,}}

\newcommand{\sig}
{\ensuremath{\sigma/f_0(500)}\xspace}

\newcommand{\kap}
{\ensuremath{\kappa/K_0^*(700)}\xspace}

\newcommand{\pipi}
{\ensuremath{\pi \pi \rightarrow \pi \pi}\xspace}

\newcommand{\pik}
{\ensuremath{\pi K \rightarrow \pi K}\xspace}

\newcommand{\pipikk}
{\ensuremath{\pi \pi \rightarrow K \bar K}\xspace}

\newcommand{\addReviewer}[2]{
  \expandafter\newcommand\csname #1\endcsname[1]{{\bf \color{#2} \capitalisewords{#1}:\,##1}}
  \expandafter\newcommand\csname #1cor\endcsname[2]{{\color{#2} \capitalisewords{#1}:\,\st{##1}{\bf ##2}}}
  \expandafter\newcommand\csname #1color\endcsname{#2}
}

\definecolor{chromeyellow}{rgb}{1.0, 0.65, 0.0}
\definecolor{DodgeBlue}{rgb}{0.118, 0.565,1.000}
\definecolor{asparagus}{rgb}{0.53, 0.66, 0.42}
\definecolor{cadmiumgreen}{rgb}{0.0, 0.42, 0.24}

\addReviewer{CP}{orange}
\addReviewer{AR}{asparagus}
\addReviewer{JR}{red}

\begin{document}

\hypersetup{allcolors=SeaGreen}

\hfill{
\vbox{
\hbox{JLAB-THY-20-3276}
}}

\begin{frontmatter}



\title{Dispersive $\pik$ and $\pipikk$ amplitudes from scattering data, threshold parameters, and the lightest strange resonance $\kappa$ or $K^*_0(700)$.}



\author{Jos\'e R. Pel\'aez}

\address{Departamento de F\'{\i}sica Te\'orica. Universidad Complutense and IPARCOS. 28040 Madrid. SPAIN}

\author{Arkaitz Rodas}

\address{Department  of  Physics, College  of  William  and  Mary,  Williamsburg,  VA  23187,  USA and Thomas Jefferson National Accelerator Facility, 12000 Jefferson Avenue, Newport News, VA 23606, USA}

\begin{abstract}
We discuss the simultaneous dispersive analyses of $\pik$ and $\pipikk$ scattering data and the $\kap$ resonance. The unprecedented statistics of present and future hadron experiments, modern lattice QCD calculations, and the wealth of new states and decays require such precise and model-independent analyses to describe final state interactions. We review the existing and often conflicting data and explain in detail the derivation of the relevant dispersion relations, maximizing their applicability range. Next, we review and extend the caveats on some data, showing their inconsistency with dispersion relations. Our main result is the derivation and compilation of precise amplitude parameterizations constrained by several $\pik$ and $\pipikk$ dispersion relations. These constrained parameterizations are easily implementable and provide the rigor and accuracy needed for modern experimental and phenomenological hadron physics. As applications, after reviewing their status and interest, we will provide new precise threshold and subthreshold parameters and review our dispersive determination of the controversial $\kap$ resonance and other light-strange resonances.
\end{abstract}


\end{frontmatter}



\tableofcontents
\newpage
\section{Introduction}
\label{sec:intro}

\subsection{Motivation}
\label{sec:Motivation}
\begin{flushright} {\it ``There is no excellent beauty that hath not some strangeness in the proportion.''\\  
{\footnotesize Francis Bacon. ``Of Beauty'', in Essays (1625)}}
\end{flushright}

Even though the existing data on \pik and \pipikk scattering below 2 GeV were obtained between 30 and 50 years ago, there is a longstanding interest in them, which has been renewed over the last years motivated by the following reasons: First, they are relevant by themselves because they test hadron physics and particularly the low-energy realm. 
Second, because pions and kaons are the lightest hadrons and then they appear in the final state of most hadronic processes, which thus require a description of their interactions. Third, they are of interest for Hadron Spectroscopy, due to the resonances that can be identified in these processes. Finally,  on the theoretical side, since  the low-energy regime lies out of reach for standard QCD perturbation theory, they test the most recent developments on non-perturbative QCD techniques, like lattice QCD calculations and the low-energy QCD effective theory, known as Chiral Perturbation Theory (ChPT).  

\begin{figure}[b]
\begin{center}
\includegraphics[width=0.48\textwidth]{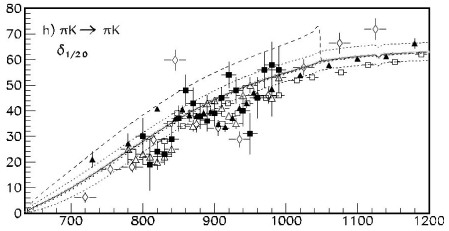}
\includegraphics[width=0.5\textwidth]{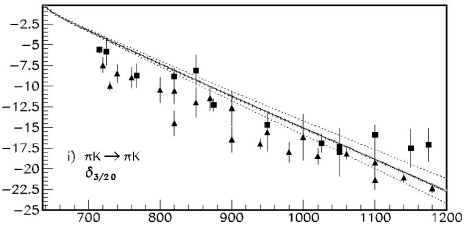}
 \caption{\rm \label{fig:chaos}
Plots are taken from \cite{GomezNicola:2001as} to illustrate the conflicting data sets and a rather popular model (unitarized Chiral Perturbation Theory).
They represent the phase shifts of the isospin 1/2 (Left) and 3/2 (Right) scalar partial waves. As we will discuss in the main text, the 1/2 phase shifts are not measured directly. See \cite{GomezNicola:2001as} for experimental references.}
\end{center}
\end{figure}

Given their relevance and the many years passed since these interactions were measured, readers outside the field may assume that a rigorous and accurate description should have been found many years ago. 
However, this is not the case and, in brief, this is due to what we could call the ``data problem'' and the ``model-dependence problem''.
The data problem is that the \pik~\cite{Cho:1970fb,Bakker:1970wg,Linglin:1973ci,Jongejans:1973pn,Estabrooks:1977xe,Aston:1987ir}
and \pipikk  \cite{Cohen:1980cq,Etkin:1981sg,Longacre:1986fh} scattering data, obtained in the '70s and '80s,  were
extracted indirectly from meson--nucleon scattering experiments. Unfortunately, this technique is plagued with systematic uncertainties, leading to conflict within or between data sets, particularly for the controversial $S$-wave, as we will review here in section \ref{sec:Data}. 
The model-dependence problem arises because, since we cannot use a systematic approach like QCD,  for many years simple models and fits  were considered good enough to describe such data. Moreover, given that conflicting data exist, any model providing a mere qualitative description was also considered acceptable, even good if the description was semiquantitative.
This situation with data and models is illustrated in Fig.~\ref{fig:chaos}. Needless to say that this state of affairs has led, for instance, to different resonance contents in different models, or different parameters for the same resonance, to scarce robust and precise determinations of low-energy parameters, etc... 
Of course, the use of models may have been justified at first, but modern hadron physics demands a more precise, rigorous, and model-independent description.

Dispersion relations, which, as we will also review in section \ref{sec:DR}, are a consequence of fundamental properties like relativistic causality and crossing symmetry, provide a solution to these two problems. First, concerning the data problem, they provide stringent constraints on  amplitudes, which allow us to neglect or identify inconsistent data and restrict the fits that can be acceptable. We will review in section \ref{sec:UFD} how simple and very nice-looking data  fits fail to satisfy the dispersive representation. Sometimes, dispersion relations can be solved in a given regime and provide information without using data there. However, in this report, we want to analyze the data and our main result will be to provide in section \ref{sec:CFD} a Constrained Fit to Data (CFD) that satisfies all the dispersive constraints. Second,  since the dispersive constraints are written in terms of integral relations, they wash out any dependence on the details of input parameterizations. 
Moreover, dispersion relations provide the correct and model-independent analytic continuation to the complex plane that allows for a rigorous determination of the existence and parameters of resonances, to which section \ref{sec:kappa} is dedicated. In particular, after reviewing the general state of the art in \ref{sec:kapreview}, we will dedicate subsection \ref{sec:pades} to the use of analytic methods to minimize model dependencies  when extracting parameters of strange resonances and
 subsection \ref{sec:kapdr} to the dispersive determinations of the still debated \kap meson. In addition, a dispersive determination of the non-ordinary Regge trajectory of the \kap, using as input its pole parameters, is presented in subsection \ref{subsec:nature}.

Finally,  apart from their model independence, the use of integrals increases the precision when calculating certain observables, like the threshold or subthreshold parameters. In section \ref{sec:sumrules} we will thus review these so-called sum rule determinations, also providing new results or updating old ones. Further applications related to the  $\pi K$ sigma term and a brief review of the $\pipikk$ contribution to the calculation of the anomalous magnetic moment of the muon are also presented in subsections \ref{sec:sigterm} and \ref{sec:g-2}, respectively.

All the advantages of dispersion theory listed above are common to many other processes where they have provided successful descriptions highly demanded by modern developments. The relatively recent and comprehensive reports on their application  to $\pi\pi\rightarrow\pi\pi$ \cite{Ananthanarayan:2000ht,Colangelo:2001df,Pelaez:2015qba} and $\pi N\rightarrow\pi N$ \cite{Hoferichter:2015hva} are illustrative of this demand. Our aim here is to provide such a comprehensive report for \pik and \pipikk.

\subsection{State of the art and goals}

Let us then describe in detail the present situation of the pieces of motivation we have enumerated above and the objectives we want to address in this report.

\subsubsection{The interest in \pik and \pipikk interactions by themselves}

First of all, these processes are interesting by themselves, since, together with $\pi\pi\rightarrow \pi\pi$ scattering, they are the simplest two-body interactions of hadrons. Moreover, these are the lightest mesons available, and, being pseudoscalar particles, they do not have the complications associated with their spin. Thus, one would expect that any basic understanding of hadronic interactions should be able to describe these processes.  

Indeed, over almost five decades, a lot of work has been devoted to building phenomenological models to describe \pik and/or \pipikk scattering. Until the late '70s, we refer the reader to the excellent review in \cite{Lang:1978fk} and after that a wide variety of these models can be found in \cite{vanBeveren:1986ea,Au:1986vs,Kaminski:1993zb,Bugg:1996ki,Ishida:1997wn,Kaminski:1997gc,Bajc:1997nx,Oller:1997ng,Oller:1997ti,Oller:1998hw,Oller:1998zr,Guerrero:1998ei,Minkowski:1998mf,Black:1998zc,Locher:1997gr,GomezNicola:2001as,Close:2002zu,Kelkar:2003iv,Maiani:2004uc,Pelaez:2004xp,Amsler:2004ps,vanBeveren:2005ha,Jaffe:2007id,Albaladejo:2008qa,Guo:2011aa,Fariborz:2015bff}. Many of them will be discussed below.  However, in this report, we will focus on the data analysis within the model-independent dispersive approach, which yields mathematical robust constraints and results, although it may also be limited in its applicability conditions.

\begin{figure}[!htb]
\includegraphics[width=0.49\textwidth]{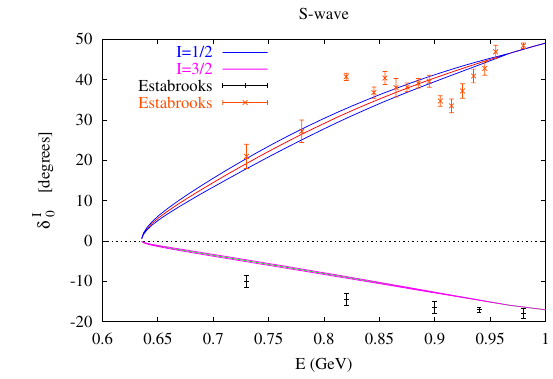}
\includegraphics[width=0.49\textwidth]{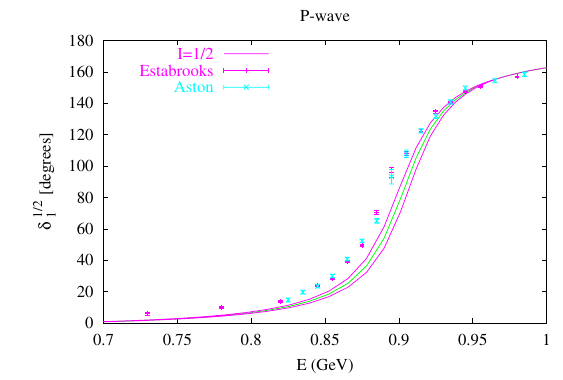}
 \caption{\rm \label{fig:BDGM}
The figure is taken from~\cite{Buettiker:2003pp}.
Solutions of Roy-Steiner equations obtained in \cite{Buettiker:2003pp} for the phase shifts of 
$\pik$ scattering partial waves. Left: $S$-waves with isospin 1/2 (positive curves) and 3/2 (negative curves). Right:  $P$-wave of isospin 1/2. The data come from \cite{Estabrooks:1977xe,Aston:1987ir}.
}
\end{figure}

Of course, to understand these interactions the first need is to have a reliable data analysis, consistent with fundamental constraints. Unfortunately, we have recently shown in \cite{Pelaez:2016tgi} and \cite{Pelaez:2018qny} that, respectively, the \pik and \pipikk data are inconsistent with dispersion relations, sometimes by a large amount. 
It should be pointed out that in a remarkable work \cite{Buettiker:2003pp}, numerical solutions of partial-wave projections of fixed-$t$ dispersion relations, i.e. the so-called Roy-Steiner equations \cite{Roy:1971tc,Steiner:1970fv,Steiner:1971ms}, were obtained for $\pi K$ below 1 GeV. As seen in Fig.~\ref{fig:BDGM} these solutions were consistent with the scalar-wave data from \cite{Estabrooks:1977xe},  but not quite so with the prominent $K^*(892)$ in the vector wave.
We emphasize that these are solutions to the dispersion relations, which do not use data in the elastic regions of these waves so that the curves are predictions and the results impressive. In the present work, however, we will follow a different approach, not solving Roy-Steiner dispersion relations, but using many types of them as constraints on data in an energy region as large as possible.

Thus, in this report we will explain and review the dispersive formalism for \pik and $\pi\pi\rightarrow K\bar K$, paying particular attention to our series of works \cite{Pelaez:2016tgi,Pelaez:2018qny,Pelaez:2020uiw}.
Besides reviewing those results, we will complete the Roy-Steiner analysis by applying it simultaneously  to \pik and \pipikk, which so far had been analyzed considering the other one as a fixed input. Moreover, we will derive and use Roy-Steiner relations with various numbers of subtractions, which weight differently different energy regions, and we will use them to determine several threshold parameters. Finally, forward dispersion relations will be used to constrain $\pi K$ amplitudes up to 1.7 GeV. 
 This defines our main goal, which is to present here simple parameterizations of both \pik up to 1.8 GeV and \pipikk up to 2 GeV that describe the data, {\it and their uncertainties}, while simultaneously satisfying an ample set of dispersion relations covering different regions.  This result will be called Constrained Fit to Data (CFD) in contrast to other Unconstrained Fits to Data (UFD) that we will also analyze here. 
 Similarly constrained parameterizations were obtained for $\pi\pi\rightarrow\pi\pi$ in a series of works  \cite{GarciaMartin:2011cn,Kaminski:2006qe,Kaminski:2006yv,Pelaez:2004vs} by one of the authors together with the Madrid-Krakow group and they have become  widely used both in theoretical and experimental studies.
To provide the hadron community with similar results for \pik and \pipikk, this review will therefore deliver the most constrained set of \pik and \pipikk partial-wave parameterizations, together with accurate values of threshold and subthreshold parameters as well as a review of the rigorous determination of the \kap parameters and other heavier strange resonances.

The simplicity of our parameterizations and their uncertainties is a goal we imposed on ourselves so that they are easy to implement in future works, either of phenomenological or experimental character, whose interest on \pik and \pipikk we detail in the next subsections.

\subsubsection{\pik and \pipikk scattering for final state interactions.}

On the experimental and phenomenological side,  we remark once again that a piece of motivation to look for such a CFD parameterization of \pik and \pipikk interactions is because, being much lighter than similar hadrons, pions and kaons are ubiquitous in final states of hadron processes. Once they appear in a final state they re-scatter strongly again. These are known as final state interactions (FSI) and are a very relevant  contribution to the description of many hadronic processes. Moreover, the interest for a precise and rigorous \pik and \pipikk description has increased  due to the high statistics at B-meson factories (BaBar, Belle, LHCb), the wealth of new hadron states, their many decay channels and the CP violation studies, which need reliable input from intermediate $\pi\pi,\pi K,K \bar K $ states and their FSI. As illustrative examples of these processes, let us mention: multi-body heavy-meson decays like $D\rightarrow K\pi\pi$ \cite{Aitala:2002kr,Aitala:2005yh,Link:2009ng}, $B^\pm\rightarrow D K^\pm$ with $D\rightarrow K\pi\pi$ or $D\rightarrow KK\pi$ \cite{Aubert:2008bd,Poluektov:2010wz,Aaij:2014iba,Aaij:2014dia}, or $B\rightarrow 3M$, with $M=\pi, K$ \cite{Aubert:2005ce}, CP violation in $D,B\rightarrow K^+ \pi^- K^- \pi^+$ \cite{Aaij:2013swa,Aaij:2017wgt,Aaij:2018nis}, and the enhancement of CP violation by meson-meson FSI in three-body charmless B decays \cite{Aaij:2013sfa,Aaij:2014iva}, particularly through \pipikk rescattering \cite{Bediaga:2013ela,Nogueira:2015tsa} (see the recent review in \cite{Bediaga:2020qxg}).
Very often, these FSI are described with simple models fitted to data, which, as we will show here, fail to satisfy the fundamental constraints encoded in dispersion relations. The precision achieved by the already existing data thus asks for model-independent parameterizations and a realistic assessment of their uncertainties.
Let us note that such model-independent dispersively constrained parameterizations for $\pi\pi\rightarrow\pi\pi$ coauthored by one of us \cite{GarciaMartin:2011cn} are widely used both by phenomenological and experimental studies. Moreover
several of these experimental groups have asked us for our constrained \pik and \pipikk parameterizations \cite{Pelaez:2016tgi,Pelaez:2018qny} too. This includes the recently accepted KLF proposal 
\cite{Amaryan:2017ldw,Amaryan:2020xhw} to use a neutral $K_L$ beam at Jefferson Lab with the Gluex experimental setup, to study strange spectroscopy and the $\pi K$ final state system up to 2 GeV.
Further developments in existing experiments and future Hadron-Physics facilities will be even more demanding for precise and model-independent meson-meson amplitudes like those reviewed in this manuscript.

\subsubsection{Lattice QCD and Chiral Perturbation Theory}

\begin{figure}[!htb]
\centerline{\includegraphics[width=0.5\textwidth]{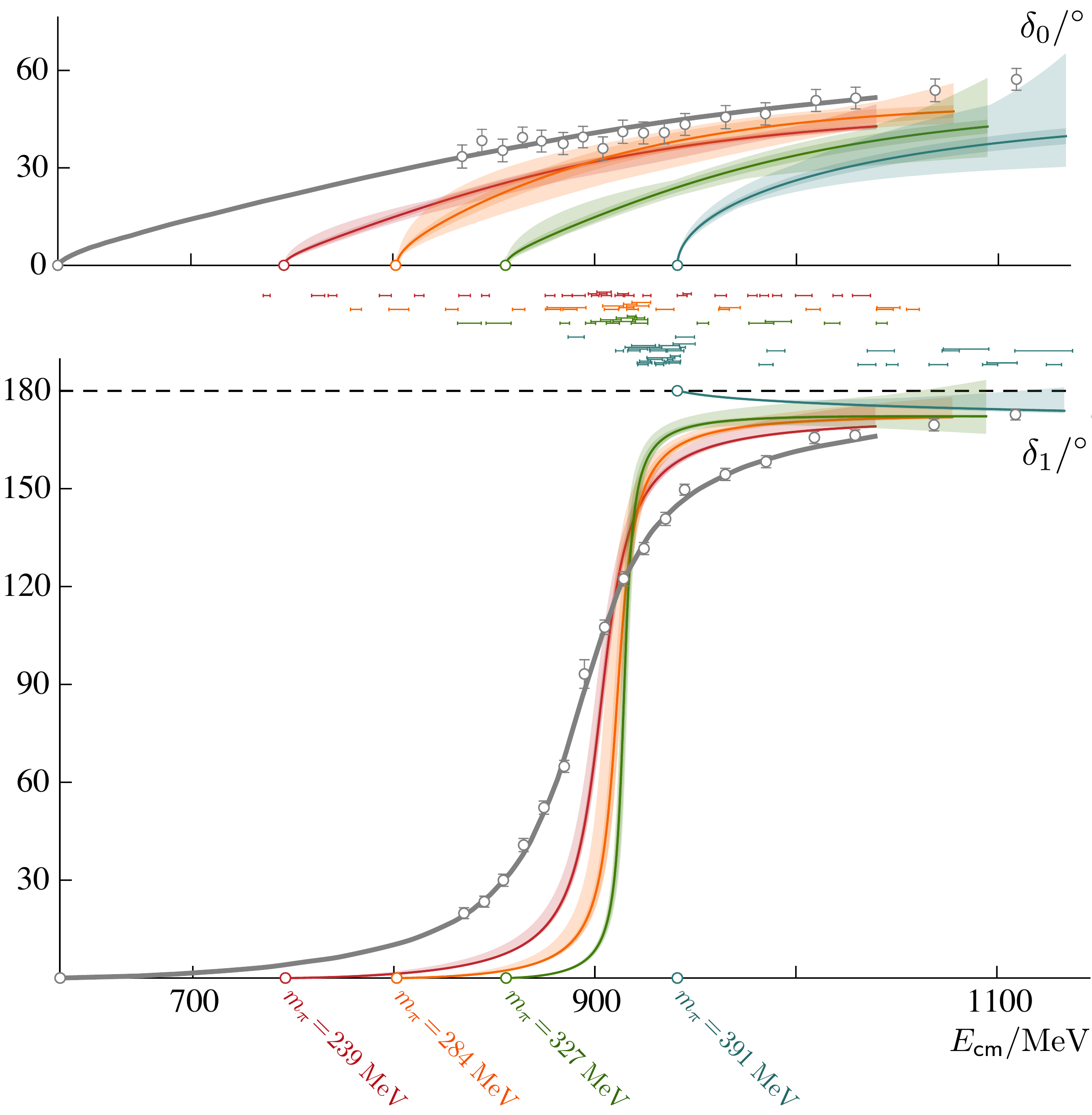}}
 \caption{\rm \label{fig:hadspeckap}
The figure is taken from~\cite{Wilson:2019wfr}. Comparison between the fitted low energy phase shifts $\delta_\ell$ obtained from lattice QCD calculations for the scalar $f^{1/2}_0(s)$ (top) and vector $f^{1/2}_1(s)$ (bottom) $\pik$ scattering partial waves, at four different quark masses. Notice how, as the lattice pion mass tends to the physical pion mass, both partial waves approach the grey curves, which represent our UFD fit~\cite{Pelaez:2016tgi} to real data. Note also that the vector channel at $m_\pi=391$ MeV has been plotted starting at 180 degrees as the $K^*(892)$ becomes a bound state.}
\end{figure}

On the theoretical side, unfortunately, the energy region below 1.5 or 2 GeV lies beyond the applicability realm of perturbative QCD. Nevertheless, unquenched lattice QCD calculations on \pik \cite{Prelovsek:2013ela,Wilson:2014cna,Wilson:2019wfr}
phase shifts and \pipikk \cite{Wilson:2015dqa,Briceno:2017qmb}, still at unphysical masses, provide scattering information at several different energies. This is illustrated in Fig.~\ref{fig:hadspeckap}, where we can see that the main features of the scalar and vector partial waves in the elastic regime are clearly visible, even if for unphysical masses. For a review on scattering processes and resonances from lattice QCD see \cite{Briceno:2017max}. We consider it very likely that precise lattice QCD calculations with close to physical pion masses may be available soon. These will also require consistent and precise data analyses, like the one we pursued here, to compare with.

In addition, since the advent of QCD, we know that pions and kaons  (together with the eta meson) can be identified with the Nambu-Goldstone-Bosons (NGB) \cite{Nambu:1960xd,Nambu:1960tm,Goldstone:1961eq,Goldstone:1962es} associated with the spontaneously broken SU(3) chiral symmetry generators of QCD. In the massless quark limit, these NGB would be massless and separated from all other hadrons  by a mass gap of $O(1 \gev)$.  Note, however, that quarks have a small non-zero mass so that pions and kaons are indeed massive, but they are still much lighter than hadrons with similar quantum numbers.
This is why, in purity, they should be called ``pseudo-NGB''. In any case, the mass gap exists even if QCD is not massless. This means that pions, kaons, and etas are the only degrees of freedom of the strong interaction at low energies and the main decay products at all energies.

Perturbative QCD may not be applicable at low energies, but, together with the observed chirally-broken spectrum, it still dictates that the spontaneous chiral symmetry pattern is that of an SU(3)$_L\times$ SU(3)$_R$ group, broken down to an SU(3)$_{L+R}$ group (for introductory texts see \cite{Donoghue:1992dd,Dobado:1997jx,Scherer:2012xha}). 
Here $R,L$ refer to right and left quark chiralities.
It is then possible to formulate the low-energy effective field theory of QCD reproducing this symmetry breaking pattern in terms of pions, kaons, and the eta. This is called 
Chiral Perturbation Theory (ChPT) \cite{Gasser:1983yg,Gasser:1984gg} and provides a rigorous and systematic perturbative treatment of hadron physics in the low-energy regime. This alternative perturbative expansion is organized in even powers of masses or momenta and can be tested against experiments
by studying meson-meson interactions at very low energies.  Consequently, the so-called threshold or sub-threshold parameters become a central object of study. In particular, the ChPT program has been carried out to NNLO for $\pik$ scattering in \cite{Bijnens:2004bu}, providing rather accurate predictions.
Let us note that the pion-only version of ChPT is known to work very well since the pion mass is very small $\sim 0.14\,$GeV. However, the SU(3) ChPT convergence is not so nice \cite{Bijnens:2014lea} when dealing with kaons, whose mass is $\sim 0.5\,$GeV.

\begin{figure}[!ht]
    \centering
    \resizebox{0.8\textwidth}{!}{\input{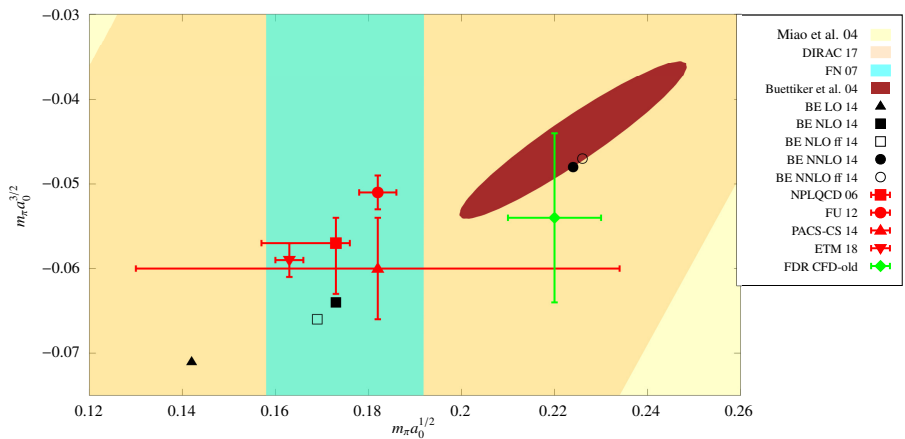}}
    \caption{Comparison between various existing  determinations of the \pik scalar scattering lengths $a_0^{1/2}$ and $a_0^{3/2}$. The only experimental value shown is the beige band of the $a_0^-$ DIRAC measurement \cite{Adeva:2014xtx}. The first lattice result (Miao et al. \cite{Miao:2004gy}) yields the rather large yellow band, whereas higher precision is claimed in more recent lattice calculations (In red: NPLQCD \cite{Beane:2006gj},
    Fu \cite{Fu:2011wc}, PACS-CS \cite{Sasaki:2013vxa}, ETM \cite{Helmes:2018nug}) as well as when lattice was combined within an Omn\`{e}s dispersion relation (Flynn-Nieves \cite{Flynn:2007ki}). Note that dispersive determinations yield somewhat larger values for both scattering lengths, as in B{\"u}ttiker et al.~\cite{Buettiker:2003pp}, or Peláez-Rodas \cite{Pelaez:2016tgi} (FDR CFD-old). ChPT calculations to LO, NLO and NNLO come from the Bijnens-Ecker review \cite{Bijnens:2014lea}, but note the NNLO parameters are obtained by including in the fit the value of \cite{Buettiker:2003pp}.}
    \label{fig:scl}
\end{figure}

  Interestingly, at present, there is some tension between the scalar $\pi K$ scattering lengths from  QCD \cite{Miao:2004gy,Beane:2006gj,Flynn:2007ki,Nagata:2008wk,Fu:2011wc,Sasaki:2013vxa,Helmes:2018nug} and the rigorous dispersive determination we mentioned above \cite{Buettiker:2003pp} and our previous dispersively constrained fits to data \cite{Pelaez:2016tgi}. 
  This situation is summarized in Fig.~\ref{fig:scl}. Only with NNLO ChPT \cite{Bijnens:2014lea} it is possible to come close to these dispersive values, but then the NNLO lies more than two standard deviations off the bulk of  lattice values. Moreover, in such case the scalar $\pi K$ scattering lengths show “the worst convergence of all” \cite{Bijnens:2014lea}.  
  
  Thus, another goal of this report is to provide accurate and rigorous determinations of $\pi K$ threshold and subthreshold parameters, with particular attention to the scalar ones. These will be  obtained in subsections \ref{subsec:SRth} and \ref{subsec:subth}, respectively, from sum rules derived from dispersion theory, and, once again, we will pay special attention to uncertainties, including those from \pipikk. 

\subsubsection{Strange meson spectroscopy and the \kap}

Finally, the fourth feature of \pik and \pipikk interactions  motivating this work is that they provide important information on light meson spectroscopy. One aim of spectroscopy is to identify as mesons the bound states of quarks and antiquarks---or, more often, as resonances since we deal with strong interactions that make most of these states rather unstable. For a compilation of our present knowledge about these bound and resonant states, we refer the reader to the Review of Particle Physics (RPP) \cite{pdg}.
Once these states are established, another aim of spectroscopy is to classify them in multiplets  related by some symmetry transformation: in our case, isospin, flavor SU(3), and spin multiplets, as well as to identify their parity, etc... Let us remark that SU(3) flavor multiplets require the presence of strange resonances, and it can be checked that the information on many of these strange resonances is dominated by \pik scattering data.

One area of very active research in Hadron Spectroscopy is the search for non-ordinary mesons, namely, those that are not made of just a valence quark and an antiquark. Probably the most interesting non-ordinary configuration is the glueball. This is a meson made of bound gluons, which contrary to photons can interact with themselves due to QCD being a non-abelian gauge theory.   Having no quarks in its composition, a glueball should have zero charge and no flavor, so that it forms a flavor singlet. The lightest one is also expected to be a scalar. This is one of the reasons why understanding light scalar mesons is very important. However, mesons made of quarks should form nonets, i.e. an octet and a singlet. Thus, naively, any singlet not associated with an octet is a glueball candidate. Unfortunately, in practice, this simple picture is complicated by the mixing of resonances with the same quantum numbers. However, the simplest way to identify how many octets should be and where their masses are is by looking at their strange members, which makes the identification of strange resonances even more relevant. As we commented above, a great deal of the information used to determine the existence and properties of strange resonances below 2 GeV comes from \pik scattering, with all the caveats about data inconsistencies and model dependencies that we have already emphasized. Hence, another goal of this report is to review the recent application of analytic techniques, using our CFD as input, to obtain model-independent determinations of strange resonances below 2 GeV.

Furthermore, the other reason why light scalar mesons are interesting is that the exchange of the lightest one, known as the $\sigma$ or $f_0(500)$ resonance, is responsible for most of the attractive part of the nucleon-nucleon potential, without which nuclei would not be formed and we would not even exist. The \sig is very  difficult to observe clearly in experiments, since it is very wide and short-lived, with no charge, isospin zero, and no strangeness, i.e. the vacuum quantum numbers. It has  very many properties that make it a robust candidate for a non-ordinary meson. Over the last years there is growing agreement that this state is dominated by some kind of four-quark or two-meson configuration (see \cite{Pelaez:2015qba} for a recent review). Nevertheless, since the \sig has the same quantum numbers of the lightest  glueball, some works have postulated that it is actually the lightest one of them \cite{Minkowski:1998mf,Ochs:2013gi}.
However, if it were made of quarks it would necessarily have a strange partner, known as the \kap, which once again is a wide and very controversial state. There has been a longstanding debate about the existence and properties of this light strange resonance and, as of today, it still ``Needs Confirmation'' in the RPP \cite{pdg}. Once again, the determination of this state is hindered by the data and model-dependence problems commented above. In this case, the model-dependence problem is aggravated because the analytic extension to the complex plane is a very unstable procedure when models are used to extrapolate to the complex plane.
Once more, the solution comes from dispersion theory, this time in the form of dispersion relations projected into partial waves.
However, we will show here that the problem is so unstable that even using naively the same partial-wave data as input, different dispersion relations can yield different poles. A unique pole is obtained only if the data description in the real axis satisfies dispersion theory.
It should be remarked that a rigorous  dispersive determination of the \kap, although without using data on the nominal \kap region, was obtained in \cite{DescotesGenon:2006uk}. Interestingly, the authors were able to prove that the \kap pole lies within the  applicability domain of  partial-wave projected hyperbolic dispersion relations. Then, by
using inside the latter the solution predicted from partial-wave projected fixed-$t$ dispersion relations obtained in \cite{Buettiker:2003pp}, they obtained their very sound prediction for the \kap pole. Since this work does not use \pik data in the \kap region, in a sense, their resonance is a prediction.
Nevertheless,  even after that work, the \kap still ``Needs Confirmation'' according to the RPP \cite{pdg}. 

 Within our approach, we are looking for a complementary determination based on data instead.
Actually, our recent determination in \cite{Pelaez:2019rwt} using analytic techniques based on truncated series of Padé approximants obtained from the derivatives of our dispersively constrained fit to data \cite{Pelaez:2016klv}, triggered the 2018 change of the \kap name and parameters in the Review of Particle Physics (it was $K^*(800)$ until then). 
It is only recently that we have completed a precise and rigorous dispersive determination of the \kap pole using partial wave hyperbolic dispersion relations \cite{Pelaez:2020uiw}. The final result for our pole  and a sketch of the constrained fit to data was given in \cite{Pelaez:2020uiw}, but one of our goals here is to provide the details of the full calculations.

Let us remark that a recent lattice study \cite{Dudek:2014qha,Rendon:2020rtw} at unphysical masses already finds a virtual pole that can be associated with the \kap. This behavior was already predicted within unitarized ChPT
\cite{Nebreda:2010wv}. Unfortunately, when using more realistic masses, even with the phase shifts shown in Fig.~\ref{fig:hadspeckap}, the extraction of the \kap pole is not so straightforward. When employing models to obtain the analytic continuation, the poles are, once again, rather unstable \cite{Wilson:2019wfr,Rendon:2020rtw}. It seems that a rigorous determination of the \kap from lattice will also require the analysis of lattice data using analytic and dispersive techniques like the ones we will describe in this report. Note that in this case  dispersion relations will not be ``solved'' but instead used to ``constrain'' the lattice data, which is the main approach we will follow here.

\subsubsection{Other applications}

Finally, we will also review other applications of dispersion relations to \pik or \pipikk scattering, as well as further applications where, as a part of a larger calculation, a precise knowledge of these interactions may be of interest.

Thus, in section \ref{subsec:nature}, we will illustrate an additional use of dispersion theory for \pik scattering.  The relevant observation is that the analytic properties of Regge trajectories are determined by the analytic properties of the partial wave where they appear (for a textbook introduction to Regge theory see  \cite{Collins:1971ff}). In the case of an ``elastic resonance'', i.e. with just one decay channel, these analytic constraints can be combined with elastic unitarity into a dispersion relation for its trajectory, which in turn determines, up to a few subtraction constants, the partial-wave in the energy region dominated by that resonance. Adjusting these constants so that the partial-wave has a pole with the given resonance parameters, one then determines its Regge trajectory.
It is here that the poles obtained from dispersive approaches can also be used as input.
When this method is applied to relatively narrow resonances like the $\rho(770)$ or  $f_2(1270)$ 
\cite{Londergan:2013dza,Carrasco:2015fva}, their trajectories come almost real and in the form of straight lines in the $(M^2,\,J)$ plane, with a slope of $\sim$ 1 GeV$^{-2}$. These are the ordinary resonances associated with quark-antiquark states bound by QCD dynamics. In section \ref{subsec:nature} we will review how, when this method has been applied to resonance poles in \pik scattering \cite{Pelaez:2017sit}, we have found the expected ordinary trajectories for the $K^*(892)$ and $K^*_0(1430)$. In contrast, the trajectory of the \kap comes out different, providing strong evidence for the non-ordinary nature of this controversial state.
Moreover, we will review how this trajectory is very similar to that of the $\sigma/f_0(500)$ meson, showing once more the striking similarities of these two non-ordinary states.

In the final section \ref{sec:applications} 
we supply examples of applications where our CFD can be used as input. We have already commented that in this section we provide precise and model-independent values of threshold and subthreshold parameters obtained using our CFD as input. For these we review and also derive  a large collection of sum rules from dispersion relations, providing many of their explicit expressions. We hope these are of relevance to test ChPT and low-energy lattice calculations.

In addition, we discuss two more applications where a precise and model-independent description of \pik or \pipikk data is of relevance. Regarding the former case, we aim at providing a value of the so-called $\pi K$ $\sigma$-term---a quantity of relevance to understand the inner structure of the kaon. 
The calculation of one of its contributions requires the value of the $\pi K$ amplitude at the unphysical Cheng-Dashen point  $t=2m_\pi^2$, $\nu=0$. For this, we have to use a dispersive sum rule evaluated with our CFD as input, as explained in section \ref{sec:sigterm}.

And finally, regarding the need for \pipikk, we review its contribution for the calculation of $(g-2)_\mu$. We comment on how our updated CFD parameterization could be of use since some of our previous parameterizations were used in the past for this purpose. Nevertheless, although it constitutes a nice possible application, the overall effect of $ K \bar K$ states in intermediate states is always found to be very small, and will not change the status of the present conflict between theoretical calculations and experiment.

\subsection{Notation}
\label{sec:Notation}

Throughout this work we will be working in the isospin limit
of equal mass for all pions,
$m_\pi\equiv m_{\pi^+}=139.57\,$MeV,  kaons, $m_K=496\,$MeV, and etas, $m_\eta=547\,$MeV, $m_{\eta'}=957\,$MeV.
It is also convenient to define $m_\pm=m_K\pm m_\pi$, $\Sigma_{12}=m_1^2+m_2^2$ and $\Delta_{12}=m_1^2-m_2^2$, as well as $t_\pi=4m_\pi^2$, $t_K=4m_K^2$.
In the rest of this work, and unless  
stated otherwise,  $m_1=m_K$, $m_2=m_\pi$, $\Delta=\Delta_{K\pi}$, $\Sigma=\Sigma_{K\pi}$.
It is also customary to use the standard Mandelstam variables $s,t,u$
for $\pi K$ scattering, satisfying $s+t+u=2\Sigma$.
The center-of-mass (CM) momentum of the $s$-channel $\pi K$ system will be called
$q=q_{\pi K}(s)$, whereas  
$q_{\pi}=q_{\pi\pi}(t)=\sqrt{t-t_\pi}/2,q_K=q_{KK}(t)=\sqrt{t-t_K}/2$
 will be the CM momenta of the respective $\pi \pi$ and $K\bar{K}$ states in the $t$-channel, i.e. for $\pi\pi \rightarrow K\bar K$ scattering. Here
\begin{equation}
q_{12}(s)=\frac{1}{2\sqrt{s}}\sqrt{(s-(m_1+m_2)^2)(s-(m_1-m_2)^2)}=
\frac{1}{2\sqrt{s}}\sqrt{s^2-2s\Sigma_{12}+\Delta_{12}^2}.
\end{equation}
Sometimes it will also be convenient to use the so-called phase-space:
\begin{equation}
    \sigma_{\pi K}(s)=\frac{2q}{\sqrt{s}}, \quad\sigma_{\pi}(t)=\frac{2q_{\pi}}{\sqrt{t}}, \quad\sigma_{K}(t)=\frac{2q_{K}}{\sqrt{t}}.
    \label{eq:phasespace}
\end{equation}
Since we will be working in the isospin limit, we will make extensive use of the isospin-defined $\pi K$ scattering amplitudes, denoted by 
$F^I(s,t,u)$, where $I=1/2,3/2$ is the total isospin of the process. 
They are related by $s\leftrightarrow u$ crossing as follows:
\begin{equation}
F^{1/2}(s,t,u)=\frac{3}{2} F^{3/2}(u,t,s)-\frac{1}{2}F^{3/2}(s,t,u).
\end{equation}
For convenience we will frequently use the $s\leftrightarrow u$ symmetric and antisymmetric $\pi K$ amplitudes, denoted $F^{\pm}$ respectively. These can be recast as
\begin{eqnarray}
F^+(s,t,u)&=&\frac{1}{3}F^{1/2}(s,t,u)+\frac{2}{3}F^{3/2}(s,t,u),\label{eq:Fplusdef}\\
F^-(s,t,u)&=&\frac{1}{3}F^{1/2}(s,t,u)-\frac{1}{3}F^{3/2}(s,t,u).\label{eq:Fminusdef}
\end{eqnarray}

In addition, using $s \leftrightarrow t$ crossing symmetry, the
$\pi\pi\rightarrow K\bar{K}$ amplitudes, denoted $G^I$ for isospin $I=0,1$,  are related to those of $\pi K$ scattering as follows:
\begin{eqnarray}
G^0(t,s,u)&=&\sqrt{6}F^+(s,t,u),\nonumber\\
G^1(t,s,u)&=&2F^-(s,t,u). 
\label{Fpm-GI}
\end{eqnarray}

In this work we will also use the partial-wave decomposition of both the $\pi K$
and $\pi\pi\rightarrow K\bar{K}$ scattering amplitudes, defined\footnotetext{Please note that in our Ref.~\cite{Pelaez:2016klv} we used the notation $t_\ell^I(s)$ instead of $f_\ell^I(s)$. In addition, we used $F(s,t,u)$ which corresponds to what we would call $F(s,t,u)/4\pi^2$ here. The present notation is the one we used in \cite{Pelaez:2018qny}.} as:
\begin{eqnarray}
&&F^I(s,t,u)=16\pi \sum_\ell{(2\ell+1)P_\ell(z_s) f^I_\ell(s)},\label{ec:pwexpansion} \\
&&G^I(t,s,u)=16\pi\sqrt{2} \sum_\ell{(2\ell+1)(q_{\pi}q_{K})^\ell P_\ell(z_t) g^I_\ell(t)}.\nonumber
\end{eqnarray}
Let us not forget that, since in the isospin limit pions are identical particles from the point of view of hadronic interactions, irrespective of their charge, then, being bosons, the  $\pi\pi$
state must be fully symmetric. Thus, for even (odd) isospin, only even (odd) angular momenta should be considered in the above partial-wave expansion of $G^I(t,s,u)$. Let us remark that it is customary to extract explicitly the $(q_\pi q_K)^\ell$ factors in the partial waves of the $t$-channel, to ensure good analytic properties for $g_\ell(t)$ 
(see \cite{Frazer:1960zza} in the $\pi\pi\rightarrow N\bar{N}$ context).
The scattering angles in the $s$ and $t$ channels are given by:
\begin{equation}
z_s=\cos\theta_s=1+\frac{2st}{\lambda_s},\qquad z_t=\cos\theta_t=\frac{s-u}{4q_{\pi}q_K}=\frac{\nu}{4q_{\pi}q_K}=\frac{\nu}{\sqrt{(t-t_\pi)(t-t_K)}},
\label{eq:anglet}
\end{equation}
where 
\begin{equation}
    \lambda_s=\lambda(s,m_\pi^2,m_K^2)=(s-m_+^2)(s-m_-^2)=s^2-2s\Sigma+\Delta^2=4s\,q_{K\pi}^2(s),
\end{equation} 
and $\nu\equiv s-u$ is the antisymmetric variable under $s\leftrightarrow u$ crossing.
For convenience, we have also defined the K\"all\'en function
\begin{equation}
    \lambda(x,y,z)=x^2+y^2+z^2-2xy-2xz-2yz.
    \label{eq:Kallenfunction}
\end{equation}

Let us now recall that in the $s$-channel partial waves are  projected using
\begin{equation}
f^I_{\ell}(s)=\frac{1}{32\pi}\int^1_{-1}dz_sP_{\ell}(z_s)\, F^I(s,t(z_s)),
\label{eq:projs}    
\end{equation}
whereas for the $t$-channel partial waves are obtained from
\begin{equation}
g^I_\ell(t)=\frac{\sqrt{2}}{32\pi(q_{\pi}q_{K})^\ell}\int_{0}^{1}{dz_t P_\ell(z_t)\, G^I(t,s(z_t))},
\label{eq:gpw}
\end{equation}
since now  we have two identical particles in the initial state, i.e. the  
two pions in the isospin conserving formalism. It is worth noticing that very often experimentalists in their partial-wave definitions included
an $\sqrt{2\ell+1}$ factor, which we will have to take  into account when comparing with data.

For later use we define the scalar $\pi K$ scattering lengths as follows:
\begin{equation}
a_0^I=\frac{2}{m_+}f_0^I(m_+^2),
\label{eq:scldef}
\end{equation}
and similarly for $a_0^-=(a_0^{1/2}-a_0^{3/2})/3$
and $a_0^+=(a_0^{1/2}+2a_0^{3/2})/3$. These parameters come from the low-energy effective range expansion for partial waves
\begin{equation}
    \frac{2}{\sqrt{s}}\re f^I_\ell(s)\simeq q^{2\ell}\left(a^I_\ell+b^I_\ell q^2+c^I_\ell q^4 +...\right),
    \label{eq:lowex}
\end{equation}
whose coefficients are called scattering lengths ($a^I_\ell$), effective ranges ($b^I_\ell$), shape parameters ($c^I_\ell$), etc... and are generically referred to as threshold parameters. All them are similarly defined for the expansion of the $f^\pm_\ell$  combinations.

The relation with the $S$-matrix partial waves, which allows for a direct comparison with some experiments, is: 
 \begin{eqnarray}
 S^{I}_\ell(s)_{\pi K\rightarrow \pi K}&=&1+i\frac{4q}{\sqrt{s}}f_\ell^I(s)\theta(s-m_+^2), \\
 S^{I}_\ell(t)_{\pi \pi\rightarrow K \bar K}&=&i \frac{4(q_\pi q_K)^{\ell+1/2}}{\sqrt{t}}g^{I}_\ell(t)\theta(t-t_K). \nonumber
 \end{eqnarray}
 When it is clear from the context that the energy under consideration is above the corresponding physical threshold it is also usual to write these equations omitting the step function. 
 
Let us now remind that the $S$-matrix should be unitary, which means that if, for a given energy, the initial state $i$ can evolve not only into a given final state $f$ but also to many other states $n$, then $\sum_n S^I_\ell(s)_{in}S^I_\ell(s)_{nf}^\dagger=\delta_{if}$. 

In case only one state is available, as for $\pi K$ at sufficiently low energies, we say the reaction is elastic. 
The elastic unitarity condition translates into the following algebraic relation for $\pi K$ partial waves 
\begin{equation}
    \im f^I_\ell(s)=\sigma_{\pi K}(s)\vert f^I_\ell(s)\vert^2,
    \label{eq:elunit}
\end{equation}
which implies that the elastic \pik 
partial wave can be recast in terms of a real phase shift, called $\delta^I_\ell(s)$, as follows:
\begin{equation}
f^I_\ell(s)=\frac{\hat f^I_\ell(s)}{\sigma_{\pi K}(s)}=\frac{e^{i\delta^I_\ell(s)}\sin\delta^I_\ell(s)}{\sigma_{\pi K}(s)}=
\frac{1}{\sigma_{\pi K}(s)}\frac{1}{\cot\delta^I_\ell(s)-i},
\label{eq:elasticpw}
\end{equation}
where we have introduced the ``Argand'' partial wave $\hat f (s)$ for later convenience. 
In practice, $\pi K$ scattering is elastic below approximately 1 GeV for all waves, and so are the maximal isospin waves within the whole region of interest in this review. In such cases, the knowledge of $\delta^I_\ell(s)$ is enough to characterize the full complex amplitude. We detail in~\ref{app:conformal} the explicit functional form used in this work when describing this elastic region. 

In contrast, in the inelastic regime, and like any other complex function, the description of a partial wave requires the knowledge of two real functions, i.e., the phase and the modulus. We will often use those quantities, but sometimes it is also convenient to define an elasticity function $0\leq\eta^I_\ell\leq 1$ to write:
\begin{equation}
f^I_\ell(s)=\frac{\hat f^I_\ell(s)}{\sigma_{\pi K}(s)}=\frac{\eta^I_\ell(s)e^{2i\delta^I_\ell(s)}-1}{2i\sigma_{\pi K}(s)}.
\label{eq:inelpw}
\end{equation}
Note that when the elasticity is one, we recover the elastic formalism.

When partial waves are considered as analytic functions of the $s$ variable, they have a complicated cut structure in the complex plane that will be explained in detail in section \ref{sec:DR}. However,
the so-called ``physical'' or ``unitarity'' cut can already be observed in Eq.~\eqref{eq:elunit} from the $\sigma_{\pi K}(s)$ factor. It starts at threshold and extends to infinity, and produces two Riemann sheets, each for a different sign of the imaginary part of the momentum. So far we have been dealing with the expressions in the first or ``physical'' Riemann sheet, so that our $f(s)$ should have been called $f^{(I)}(s)$.
The partial wave in the second Riemann sheet. i.e., $f^{(II)}(s)$, is accessible by crossing continuously the unitarity cut. 

Let us now recall 
that, for elastic scattering, the $S$-matrix in the second Riemann sheet is the inverse of the 
$S$-matrix on the first. 
But then, since the $f$-matrix partial 
waves are related to the $S$-matrix partial waves by 
$S(s)=1-2 i \sigma_{\pi K}(s) f(s)$, where we have suppressed the isospin and angular momentum indices for simplicity, we can write 
the amplitude in the second Riemann sheet $f^{(II)}(s)$
in terms of the one in the first Riemann sheet $f^{(I)}(s)$, as follows:
\begin{equation}
f^{(II)}(s)=\frac{f^{(I)}(s)}{1+2 i \sigma_{\pi K}(s)f^{(I)}(s) }.
\label{ec:firsttosecondsheet}
\end{equation}
Note that in the equation above the determination of $\sigma_{\pi K}(s)$ is chosen such that  $\sigma_{\pi K}(s^*)=-\sigma_{\pi K}(s)^*$ to ensure the Schwarz reflection
symmetry of the amplitude.  In other words, on the upper half $s$ plane we can take $\sigma_{\pi K}=+2q_{\pi K}/\sqrt{s}$ as usual,
 whereas on the lower half $s$ plane we must then take $\sigma_{\pi K}(s)=-\sigma_{\pi K}(s^*)^*$.

The above equation is useful to look for poles associated with elastic resonances, which are found in the second sheet. In particular,
a pole 
at $\sqrt{s_R}=M_R-\,i\, \Gamma_R/2$ in the second sheet,
 corresponds exactly with the position of a {\sl zero} 
of the $S$-matrix partial wave in the first Riemann sheet. Restoring the isospin and angular momentum indices back, this reads:
\begin{equation}
S_\ell^{I}(s)=1+2\,i\, \hat{f}_\ell^{I}(s)=e^{2\,i\,\delta_\ell^{I}}=0.
\label{ec:resonancecondition}
\end{equation}
This zero condition may be recast as
\begin{equation}
\cot\delta_\ell^{I}(s_R)=-\,i\,,
\label{eq:rescond}
\end{equation}
where, of course, $\cot\delta_\ell^{I}(s)$ now corresponds to a function in the complex plane
that has all the singularities
of the amplitude, except for the cut along the real axis above threshold,
where it coincides with the physical $\cot\delta_\ell^{I}(s)$. We will use this method to find elastic resonances from a given parameterization of data in section \ref{sec:kappa}.  The inelastic case is not so simple and specific analytic methods to find resonances will be discussed in section \ref{sec:pades}.

Finally, once the pole position $s_p$ of a resonance is found, another parameter of relevance is its coupling to the partial wave, defined as
\begin{equation}
g^{2}=-16 \pi (2 \ell+1)\lim _{s \rightarrow s_p}\left(s-s_{p}\right) \frac{f^{RS}_{\ell}(s)}{(2 q_{\pi K}(s))^{2 \ell}},
\label{eq:coupling}
\end{equation}
where $f^{RS}_{\ell}(s)$ represents the partial wave in the contiguous  Riemann sheet.


\section{The data}
\label{sec:Data}
\subsection{\pik scattering data }
\label{sec:piKData}

\subsubsection{Introduction}
\begin{figure}[ht]
\begin{center}
\centerline{\includegraphics[width=.45\textwidth]{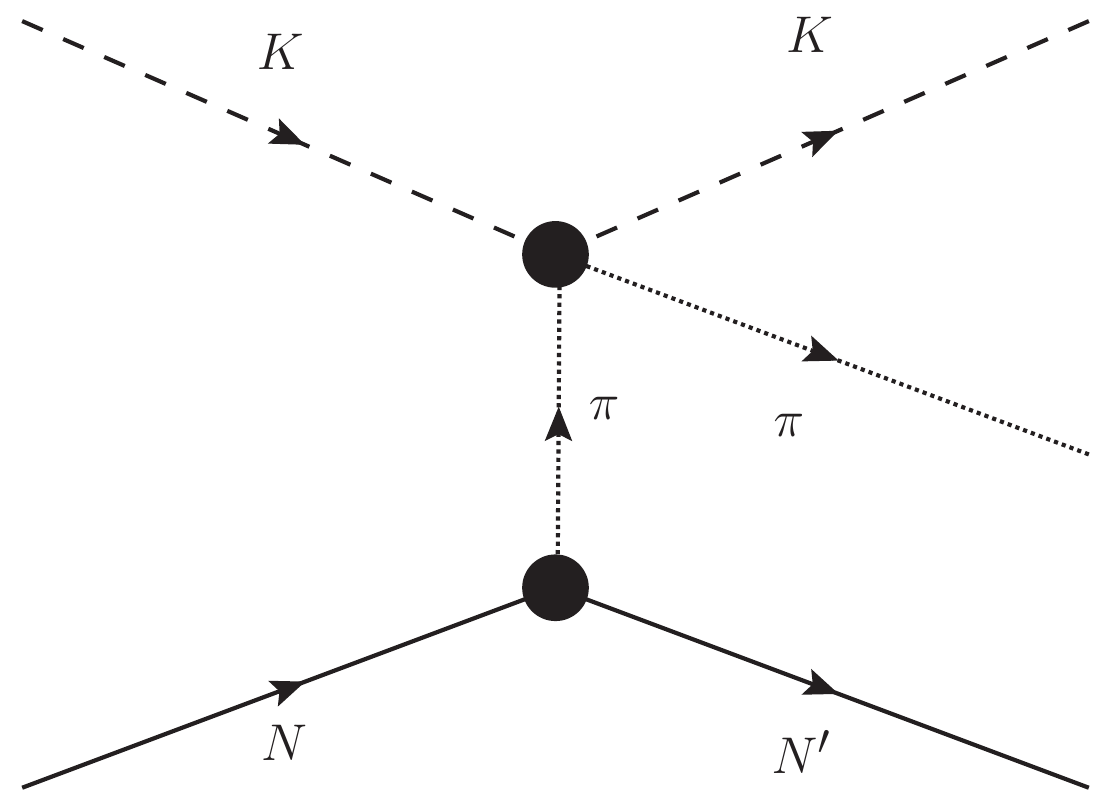} \hspace{0.5 cm}\includegraphics[width=.45\textwidth]{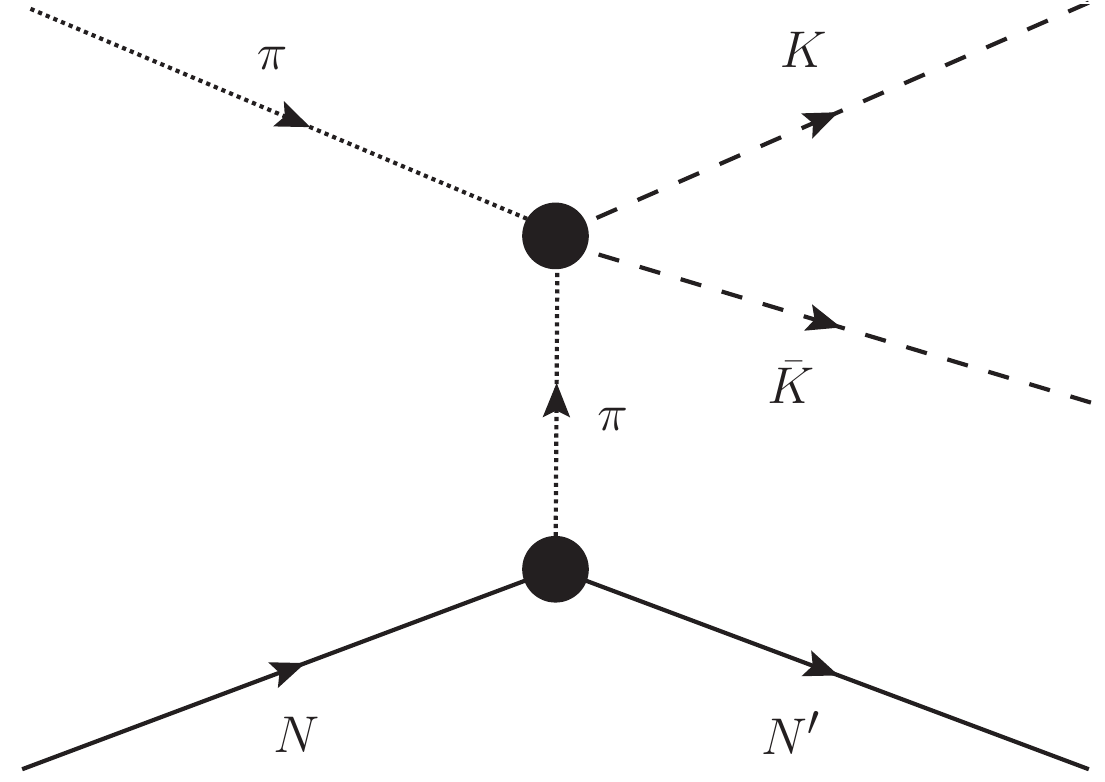}}
\caption{One-pion exchange diagrams for the processes used to extract meson-meson scattering data. It is then assumed that, when the exchanged pion is nearly on-shell, the meson-meson interaction factorizes out in the whole process description. Left panel: \pik case. Right panel \pipikk case. \label{fig:processes}}
\end{center}
\end{figure}

Most of the data on $\pi K$ scattering were obtained during the '70s and the '80s. Due to the practical impossibility of making pion or kaon beams sufficiently luminous to measure these collisions directly, data are measured indirectly in fixed-target $KN\rightarrow K\pi N'$ experiments
assuming they are dominated by the exchange of a single pion. This generic mechanism is illustrated in the left panel of Fig.~\ref{fig:processes}.  Here $N$ is a nucleon and $N'$ can either be a nucleon or a $\Delta$ resonance (although some experiments used deuterium in the initial state and $pp$ in the final one). Experimentally, events whose momentum transfer is as close as possible to the pion pole are selected and then approximated by considering that the exchanged pion is on-shell. This technique was first proposed for $\pi\pi$ scattering \cite{Goebel:1958zz,Chew:1958wd} and later extended to \pik in  \cite{Bingham:1972vy,Baker:1974kr,Estabrooks:1975zw,Estabrooks:1976hc,Estabrooks:1976sp}, see \cite{Martin:1976mb} for a textbook introduction.
Unfortunately 
this one-pion-exchange formalism 
needed to extract $\pi K$ scattering amplitudes from $KN\rightarrow K\pi N'$ is just an approximation  and has 
several sources of large systematic uncertainties. Namely: 
corrections to
the on-shell extrapolation of the exchanged pion, 
rescattering effects, absorption, exchange of other resonances, etc... (see~\cite{Nys:2017xko}). However, most experimental works only quote statistical uncertainties for each solution and it is therefore rather usual that different experiments disagree within their quoted experimental errors, which  do not include these sources of systematic uncertainties.  This will be clearly seen in the figures below.
One of our main tasks in later sections  will be to estimate the systematic uncertainty for different sets, or data points within the same set, in conflict within a specific energy region.

Before presenting the data on each partial-wave, let us recall that isospin is conserved to a very good approximation and we will thus work in the isospin limit. Then, there are two possible isospins for a $\pi K$ state, i.e. $I=1/2$ and $I=3/2$. 
In particular, cross-section measurements 
for the $I=3/2$ isospin channel
were obtained in the early '70s using different reactions.
These include $K^- d \rightarrow K^- \pi^- p p$ in Y. Cho et al. \cite{Cho:1970fb}, 
$K^- n \rightarrow K^- \pi^- p$ in A.M. Bakker et al. \cite{Bakker:1970wg} as well as 
$K^{\pm}p\rightarrow K^{\pm}\pi^{-}\Delta^{++}$
in B. Jongejans et al. \cite{Jongejans:1973pn}.
In practice,  this isospin channel is elastic up to at least 1.8 GeV and then it is straightforward to obtain its phase shift from the cross-section.
This was done by D. Linglin et al. in \cite{Linglin:1973ci}
using their $K^- p\rightarrow K^- \pi^- \Delta^{++}$ data.
Generically, experiments in the earlier '70s have low statistics. An excellent review on the experimental and phenomenological situation until 1978, which also comments on dispersive approaches, can be found in \cite{Lang:1978fk}. Fortunately, such a situation was improved by the end of the decade.
Indeed, in 1978  P. Estabrooks et al. \cite{Estabrooks:1977xe}
presented an analysis of $K^{\pm}p\rightarrow K^{\pm}\pi^{+}n$ 
and $K^{\pm}p\rightarrow K^{\pm}\pi^{-}\Delta^{++}$ at 13 GeV with relatively high statistics
and obtained the $I=3/2$ $\pi K$ scattering contribution,
without any evidence of inelasticity up to 1.8 GeV.

Concerning isospin $I=1/2$ scattering, let us first of all note that, when extracted from $KN\rightarrow K\pi N'$,
it always appears mixed with $I=3/2$ in the following isospin combination: $f_l=f^{1/2}_l+f^{3/2}_l/2$.
This was the case, for example, of the first experiments by
R. Mercer et al. in \cite{Mercer:1971kn} using
$K^+ p \rightarrow K^+ \pi^- \Delta^{++}$ and $K^+ p \rightarrow K^0 \pi^0 \Delta^{++}$
reactions.  In practice, they separated different isospins by combining their results  with a heterogeneous and not very precise data collection that existed at the time, which was called the ``World Data Summary Tape''.
Together with their low statistics, this means that their results for both the $I=1/2$ and $3/2$ waves have huge uncertainties. For this reason  these data
are usually neglected against later and more precise experiments.

The first study of \pik scattering with relatively high statistics for the $f_l\equiv f^{1/2}_l+f^{3/2}_l/2$ isospin combination,  up to 1.85 GeV, was published in 1978 by Estabrooks et al. \cite{Estabrooks:1977xe}, using the SLAC 13 GeV spectrometer.
However, the experiment with the highest statistics so far was performed about a decade later by Aston et al.~\cite{Aston:1987ir}, using the Large Acceptance Superconducting Solenoid (LASS) Spectrometer 
at SLAC. This LASS experiment studied the $K^{-}p\rightarrow K^{-}\pi^{+}n$ reaction at 11 GeV
and obtained the same $\pi K$ partial-wave combination up to 2.6 GeV. These two experiments are the most widely used in the literature, particularly the latter.

Apart from the large, but frequently omitted, systematic uncertainties, an additional problem affects \pik scattering data phases. Namely, ambiguities
appear in the determination of the phase
leading to different solutions even when extracted from the same $KN\rightarrow K\pi N$ experiment. 
In the case of Aston et al. \cite{Aston:1987ir}, they appear above the region of interest for this review. However, Estabrooks et al. \cite{Estabrooks:1977xe} presented four solutions above 1.5 GeV.
We will only consider Solution B since it is the one qualitatively closer to the LASS results.

So far we have discussed fixed-target experiments, from which  data on both the modulus and the phase of the amplitude can be obtained. However, it is also possible to gather experimental information on the meson-meson scattering phase by measuring processes where two mesons appear in the final state, as long as the other particles interact very weakly with the two mesons and among themselves. Then, Watson's theorem \cite{Watson:1952ji} implies that the phase of the whole process is the same as the phase of its two-meson rescattering.
This technique has been applied using $D$ or $\eta_c$  meson decays. Unfortunately, the uncertainties are much bigger than those from fixed-target experiments. Nevertheless, they are useful because they 
provide direct access to $I=1/2$. We will review them here too.

Finally, let us recall that the $I=3/2$ data ends at 1.74 GeV. 
That is the reason why all our plots in this subsection end at 1.74 or 1.8 GeV at most.
Nevertheless, above that energy, we will not use the partial-wave expansion but  Regge parameterizations obtained from the factorization of nucleon-nucleon and meson-nucleon cross-sections.
Since these are not \pik data but phenomenological parameterizations, they will be described in section \ref{sec:ufdregge}.

Let us then discuss the data on each partial wave in detail.

\subsubsection{$S$-wave data}
\label{subsec:Sdata}

\vspace{.5cm}
{\em $I=3/2$ $S$-wave data  }\\

\begin{figure}[ht]
\centering
\resizebox{\textwidth}{!}{\input{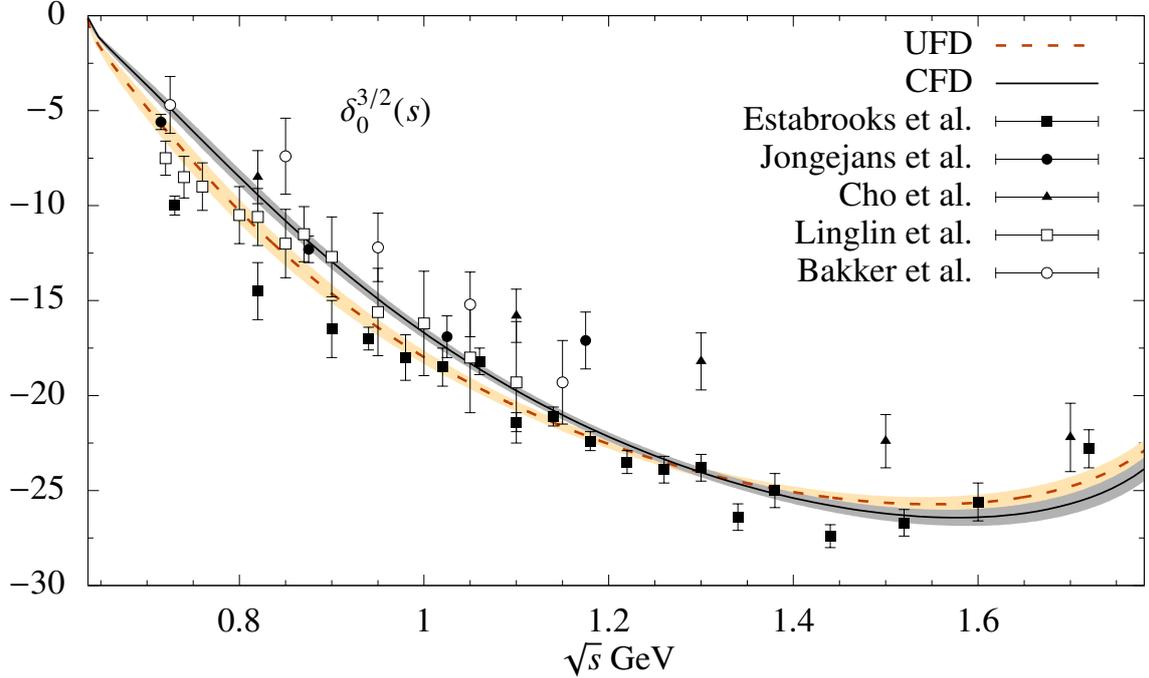}}\\
\caption{Phase-shift data on  the $S^{3/2}(s)$ \pik  from \cite{Estabrooks:1977xe} (solid squares), \cite{Jongejans:1973pn} (solid circles), \cite{Cho:1970fb} (solid triangles), \cite{Linglin:1973ci} (empty squares) and \cite{Bakker:1970wg} (empty circles).
We also show the results of our Unconstrained and Constrained Fits to Data (UFD and CFD respectively).
\label{fig:s32data}}
\end{figure}

We will first describe the $I=3/2$ data since it is rather large and needed to extract the $I=1/2$.
As commented above, several experiments measured first this wave in the early '70s: Y. Cho et al. \cite{Cho:1970fb},  A.M. Bakker et al. \cite{Bakker:1970wg}, D. Linglin et al. \cite{Linglin:1973ci} and B. Jongejans et al. in \cite{Jongejans:1973pn} . None of them found any evidence of inelasticity up to 1.8 GeV so that the amplitude is fully determined in terms of the phase shift, which is shown in Fig.~\ref{fig:s32data} as a function of the CM energy. 
The first observation is that these phase shifts are negative, which means that the interaction in this channel is repulsive, and no resonances occur.
Second, it can be seen that most of these experiments have relatively large uncertainties and provide data between 800 and 1100 MeV, except for \cite{Cho:1970fb}, which reaches up to 1.7 GeV. In contrast, the data from Estabrooks et al. \cite{Estabrooks:1977xe}, obtained in 1977 and also reaching 1.75 GeV,  quotes much smaller uncertainties. It will therefore dominate any fit, particularly at high energies.
Although there is a crude agreement at low-energies, the conflict between different experimental sets when taking the uncertainties at face value is evident from the plot, particularly between Estabrooks et al. and Bakker et al. at low energies or Cho et al. at higher energies. It is also important to remark that the lowest data points sit around 700 MeV, and therefore somewhat far from threshold at 636 MeV. Hence, the direct extraction of threshold parameters, like scattering lengths, from a fit to data requires an  extrapolation that produces large uncertainties.
For example, although in \cite{Estabrooks:1977xe} a value of $a_0^{3/2}=-0.14\pm0.07$ is quoted (in $m_\pi^{-1}$ units), the compilation of low-energy parameters in 1983 \cite{Dumbrajs:1983jd} provided five values for this
scattering length ranging from -0.14 to -0.05. We already commented that this observable is relevant for chiral perturbation theory and
is the subject of a renewed interest from lattice QCD. It will be discussed in section
\ref{sec:sumrules}, where we will provide more precise and robust determinations from sum rules using our constrained parameterizations.

\vspace{.5cm}
{\em $I=1/2$ $S$-wave data  }\\

\begin{figure}[ht]
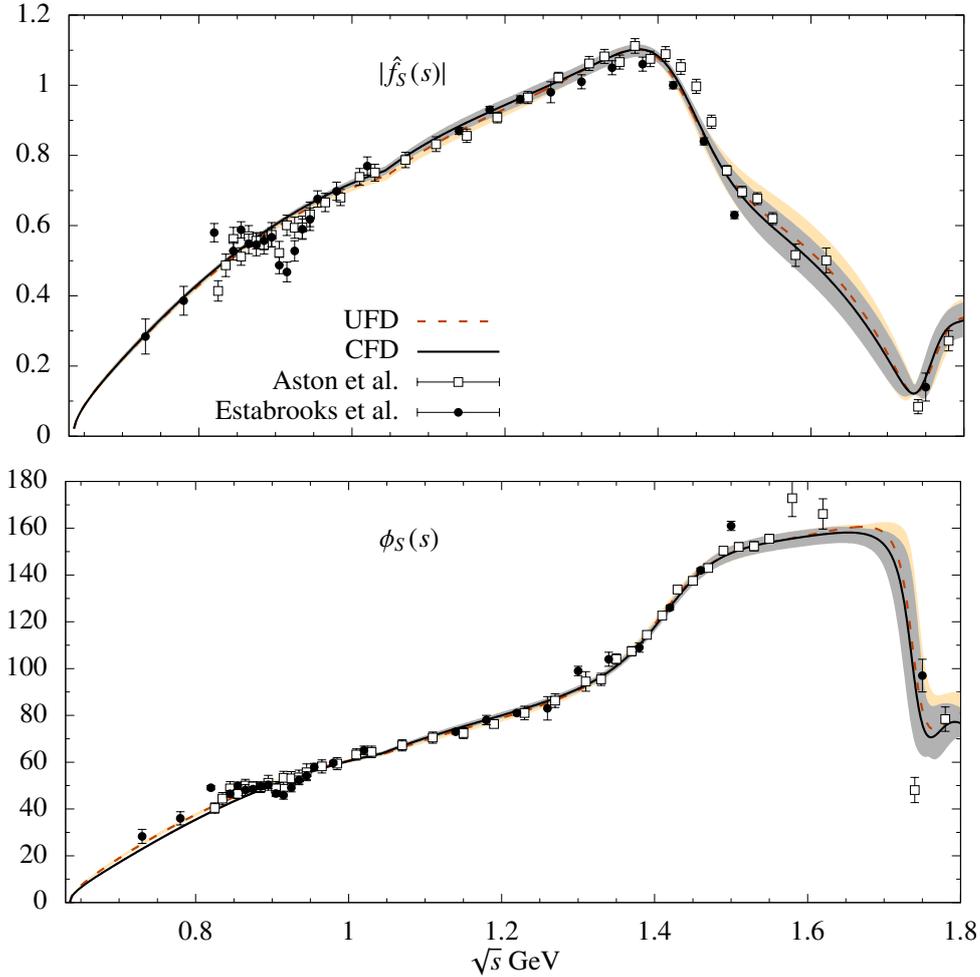

\begin{center}
\resizebox{0.8\textwidth}{!}{\input{figures/smod.tex}}\\
\vspace{-0.3cm}
\resizebox{0.8\textwidth}{!}{\input{figures/sphase.tex}}
\vspace{.2cm}
\caption{Data on the $S$-wave, measured by Estabrooks et al. \cite{Estabrooks:1977xe} and Aston et al. \cite{Aston:1987ir}. The upper panel shows the modulus and the lower panel the phase. 
We also show the results of our Unconstrained and Constrained Fits to Data (UFD and CFD respectively).
\label{fig:S12data} 
}
\end{center}
\end{figure}

This is the wave with the most structure and the most interesting for spectroscopy.
Unfortunately, as already discussed,  in fixed-target experiments 
the data on this isospin always appears mixed with $I=3/2$ in the following combination:
$f_S\equiv f^{1/2}_0+f^{3/2}_0/2$. Its modulus and phase are defined as follows:
\begin{equation}
    f_S(s)=\vert f_S(s)\vert \, e^{i \Phi_S(s)},
\end{equation}
which, in principle, were measured independently and therefore will be fitted separately. For comparison with data, the following normalization is used:
\begin{equation}
    \hat f_S(s)=f_S(s)\,\sigma_{\pi K}(s),
\end{equation}
where the phase space $\sigma_{\pi K}(s)$ was defined in Eq.~\eqref{eq:phasespace}.
With this notation, we show in Fig.~\ref{fig:S12data} the high-statistics data from the fixed-target experiments \cite{Estabrooks:1977xe,Aston:1987ir}.
Taking into account that the $I=3/2$ has a negative phase decreasing smoothly from zero to about -25 degrees, then the structure that is observed in $f_S$ is mostly due to $I=1/2$. In particular, there is a peak in the modulus and a simultaneous rapid increase of the phase  around 1430 MeV. This is a scalar strange resonance, nowadays called $K^*(1430)$, whose existence was supported by both experiments. In addition, there is a considerable increase in the modulus and the phase from threshold to 1.2 GeV, but not clearly resonant, which is the origin of the long-standing controversy about the existence of the \kap meson, which still ``Needs confirmation'' according to the present edition of the Review of Particle Physics \cite{pdg}. We will dedicate section~\ref{sec:kappa} to review the dispersive determination of this resonance.

Let us note the discrepancies between both sets of data in the whole energy region. Since the quoted errors are purely statistical, it is evident that there are systematic effects that we will have to estimate and consider when fitting the data.
In addition, it is important to remark that there are only two points below 800 MeV,
coming from \cite{Estabrooks:1977xe},
and thus the extraction of the $I=1/2$ scattering length from fits to data requires a large extrapolation, which yields large uncertainties. For illustration, the compilation of low-energy parameters from data in 1983 \cite{Dumbrajs:1983jd} provided five values for this
scattering length ranging
from $0.13\pm0.09$ to $0.33\pm0.01$ (in $m_\pi$ units).
The latter is from \cite{Estabrooks:1977xe}. The LASS data \cite{Aston:1987ir} starts even higher, at 825 MeV.
Once again, we recall that this scattering length is relevant for chiral perturbation theory and has been the subject of a renewed interest from lattice QCD. We will provide robust results from sum rules using our constrained parameterizations in section \ref{sec:sumrules}.

Up to here, we have discussed scattering data 
from fixed-target experiments on nucleons. 
However, it is also possible to extract $I=1/2$ data
from heavy meson decays.
In particular, when $\pi K$ are the only strongly interacting products
in the decay, Watson's theorem implies \cite{Watson:1952ji} that, in the $\pi K$ elastic region,
the phase of the whole process should be the same as the \pik scattering phase shift. 

\begin{figure}[ht]
\centering
 \includegraphics[width=.32\textwidth]{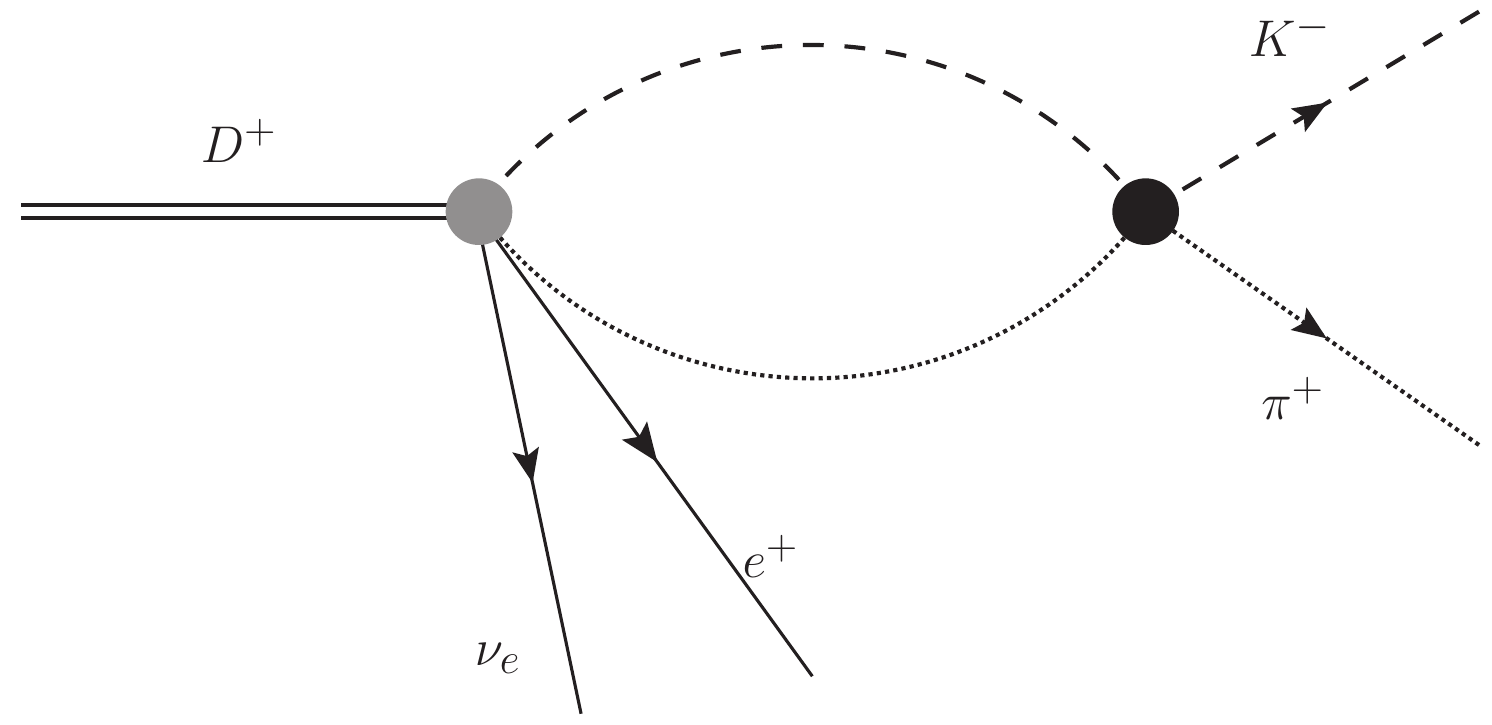} \hspace{0.1 cm}
\includegraphics[width=.32\textwidth]{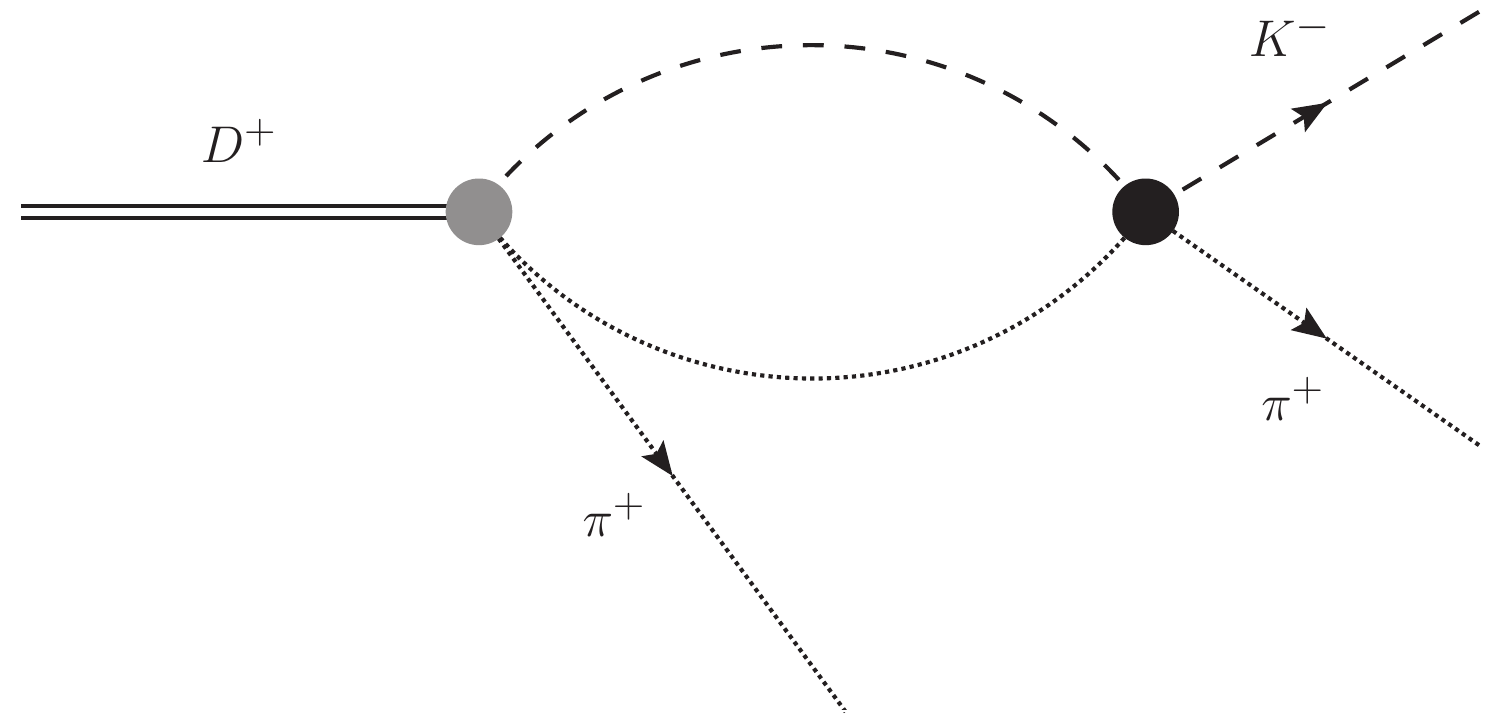} \hspace{0.1 cm} \includegraphics[width=.32\textwidth]{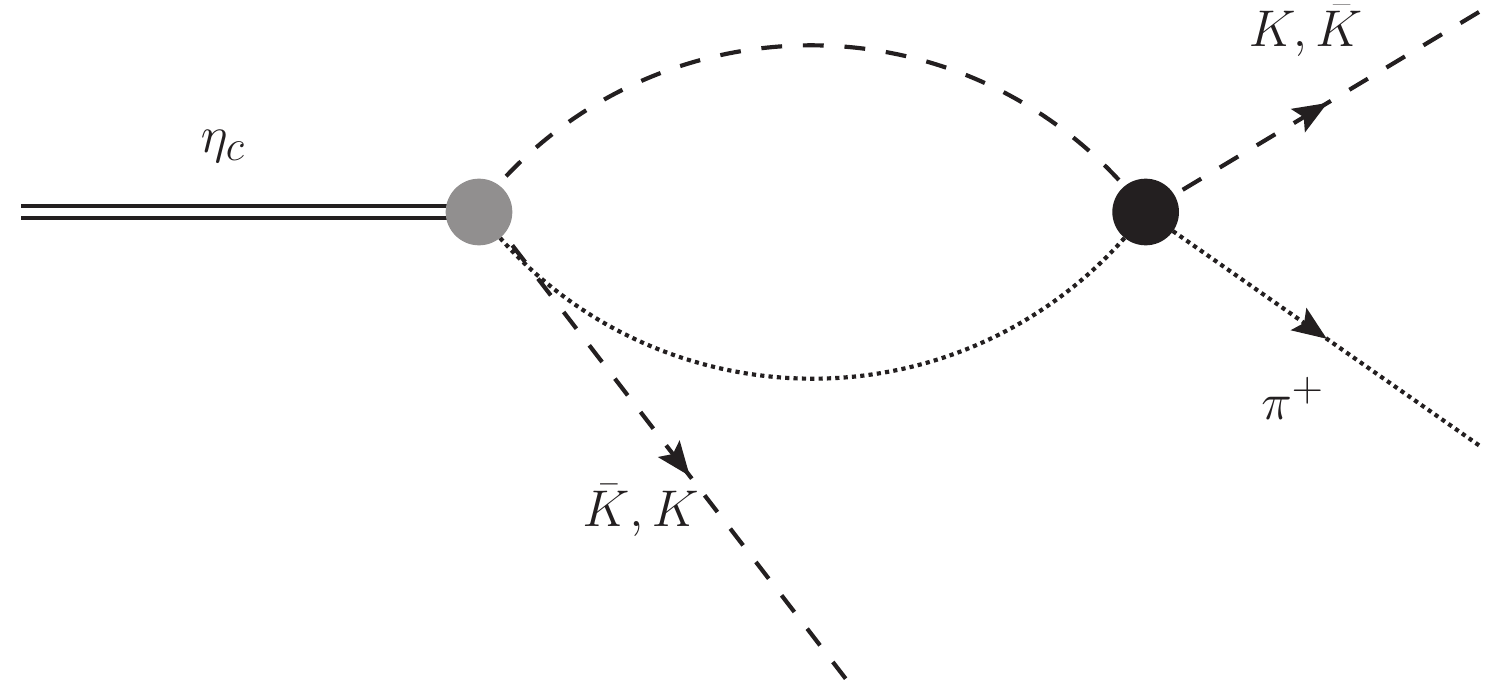}
\caption{ Other mechanisms that are sensitive to $\pi K$ scattering through its appearance as part of the final state. Assuming the other particles in the final state are spectators, Watson's theorem implies that the phase of the whole process is that of $\pi K$ scattering. Left: The cleanest case is when the other particles do not interact strongly as the lepton and neutrino produced in $D^+\rightarrow K^-\pi^+e^+\nu_e$ decays observed by the BaBar and BESIII Collaborations \cite{Ablikim:2015mjo, delAmoSanchez:2010fd}.  Center: $D^+\rightarrow K^-\pi^+\pi^+$ measured by the by the E791 \cite{Aitala:2005yh}, FOCUS \cite{Pennington:2007se,Link:2009ng} and CLEO-c \cite{Bonvicini:2008jw} collaborations. 
Right: $\eta_c\rightarrow K\bar K \pi$ measured by the BaBar Collaboration \cite{Lees:2015zzr}.
In the last two processes, the $\pi K$ scattering phase is obtained assuming the effect of the spectator particle plays a minor role in the energy dependence of the phase-shift, which is then observed displaced by a constant phase.}
\label{fig:decays}
\end{figure}

\begin{figure}[htb]
\begin{center}
\includegraphics[width=0.5\textwidth]{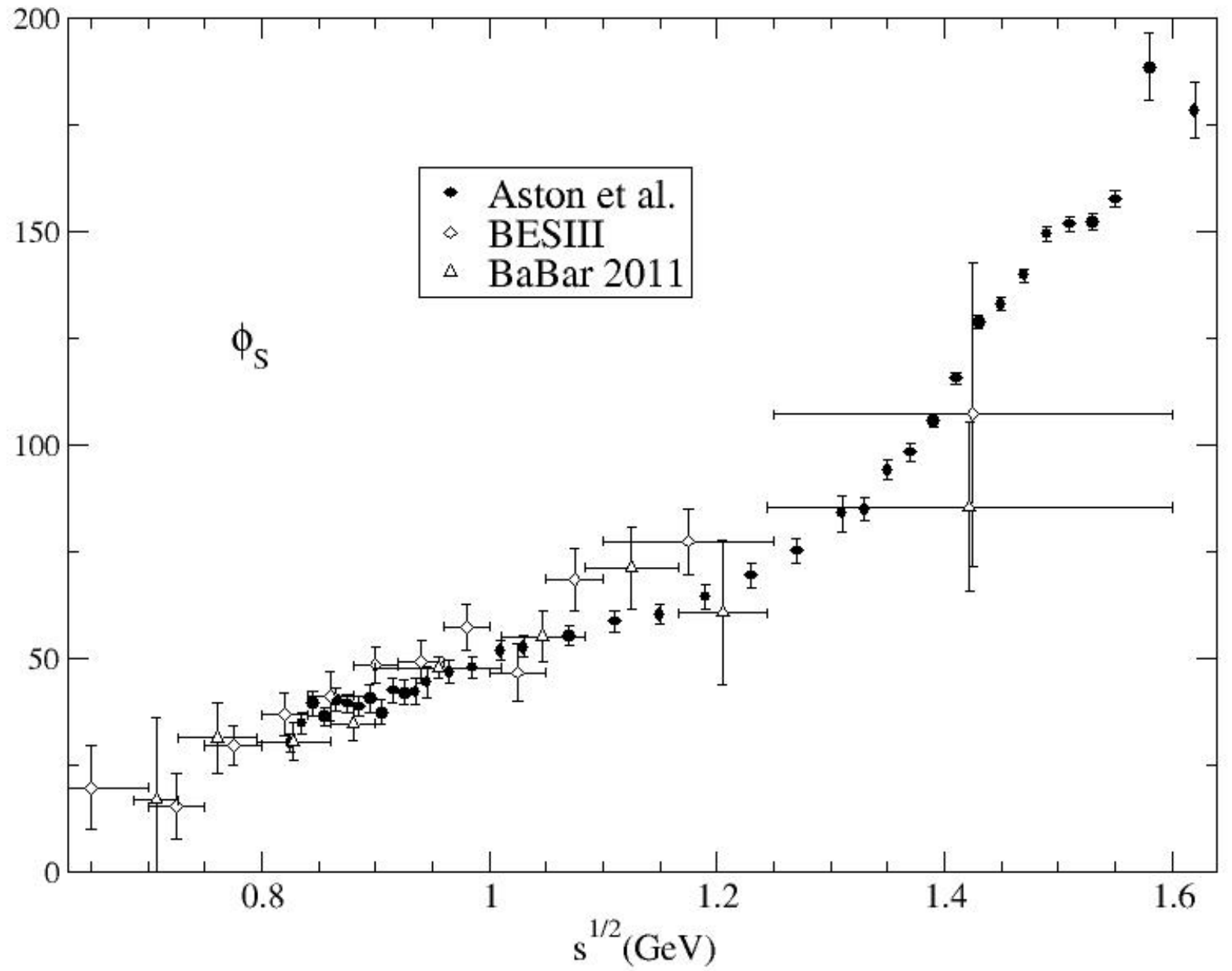}
\hspace*{-.3cm}
\includegraphics[width=0.5\textwidth]{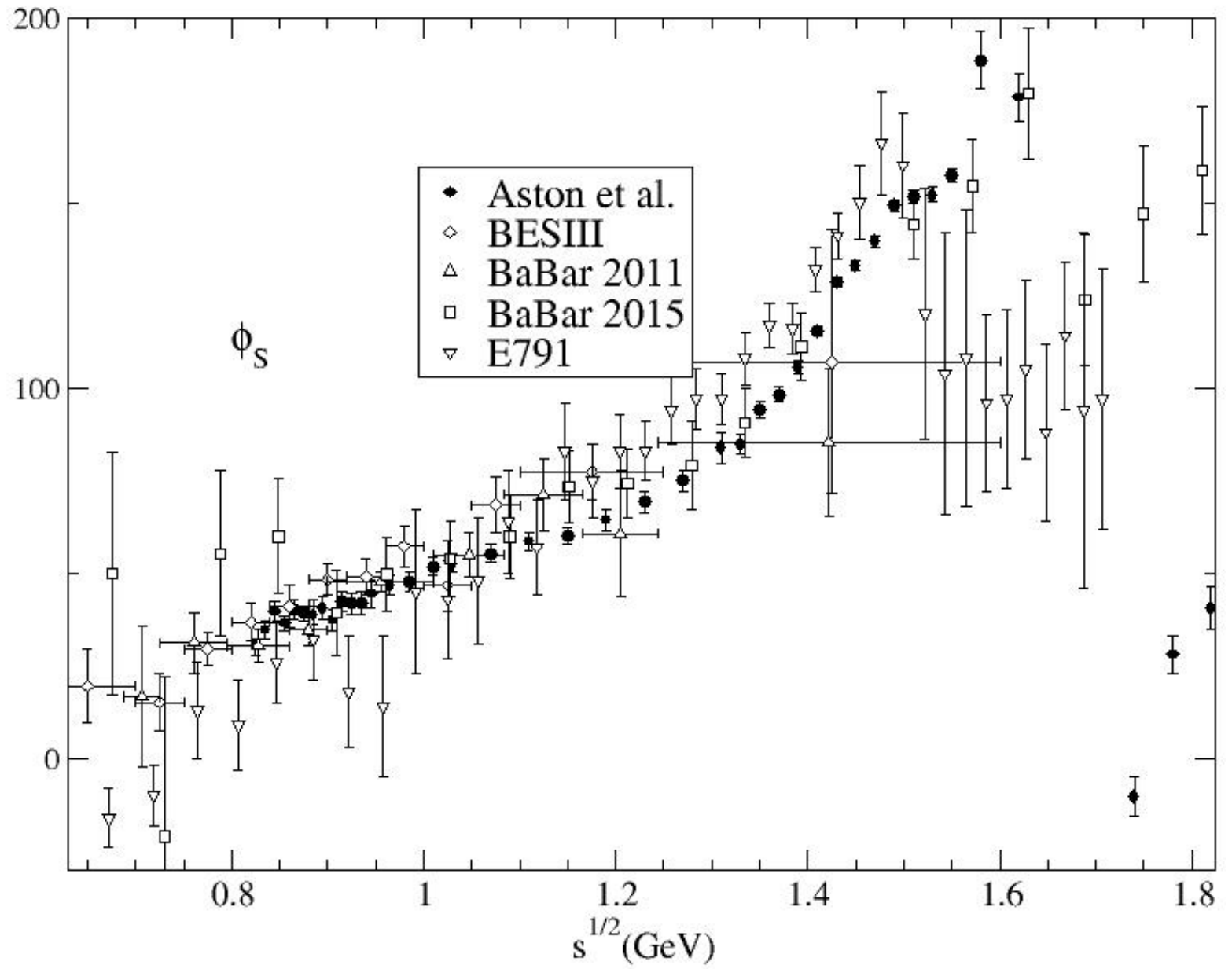}
\caption{$I=1/2$ $S$-wave phases obtained from the decay of heavy mesons versus the LASS data (using their own $I=3/2$ model to separate the $I=3/2$ component).
Left: From the semileptonic $D^+\rightarrow K^-\pi^+e^+\nu_e$ decay measured by BaBar and BESIII Collaborations \cite{Ablikim:2015mjo,delAmoSanchez:2010fd}, which used a simple $P$-wave model to separate the $P$ wave component. 
Right:  From Dalitz
plot analyses of $D^+\rightarrow K^-\pi^+\pi^+$  by the E791 Collaboration \cite{Aitala:2005yh} and
of $\eta_c\rightarrow K\bar K \pi$ by the BaBar Collaboration \cite{Lees:2015zzr}. Statistical plus systematic uncertainties are plotted for \cite{Aitala:2005yh}. As explained in the main text, the data from BaBar are displaced by 34$^\degree$ while those from E791 are displaced by 86$^\degree$, to ease the comparison with $\pi K$ data. 
\label{fig:decaydata} }
\end{center}
\end{figure}

The ideal situation is when the other particles in the decay are weakly interacting, as in so-called semileptonic decays. For example, the $I=1/2$ phase-shift difference between S and P waves has been measured from $D^+\rightarrow K^-\pi^+e^+\nu_e$ decays by the BaBar and BESIII Collaborations \cite{Ablikim:2015mjo,delAmoSanchez:2010fd}.  In the left panel of Fig.~\ref{fig:decays} we illustrate how \pik rescattering  (represented as a black disk) appears in this process. Note that the lepton and neutrino come from a weakly interacting $W$ boson, represented by a grey disk. 
Thus, in the left panel of Fig.~\ref{fig:decaydata} 
we show the $S$-wave phase, extracted from the measured $S-P$ phase difference with a simple model for the $P$-wave, whose uncertainties are much smaller than those of the $S$-wave and therefore can be neglected. In that plot, it can be seen that both the BaBar and BESIII results are quite consistent with those of the LASS experiment (separated from the $I=3/2$ component using their parameterization). However, they will not be included in our fits because their uncertainties are much larger than those from fixed-target experiments and  they should only be the same in the elastic region. Nevertheless, they provide a nice check of consistency.

Furthermore, the $I=1/2$ phase of the 
$\pi K$ $S$-wave amplitude has also been obtained from Dalitz plot analyses 
of $D^+\rightarrow K^-\pi^+\pi^+$ by the E791 \cite{Aitala:2005yh}, FOCUS \cite{Pennington:2007se,Link:2009ng} and CLEO-c \cite{Bonvicini:2008jw} collaborations, as well as a recent similar analysis 
of $\eta_c\rightarrow K\bar K \pi$ by the BaBar Collaboration \cite{Lees:2015zzr}.  The illustration of how \pik rescattering appears in these processes is shown in the center and right panels of Fig.~\ref{fig:decays}, respectively.
In principle, these phases (and amplitudes) are not necessarily those of $\pi K$ scattering due to the presence of a third meson that could also interact strongly.
However, a comparison with
the scattering data has shown that, within the large uncertainties and
at least in the elastic region, 
the resulting phase (but not the amplitude) still bears some similarity to that of LASS. It seems that, to a good degree of approximation, the third meson acts as a spectator and its effect on the phase can be recast as a global constant shift. 
Thus, we show in the right panel of Fig.~\ref{fig:decaydata}
that, up to 1.5 GeV and mostly due to their large uncertainties,
the phases obtained from E791 and BaBar \cite{Aitala:2005yh,Lees:2015zzr} 
are fairly compatible with those of LASS (extracting once again their $I=1/2$ with their own $I=3/2$ parameterization).
Note, however, that the data from BaBar are displaced by 34$^\degree$ while
those from E791 are displaced by 86$^\degree$. We do not show FOCUS and CLEO-c data, because they only provide some curves coming from their model for the phase, which, once displaced by a similar constant phase are relatively similar to that of LASS.
The additional phase is originated in the production process, represented by the grey disk in Fig.~\ref{fig:decays}, as well as in the interaction with the additional pion.  Within the range of interest, these two energy dependencies are expected to be very mild compared to that of $\pi K$ scattering itself and therefore crudely approximated by a constant.
Nevertheless, apart from their huge uncertainty, which makes them of little use,  these data cannot be interpreted as a scattering phase beyond this simplistic approximation. As a matter of fact, the effect of the interaction with the second pion has been investigated \cite{Magalhaes:2011sh,Guimaraes:2014kor,Nakamura:2015qga,Magalhaes:2015fva,Niecknig:2015ija,Niecknig:2017ylb} within several theoretical frameworks implementing
rescattering beyond what is typically called the isobar model and, to a varying degree, they explain why the $I=1/2$ $S$-wave phase shift extracted from D-decays should not be expected
to agree with the scattering data.
Therefore, $D$-meson decay data are not included in our fits, although they still provide a qualitative check of consistency, at least in the elastic region.

\subsubsection{$P$-wave data}
\label{subsec:Pdata}

\vspace{.5cm}
{\em I=3/2 $P$-wave data  }\\

\begin{figure}[ht]
\centering
\resizebox{0.8\textwidth}{!}{\input{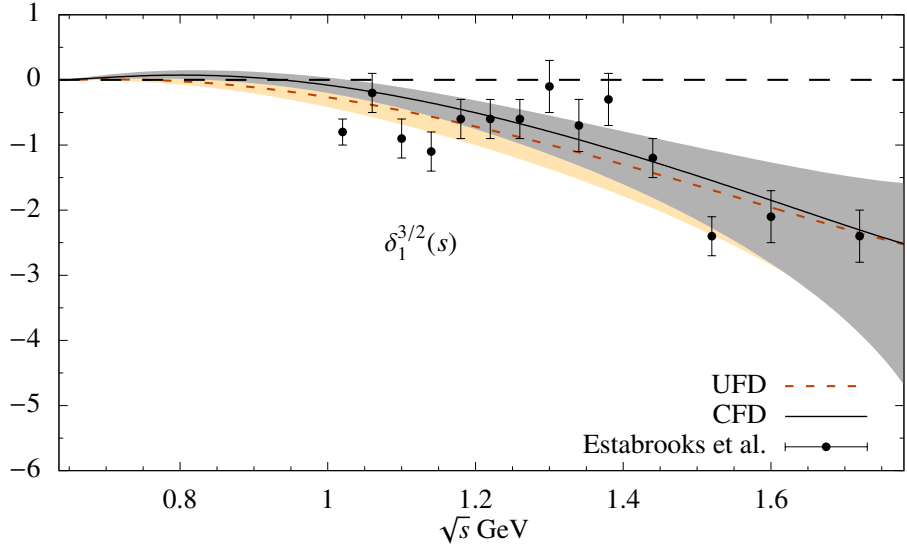}}\\
 \caption{\rm \label{fig:p32wavephase} 
Data on the $I=3/2$ $P$-wave from Estabrooks et al. \cite{Estabrooks:1977xe}.
We also show our CFD result as a solid line with a grey uncertainty band, 
which is obtained by fitting these data 
together with the data on the $f_P=f^{1/2}_1+f^{3/2}_1/2$ combination.
For comparison, we show with a dashed line the unconstrained fit
to the data in this figure, whose uncertainty is
    delimited by the orange band. }
\end{figure}

Only Estabrooks et al.~\cite{Estabrooks:1977xe} provide data for the $I={3/2}$ $P$-wave
phase-shift up to 1.74 GeV, which we show in Fig.~\ref{fig:p32wavephase}. 
No inelasticity is observed.
As it happened in the scalar case, this isospin wave is negative and therefore also repulsive. However,
the phase is an order of magnitude smaller. Actually, below 1.1 GeV the modulus of the phase shift is less 
than $1^\degree$, below 1.4 GeV is less than $2^\degree$, and below 1.74 GeV it is less than $3^\degree$. For this reason, is very frequently neglected in many analyses.
However, it should be considered for precision studies, and in particular to separate its contribution from that of $I=1/2$ in fixed-target experiments. Finally, it should be noted that data starts at 1 GeV and has huge oscillations. Therefore there is no information on its behavior near threshold, although NLO and NNLO ChPT
\cite{Bernard:1990kx,Bijnens:2004bu} and sum rules \cite{Buettiker:2003pp} predict that the scattering length should be positive. This suggests that the phase might be positive close to threshold and below 1 GeV, which will be confirmed by our dispersive analysis.

\vspace{.5cm}
{\em $I=1/2$ $P$-wave data  }\\

\begin{figure}[ht]
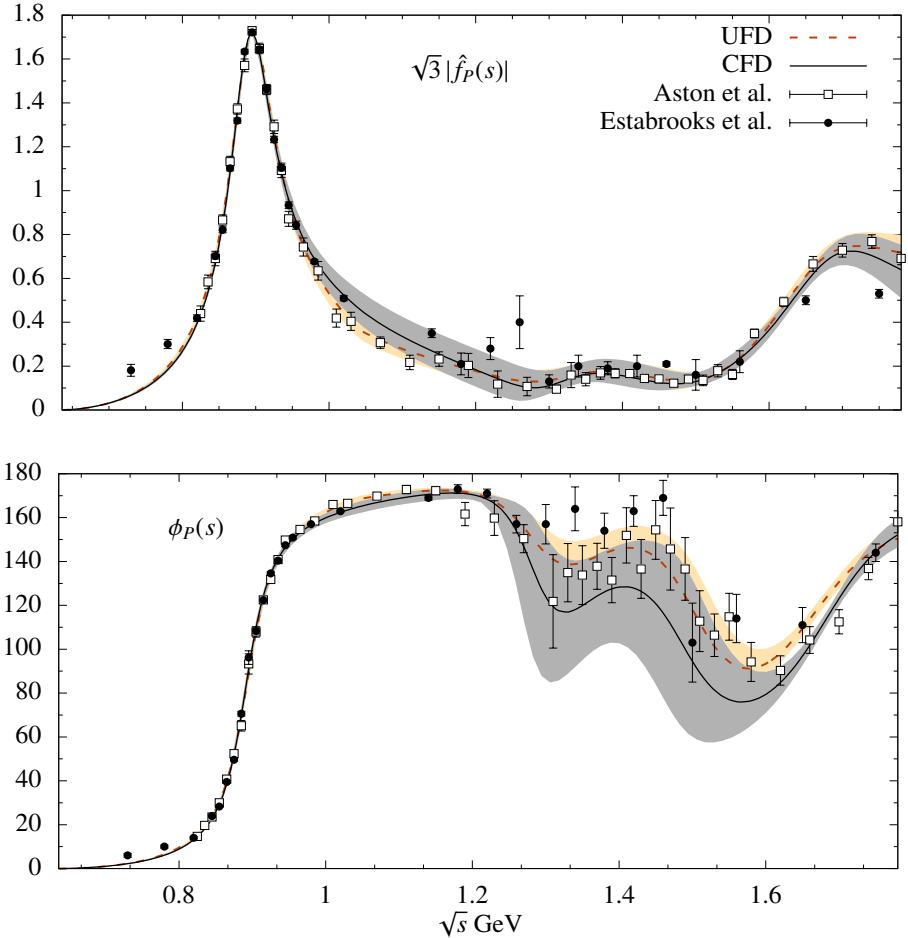

\begin{center}
\resizebox{0.75\textwidth}{!}{\input{figures/pmod.tex}}\\
\resizebox{0.75\textwidth}{!}{\input{figures/pphase.tex}}
\vspace{.2cm}
\caption{ Experimental results on $P$-wave \pik scattering from 
Estabrooks et al. \cite{Estabrooks:1977xe} and Aston et al.
 \cite{Aston:1987ir}.
Top: Data on $\vert \hat f_P(s)\vert$.
Bottom: Data on $\phi_P(s)$ . The continuous line is our constrained fit (CFD), 
whose uncertainties are covered by the grey band, whereas the dashed line is the Unconstrained Fit to Data (UFD),
whose corresponding uncertainties
are delimited by the orange band.
\label{fig:pwavedata} }
\end{center}
\end{figure}

Once again this wave is measured mixed with the $I=3/2$ component in the combination
$f_P\equiv f^{1/2}_1+f^{3/2}_1/2$, whose modulus and phase we define as follows:
\begin{equation}
    f_P(s)=\vert f_P(s)\vert\, e^{i \Phi_P(s)},
\end{equation}
although for comparison with data, the following normalization is used:
\begin{equation}
\hat f_P(s)=f_P(s)\,\sigma_{\pi K}(s),
\end{equation}
where the phase space $\sigma_{\pi K}(s)$ was defined in Eq.~\eqref{eq:phasespace}.
With this notation, we show in Fig.~\ref{fig:pwavedata}
the data measured by Estabrooks et al. \cite{Estabrooks:1977xe} and the LASS Collaboration \cite{Aston:1987ir}. There is a previous production experiment that was able to extract the $P$-wave \cite{Mercer:1971kn}. However, its statistics are low compared to LASS experiments, yielding  remarkably larger uncertainties, for which we consider it superseded by later experiments.

Then, taking into account that the $I=3/2$ contribution is tiny, one can see the general features of this $I=1/2$ $P$-wave. The low-energy part below 1.2 GeV is completely dominated by the presence of the $K^*(892)$ resonance peak  and its very
rapid associated increase of 180$^\degree$ in the phase. 
This is a very well-established resonance, measured in many other processes. For the neutral case, which is the one measured in \cite{Estabrooks:1977xe,Aston:1987ir}, the Review of Particle Physics quotes a mass of $895.55\pm0.20\,$MeV and a width of $47.3\pm0.5\,$MeV.
These features are usually described by means of some sort of Breit-Wigner parameterization, which may be justified when high precision is not required. Above  1.2 GeV,
there are two other strange resonances:
First, the  $K^*(1410)$ whose mass and width averages in the RPP are $1403\pm7\,$MeV and
$174\pm13\,$MeV, respectively. This resonance couples very little to the $\pi K$ channel and, accordingly, its associated peak is very small in the modulus. 
In contrast, the $K^*(1680)$, whose mass and width averages at the RPP are $1718\pm18\,$MeV and
$322\pm110\,$MeV, has a $\simeq 40 \%$ branching fraction to $\pi K$ and its peak is more visible.
The LASS experiment, using different decays channels, plays a very relevant role in the determination of the parameters of these resonances.
Note that, being so close and wide, they largely overlap and interfere, giving rise to the complicated behavior observed in the phase. Obviously, since these resonances decay predominantly to other channels, the \pik inelasticity becomes large in some parts of this region.
Concerning the threshold parameters, 
note once more that there are only two points below 800 MeV so that simple extrapolations of data down to threshold are rather unstable. Moreover, these two points are at odds with the rest of the data and the dispersion relations, so we do not include them in our fits. We will provide sum-rule determinations of these threshold parameters in section \ref{sec:sumrules}.

\begin{figure}[ht]
\begin{center}
\resizebox{.95\textwidth}{!}{
\includegraphics[height=4cm]{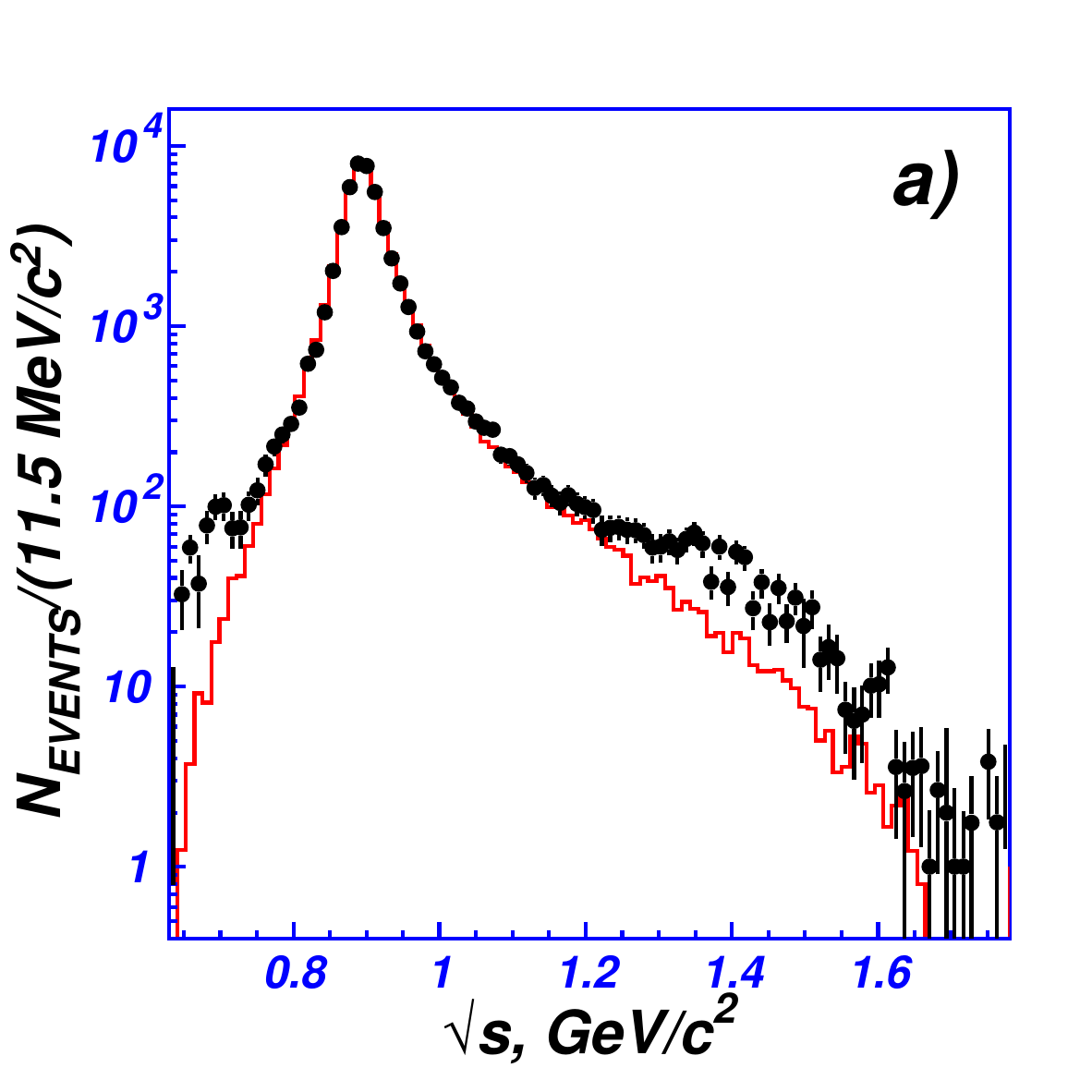}
\hspace{.2cm}
\includegraphics[height=4.2cm]{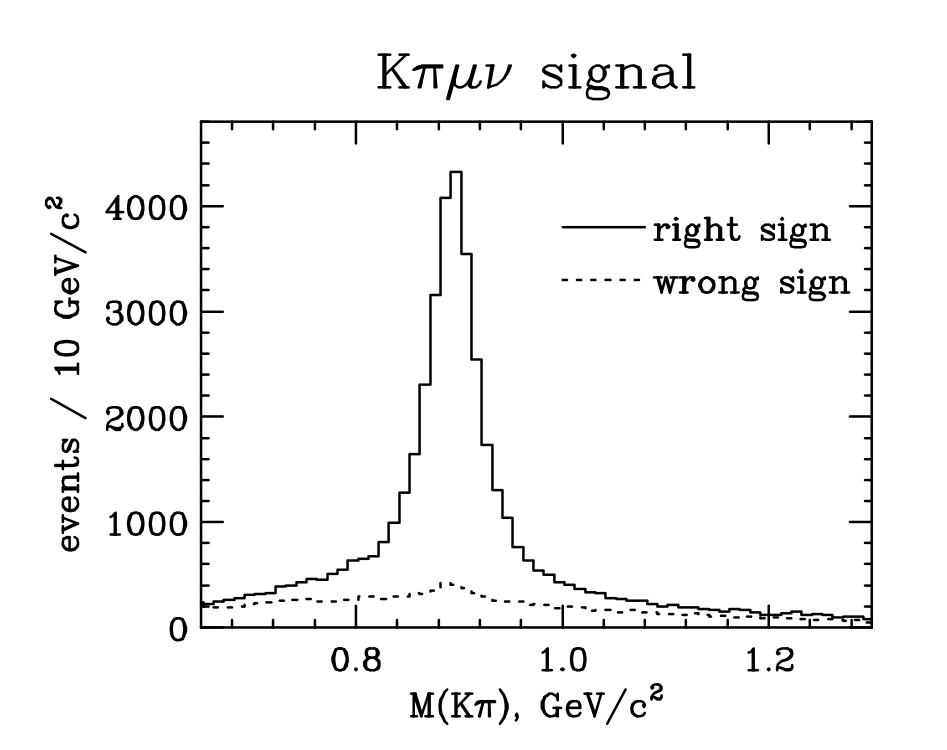}
}
\caption{$\tau\to K \pi \nu_{\tau}$ data measured by the BELLE collaboration (left panel) \cite{Epifanov:2007rf} and $D^+\to K^+ \pi^-\mu^+\nu$ data measured by the FOCUS collaboration (right panel) \cite{Link:2002ev}.
\label{fig:decayspwave}}
\end{center}
\end{figure}

Besides scattering experiments on nuclei,  other ways of extracting information on $\pi K$ scattering in the $P$-wave are possible as shown in Fig.~\ref{fig:decayspwave}. In particular, according to \cite{Ananthanarayan:2005us}, data measured on $D^+\to K^+ \pi^-\mu^+\nu$ by the FOCUS collaboration \cite{Link:2002ev} could provide stringent constraints on the $P$-wave phase shifts. In addition, there are two measurements of $\tau\to K \pi \nu_{\tau}$ decays \cite{Epifanov:2007rf,Paramesvaran:2009ec}. These observations, together with information on other decays, have been recently used to improve our knowledge of the $\pi K$ form factors \cite{Boito:2010me,Bernard:2011ae,Escribano:2014joa,Gonzalez-Solis:2019owk,Rendon:2019awg}. Furthermore, the first extraction of the $f_+(0)|V_{us}|$ term coming from these combined analyses was performed in \cite{Bernard:2013jxa}. Due to the fact that the $\pi K$ vector partial wave is the dominant contribution in all these processes, the $K^*(892)$ plays a very relevant role in all these works. Therefore, in~\ref{app:palt} we describe a new alternative elastic $P$-wave obtained by using the fits of the FOCUS collaboration~\cite{Link:2002ev}, and we also perform there its dispersive study. Fortunately, although starting from different data, the dispersively constrained final result of the alternative  description and the one we will discuss in the main text, using scattering data only, turn out to be very similar and compatible, which is why we have relegated the alternative one to the appendix.

\subsubsection{$D$-waves data}
\label{subsec:Ddata}

\vspace{.5cm}
{\em $I=3/2$ $D$-wave data  }\\

\begin{figure}[ht]
\begin{center}
\resizebox{0.8\textwidth}{!}{\input{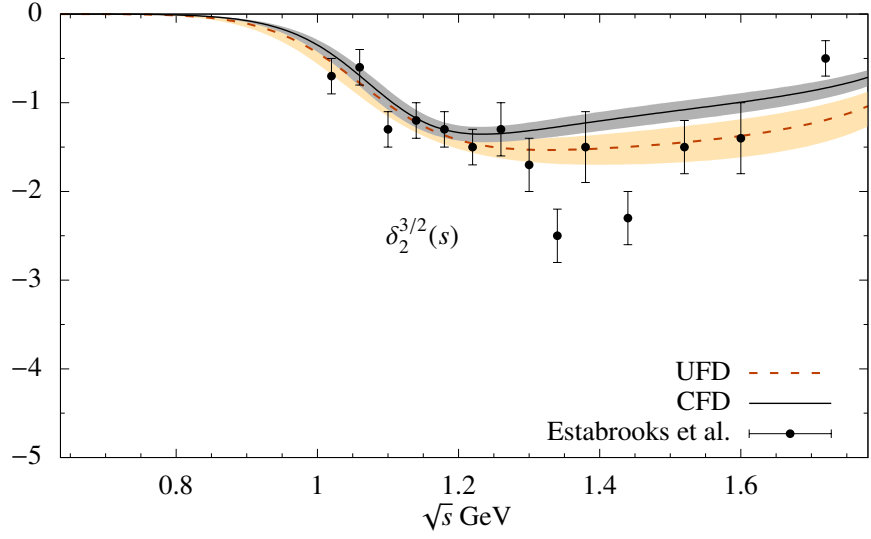}}
\caption{Data on the $I=3/2$ $D$-wave from Estabrooks et al. \cite{Estabrooks:1977xe}.
 We also show our UFD and CFD parameterizations (dotted and continuous lines, respectively).
\label{fig:D32phase}}
\end{center}
\end{figure}

As with the $I=3/2$ $P$-wave, only Estabrooks et al. \cite{Estabrooks:1977xe} provide data for the $I=3/2$ $D$-wave phase shift up to 1.74 GeV.  Since no inelasticity has been measured, the phase shift determines completely the amplitude. 
The data for the phase shift are shown in Fig.~\ref{fig:D32phase} and are very small in the whole energy region, not even reaching 3$^\degree$. Note there are no data below 1 GeV so that there is no information on threshold, however, in this case, both NNLO ChPT 
\cite{Bijnens:2004bu} and sum rules \cite{Buettiker:2003pp} yield a negative scattering length, and then it is natural to assume that the phase shift is also negative from threshold up to the first data point. We will confirm this with our dispersive analysis.

\vspace{.5cm}
{\em $I=1/2$ $D$-wave data  }\\

\begin{figure}[ht]
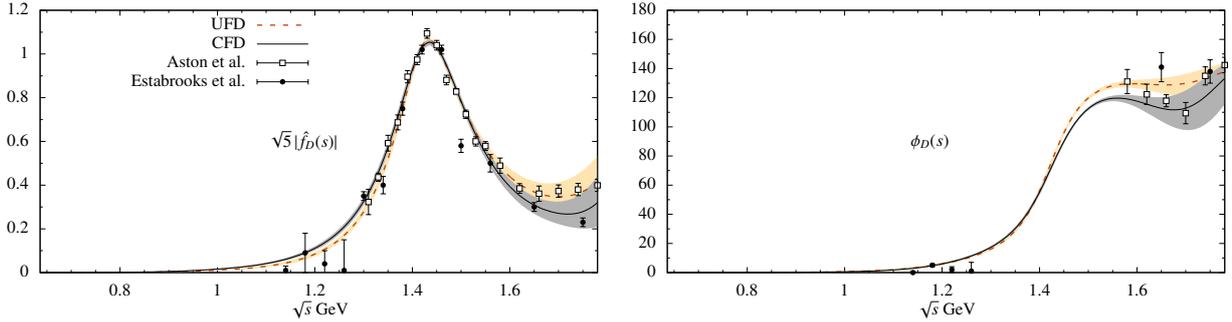

\begin{center}
\resizebox{\textwidth}{!}{\input{figures/dmod.tex} \input{figures/dphase.tex}}
\caption{Data on the $D$-wave isospin combination $f_D$ measured by Estabrooks et al.~\cite{Estabrooks:1977xe} and Aston et al. \cite{Aston:1987ir}. Left: Modulus of $\vert \hat f_D\vert$. Right: Phase $\Phi_D$.  We also show our UFD and CFD parameterizations (dotted and continuous lines, respectively).
\label{fig:Dwavedata}
}
\end{center}
\end{figure}

As it happened with the $S$ and $P$-waves, 
the $I$=1/2 $D$-wave is only measured together
with the $I$=3/2-wave in the 
$f_D \equiv f^{1/2}_2+f^{3/2}_2/2$
combination, for which, once again, we define the modulus and the phase
\begin{equation}
    f_D(s)=\vert f_D(s)\vert\, e^{i \Phi_D(s)},
\end{equation}
as well as the usual normalization to compare with data:
\begin{equation}
\hat f_D(s)=f_D(s)\,\sigma_{\pi K}(s).
\end{equation}
With this notation we show in Fig.~\ref{fig:Dwavedata} the experimental results of \cite{Estabrooks:1977xe} and \cite{Aston:1987ir}. Since the $I=3/2$ component is so small and featureless, all the features seen in that figure correspond to the $I=1/2$ channel.
In particular, below 1.8 GeV there is just a clear peak and phase motion corresponding to the well-established $K^*_2(1430)$ strange resonance. This resonance is seen in many other processes, but note that the averaged mass of the neutral case is dominated in the RPP by the results of \cite{Estabrooks:1977xe} and LASS.  Its decay branching ratio to $\pi K$ is approximately 50\% and an inelastic formalism will be needed. Let us remark that the data starts above 1.1 GeV so that there is no real experimental information on threshold and extrapolations are unstable.
Actually,  although NNLO ChPT \cite{Bijnens:2004bu} and sum rules \cite{Buettiker:2003pp} predict a positive scattering length, they are not very consistent with each other. We will review this situation and provide sum rule determinations from our constrained dispersive fits in section \ref{sec:sumrules}.

\subsubsection{$I=1/2$ $F$-wave data}
\label{subsec:Fdata}
\begin{figure}[ht]
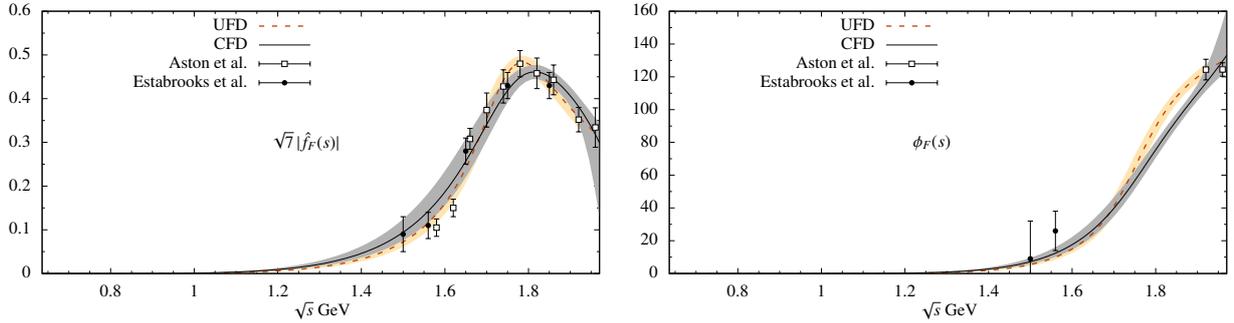

\begin{center}
\resizebox{\textwidth}{!}{\input{figures/fmod.tex} \input{figures/fphase.tex}}\\
\caption{Scattering data on the \pik $F$-wave \label{fig:Fwavedata}
measured by Estabrooks et al.~\cite{Estabrooks:1977xe} and Aston et al. \cite{Aston:1987ir}. Left: Modulus of $\vert \hat f_F\vert$. Right: Phase $\Phi_F$.  We also show our UFD and CFD parameterizations  as dashed and continuous lines, respectively, with their corresponding uncertainty bands (grey and orange respectively). }
\end{center}
\end{figure}

As usual with other waves we define 
\begin{equation}
f_F(s)\equiv \vert f_F(s)\vert\, e^{i\phi_F(s)},
\quad \hat f_F(s)=f_F(s)\, \sigma_{\pi K}(s). 
\end{equation}
However, for this wave, there are 
no observations of $I=3/2$ scattering, which is therefore neglected in 
the literature. We can thus consider that the whole $f_F$ is just $I=1/2$. We show in Fig.~\ref{fig:Fwavedata} the data obtained in \cite{Estabrooks:1977xe,Aston:1987ir}.
Note that the threshold suppression is so large that there are no data below 1.5 GeV. We will provide sum-rule results for the scattering length.

The most salient feature of this wave is the peak of the 
$K_3^*(1780)$ resonance, whose branching ratio to $\pi K$ is slightly less than 20\%. We will thus need an inelastic formalism. In the RPP, the parameters of this wave are completely dominated by the LASS Collaboration results \cite{Aston:1987ir}.

\subsection{$\pipikk$ scattering data}
\label{sec:pipiKKData}

Let us recall that, in the isospin limit, all pions are identical particles, and, being bosons, 
the $\pi\pi$ state must be fully symmetric. Thus, the two possible isospin states that couple to $K\bar K$, which are $I=0$ and $I=1$, are expanded in terms of only-even or only-odd partial waves, respectively. For all  of them, we define a modulus and a phase $g^I_\ell=|g^I_\ell|e^{i\phi^I_\ell}$.

The experimental results on $\pi\pi\rightarrow K\bar{K}$ partial waves that will be reviewed next 
were obtained in the early eighties  \cite{Cohen:1980cq,Etkin:1981sg}, indirectly
from fixed-target experiments on $\pi N\rightarrow K\bar{K} N'$. In order to extract the meson-meson amplitude, it is then assumed that the one-pion-exchange mechanism illustrated in the right panel of Fig.~\ref{fig:processes} dominates the whole process and that the meson-meson sub-process is factorizable. This is a fairly good approximation if the events are selected with the exchanged pion momenta close to the pion mass shell, but as commented in the \pik case, and as illustrated below, the final result is plagued with systematic uncertainties. It is therefore usual to find that different experiments do not agree within their statistical uncertainties, and a systematic uncertainty will have to be considered.
For our purposes, the data can be grouped in
four different types. First,
we will use data on phases and modulus for the $g^0_0, g^1_1$
partial waves extracted from  $\pi^- p\rightarrow K^-K^+ n$ and 
$\pi^+ n\rightarrow K^-K^+ p$ at the Argonne National Laboratory  \cite{Cohen:1980cq}
and from $\pi^- p\rightarrow K^0_sK^0_s n$ at the Brookhaven National Laboratory in a series of three works 
\cite{Etkin:1981sg,Longacre:1986fh,Lindenbaum:1991tq}. The latter will be called Brookhaven-I, Brookhaven-II, and Brookhaven-III, respectively. 
Second, although 
data for the modulus of the tensor $g^0_2$ wave was obtained in
Brookhaven-II and Brookhaven-III, we will see that the old experimental parameterizations are not
quite compatible with the  resonance parameters presently compiled in the RPP. 
Third, for higher partial waves, which play a very minor role in the dispersive analysis of the lower waves and have no scattering data, we use 
simple resonance parameterizations adjusting their parameters
to those in the RPP. 

Finally, let us remark that in the high-energy region above 2 GeV there are no data on all the partial waves we need for our dispersive integrals. It is for this reason that our plots in this subsection will end at that energy.
Nevertheless,  we will follow closely our approach in 
\cite{Pelaez:2018qny} and
rely on recent updates \cite{DescotesGenon:2006uk,GarciaMartin:2011cn,Pelaez:2016tgi}
of Regge parameterizations 
\cite{Pelaez:2003ky}
obtained from factorization from nucleon-nucleon and meson-nucleon processes and the phenomenological observations of Regge trajectories
or the Veneziano model \cite{Veneziano:1968yb}.
All this will be discussed in section \ref{sec:ufdregge}.

Let us then describe the \pipikk data for each partial wave in detail.

\subsubsection{$I=0$ $S$-wave}
\label{subsec:S0data}

\begin{figure}[!ht]
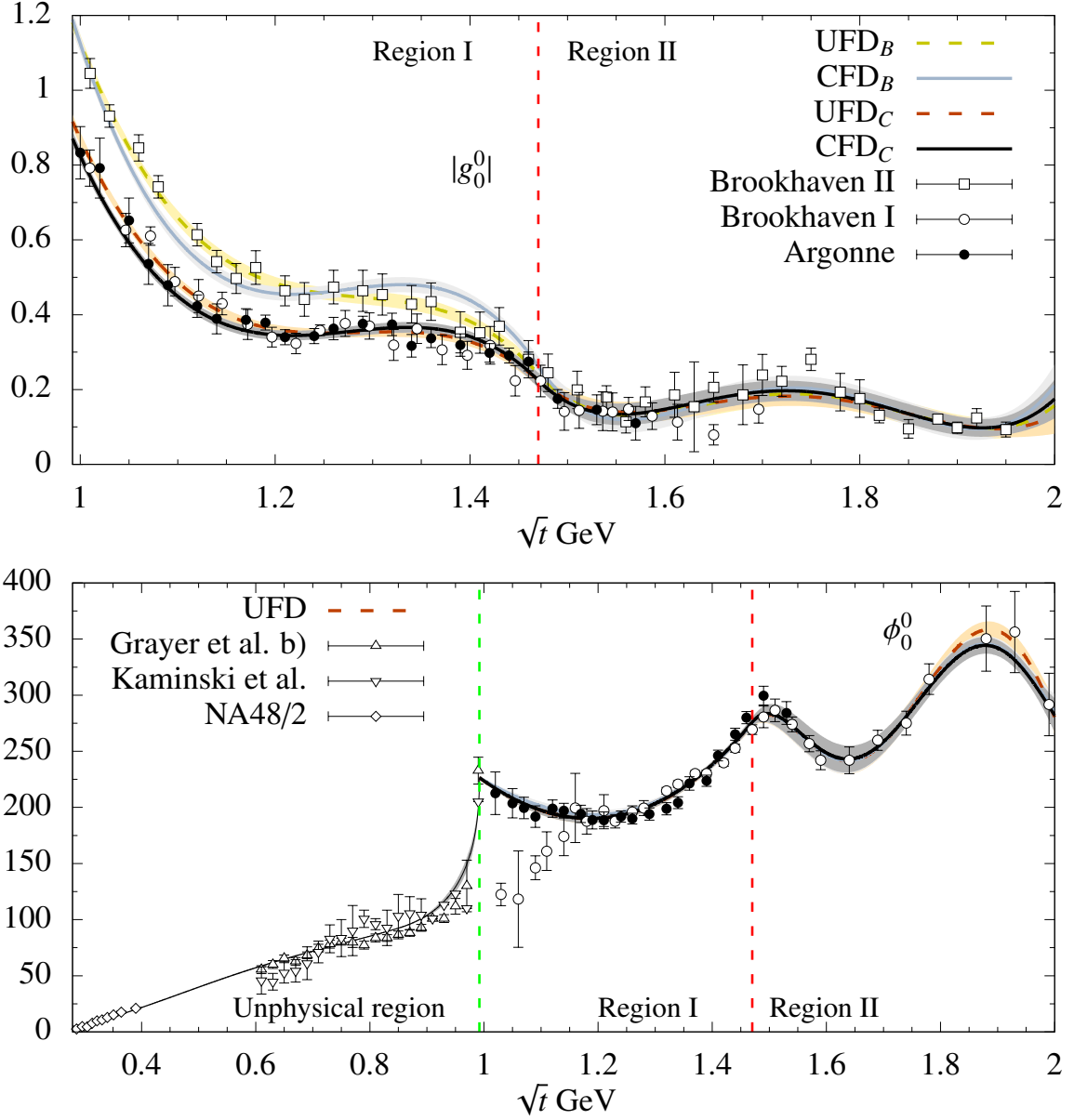

\vspace{-0.5cm}
\begin{center}
\resizebox{\textwidth}{!}{\input{figures/g00mod.tex}}\\
\resizebox{\textwidth}{!}{\input{figures/g00phase.tex}}\\
\caption{\label{fig:g00data}\pipikk scattering data on the scalar-isoscalar partial wave $g^0_0$, coming from  \cite{Cohen:1980cq} (Argonne), \cite{Etkin:1981sg} (Brookhaven-I) and \cite{Longacre:1986fh} (Brookhaven-II). As explained in the main text, in the ``unphysical region'' below $K\bar K$ threshold, due to Watson's theorem and the fact that no multi-pion states are observed, 
the $\pipikk$ phase shift is precisely that of $\pi\pi\rightarrow\pi\pi$ scattering.  Thus in that region, we provide a representative sample of such $\pi\pi$ scattering data  \cite{Grayer:1974cr} (Grayer et al., solution b), \cite{Kaminski:1996da,Kaminski:2001hv} (Kaminski et al.),
or the very precise $K_{\ell4}$ decays from \cite{Batley:2005ax} (NA48/2). We have also separated Region I, where we will be able to apply dispersion relations as tests or constraints, from Region II, where these relations are not applicable. }
\end{center}
\end{figure}

This $g^0_0(t)$ wave is quite complicated but also a very interesting 
one for hadron spectroscopy, since it couples to the much-debated scalar-isoscalar 
resonances.
Data for both the modulus $\vert g_0^0\vert$ and the phase
$\phi^0_0$ exist in the physical region and
are shown in  Fig.~\ref{fig:g00data}. 
Although data for both observables extend up to 2.4 GeV, we only show them up to 2 GeV, since above that energy we will use Regge parameterizations.

Note that in the lower panel of Fig.~\ref{fig:g00data} we also show 
data on the phase below $K \bar K$ threshold, coming from $\pi\pi$ elastic scattering in the scalar-isoscalar channel.
Watson's theorem \cite{Watson:1952ji} tells us that
in the $\pi\pi$ scattering elastic regime this is also the \pipikk phase.  Since multi-pion states have not been observed
in $\pi\pi$ scattering below the $K \bar K$ threshold, this phase shift 
is in practice the same as that of \pipikk below threshold.

There are several incompatible data sets in different parts of the inelastic regime and some of them will be discarded based on physical shortcomings, like Watson's theorem. However, for the modulus two distinct sets can be seen:
on the one hand the Argonne and Brookhaven I data, which are roughly compatible, and, on the other hand, the Brookhaven II set.
In \cite{Pelaez:2016tgi} we studied them both and we were able to find constrained fits describing either one of them while
satisfying simultaneously the dispersive representation, although the combined solution of Argonne and Brookhaven I seemed slightly favored. 
Of course, that was done with a fixed \pik input, and here we will complete that analysis constraining simultaneously  \pik and \pipikk. Thus we will analyze dispersively again the two different $g^0_0$ incompatible sets, referring by default to the combined solution in the main text, and explaining the alternative one in~\ref{app:g00alt}. To this end it is more convenient to separate the whole inelastic fit range into two regions, as follows:

\begin{enumerate}
\item[I)] Region I, extending from $\sqrt{t_{min,I}}=2m_K$ up to $\sqrt{t_{max,I}}=1.47\,$GeV. As shown in \ref{app:Applicability}, this region lies within the applicability domain of Roy-Steiner equations and will be constrained to satisfy dispersion relations in section \ref{sec:CFDpipiKK} below.

It is clearly seen in the lower panel of Fig.~\ref{fig:g00data}
that from $2m_K$ up to $1.2\,$GeV, the phase $\phi^0_0$ from the Argonne \cite{Cohen:1980cq} and Brookhaven-I
\cite{Etkin:1981sg} collaborations are incompatible. 
However, by Watson's theorem, $\phi^0_0$
at $K \bar{K}$ threshold should be the same as that of the scalar-isoscalar
$\pi\pi\rightarrow\pi\pi$ phase shift $\delta_0^{(0)}$. Here one should notice that,
as seen in the figure in the unphysical region, both the $\pi\pi\rightarrow\pi\pi$ data and their dispersive analyses with Roy and GKPY equations \cite{GarciaMartin:2011cn,Moussallam:2011zg},  
which extend up to or beyond $K \bar K$ threshold,
find $\delta_0^{(0)}>200^\degree$. The huge rise of the phase right below $K\bar K$ threshold is due to the presence of the well-known $f_0(980)$ resonance.
Thus, it seems that the phase of Brookhaven-I \cite{Etkin:1981sg} right above $K \bar K$ threshold is inconsistent with Watson's theorem. Moreover, this phase was extracted using a $g_2^0$ wave that also violates Watson's theorem, as we will see soon below.
Hence, for our fits, we will discard the  
Brookhaven-I phase data
\cite{Etkin:1981sg} below $\sim$1.15 GeV, i.e. until it agrees again with that of 
Argonne \cite{Cohen:1980cq}. 

Concerning the data on $\vert g_0^0\vert$,
shown in the upper panel of Fig.~\ref{fig:g00data},  we can see that, up to roughly 1.4 GeV, the sets of
Argonne and Brookhaven-I  are consistent among themselves
but not with Brookhaven-II. 
For later purposes, it is relevant to remark that the latter is consistent up to 1.2 GeV with the so-called ``dip solution''
of the elasticity favored from dispersive
 $\pi\pi\rightarrow\pi\pi$ analyses \cite{GarciaMartin:2011cn,Moussallam:2011zg}, assuming that only $\pi\pi$ and $K\bar K$ states are relevant. In contrast, Brookhaven-II would require the presence of some non-negligible coupling to another state, possibly four pions. 
 As we did in \cite{Pelaez:2018qny} we will  consider both alternative possibilities in our fits.
Finally, in the 1.2 GeV to 1.47 region the ``dip'' solution from $\pi\pi$ scattering has such large uncertainties that
it is roughly consistent with the three data sets.

\item[(II)] Region II, extending from
 $\sqrt{t_{min,II}}=1.47$ GeV up to $\sqrt{t_{max,II}}=2$ GeV. We only use this region 
as input for our dispersive calculations for lower energies, since Roy-Steiner equations are not applicable here (see \ref{app:Applicability}).
Above 1.4 GeV all sets seem compatible again although the data from Argonne finishes around 1.5 GeV, whereas the Brookhaven-I set reaches up to $\sim$1.7 GeV
and only Brookhaven-II reaches up to 2 GeV.
It should be noticed that above 1.5 GeV this wave is rather small and possible resonance shapes are not evident. Nevertheless, several scalar isoscalar resonances are claimed to exist above $K\bar K$ threshold: the $f_0(1370)$, $f_0(1500)$ and $f_0(1710)$. They enjoy different statuses: from still some debate about the existence and parameters of the first, to well established for the $f_0(1500)$. In general, their parameters are not determined very precisely. However, for all three, both their couplings to $\pi\pi$ and $K\bar K$ are small and they do not appear as clear peaks in the plot of $\vert g^0_0 \vert$. Also, they seem to be fairly wide and there should be a significant overlap between them.

\end{enumerate}
\subsubsection{$I=1$ $P$-wave data}

\begin{figure}[!ht]
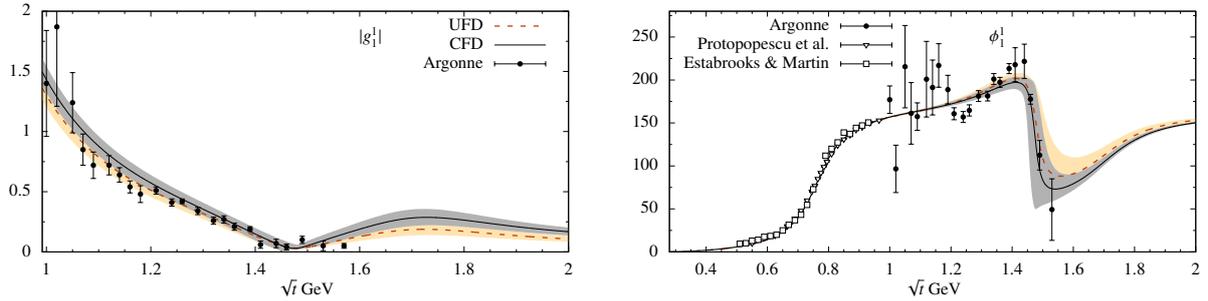

\begin{center}
\resizebox{\textwidth}{!}{\input{figures/g11mod.tex} \input{figures/g11phase.tex}}\\
\caption{
Modulus (Left) and phase (Right)  of the $g_1^1(t)$ $\pi\pi\rightarrow K \bar K$ partial wave. 
The continuous line 
and the grey uncertainty band correspond to the CFD 
parameterization described in the text, whereas the dashed line and orange band correspond to the UFD. Note that due to Watson's theorem, the phase below $K \bar K$ 
follows that of $I=1,\ell=1$  elastic $\pi\pi$ scattering \cite{GarciaMartin:2011cn}. 
The white circles and squares come from the $\pi\pi$ scattering experiments of
Protopopescu et al. 
\cite{Protopopescu:1973sh} and Estabrooks and Martin \cite{Estabrooks:1974vu}, respectively.
Let us remark that this is the only case for $\ell\geq 1$ when the data on the modulus was provided  with the same normalization as ours, i.e. without a $\sqrt{2\ell+1}$ factor.
\label{fig:g11data}}
\end{center}
\end{figure}

In the physical \pipikk region, only the Argonne Collaboration 
(Cohen et al. \cite{Cohen:1980cq}), 
has provided \pipikk scattering data on the $g_1^1$ partial wave. They reach
up to $\sim1.55\,$GeV for both the modulus $\vert g_1^1\vert$ and its phase 
$\phi^1_1$ and are shown in Fig.~\ref{fig:g11data}. It can be noticed that, for the phase, the error bars are very large in the 1 to 1.2 GeV region as well as in the last two data points above 1.45 GeV.

In addition, in Fig.~\ref{fig:g11data} we also show below $K \bar K$ threshold, the data on the phase coming from $\pi\pi$ elastic scattering with these quantum numbers.  
Note that the very rapid increase of the $\rho(770)$ meson is clearly seen. Since multiple pion states have not been observed in $\pi\pi$ scattering below the two-kaon threshold, 
Watson's theorem tells us that this is also the \pipikk phase in this pseudo-physical region. We will need this phase later on for our dispersive representation. Let us remark that the uncertainties here are much smaller than in the physical region.

Finally, the fact that there are no scattering data above 1.6 GeV and very poor information above 1.4 GeV, forces us to consider the information on resonances  measured in other processes. 
Apart from the $\rho(770)$ that dominates completely the region below 1 GeV, there are three other resonances below 2 GeV listed in the RPP with $J^{PC}=1^{--}$ quantum numbers.
However, only two of them are relevant for us. Namely, the $\rho'=\rho(1450)$
and $\rho''=\rho(1700)$, which have sizable couplings to both the $\pi\pi$ and $K\bar K$ channels.  The values of these parameters will be reviewed and used in section \ref{sec:UFDpipiKK} for our fits. In contrast, the couplings of these two channels to the $\rho(1570)$, whose existence is less certain (according to the RPP it `` may be an OZI-violating decay mode of the $\rho(1700)$''), have not been seen. We will therefore neglect it in our analysis.

\subsubsection{$I=0$ $D$-wave data}

\begin{figure}[!ht]
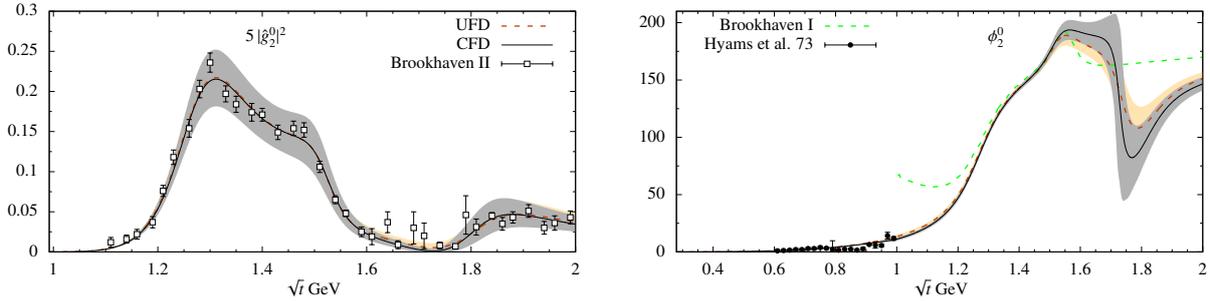

\begin{center}
\resizebox{\textwidth}{!}{\input{figures/g02mod.tex} \input{figures/g02phase.tex}}
\caption{\label{fig:g02data} Left: Data on the modulus of $\hat g^0_2(t)$
from the Brookhaven-II analysis \cite{Longacre:1986fh} together with our UFD and CFD parameterizations,
described in the text. Right: Phase of the $g^0_2(t)$
partial wave. Watson's theorem implies that in the ``unphysical'' region below $K \bar K$ scattering this phase should be the same as for elastic $\pi\pi$ scattering with the same quantum numbers because an inelasticity to other states has not been observed. Note that both our UFD and CFD satisfy this theorem since they match the 
$\pi\pi$ scattering data below 1 GeV \cite{Hyams:1973zf} and the dispersive analysis in \cite{GarciaMartin:2011cn}, whereas the Brookhaven-I model \cite{Etkin:1981sg}, does not.}
\end{center}
\end{figure}

We show in Fig.~\ref{fig:g02data} the data for this wave in the physical region.
These data were obtained in the
Brookhaven-II analysis \cite{Longacre:1986fh},
which was published 6 years after Brookhaven-I. 
Brookhaven-II was a study of  $\pi\pi\rightarrow\bar KK$ scattering
in the $I=0$, $J^{PC}=2^{++}$ channel employing 
a coupled-channel formalism, which 
included data from other reactions.
Later on, some members of that collaboration published  in \cite{Lindenbaum:1991tq}
a re-analysis, that we call Brookhaven-III, 
including even further information on other processes. 
Note that our normalization differs from that used by the experimental collaborations
and this is why we are plotting $\vert \hat g^0_2\vert$, defined as:
\begin{equation}
\hat g^0_2(t)\equiv\frac{2(q_\pi q_K)^{5/2}}{\sqrt{t}} g^0_2(t)\equiv
\vert \hat g^0_2(t) \vert e^{i\phi^0_2(t)}.
\end{equation}

Concerning the phase, below the $K \bar K$ threshold we will use Watson's theorem and the elastic data on $\pi\pi$ scattering. However, the relevant observation here is that for this partial wave there are no data on the \pipikk phase in the physical region. Thus, we need to look at the information on resonances with these quantum numbers observed in other processes.
According to the RPP, there are eight possible $J^{PC}=2^{++}$ resonances below 2 GeV.
These are the $f_2(1270)$, $f_2(1430)$ $f'_2(1525)$, $f_2(1565)$, $f_2(1640)$, $f_2(1810)$, $f_2(1910)$ and $f_2(1950)$. 
The only really well-established and clearly seen in many different processes are the $f_2(1270)$, $f'_2(1525)$ and $f_2(1950)$. Although the first couples predominantly to 
$\pi\pi$ and the second to $K\bar K$, their decays to both states have been measured
and therefore they do couple significantly to \pipikk.  
Actually, the peak of the $f_2(1270)$ is the most prominent feature in Fig.~\ref{fig:g02data}
and there is a hint of a second structure around 1.5 GeV.
However, the existence of the other five resonances is much into question, and either they
``Need confirmation'' or they are omitted from the summary RPP tables. 
Still, in Fig.~\ref{fig:g02data} there is some hint of a raise in the modulus above 1.8 GeV, and the $f_2(1810)$ resonance was considered both by Brookhaven-II and Brookhaven-III \cite{Longacre:1986fh,Lindenbaum:1991tq} in their phenomenological fits.

\subsubsection{Data on higher \pipikk  partial waves}
\label{subsec:higherpw}

There are no data on \pipikk scattering for partial-waves with an angular momentum higher than $\ell>2$.
We thus have to resort to the information on resonances below 2 GeV from the RPP. For $J^{PC}=3^{--}$ there is one well-established resonance, the $\rho_3(1690)$ with a $\sim160\,$MeV width, whose decays to $\pi\pi$ and $K\bar K$ are both measured and should therefore couple significantly to \pipikk. A not-so-well-established $\rho_3(1990)$ candidate is too high for our purposes and no decays to $K\bar K$ have been reported. Concerning $J^{PC}=4^{++}$, an $f_4(2050)$ resonance is listed in the RPP, with sizable decays to $\pi\pi$ and $K \bar K$. Its width is $\sim240\,$MeV so that its tail should affect somewhat \pipikk below 2 GeV, but we have checked that its effect is negligible in our calculations.

\section{Dispersion relations for $\pik$ and $\pipikk$}
\label{sec:DR}

Dispersion relations are the mathematical consequence of causality once we consider that pions and kaons have a sufficiently long life that they can propagate to infinity, which is a remarkably good approximation compared to the size of typical hadronic interactions.
Causality implies that the amplitude has to be analytic in the first Riemann sheet of the complex plane except for cuts due to the presence of thresholds (see \cite{Nussenzveig:1972ca,Pelaez:2015qba} for introductory texts) in the direct or crossed channels. Poles associated with bound states can only appear in the real axis below threshold, but this does not occur in \pik nor \pipikk scattering and we can thus ignore them. Rigorous proof of this connection between causality and
analyticity only exists within non-relativistic scattering \cite{Nussenzveig:1972ca}, but for relativistic scattering there is no general proof beyond axiomatic field theory or perturbation
theory~\cite{Eden:1966dnq}. For the general non-perturbative case we resort to the so-called Mandelstam hypothesis \cite{Mandelstam:1958xc,Mandelstam_1962} or ``maximal analyticity'', which we will assume throughout this review.

Dispersion relations take the form of integral equations in which the amplitude is represented as an integral over its imaginary part. 
In section \ref{sec:UFD}, we will review and update 
tests showing that the data on both \pik and \pipikk do not satisfy different dispersive representations.
One of the aims of this review, attained in section \ref{sec:CFD}, is to provide an update of 
data parameterizations that satisfy the different kinds of dispersion relations that we will present below, by imposing them as constraints on the fits to data.

The derivation of dispersion relations requires the use of Cauchy's theorem for a single complex variable. Since two-body scattering depends on two variables, different kinds of dispersion relations are obtained depending on 
whether we fix one variable, we relate one variable to the other, or whether we integrate one variable 
leaving just the explicit dependence on the other one. Respectively, these cases correspond in this review to fixed-$t$, hyperbolic, and partial-wave dispersion relations that we describe in detail next.

\begin{figure}[!ht]
\centering
\includegraphics[scale=0.65]{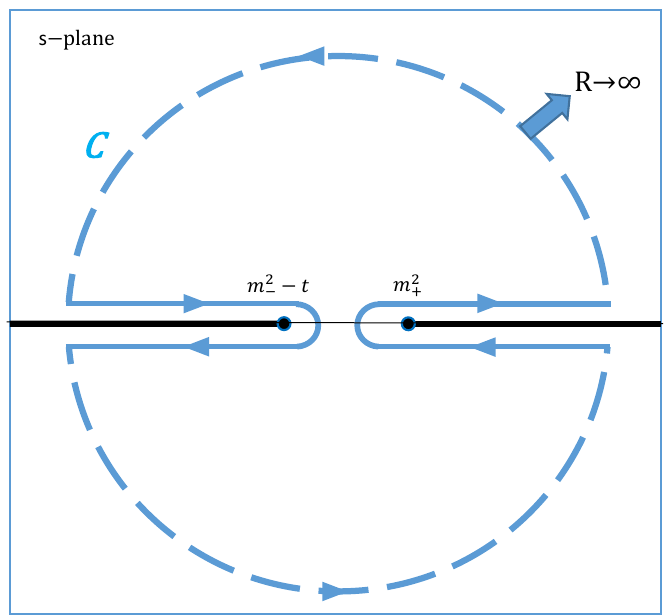}
 \caption{\rm \label{fig:Cauchy} Analytic structure of the fixed-$t$ \pik amplitude in the complex $s$-plane. We show as continuous thick black lines the ``right'' or ``physical'' cut, extending from $m_+^2$ to $+\infty$, as well as the ``left'' cut, from $m_-^2-t$ to $-\infty$.
 In blue we show the typical contour $C$ used to obtain a dispersion relation from Cauchy's theorem, enclosing the complex plane but avoiding the two cuts and sending the radius of its curved part (dashed) to infinity. Its straight sections are infinitesimally close to the cuts on the real axis.
}
\end{figure}

In Fig.~\ref{fig:Cauchy} we illustrate how a fixed-$t$ dispersion relation is obtained by applying Cauchy's theorem to the integral over the contour  $C$ (in blue) that encloses the complex plane except for the cuts (shown in black). The right cut corresponds to the opening of the $s$  channel threshold at $s=m_+^2$, which then extends to $+\infty$, whereas the left cut corresponds to the opening of the \pik $u$-channel, starting at $s=m_-^2-t$ and extending to $-\infty$. Then, the value of the amplitude at any point $s$ inside this contour is given by:
\begin{equation}
F(s,t,u)= \frac{1}{2\pi i}\oint_C \frac{F(s',t,u')}{s'-s}ds'.
\end{equation}
If the contribution of the amplitude in the curved part vanishes as its radius $R$ is taken to infinity, we are left only with the straight contours, separated by an infinitesimal distance $\epsilon$ from the real axis. Since amplitudes satisfy the reflection symmetry $F(s'+i\epsilon',t,u'-i\epsilon')=F^*(s'-i\epsilon',t,u'+i\epsilon')$, in the $\epsilon\rightarrow0$ limit, and given that the straight contours above and below the real axis run in opposite sense, we are left with
\begin{eqnarray}\label{Tcauchy-2}
F(s,t,u)
&=&\frac{1}{\pi}\int_{m_+^2}^{\infty}{ds^\prime 
\frac{\mathrm{Im}F(s^\prime,t,u')}{s^\prime-s}}+\frac{1}{\pi}\int_{-\infty}^{m_-^2-t}{ds^\prime \frac{\mathrm{Im}F(s^\prime,t,u')}{s^\prime-s}}.
\label{ec:predisp0}
\end{eqnarray}
This is a fixed-$t$ dispersion relation  valid everywhere in the $s$-complex plane
except on the singularities. If we want the amplitude from the dispersion relation in the real axis over the cut singularities, we must then consider the amplitude at $s+i\epsilon$ 
with $s$ real, 
and use the relation:
\begin{equation}
\frac{1}{s^\prime-s-i\epsilon}=PV\frac{1}{s^\prime-s}+i\pi\delta(s^\prime-s),
\label{eq:PVdistribution}
\end{equation}
where $PV$ denotes the principal value. Note that the effect of $i\pi\delta(s^\prime-s)$ on Eq.~\eqref{ec:predisp0}
is to extract $i\, \im T (s,t,u)$ out of the first integral, which cancels out the imaginary part on the left side. Hence on the real axis, we find:
\begin{equation}\label{Tcauchy-real}
\mathrm{Re}F(s,t,u)=\frac{1}{\pi}PV\int_{m_+^2}^{\infty}{ds^\prime \frac{\mathrm{Im}F(s^\prime,t,u^\prime)}{s^\prime-s}}+\frac{1}{\pi}\int_{-\infty}^{m_-^2-t}{ds^\prime \frac{\mathrm{Im}F(s^\prime,t,u^\prime)}{s^\prime-s}}.
\end{equation}
Therefore, for real values of $s$  {\it dispersion relations provide the real part of the amplitude from its imaginary part}.

When the amplitude $F(s,t,u)$ does not tend to zero 
fast enough at $\infty$, 
the circular contribution of the contour
will not vanish.  In such cases, we can try applying the theorem
to the ``subtracted" function $\big(F(s,t,u)-F(s_0,t,u)\big)/(s-s_0)$, to write
\begin{equation}\label{Tcauchy-1s}
F(s,t,u)-F(s_0,t,u)=\frac{1}{2\pi i}(s-s_0)\oint{ds^\prime \frac{F(s^\prime,t,u^\prime)}{(s^\prime-s)(s^\prime-s_0)}},
\end{equation}
and for the circular part of $C$ to vanish it is enough to demand $F(s,t,u)/s$
to tend to zero at $\infty$ faster than $1/s$. This yields a so-called ``once subtracted''
dispersion relation, which now reads:
\begin{equation}\label{Tcauchy-onesubstracted}
F(s,t,u)=F(s_0,t,u)+\frac{s-s_0}{\pi}\int_{m_+^2}^{\infty}{ds^\prime \frac{F(s^\prime,t,u^\prime)}{(s^\prime-s)(s^\prime-s_0)}}+\frac{s-s_0}{\pi}\int_{-\infty}^{m_-^2-t}{ds^\prime \frac{F(s^\prime,t,u^\prime)}{(s^\prime-s)(s^\prime-s_0)}}.
\end{equation}
The price to pay is that the amplitude at the subtraction point $s_0$ is now required as input. If that is still not enough to ensure the vanishing of the circular part of $C$, one can make another subtraction, typically at the same point, finding
\begin{eqnarray}\label{Tcauchy-two-substracted}
F(s,t,u)&=&F(s_0,t,u)+(s-s_0)\left(\frac{\partial}{\partial s}F(s,t,u)\right)_{s=s_0}+\frac{1}{2\pi i}(s-s_0)^2\oint{ds^\prime \frac{F(s^\prime,t,u^\prime)}{(s^\prime-s)(s^\prime-s_0)^2}}. \nonumber
\end{eqnarray}
This is now called  a ``twice subtracted'' dispersion relation, which requires the knowledge of two subtraction constants. 
In principle, one can calculate dispersion relations with an arbitrary number of subtractions. However, due to the Froissart bound \cite{Froissart:1961ux}, i.e. $\sigma_{tot}(s)<c (\log s)^2$, two subtractions
are enough to ensure convergence. Nevertheless, more subtractions than strictly needed can be made in order to weigh some regions of the integrand more than others. Actually, in this review, we will use the same dispersion relation with different numbers of subtractions for that purpose.

\begin{figure}[ht]
\centering
\includegraphics[scale=0.65]{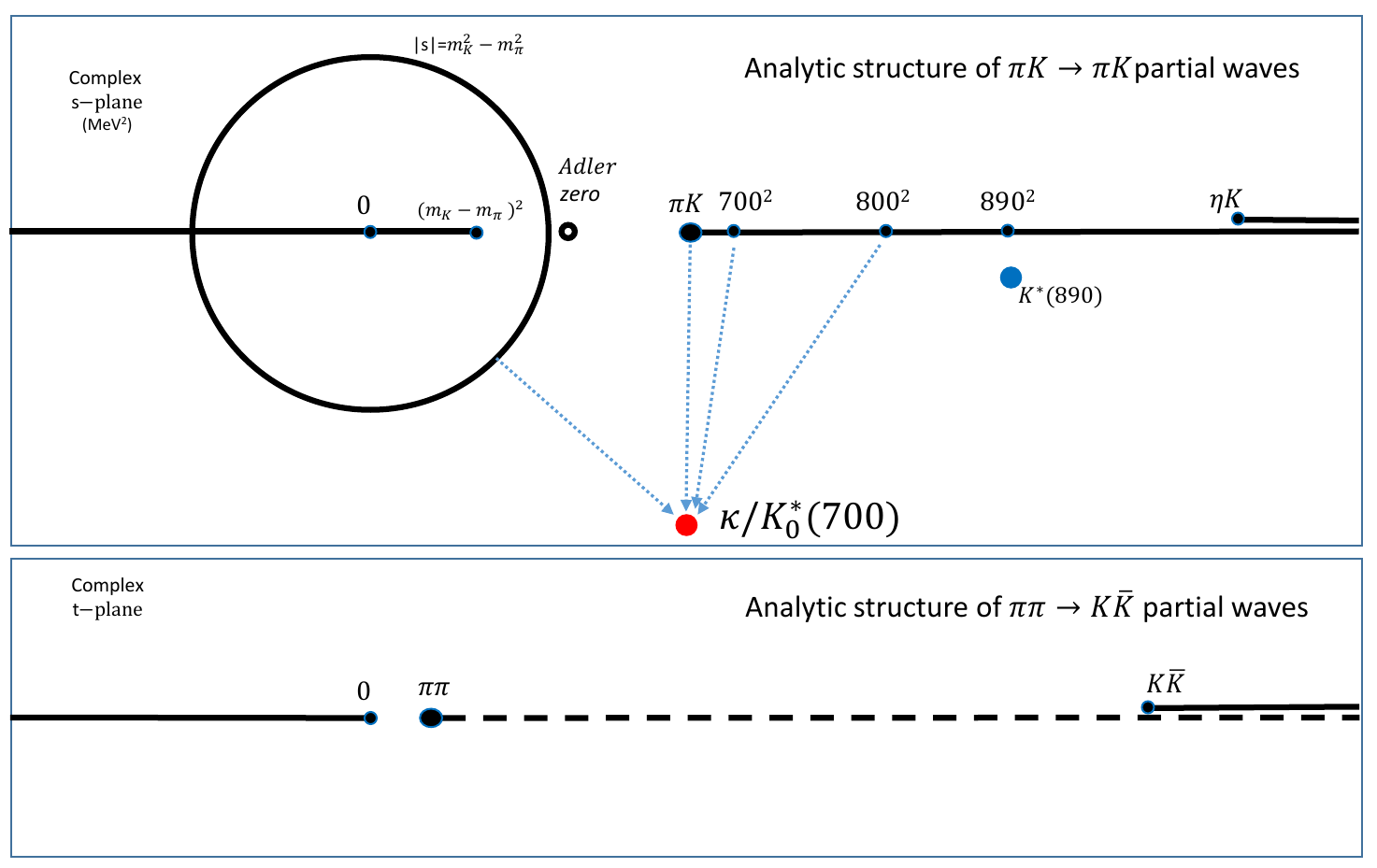}
 \caption{\rm \label{fig:anstrucpw} 
Analytic structure of partial-wave amplitudes in the complex plane for \pik scattering ($f^I_\ell(s)$, top) and \pipikk scattering ($g^I_\ell(t)$, bottom). Note that both present ``right cuts'' from each threshold to infinity. We only show two-particle thresholds.
In the \pipikk case, the real axis segment from the two-pion threshold to the $K\bar K$ threshold is below the physical region, but the imaginary part of the amplitudes is non-zero. This is called the pseudo-physical region.
In addition, both types of partial waves have ``left cuts'' for negative values of their center-of-mass Mandelstam variable. However, \pik partial waves also have  a circular cut, centered at $s=0$ with radius $m_K^2-m_\pi^2$, and an additional piece of left cut on the real axis from $s=0$ to $(m_K-m_\pi)^2$. We also show the position of several dynamical features of \pik scattering, like the Adler zero required by spontaneous chiral symmetry breaking (we show that of the $f^{1/2}_0$ wave), and the position of the \kap and $K^*(892)$ poles in the second Riemann sheet of the $f^{1/2}_0$ and $f^{1/2}_1$, respectively. Note that while the latter is much closer to its nominal mass in the real axis, the \kap is as close to its nominal mass region in the real axis as it is to the \pik threshold, or the left and circular cuts.
Adler zeros \cite{Adler:1964um} appear in scalar partial waves only as a consequence of spontaneous chiral symmetry breaking. To leading order in ChPT, they are located in the real axis slightly above the circular cut, roughly as depicted in the upper panel.
}
\end{figure}

The previous discussion was made in terms of a fixed-$t$ dispersion relation, but for partial waves one can proceed similarly once the analytic structure is known. Thus we show in Fig.~\ref{fig:anstrucpw}  the analytic structure in the complex plane for the \pik scattering $f^I_\ell(s)$ partial waves (top panel) and the $g^I_\ell(t)$  \pipikk partial waves (bottom panel). Due to the partial-wave integration and the two different masses involved in the process, the \pik partial waves have an additional circular cut. It is also important to notice that, in the \pipikk case, the right cut extends below the physical $K\bar K$ threshold down to the two-pion threshold, which is the so-called ``pseudo-physical'' or ``unphysical'' region.

The main problem with dispersion relations is to recast all the integrals in terms of the amplitudes in physical regions. For this, crossing symmetry is essential. It is particularly easy to implement for fixed-$t$ amplitude dispersion relations, giving rise to closed and simple expressions. In contrast, it is more cumbersome for partial-wave relations, since for each partial wave in a given channel they might involve the infinite tower of partial waves in the crossed channel. We will derive expressions for all the dispersion relations of interest in the next subsections.

However, let us remark that the previous derivation of a dispersion relation and the derivations in the next subsections are purely formal. For the sake of brevity, we will proceed as if our manipulations on integrals, series, etc... are always well justified. For instance,
we will assume that $\im F(s,t,u)$ is a real function, or that the partial-wave expansions converge. However, these conditions are only met in certain regions of the $(s,t)$ Mandelstam plane. The applicability domain of all dispersion relations described next is studied in rigor and detail in \ref{app:Applicability}. 

\subsection{State of the art}
\label{sec:Stateofart}

 As explained above, a dispersion relation can always be written for a given amplitude  in a certain region of the Mandelstam plane. These integral equations enforce not only analyticity, but also crossing, and thus, when constrained with unitary partial waves, produce as a result a system of scattering amplitudes that fulfill all three first principles. On top of that, it has been proven for two-hadron elastic scattering that dispersive solutions are unique under certain conditions~\cite{Gasser:1999hz}. Considering, as explained in section~\ref{sec:Data}, that experimental data are usually plagued with systematic uncertainties, dispersion relations offer the possibility to constrain the desired amplitudes and their data description, without including any model dependencies. As a result, such dispersive analyses are considered very robust results within the hadron physics community and, in practice, ``model-independent'' studies, up to minor simplifying assumptions like isospin conservation, or that hadronic states are truly asymptotic, etc. For studies of isospin violation in $\pi K$ scattering, we refer the reader to~\cite{Nehme:2001wf,Nehme:2001wa,Kubis:2001bx,Kubis:2001ij}.
 
 Furthermore, one of the most relevant topics in Hadron  Physics is the extraction of resonances. 
 The most interesting resonances nowadays are not those easily identified by nice peaks in cross-sections, but those which are very wide and/or masked by thresholds or other nearby  resonances and dynamical features. An example is the much-debated  lowest-lying scalar-nonet $\sig,f_0(980),a_0(980)$ and $\kap$. We will dedicate the whole section \ref{sec:kappa} to the latter.
 When these complications occur, one has to resort to the rigorous resonance definition in terms of its associated pole in the complex plane, which is a feature that cannot be removed by any other nearby effect. The pole position of these resonances is usually far from the real axis, or surrounded by nearby thresholds, which produce a very unstable analytic extrapolation.
 Actually, different models providing similar-quality descriptions of data may yield rather different poles. Dispersion relations can also overcome this problem, since they are derived from Cauchy's theorem, thus supplying the correct analytic continuation to the complex plane, which is also  very  stable. Last but not least, apart from resonance poles,  several interesting parameters or quantities can only be determined far from the physical region, like the renowned $\pi N$ $\sigma$ term, Adler zeros, etc... which can only be extracted with high precision if a stable analytic extrapolation is performed.

The formulation of dispersion relations using crossing constraints and coupling partial waves from crossed channels appeared in the '70s. 
These lead to the so-called Roy equations \cite{Roy:1971tc}  for $\pi\pi$ scattering and Steiner-equations \cite{Steiner:1971ms,Hite:1973pm} for $\pi N$ scattering. Other phenomenological works along these lines
where developed during that decade on $\pi\pi$ scattering  \cite{Pennington:1973hs,Basdevant:1972gk,Basdevant:1972uu,Basdevant:1972uv, Basdevant:1973ru,Froggatt:1977hu} together with the developments of 
sum-rules for the determination of low-energy $\pi\pi$ parameters, also  using analyticity and crossing \cite{Palou:1974ma}.
These techniques were soon applied to \pik and \pipikk scattering in
\cite{HedegaardJensen:1974ta,Johannesson:1974ma,Palou:1975uu,Bonnier:1975ne,Johannesson:1976qp,Johannesson:1977zu} and, although these works were mostly focused on the determination of low-energy parameters, they laid the ground for the formalism we will use here. For a nice description of dispersive \pik and \pipikk analyses at the end of that decade, we refer the reader once again to the excellent review in \cite{Lang:1978fk}.

With the advent of QCD, the interest in these techniques faded partially, but a renewed interest arose around the '90s, which continues to our days, due, first, to the appearance of ChPT and its need for very precise and rigorous low-energy observables to which it could be compared, and, second, to the efforts to determine the existence and properties of the much-debated lightest scalar resonances. This has given rise to extensive studies of $\pi\pi$ scattering  with Roy or similar equations  \cite{Ananthanarayan:2000ht,Colangelo:2001df,DescotesGenon:2001tn,Kaminski:2002pe,Kaminski:2008fu,GarciaMartin:2011cn,Kaminski:2011vj,Caprini:2011ky,Nebreda:2012ve,Bydzovsky:2016vdx,Albaladejo:2018gif}, so that very high-precision amplitude parameterizations have been built  over the years, with various dispersive solutions converging to very similar
 final results. Furthermore, these techniques also produced a very precise extraction of the $\sig$ and $f_0(980)$ resonances~\cite{Caprini:2005zr,GarciaMartin:2011jx,Moussallam:2011zg},
 which lead to a major revision of their status in the 2012 Review of Particle Physics, reducing the  quoted uncertainty for the former by a factor of 5 and even changing its name to the present one, i.e., $f_0(500)$.
 On the $\pi N$ front, similar dispersive analyses \cite{Ditsche:2012fv, Hoferichter:2015hva} exist, that will be of relevance for our $\pik$ system. Moreover, besides determining the amplitudes on the real axis, all these dispersive techniques  allow to extract information on different observables like the $\pi N$ $\sigma$-term~\cite{Gasser:1988jt,Gasser:1990ce,Gasser:1990ap,Hoferichter:2015dsa,Hoferichter:2016ocj,RuizdeElvira:2017stg}, the matching to ChPT~\cite{Hoferichter:2015tha,Siemens:2016jwj}
  or several other relevant quantities associated with the nucleon~\cite{Hoferichter:2016duk, Hoferichter:2018zwu}.
  We will also follow this path here for $\pi K$ when discussing applications in section \ref{sec:applications}. 
Finally, we would like to point out that, given the success of these techniques, they have also been applied recently to other processes like:
 $e^+ e^- \to \pi^+ \pi^-$ \cite{Colangelo:2018mtw} and $\gamma^{(*)} \gamma^{(*)} \to \pi \pi$ \cite{GarciaMartin:2010cw,Hoferichter:2011wk,Moussallam:2013una, Danilkin:2018qfn, Hoferichter:2019nlq}. 
 
Concerning $\pik$  and $\pipikk$ scattering, several previous dispersive studies must be recalled before addressing the details of our calculations. Seminal works using fixed-$t$ \cite{Nielsen:1973au} or hyperbolic dispersion relations can be found in \cite{HedegaardJensen:1974ta,Johannesson:1974ma,Bonnier:1975ne,Johannesson:1976qp,Johannesson:1977zu}, although they mostly focused on the determination of scattering lengths and very low-energy or even sub-threshold regions. A contemporary study on $\pipikk$ scattering lengths \footnote{For this inelastic process, scattering lengths are defined as $\tilde a^I_\ell=\lim_{q_K\rightarrow 0^+} g^I_\ell (s)$.} using an alternative dispersive technique is also found in \cite{Palou:1975uu}. Unfortunately, the data at that time was not so good and the precision attained by all those works is comparatively poorer than more recent developments.
In particular, more modern works used dispersion theory to obtain sum rules for threshold parameters to compare with SU(3) ChPT ~\cite{Ananthanarayan:2000cp,Ananthanarayan:2001uy}. 
 Finally, the most rigorous and robust dispersive analysis of $\pik$ scattering was carried out in a series of works by the Paris group \cite{Buettiker:2003pp,DescotesGenon:2006uk}.
 In the first work \cite{Buettiker:2003pp} they {\it solved} the partial-wave projected fixed-$t$  dispersion relations, to obtain a prediction of low-energy $S$ and $P$ partial waves, below $\sim$1 GeV, using as input higher partial waves, data above 1 GeV and phenomenological fits to $\pipikk$ scattering. The results describe well the data on the $S$-wave and just qualitatively the data on the $P$-wave, particularly around the $K_0^*(892)$ resonance, see Fig.\ref{fig:BDGM}.
 Later on, they used their solutions inside partial-wave projected hyperbolic dispersion relations to determine with high accuracy the pole position of the light scalar \kap resonance~\cite{DescotesGenon:2006uk}. We definitely consider this work very robust.
 However, note that  the dispersion relations used for this pole extraction were not the same ones solved in the first work. Moreover, the \pipikk input was kept fixed from simple fits to data, and not all the possible sources of uncertainty were considered. Despite the existence of this rigorous determination of  the \kap pole, it still ``Needs Confirmation'' in the Review of Particle Physics. 
 
 Thus, even more recently, we have pursued an alternative ``data-driven'' dispersive program, whose completion we finally present in this report. We have aimed at using several different dispersion relations, including  hyperbolic ones, to constrain data parameterizations up to as high as possible energies, simultaneously for $\pik$ and \pipikk and to use the resulting amplitudes also to determine the \kap pole.  In particular, we first obtained \cite{Pelaez:2016tgi} a precise description of $\pik$ scattering consistent with forward dispersion relations (FDRs) up to $\sim$1.6 GeV. Next, keeping those $\pi K$ amplitudes fixed, we obtained \cite{Pelaez:2018qny} a precise description of $\pipikk$ scattering up to 1.47 GeV, consistent with hyperbolic dispersion relations. In all these cases we paid particular attention to uncertainties, both statistical and systematic. Our final result, that we will present below in full detail, is to use all dispersion relations together to constrain  simultaneously the data description of both \pik and \pipikk. In \cite{Pelaez:2020uiw}, we already advanced the result of this analysis for the \kap resonance, quite consistent with the result in \cite{DescotesGenon:2006uk}. 
 
 We would not like to finish this state-of-the-art section without mentioning that there are several other applications of dispersion relations related to \pik interactions~\cite{Zheng:2003rw,Albaladejo:2016mad,Danilkin:2017lyn,Ropertz:2018stk,Danilkin:2018qfn,Danilkin:2019opj,Deineka:2019bey}, in which some simplifying approximations are made, or cutoffs are used, etc...,  which lie outside the scope of this manuscript. These include ``unitarized'' ChPT for $\pik$ or $\pipikk$  \cite{Dobado:1992ha,Dobado:1996ps,Oller:1997ng,Oller:1997ti,Oller:1998hw,Oller:1998zr,Jamin:2000wn,GomezNicola:2001as,Pelaez:2004xp,GomezNicola:2007qj, Pelaez:2015qba,Yao:2020bxx} and in general chiral unitary approaches. In the elastic case they can be justified from a dispersion relation for the inverse of the partial wave (since its imaginary part is known exactly from elastic unitarity), or in the coupled-channel case can be derived from Bethe-Salpeter equations \cite{Nieves:1999bx} or the
 so-called N/D method \cite{Chew:1960iv}
 (see also  \cite{Oller:2019rej} for a recent review). The latter assumes that, similarly to the elastic case, the right cuts, which are common to all coupled partial waves, can be obtained from unitarity in coupled channels  together with a dispersion relation for the denominator. Such a dispersion relation is coupled to another one for the numerator that carries the other cuts (left cut, circular, etc...). 
 In both the elastic and inelastic cases, subtractions are needed, whose values can be obtained from Chiral Perturbation Theory (ChPT) up to a specific order. These methods are very successful phenomenologically, although, due to the approximations they use, not so suited for the kind of precision we are after in this report. For this  reason, their success and  details lie beyond the scope of this work.
For a thorough discussion of the advantages and disadvantages of these unitarization methods versus fully dispersive approaches, we refer the reader to the recent extensive review in \cite{Pelaez:2021dak}.

 Thus, in the next sections we will summarize the dispersion relations we will implement for our full dispersive analysis here. These are a total of {\it sixteen} dispersion relations that comprise:
 \begin{itemize}
     \item {\it Two} forward dispersion relations (FDRs), useful to constrain the $\pik$ scattering amplitudes up to roughly 1.7 GeV. Contrary to partial-wave dispersion relations (PWDR),  FDRs constrain the full partial-wave series instead of individual partial waves. These dispersion relations are the ones reaching the highest in applicability and, in addition, they help to obtain an improved description of the high-energy asymptotic region described in section~\ref{sec:ufdregge}. One obvious advantage is that they do not need \pipikk input. We already made use of them in \cite{Pelaez:2016tgi}. Unfortunately, they alone cannot reach the second Riemann sheet, in search of resonances.
     
     \item {\it Four} Hyperbolic partial-wave projected dispersion relations (HPWDR) for the crossed channel $\pipikk$. For our purposes, it is enough to study dispersively the lowest three partial waves in this channel, for which data exist. Note we will make use of an extra dispersion relation for $g^1_1$. The reason is that we found inconsistencies in the continuation to the pseudo-physical region between subtracted and unsubtracted dispersion relations when using unconstrained fits as input. This region of the $P$-wave is relevant for the determination of the \kap resonance and we want to make sure both solutions are  consistent in the end. We already enlarged their applicability region to 1.47 GeV and made use of them in \cite{Pelaez:2018qny}, but keeping fixed the $\pi K$ input.
     
     \item {\it Four} Fixed-$t$ partial-wave projected dispersion relations (FTPWDR) for the four $\pik$ partial waves $f^\pm_{0,1}$. These are mostly dominated by their own inputs, and thus offer a simpler dispersive constraint than the HPWDR. Unfortunately, as explained in detail in~\ref{app:Applicability}, their applicability region  ends slightly above 1 GeV in the real axis and in the complex plane does not reach the \kap pole. These were solved in the \cite{Buettiker:2003pp} analysis, keeping the \pipikk input fixed, whose influence is small.
     
     \item {\it Six} Hyperbolic partial-wave projected  dispersion relations (HPWDR) for the four $\pik$ partial waves $f^\pm_{0,1}$. Again as in the case of the crossed channel, extra dispersion relations will be used to obtain consistency between the unsubtracted and once-subtracted anti-symmetric dispersion relations. These produce a more intricate realization of the crossing between the two channels and this time the \pipikk crossed channel provides a quite significant contribution, which therefore really links together the two channels. Although, as the FTPWDR, they reach only up to $\sim$ 1 GeV in the real axis, we need these HPWDR because the \kap pole lies inside their applicability region.
     
 \end{itemize}
 
As detailed in section~\ref{sec:CFD} these constraints are enough to obtain a very consistent result not only concerning dispersion relations, but also with the existing \pik and \pipikk data within uncertainties.

 Let us now describe in detail the derivation and subtleties of each one of these dispersion relations we have just enumerated.

\subsection{Fixed-t dispersion relations}
\label{sec:fixedtDR}

In this review, we will show three relevant uses of fixed-$t$ dispersion relations for $F^\pm(s)$:
First, they will be an intermediate step for
the derivation of more elaborated dispersion relations for partial waves \cite{Ader:1974yk,Steiner:1971ms,Buettiker:2003pp}. 
Second,  they will provide sum rules for low-energy parameters \cite{Palou:1974ma,Palou:1975uu,Ananthanarayan:2001uy}.
Third, FDRs, which correspond to the fixed $t=0$ case,  will be used as checks and constraints on amplitudes \cite{Pelaez:2016tgi}.
FDRs have the advantage that they converge for any value of $s$ (see \ref{app:Applicability}).
In contrast, partial-wave dispersion relations obtained from fixed-$t$ dispersion relations have a limited applicability range,  for \pik scattering amplitudes, up to roughly $1.05 \,\gev$. 
Finally, one might wonder why we consider fixed-$t$ for $\pik$ amplitudes but not  fixed-$u$ or fixed-$s$ dispersion relations for $\pi\pi\rightarrow \bar KK$ amplitudes. The reason is that, as shown in \ref{app:Applicability}, their applicability region does not reach the physical region where data exist.

According to Eqs.~\eqref{Fpm-GI}, $F^+$ and $F^{-}$
correspond, by crossing, to the exchange of isospin 0 or 1 in the $t$-channel, respectively. This means that, at high energies, $F^+$ is dominated 
by the $t$-channel exchanges
of the Pomeron and $P'$ trajectories, with no $\rho$ trajectory contribution, 
whereas the opposite occurs for $F^-$.
If one looks only at either the right or the left-hand-side integral, one would then need 
two subtractions to ensure the convergence of the Pomeron contribution and one for that of the $\rho$ trajectory. Thus, when used as intermediate steps for the derivation of other dispersion relations, fixed-$t$ dispersion relations for $F^+$ are customarily written with two subtractions and those for $F^-$ at least with one.
However, the $s\leftrightarrow u$ behavior of the $F^\pm$ amplitudes implies that the leading Regge contributions in the right-hand and left-hand cuts cancel against each other. As a consequence, it is enough to have one or no subtraction to ensure the convergence of the $F^+$ 
and $F^-$ fixed-$t$ dispersion relations, respectively. These minimally subtracted fixed-$t$ dispersion relations
have been recently used for $\pi\pi$ scattering FDR in \cite{GarciaMartin:2011cn,Pelaez:2004vs,Kaminski:2006qe,Kaminski:2006yv} and for $\pi K$ in \cite{Pelaez:2016tgi}. Generically, fewer subtractions 
are convenient to avoid the propagation of the uncertainties in the subtraction constants
becoming too large in the resonance region, whereas more subtractions are useful when concentrating on the threshold region. 

Thus, using crossing symmetry to write the left-hand cut contribution as an integral over the right one, the unsubtracted fixed-$t$ dispersion relation for $F^-$ reads:
\begin{equation}
 F^-(s,t)=
\frac{1}{\pi}\int^{\infty}_{m_+^2} ds'\im F^-(s',t)\left[\frac{1}{s'-s}-\frac{1}{s'-u}\right].
\label{eq:F-fixedt}
\end{equation}

One of the main advantages of using this non-subtracted relation
is that it provides a sum rule for the scattering length $a^-_0$, which therefore is not input. Another relevant advantage is that it does not depend on the crossed channel $\pi \pi \rightarrow K \bar K$, thus avoiding an additional source of uncertainty. 

In contrast, for the fixed-$t$ dispersion relation of $F^+$, we need two subtractions. For the moment and later convenience, we choose the subtraction point at $s=0$.
Once again we can use crossing symmetry to rewrite the left cut in terms of an integral over the right one.
It can be shown \cite{Roy:1971tc,Johannesson:1976qp} that it can always be recast into the following simple form:
\begin{align}
 F^+(s,t)=c^+(t)
&+\frac{1}{\pi}\int^{\infty}_{m_+^2} ds' \frac{\im F^+(s',t)}{s'^2} \left[\frac{s^2}{s'-s}+\frac{u^2}{s'-u}\right], 
\label{eq:F+t}
\end{align}
where  $c^+(t)$ is a subtraction constant once each value of $t$ is fixed. In section \ref{sec:HDR} below, we will show how this subtraction constant can be determined using a hyperbolic dispersion relation for $F^+$ and the resulting fixed-$t$ dispersion relation, Eq.~\eqref{eq:F+fixedt}, will therefore have some dependence on \pipikk.

Fixed-$t$ dispersion relations will be used as intermediate steps for partial wave dispersion relations in later sections where this subtraction constant will be rewritten using a hyperbolic dispersion relation. At that point, we will introduce a mild dependence on the crossed channel.

Furthermore, we will also use the particular case when we fix $t=0$, i.e., FDRs, that we describe in detail next.

\subsection{Forward dispersion relations for $\pik$}
\label{sec:FDR}

The $t=0$ case is useful because
forward scattering is related  by the optical theorem to the total cross-section, for which 
there is experimental information at high energies for many hadron scattering processes.
Moreover, this is the only fixed value of $t$ for which the integrands in the dispersion relation are given directly in terms of the imaginary part of a physical amplitude.
FDRs are applicable at any energy, in contrast to Roy-like equations which, in practice, 
have a limited applicability energy range of $O(1 \gev)$ due to the projection in partial waves (see  \ref{app:Applicability}).
Actually, we recently used FDR to check different sets of $\pi K$ data and to obtain a set of constrained data parameterizations of $\pi K$ that satisfy well these FDRs up to 1.6 GeV
\cite{Pelaez:2016tgi}. In this report, we go one step forward and we will impose them together with Roy-Steiner-like equations, the latter both for $\pi K$ and $\pi\pi\rightarrow \bar KK$ channels.

The unsubtracted FDR for $F^-$ can be obtained from Eq.~\eqref{eq:F-fixedt} just by setting $t=0$.
The only subtlety is that, since we will use the FDRs as checks or constraints on the amplitudes for physical values of $s$,  we need the $s$ variable to lie on the real positive axis above threshold. Therefore, we have to use the principal value as in 
Eqs.~\eqref{eq:PVdistribution} and \eqref{Tcauchy-real}
\begin{equation}
\re F^-(s)=
\frac{2(s-\Sigma)}{\pi}PV\!\!\int^{\infty}_{m_+^2}\!\!\!\! ds'
\frac{\im F^-(s')}{(s'-s)(s'+s-2\Sigma)}.
\label{eq:FDRTan}
\end{equation}

For the $F^+$ FDR, however, it is convenient to make the subtraction at threshold, instead of $s=0$, so that the subtraction constant can be recast later in terms of $\pi K$ scattering lengths. For this reason
the FDR is not obtained directly by setting $t=0$ in Eq.~\eqref{eq:F+t}, but can be recast as:
\begin{align}
\re F^+(s)=F^+(m_+^2)
&+\frac{(s-m_+^2)}{\pi}PV\!\int^{\infty}_{m_+^2}\!ds'\left[\frac{\im F^+(s')}{(s'-s)(s'-m_+^2)} -\frac{\im F^+(s')}{(s'+s-2\Sigma)(s'+m_+^2-2\Sigma)}\right].
\label{eq:FDRTsym}
\end{align}

Their main drawback is that when looking for resonances, their associated poles appear in the second sheet of partial waves with the same quantum numbers of each resonance. FDRs do not deal with partial waves, and even worse, just by themselves, they do not provide access to the second Riemann sheet. These two drawbacks can be overcome by the use of partial-wave dispersion relations, although then the advantages of the FDRs are lost.

\subsection{Hyperbolic dispersion relations}
\label{sec:HDR}

In this kind of dispersion relations, abbreviated as HDR, the $t$ variable is not fixed to the same value for all $s$,
but by imposing that $u$ and $s$ should lie in a hyperbola $(s-a)(u-a)=b$. In the literature, it has been usual to set $a=0$, but we have recently obtained in \cite{Pelaez:2018qny} HDRs for arbitrary $a$. As shown in \ref{app:Applicability}, the advantage of such generalization is that 
the $a$ value can then be chosen to modify the HDR applicability
domain, as well as the weight of different parts of the integral.
In particular, in \cite{Pelaez:2018qny} we chose $a$ to maximize 
the applicability of HDR for either $\pi K$ or \pipikk  in their respective physical regions. Since in this report we are also interested in the precise determination of the $\kap$ resonance, in \ref{app:Applicability} we have determined the $a$ value that maximizes the partial-wave HDR applicability reach in the $\pi K$ physical region while enclosing the $\kap$ pole region in the complex plane.

In this section we will briefly review our results in \cite{Pelaez:2018qny}. In addition, we will also provide the expression for the subtraction constant of the fixed-$t$ $F^+$ dispersion relation in Eq.~\eqref{eq:F+t} and the once-subtracted expression for the $F^-$ HDR.

Thus, in what follows  $s_b$ and $u_b$ will be the values of $s$ and $u$ that lie in the hyperbola $(s-a)(u-a)=b$ for a given value of $t$. Together with the condition $s+t+u=2\Sigma$, these values can be recast as:
\begin{eqnarray}
&&s_b\equiv s_b(t)= \frac{1}{2}\left(2\Sigma -t+\sqrt{(t+2a-2\Sigma)^2-4b} \right), \nonumber \\
&&u_b\equiv u_b(t)=\frac{1}{2}\left(2\Sigma -t-\sqrt{(t+2a-2\Sigma)^2-4b} \right).
\label{eq:sboft}
\end{eqnarray}

Then we can write the following once-subtracted hyperbolic dispersion relation for $F^+$
\begin{align}
F^+(t,b,a)\!=\!h^+(b,a)\!+\!\frac{t}{\pi}\!\int^{\infty}_{4m_{\pi}^2}\frac{\im G^0(t',s'_b)}{\sqrt{6}\,t'(t'-t)}dt'
\!+\!\frac{1}{\pi}\!\int^{\infty}_{m_+^2}\!ds'
\frac{\im F^+(s',t'_{b})}{s'}\Big(\frac{s}{s'-s}+\frac{u}{s'-u}\Big),
\label{eq:preshdr}
\end{align}
where
\begin{equation}
s'_b\equiv s_b(t'), \quad u'_b \equiv u_b(t'),\quad
t'_b=2\Sigma-s'-\frac{b}{s'-a}-a.
\end{equation}

Note that one subtraction is enough in Eq.~\eqref{eq:preshdr} since the first term converges because at high energies its leading Regge exchange is the $K^*$-trajectory, whereas the second term is backward scattering.

Then, similarly to what was done in \cite{Ananthanarayan:2001uy} for $a=0$, by combining the fixed-$t$ dispersion relation in Eq.~\eqref{eq:F+t} with Eq.~\eqref{eq:preshdr} right above at $t=0$ and $b_0=a^2-2\Sigma a +\Delta^2$, we can determine both subtraction terms $c^+(t)$ and $h^+(b,a)$. Thus we can rewrite the HDR for $F^+$ as follows:
\begin{align}
F^+(s_b,t)=&8\pi m_+ a_0^+ +\frac{t}{\pi}\int^{\infty}_{4m_{\pi}^2}\frac{\im G^0(t',s'_b)}{\sqrt{6}\,t'(t'-t)}dt' 
\nonumber\\
&
+\frac{1}{\pi}\int^{\infty}_{m_+^2}ds'\frac{\im F^+(s',t'_b)}{s'}\left[h(s',t,b,a)-h(s',0,b,a) \right]
\nonumber \\
&
+\frac{1}{\pi}\int^{\infty}_{m_+^2}ds'\frac{\im F^+(s',0)}{s'^2} \left[ g(s',b,a)-g(s',\Delta^2,0) \right], 
\label{eq:shdr}
\end{align}
where we have defined
\begin{align}
&h(s',t,b,a)=\frac{s'(2\Sigma-t)-2[b-a^2+(2\Sigma-t)a]}{s'^2-s'(2\Sigma-t)+[b-a^2+(2\Sigma-t)a]},\nonumber\\
& g(s',b,a)=\frac{s'(2\Sigma)^2-2[b-a^2+2\Sigma a](s'+\Sigma)}{s'^2-s'2\Sigma+[b-a^2+2\Sigma a]}.
\end{align}

Moreover, since we have also determined $c^+(t)$, we can now
rewrite, as promised in the previous section, the fixed-$t$ dispersion relation in Eq.~\eqref{eq:F+t} as:

\begin{align}
 F^+(s,t)&=8\pi m_+ a_0^+
+\frac{1}{\pi}\int^{\infty}_{m_+^2} ds' \frac{\im F^+(s',t)}{s'^2} \left[\left(\frac{s^2}{s'-s}+\frac{u^2}{s'-u}\right)-k(s',t) \right] \nonumber \\
&+\!\frac{t}{\pi}\!\!\int^{\infty}_{4m_{\pi}^2}\frac{\im G^0(t',s'_{b_0})}{\sqrt{6}\,t'(t'-t)}dt'\! 
+\!\frac{1}{\pi}\!\!\int^{\infty}_{m_+^2} ds' \frac{\im F^+(s',t_{\Delta^2})}{s'}\left[ h(s',t,\Delta^2,0)-h(s',0,\Delta^2,0)\right],
\label{eq:F+fixedt}
\end{align}
where we have now defined
\begin{equation}
 k(s',t)=\frac{1}{s'^2}\frac{s'(2\Sigma-t)^2-2\Delta^2s'-\Delta^2(2\Sigma-t)}{s'^2-s'(2\Sigma-t)+\Delta^2}.
\end{equation}

For the  $F^-$ case, in this report
we will use both the unsubtracted and once-subtracted hyperbolic dispersion relations. This is just because, as we will show later, we have found that the unconstrained data fits do not satisfy well these two dispersion relations simultaneously. 
For convenience one has to write the HDR for the $s\leftrightarrow u$ symmetric combination $F^-/(s-u)$.
Thus, the unsubtracted relation reads:
\begin{equation}
\frac{F^-(s_b,t)}{s_b-u_b}=\frac{1}{2\pi}\int^{\infty}_{4m_{\pi}^2}dt'\frac{\im G^1(t',s'_b)}{(t'-t)(s'_b-u'_b)}+\frac{1}{\pi}\int^{\infty}_{m_+^2}ds'\frac{\im F^-(s',t'_b)}{(s'-s_b)(s'-u_b)},
\label{eq:ahdr}
\end{equation}
whereas the once-subtracted reads:
\begin{equation}
\frac{F^-(s_b,t)}{s_b-u_b}\!=\!h^-(b,a)+\frac{t}{2\pi}\!\!\int^{\infty}_{4m_{\pi}^2}\!\!\!\!dt'\frac{\im G^1(t',s'_b)}{t'(t'-t)(s'_b-u'_b)}+\frac{1}{\pi}\!\int^{\infty}_{m_+^2}\!\frac{ds'}{s'}\frac{\im F^-(s',t'_b)}{(s'_b-u'_b)}\!\left[\frac{s}{s'-s}\!+\!\frac{u}{s'-u}\right].
\label{eq:ahdr1}
\end{equation}
For the latter, the subtraction constant
can be determined, as done in \cite{Ananthanarayan:2001uy} for the $a=0$ case and for $F^+$ above. Namely,
by using the once-subtracted fixed-$t$ dispersion relation 
\begin{align}
 F^-(s,t)=c^-(t)(s-u)+
\frac{1}{\pi}\int^{\infty}_{m_+^2} \frac{ds'}{s'^2}\im F^-(s',t)\left[\frac{s^2}{s'-s}-\frac{u^2}{s'-u}\right].
\label{eq:ftmi}
\end{align}

Thus, we can rewrite Eq.~\eqref{eq:ahdr1} as
\begin{align}
\frac{F^-(s_b,t)}{s_b-u_b}=\frac{8\pi m_+ a_0^-}{m_+^2-m_-^2} 
&+\frac{t}{2\pi}\int^{\infty}_{4m_{\pi}^2}dt'\frac{\im G^1(t',s'_b)}{t'(t'-t)(s'_b-u'_b)} \nonumber \\
&+\frac{1}{\pi}\int^{\infty}_{m_+^2}ds'\im F^-(s',t'_b)\left[d(s',t,b,a)-d(s',0,b,a)\right] \nonumber \\
&+\frac{1}{\pi}\int^{\infty}_{m_+^2}ds'\im F^-(s',0)\left[f(s',b,a)-f(s',\Delta^2,0)\right],
\label{eq:mihdr}
\end{align}
with
\begin{align}
&d(s',t,b,a)=\frac{1}{s'^2-s'(2\Sigma-t)+[b-a^2+(2\Sigma-t)a]},\nonumber\\
& f(s',b,a)=\frac{1}{s'^2-s'2\Sigma+[b-a^2+2\Sigma a]}.
\end{align}

We have explicitly checked that setting $a=0$ we recover the
same expressions as in \cite{Johannesson:1974ma, Johannesson:1976qp, Ananthanarayan:2001uy} for the once-subtracted HDR. 
However, as already commented and explained in \ref{app:Applicability}, with our $a\neq0$ HDR above we will be able to choose the $a$ parameter to maximize the applicability region for the $s$ or $t$ channels of the HDR once projected into partial waves, while still reaching the \kap region.

\subsection{Partial-wave dispersion relations. Roy-Steiner equations.}
\label{sec:Roy-Steiner}

Once the $F(s,t,u)$ and $G(t,s,u)$ amplitudes are calculated dispersively, one can just project Eqs.~\eqref{eq:F-fixedt}, \eqref{eq:shdr},\eqref{eq:F+fixedt}, \eqref{eq:ahdr} and \eqref{eq:ahdr1} to 
write partial-wave dispersion relations. Of course the partial-wave projection is different for the $s$ and $t$ channels and we will examine them separately in what follows.

\subsubsection{$s$-channel partial-wave dispersion relations}

We now use Eq.~\eqref{eq:projs} to obtain $\pi K$ scattering partial waves.
First, we project the fixed-$t$ dispersion relations Eqs. \eqref{eq:F-fixedt} and \eqref{eq:F+fixedt}, obtaining:
\begin{align}
f^+_l(s)&=\frac{m_+a^+_0}{2}+\frac{1}{\pi}\sum_\ell \int^{\infty}_{m_+^2}ds' L^+_{l, \ell}(s,s') \im f^+_\ell(s')
+\frac{1}{\pi}\sum_{\ell\geq0}\int^{\infty}_{4m_{\pi}^2} dt' L^0_{l, 2\ell}(s,t') \im g^0_{2\ell}(t'), \nonumber \\
f^-_l(s)&=\frac{1}{\pi}\sum_\ell \int^{\infty}_{m_+^2}ds' L^-_{l, \ell}(s,s') \im f^-_\ell(s'),
\label{eq:sftpwdr}
\end{align}
where the kernels $L^{\pm}_{l, \ell}(s,s'),L^I_{l, 2\ell}(s,t')$ were given in
\cite{Johannesson:1976qp} and can be found here in~\ref{app:Kernels}.
Beware that when fixing $t$, the $z_s$ variable changes within the range  $[-\lambda(s)/s,0]$. This makes the $L$ kernels relatively simple. 
Note also that there are no subtractions for $F^- (s,t)$.

In \ref{app:Applicability} we show that the applicability of these partial-wave relations obtained from fixed-$t$ dispersion relations  reaches $\sqrt{s_{max}}\simeq 1.05\,$
GeV in the real axis. Unfortunately, they do not reach the \kap region in the complex plane, which is one of the reasons why we have to resort to HDR.

In contrast, the projection over an HDR is more complicated, since now $t_b=2\Sigma-s-b/(s-a)-a$, and $b$ is determined implicitly by $b=(s-a)(2\Sigma-s-t-a)$. In addition, we will now find inside the
integrands the crossed-channel partial waves $g^I_\ell(t)$, defined in Eq.~\eqref{eq:gpw}.

Then, projecting into $s$-wave partial waves the HDR for $F^+$ in Eq.~\eqref{eq:shdr}, which has one subtraction, we find:
\begin{align}
f^+_l(s)&=\frac{m_+a^+_0}{2}
+\frac{1}{\pi}\sum_{\ell\geq0}\int^{\infty}_{4m_{\pi}^2}dt' K^0_{l, 2\ell}(s,t') \im g^0_{2\ell}(t')+\frac{1}{\pi}\sum_\ell  \int^{\infty}_{m_+^2}ds' K^+_{l, \ell}(s,s') \im f^+_\ell(s'),
\label{eq:shdrfm}
\end{align}
where the kernels $K^0_{l, 2\ell}(s,t')$ and $K^+_{l, \ell}(s,s')$ were already obtained in \cite{Pelaez:2018qny} and
are collected here in \ref{app:Kernels} for completeness.

Let us recall that for the antisymmetric amplitude $F^-$ 
we are considering here both the unsubtracted HDR, Eq.~\eqref{eq:ahdr}, and the once-subtracted HDR, Eq.~\eqref{eq:ahdr1}.  When projected into $s$-channel partial waves, they read, respectively:
\begin{align}
f^-_{l}(s)&=\frac{1}{\pi}\sum_\ell \int^{\infty}_{m_+^2}ds' K^-_{l, \ell}(s,s') \im f^-_\ell(s') 
+\frac{1}{\pi}\sum_{\ell\geq1}\int^{\infty}_{4m_{\pi}^2}dt' K^1_{l, 2\ell-1}(s,t') \im g^1_{2\ell-1}(t'), \nonumber \\
f^-_{l}(s)&=\delta_{l,0}\frac{m_+ a^-_0}{2}\frac{3s^2-2s\Sigma-\Delta^2}{8 s m_{\pi}m_K}+\delta_{l,1}\frac{m_+ a^-_0}{2}\frac{m_{\pi}^4+(m_K^2-s)^2-2m_{\pi}^2(m_K^2+s)}{24 s m_{\pi} m_K} \nonumber  \\
&+\frac{1}{\pi}\sum_\ell \int^{\infty}_{m_+^2}ds' \hat K^-_{l, \ell}(s,s') \im f^-_\ell(s') 
+\frac{1}{\pi}\sum_{\ell\geq1}\int^{\infty}_{4m_{\pi}^2}dt' \hat K^1_{l, 2\ell-1}(s,t') \im g^1_{2\ell-1}(t'),
\label{eq:shdrfmi}
\end{align}
where we have used a hat to denote the once-subtracted $\hat K^-_{l, \ell}(s,s'),\hat K^1_{l, 2\ell-1}(s,t')$ kernels, which are provided in \ref{app:Kernels} for the $a\neq0$ case. For completeness, we also provide there the $K^-_{l, \ell}(s,s'), K^1_{l, 2\ell-1}(s,t')$ unsubtracted kernels that were obtained in \cite{Pelaez:2018qny} for the $a\neq0$ case.

In \ref{app:Applicability} we show how we can now choose the $a$ parameter to maximize the applicability reach of partial-wave HDR. For instance, the applicability range in the real axis is maximized with the choice  $a=-13.9m_{\pi}^2$, reaching $\sqrt{s_{max}}\simeq 0.989\,$GeV. However, this choice leads to an applicability domain that does not cover the \kap region in the complex plane, which was shown in \cite{DescotesGenon:2006uk} to be reachable with the $a=0$ choice. The latter, however, was only applicable up to $\sqrt{s_{max}}\simeq 0.934\,$GeV int the real axis. We have thus adopted an intermediate round value $a=-10m_{\pi}^2$, which yields an applicability domain that encloses the \kap region and extends up to 
$\sqrt{s_{max}}\simeq 0.976\,$GeV in the real axis.
This is only a small improvement in the applicability in the real axis for $\pi K\rightarrow \pi K$ scattering, but the choice of $a$ is much more relevant 
for the $t$-channel case that we study next.

\subsubsection{$t$-channel partial-wave dispersion relations}
For $\pi\pi\rightarrow \bar KK$, we are only 
considering HDR. Now, recalling the $s\leftrightarrow t$ crossing relation in Eq.~\eqref{Fpm-GI} we see that these HDR
are once again those in Eqs.~\eqref{eq:shdr}, \eqref{eq:ahdr} and
\eqref{eq:ahdr1}  we have just used for the $s$-channel,
but this time we will project them into $t$-channel partial waves $g^I_\ell$, using  Eq.~\eqref{eq:gpw}. Of course, since  the HDRs  also need the crossed channel, we will find the $\pi K$ partial waves $f^I_\ell(s)$ inside the integrals. 
Specifically, we get the following once-subtracted partial-wave dispersion relations
for $I=0$ and $\ell=0,2$:
\begin{align}
g^0_0(t)=&\frac{\sqrt{3}}{2}m_+a^+_0+\frac{t}{\pi}\int^{\infty}_{4m_{\pi}^2}\frac{\im g^0_0(t')}{t'(t'-t)}dt' +\frac{t}{\pi}\sum_{\ell\geq2}\int^{\infty}_{4m_{\pi}^2}\frac{dt'}{t'} G^0_{0, 2\ell-2}(t,t') \im g^0_{2\ell-2}(t') \nonumber \\
&+\frac{1}{\pi}\sum_\ell \int^{\infty}_{m_+^2}ds' G^+_{0, \ell}(t,s') \im f^+_\ell(s'),\nonumber \\
g^0_2(t)=&\frac{t}{\pi}\int^{\infty}_{4m_{\pi}^2}\frac{\im g^0_2(t')}{t'(t'-t)}dt' +\frac{t}{\pi}\sum_{\ell\geq 3 }\int^{\infty}_{4m_{\pi}^2}\frac{dt'}{t'} G'^0_{2, 2\ell-2}(t,t') \im g^0_{2\ell-2}(t') \nonumber \\
&+\frac{1}{\pi}\sum_\ell \int^{\infty}_{m_+^2}ds' G'^+_{2, \ell}(t,s') \im f^+_\ell(s').
\label{eq:pwhdr}
\end{align}
The explicit expressions of the $G^0_{\ell \ell'}(t,t'),G^+_{\ell \ell'}(t,s')$ integration kernels are given in \ref{app:Kernels} for completeness, although they were already derived in \cite{Pelaez:2018qny}.

We also write two dispersion relations for $I=1, \ell=1$,  unsubtracted 
or once-subtracted. These read, respectively:
\begin{align}
g^1_{1}(t)=&\frac{1}{\pi}\int^{\infty}_{4m_{\pi}^2}\frac{\im g^1_1(t')}{t'-t}dt' +\frac{1}{\pi}\sum_{\ell\geq2}\int^{\infty}_{4m_{\pi}^2}dt' G^1_{1, 2\ell-1}(t,t') \im g^1_{2\ell-1}(t') \nonumber \\
&+\frac{1}{\pi}\sum_\ell \int^{\infty}_{m_+^2}ds' G^-_{1, \ell}(t,s') \im f^-_\ell(s'), \nonumber \\
g^1_{1}(t)=&\frac{2\sqrt{2}m_+ a^-_0}{3(m_+^2-m_-^2)}+\frac{t}{\pi}\int^{\infty}_{4m_{\pi}^2}\frac{\im g^1_1(t')}{t'(t'-t)}dt' 
+\frac{t}{\pi}\sum_{\ell\geq2}\int^{\infty}_{4m_{\pi}^2}\frac{dt'}{t'} \hat G^1_{1, 2\ell-1}(t,t') \im g^1_{2\ell-1}(t') \nonumber \\
&+\frac{1}{\pi}\sum_\ell \int^{\infty}_{m_+^2}ds' \hat G^-_{1, \ell}(t,s') \im f^-_\ell(s'). 
\label{eq:pwhdr1}
\end{align}

The expressions for the once-subtracted $\hat G^1_{\ell \ell'}(t,t'),\hat G^-_{\ell \ell'}(t,s')$ kernels
have been obtained in this review and are given in \ref{app:Kernels},
together with the unsubtracted $G^1_{\ell \ell'}(t,t'), G^-_{\ell \ell'}(t,s')$
that were already derived in \cite{Pelaez:2018qny} but are also provided for completeness.

Since we have left  the $a$ parameter free, 
it can be used to maximize the
applicability of the equations right above. Let us remark that
constraints are coming from the applicability of the HDR
in Eqs.~\eqref{eq:shdr}, \eqref{eq:ahdr} and \eqref{eq:ahdr1} as well as from the convergence of the partial-wave expansions.
As shown in \ref{app:Applicability}, by setting $a=-10.9m_{\pi}^2$ 
the applicability range of these equations 
is $-0.286 \,\gev^2\leq t\leq 2.19 \,\gev^2$. 
In other words, we can study the physical region from the $K \bar K$ threshold $\simeq 0.992\,\gev$ up to 
$\sqrt{t}\simeq 1.47\,\gev$.
In contrast, the HDR projected into partial waves in the $a=0$ case are only valid up to $\simeq 1.3,\gev$.
Thus, with our choice of $a$, 
the applicability of the dispersive approach in the physical region, where we can test or use data as input, can be extended by 55\% in terms of the $\sqrt{t}$ variable, or 67\% in terms of $t$.

\subsection{Muskhelishvili-Omn\`{e}s method for the unphysical $\pipikk$ region.}
\label{sec:MO}

A very important complication when dealing with dispersive integrals for the \pipikk channel is that the integration region
starts at $\pi \pi$ threshold.  This implies that 
they should be calculated over an
``unphysical'' region where actual $\pi\pi\rightarrow K \bar K$ scattering data cannot occur.
Fortunately,  below $K\bar K$ threshold the inelasticity to more than two-pion states 
is  negligible in practice. Consequently, $\pi\pi$ is the only available state and then  Watson's theorem \cite{Watson:1952ji} implies that the $g_\ell^{I}$ phase in that region
is nothing but that of $\pi \pi$ scattering, i.e. $\phi^{I}_\ell(t)=\delta^{I}_{\ell,\pi \pi \rightarrow \pi \pi}(t)$. However, Watson's theorem does not provide any direct information on  the modulus of $g_\ell^{I}$.
Nevertheless, as already illustrated in \cite{Johannesson:1976qp,Ananthanarayan:2001uy,Buettiker:2003pp,Hoferichter:2011wk,Ditsche:2012fv,Pelaez:2018qny},
once the phase is known, $\vert g_\ell^{I}\vert$ in the unphysical region can be described using the rather standard
Muskhelishvili-Omn\`es method \cite{Omnes:1958hv,MuskhelishviliSingular:1953}, which we describe next.

Let us first separate the left- and right-hand contributions in  Eqs.~\eqref{eq:pwhdr} as follows:
\begin{align}
&g^0_\ell(t)=\Delta^0_\ell(t) + \frac{t}{\pi}\int^{\infty}_{4m_\pi^2}\frac{dt'}{t'}\frac{\im g^0_\ell(t)}{t'-t}, \quad \ell=0,2,\nonumber \\
&g^1_1(t)=\Delta^1_1(t) + \frac{1}{\pi}\int^{\infty}_{4m_\pi^2}dt'\frac{\im g^1_1(t)}{t'-t}, 
\end{align}
where $\Delta^{I}_\ell(t)$ contains both the left-hand cut contributions and subtraction terms.
It is important to realize that $\Delta^{I}_\ell(t)$ does not depend on $g^I_\ell$ itself, but on the
other $g^I_{\ell'}$ with $\ell'\geq\ell+2$. Fortunately, these last are much more suppressed  than $g^I_\ell$ in the unphysical region,
due to the centrifugal barrier.

In principle, the Muskhelishvili-Omn\`es method is only needed in the region between the two-pion and two-kaon thresholds. However, its solution would then depend very strongly on the value of the amplitude at $t_K=4m_K^2$. This is inconvenient in practice, because, first, the experimental data starts above that energy and, second,  this threshold is very sensitive to isospin breaking. In particular for the scalar I=0 wave. The reason is that, in reality, there are two different thresholds for $K^+K^-$ and $K^0\overline{K^0}$ separated by 8 MeV, which are felt by experiment but not accommodated by our isospin-conserving formalism.
For these reasons, following \cite{Ananthanarayan:2001uy} and our previous analysis in \cite{Pelaez:2018qny}, we apply the method up to a ``matching energy point'' $t_m>t_K$. The best choice for $t_m$ is dictated by phenomenology and will be discussed in detail below.

It is now that we introduce the Omn\`es function
\begin{equation}
\Omega^I_\ell(t)=\exp\left(\frac{t}{\pi}\int^{t_m}_{4m_\pi^2}\frac{\phi^I_\ell(t')dt'}{t'(t'-t)}\right),
\end{equation}
satisfying 
\begin{equation}
\Omega^I_\ell(t)\equiv\Omega^I_{\ell,R}(t)e^{i\phi^I_\ell(t)\theta(t-4m_\pi^2) \theta(t_m-t)}, 
\end{equation}
where, in the real axis, $\Omega^I_{\ell,R}(t)$ is now defined as:
\begin{eqnarray}
\Omega^I_{\ell,R}(t)&=&\left\vert \frac{t_m}{t_\pi}(t-t_\pi)^{-\phi^I_\ell(t)/\pi}(t_m-t)^{\phi^I_\ell(t)/\pi}\right\vert 
\exp\left(\frac{t}{\pi}\int^{t_m}_{4m_\pi^2}dt'\frac{\phi^I_\ell(t')-\phi^I_\ell(t)}{t'(t'-t)}\right).
\end{eqnarray} 
Note that, for real values of $s$, $\Omega^I_{\ell,R}$ is nothing but the modulus of $\Omega^I_{\ell}$, 
and hence a real function.

Since the  Omn\`{e}s function has the same cut 
from $4m_\pi^2$ to $t_m$ as $g^I_\ell(t)$, we can now define an auxiliary  function
\begin{align}
F^I_\ell(t)=\frac{g^I_\ell(t)-\Delta^I_\ell(t)}{\Omega^I_\ell(t)},
\end{align}
which is analytic except for a right-hand cut starting at $t_m$. 
Therefore, it is possible to write dispersion relations for $F^I_\ell(t)$ along that cut, which, recast back in terms of  $g^I_\ell(t)$, read:
\begin{eqnarray}
g^0_0(t)&=&\Delta^0_0(t)+\frac{t\,\Omega^0_0(t)}{t_m-t}\left[\raisebox{0pt}[0.6cm][0pt]{} \alpha
+\frac{t}{\pi}\int^{t_m}_{4m_\pi^2}dt'\frac{(t_m-t')\Delta^0_0(t')\sin\phi^0_0(t')}{\Omega^0_{0,R}(t')t'^2(t'-t)}
+\frac{t}{\pi}\int^{\infty}_{t_m}dt'\frac{(t_m-t')\vert g^0_0(t')\vert \sin\phi^0_0(t')}{\Omega^0_{0,R}(t')t'^2(t'-t)} \right], \nonumber  \\
g^1_{1}(t)&=&\Delta^1_{1}(t)+\Omega^1_1(t)\left[\frac{1}{\pi}\int^{t_m}_{4m_\pi^2}dt'\frac{\Delta^1_{1}(t')\sin \phi^1_1(t')}{\Omega^1_{1,R}(t')(t'-t)}\right.
\left.+\frac{1}{\pi}\int^{\infty}_{t_m}dt' \frac{\vert g^1_1(t')\vert \sin \phi^1_1(t')}{\Omega^1_{1,R}(t')(t'-t)}\right], 
\nonumber \\
g^1_{1}(t)&=&\hat \Delta^1_{1}(t)+t\, \Omega^1_1(t)\left[\frac{1}{\pi}\int^{t_m}_{4m_\pi^2}dt'\frac{\hat\Delta^1_{1}(t')\sin \phi^1_1(t')}{\Omega^1_{1,R}(t')t'(t'-t)}\right.
\left.+\frac{1}{\pi}\int^{\infty}_{t_m}dt' \frac{\vert g^1_1(t')\vert \sin \phi^1_1(t')}{\Omega^1_{1,R}(t')t'(t'-t)}\right], \nonumber
\\
g^0_2(t)&=&\Delta^0_2(t)+t\, \Omega^0_2(t)\left[\frac{1}{\pi}\int^{t_m}_{4m_\pi^2}dt'\frac{\Delta^0_2(t')\sin \phi^0_2(t')}{\Omega^0_{2,R}(t')
t'(t'-t)}\right. 
\left.+\frac{1}{\pi}\int^{\infty}_{t_m}dt' \frac{\vert g^0_2(t')\vert \sin \phi^0_2(t')}{\Omega^0_{2,R}(t')t'(t'-t)}\right].
\label{eq:MOeq}
\end{eqnarray} 
Note that, as explained above, for the $g^{1}_{1}$ wave we are interested in both the unsubtracted and once-subtracted relations, whose expressions are provided in the equations above in that respective order.
Of course, when $t$ is real and larger than the $\pi\pi$ threshold, 
a principal value must be taken on each integral, as illustrated with Eqs.~\eqref{eq:PVdistribution} and \eqref{Tcauchy-real} in the introduction of this section.
Beware that, since by construction the Omn\`{e}s function removes the phase, 
on the left-hand sides of these equations the partial wave is reduced to its modulus between $\pi\pi$ threshold and $t_m$. In contrast,
 it is reduced to its real part above $t_m$.

Note also that  we have considered a once-subtracted relation for the $g^0_0(t)$ Omn\`es solution. This is because, in what follows, we will choose $t_m$ with $\phi^0_0(t_m)\geq \pi$ and then a subtraction is  needed  to ensure the convergence when $t\rightarrow t_m$.
The corresponding subtraction constant $\alpha$ will be determined by imposing numerically a non-cusp condition for $g^0_0(t)$ at $t_m$.

Let us now explain the use of these equations. 
On the one hand, we need input   from $\pi K$ scattering, which 
in \cite{Pelaez:2016tgi} we took as fixed input from the data analysis constrained with dispersion relations in \cite{Pelaez:2016tgi}. However, in this report, we will use these partial-wave HDR to analyze first and constrain later both the \pik and \pipikk data parameterizations simultaneously.
Thus, instead of fixing one set of partial waves when studying the other, we will allow both sets to vary when considering their coupled HDR.

On the other hand, let us recall we need the Muskhelishvili-Omn\`{e}s method because
{\it in the unphysical region} there is no $\pipikk$ data and it may seem that we will not know what input to provide.  Nevertheless, 
the equations above do not need the full knowledge of $g^I_\ell(t)$ in the unphysical region, but only their phases and the $\Delta^I_\ell$. As already explained,  Watson's theorem tells us that these phases are known from 
$\pi\pi$ scattering, since in that regime other possible states made of more pions are negligible. For this purpose we will take these $\phi^{I}_\ell(t)=\delta^{I}_{\ell,\pi \pi \rightarrow \pi \pi}(t)$ phase shifts from the dispersive analysis of \cite{GarciaMartin:2011cn}.
In addition, each $\Delta^I_\ell$ does not involve $g^I_\ell(t)$ itself, but only other 
partial waves with $\ell'-\ell\geq 2$. 
However, in the unphysical region, such higher partial waves are suppressed  compared to that with $\ell$. We have  explicitly 
checked that the $\ell=3$ contribution to the  $g^1_1(t)$ 
is rather small indeed, and even higher waves are negligible. 
Thus, with our simple phenomenological description of $g^1_3(t)$ we can get a good dispersive representation of  $g^1_1(t)$. 
In addition, the $\ell=4$ contribution to $g^0_2(t)$, which we have also considered,
is almost negligible. Once we have $g^0_2(t)$ it can be  used as input for Eq.~\eqref{eq:MOeq} to obtain the dispersive representation of $g^0_0(t)$ (Note that the $\ell=4$ contributions for $g^0_0(t)$ were neglected in \cite{Pelaez:2016tgi} but are considered here.).

There is one technicality  worth mentioning here. 
Since we will parameterize the high-energy region in terms of Regge amplitudes, which in principle contain all partial waves, 
we must subtract, from the Regge contribution to $\Delta^I_\ell(t)$,
 the projection of the 
Regge amplitude itself into the $I,\ell$ partial wave under consideration. 
Fortunately, this projection is negligible, and we have explicitly checked that the integrals barely change whether we use the full Regge amplitude or the one with its own projection into $I,\ell$ subtracted.

Finally, we have to choose a value for the matching energy $t_m$, which should always be above the $K \bar K$ threshold.
It is relevant to keep in mind that the derivation
of the dispersion relations above implies  $g_{output}(t_m)=g_{input}(t_m)$.
This condition will always occur for the integral output irrespective
of whether the data at that energy has a statistical or systematic deviation
or if it is in good or bad agreement with dispersion relations.
As a consequence, if the data in that energy region were not close to the dispersive solution, the output 
will nevertheless be compelled to describe it, forcing the dispersive calculation to be distorted in other regions. 
In practice, we have found that, given the existing data, the $g^0_0$ wave is the most sensitive to this instability, the effect is more moderate on $g^0_2$ 
and almost negligible for $g^1_1$ because it is already very consistent for any $t_m$ choice.
For that reason, we studied in \cite{Pelaez:2018qny} what energy region is most consistent for $g^0_0$ when changing $t_m$.
We concluded that two regions yield systematically rather consistent results between input and output: one around $\sqrt{t_m}=1.2\, \gev$, which is also valid for $g_2^0$,
and another one around $\sqrt{t_m}=1.45\, \gev$.
For the latter choice, however, we found that the resulting uncertainty in the dispersive calculation between  $K \bar K$ threshold
and 1.2 GeV is so large that there is no dispersive constraint at all. Actually, with such large uncertainties, we could even find that the dispersive output using any of the two $g^0_0(t)$ solutions as input (see Section \ref{sec:Data}), comes out compatible with itself as well as with the other one. 
Furthermore, note that $t_m$ in Eq.~\eqref{eq:MOeq} is
the energy above which $\vert g_\ell^I\vert $ is used as input for its own equation. Since within our approach we are either
testing or constraining the data parameterizations, we are then interested in
maximizing that region by choosing the smallest possible $t_m$.
Taking into account the previous considerations we have finally chosen $\sqrt{t_m}=1.2\, \gev$ for $g^0_0$ and $g^0_2$, and $t_m=1\, \gev$ for $g^1_1$. This energy is above $K\bar K$ threshold where the two most important inelasticities, i.e.
$K\bar K$ and $\eta \eta$, show no cusps. Moreover, the $g^0_2$ data are well understood and under control
at this energy, since its largest 
contribution comes from the well-established $f_2(1270)$ resonance, which lies very near $t_m$. As a final consistency check, we have shown  in~\ref{app:differenttm} that, once a constrained solution is obtained, the different choices of $t_m$ produce negligible  variations in the dispersive output, as expected.

\section{Data parameterizations and Unconstrained Fits to Data}
\label{sec:UFD}

In this section, we will present simple but rather flexible parameterizations that can describe the data and, very importantly, a realistic estimation of the experimental uncertainties. Our aim for simplicity is to allow for easy implementation for later use in experimental or theoretical analyses. Our parameterizations will thus follow  the spirit of what was traditionally known as ``energy-dependent'' data analyses, just simple functions that describe data and we will avoid any particular model dependence. 
Of course, some approximations will be performed, like imposing isospin conservation or elastic unitarity on partial waves, not only in the strictly elastic regime but also where experiments have not found evidence of inelasticity or imposing fundamental constraints as the correct dependence on the powers of momenta near threshold. In the region where the existence of resonances is well established, we will introduce rather flexible resonant shapes, but not necessarily pure Breit-Wigner formulae.

Note that, with these parameterizations,  we will first  obtain  a set of Unconstrained Fits to Data, that we will refer to as the UFD set. This is what this section is about. 
The flexibility of the parameterizations is required so that later on we can  impose on them the dispersive constraints we have discussed in previous sections, while still describing the data. In this way, the parameterizations will stay the same but their parameters will change to define a Constrained Fit to Data (CFD) discussed in section \ref{sec:CFD}.

Concerning uncertainties, we will not only consider the statistical error provided by experiments but we will include estimates of systematic uncertainties needed to reconcile the observed incompatibilities within the same data set and/or between different experiments.

A word of caution is in order. As seen in Section~\ref{sec:Data}, data are not precise nor numerous enough to exclude large fluctuations between successive data points. And this is particularly evident in some relatively large energy regions where just a few data points may exist. In principle, we cannot exclude  complicated parameterizations that would adjust perfectly every single piece of data but produce large fluctuations or structures between points. Thus, from the outset, we make it clear that we assume such fluctuations not to occur and that the data can be fitted with simple and relatively smooth parameterizations. Obviously, the size of our uncertainties depends on this assumption. 
 We have chosen the parameterizations we provide below because they satisfy the above assumption and yield uncertainty bands that do not show wild fluctuations or become too large in the regions where data do not exist or do not require so. In addition, we have explored many alternatives: different conformal expansions (with different centers and more terms in the expansion, see~\ref{app:conformal}), simpler or more complicated polynomials in different variables, adding or removing resonant shapes, etc.  Of course, we can make them fit the data, but in so doing their central result is always very similar to our final choice. Therefore, apart from a few exceptions of interest, we spare the reader the list of pros and cons of the many other parameterizations we tried and we just present our final choice. Moreover, for a given parameterization, and once the systematic uncertainty that affects the data has been estimated, we have decided to stop adding parameters 
when the $\chi^2/dof$ has roughly ``converged'' to 1, or normalize it in case it is needed.

The existing data for the \pik and \pipikk scattering have been reviewed in detail in section \ref{sec:Data}. However, for our purposes, we will also need parameterizations of the amplitudes up to, in principle, arbitrarily high energies, where no data for strange trajectories exists. This region provides small, but non-negligible contributions to our integrals, and here we will also provide high-energy parameterizations of  $\pik$ and $\pipikk$ amplitudes in terms of Regge theory. Thus, in section \ref{sec:ufdregge} below, we will present the required Regge parameterizations that  complete what we call the UFD set since no dispersion relations will be imposed upon them in this first stage.

One might wonder why the dispersive constraints are not imposed from the very beginning. There are several reasons for this. First, in the future more or better data may appear in a particular wave and  then one may refit that wave without altering the other ones. This will not be so easy once we impose the constraints, since dispersion relations will correlate all waves among themselves.
Second, we want to check if there are data points or sets of data that fare particularly bad against the dispersive representation and then prune the experimental data before imposing any constraint. As long as experimental analyses do not incorporate these relations, it seems natural to  check the consistency of the data with respect to causality and crossing symmetry, at least within experimental uncertainties. Since we have already seen that meson-meson scattering data are affected by large systematic errors, which are not usually included in the experimental uncertainties,  it should not be surprising that unconstrained fits to $\pi\pi$ \cite{Pelaez:2004vs,Kaminski:2006yv,GarciaMartin:2011cn}, \pik \cite{DescotesGenon:2006uk,Pelaez:2016tgi} or \pipikk \cite{Pelaez:2018qny} data do not satisfy well these dispersive constraints.
 
Later on, we will construct a ``Constrained Fit to Data'' (CFD) set  by imposing the fulfillment of dispersion relations in the fits. But note that, although differing in the values of the parameters, the functional form of the UFD and CFD parameterizations are the same. Therefore, the parameterizations  will be presented in what follows, discussing for the moment just the UFD values of the fit parameters and leaving the CFD for section \ref{sec:CFD}.

\subsection{ $\pik$ Parameterizations and Unconstrained Fits to Data}
\label{sec:UFDpik}

In principle, since the quantities of interest  like resonance poles or threshold parameters, etc... will be obtained from the output of dispersion relations, any parameterization that describes data could do as input, i.e. a model, polynomials, or even splines. However, the whole approach becomes easier if some relevant physical features like cuts, unitarity, Adler zeros, or some poles associated with resonances are already implemented in our parameterizations from the very start. In addition, for many later applications, the full rigor of dispersion relations may not be needed, but it could be more interesting to use a relatively simple parameterization that is consistent with the dispersive representation, the data and that is also able to describe the most salient analytic features.
Hence, in this section we will provide parameterizations that we will use as input for our dispersion relations but that satisfy unitarity, display the required analytic structures like thresholds, or possible poles  to accommodate resonant structures. In addition, we will make a particular effort in doing this while keeping them relatively simple for later applications while describing not just the data but also their uncertainties. Moreover, we want them flexible enough to be able to satisfy the dispersive constraints when we will impose them in a later section.

Thus, following the usual conventions, when the partial-wave decomposition is possible, we will provide the phase shift (or its cotangent) and, in the inelastic regime, also  the elasticity function for each partial wave. At high energies, where no partial-wave data exist or the partial-wave expansion does not converge, we will use Regge theory.

The $\pik$ partial-wave parameterizations we will present here follow closely those we first introduced in \cite{Pelaez:2016tgi}, although we have made some slight modifications in the $S^{1/2}$, $S^{3/2}$, $P^{1/2}$ and $P^{3/2}$ waves.  Our main goal with these modifications is to improve the description of the uncertainties associated with the data, especially when estimating systematic errors.

In particular, we fit the data discussed in Section~\ref{sec:Data}, but only up to $\sim$1.7 GeV, and thus we will  only fit $S$, $P$, $D$ and $F$ waves,
since there are no data for $G$, $H$, and higher waves below 1.8 GeV.
The reason to choose a common maximum energy for all partial-wave fits is that above some energy we have to stop relying on partial waves for the input within our integrals, and we have to use total amplitudes, which are the only ones for which data exist at high energies.
We choose to parameterize them with Regge theory, which in principle 
includes all partial waves, so, once we start using high-energy parameterizations we cannot separate partial waves any longer. In addition, the convergence of the partial-wave expansion becomes less reliable as the energy increases.

\subsubsection{General form of our $\pik$ partial-wave parameterizations}
\label{subsec:generalforms}
Some generic features will be shared by most of our partial-wave parameterizations. In waves with little structure, such as those that are elastic in the whole energy region, a single functional form will be used throughout the whole energy range. However, more complicated waves will require different functional forms in different regions. Typically these piece-wise functions will be matched at thresholds demanding continuity. Otherwise, a continuous matching will be imposed and if possible even a continuous derivative at least for the central value of the fit. Let us then describe separately the elastic and inelastic generic forms:

\vspace{.3cm}
{\em \underline{Partial waves in elastic regions}}\\

Following Eq.~\eqref{eq:elasticpw} in the Notation section \ref{sec:Notation}, $\pi K$ elastic partial waves can be recast as
\begin{equation}
f_\ell(s)=\frac{1}{\sigma(s)}\frac{1}{\cot \delta_\ell(s)-i},
\end{equation}
where for simplicity we have suppressed the isospin index with respect to section \ref{sec:Notation}.
Thus, we will just provide a parametrization of $\cot \delta_\ell(s)$.

Although in principle any parameterization would do if we just want to calculate the dispersive integrals, from section \ref{sec:DR} we know that in the complex $s$-plane, partial waves for the scattering 
of two particles  with different masses have a distinct analytic structure in the first Riemann sheet,
shown for the \pik case in the upper panel of Figure.~\ref{fig:anstrucpw}.
First of all, there is
a right-hand cut, also called ``physical cut'', extending from $\pi K$ threshold to infinity. In addition, due to the thresholds in the crossed channels, there is
a left-hand cut extending from $(m_K-m_\pi)^2$ to $-\infty$, 
as well as  a circular-cut at $\vert s \vert^2=(m_K^2-m_\pi^2)^2$. For $\pi K$ scattering no other singularities appear in the complex plane since this system has no bound states that could give rise to poles in the real axis below threshold. 
The
cut singularities are reproduced in the second Riemann sheet, where poles associated with resonances can appear. 

Then, to describe the amplitude in the complex $s$-plane,
it is customary to use the elastic unitarity condition in Eq.~\eqref{eq:elasticpw} and, defining an {\it effective range function} $\Phi_\ell(s)$, recast the partial wave as
\begin{equation}
f_\ell(s)=\frac{q^{2\ell}}{\Phi_\ell(s)-iq^{2\ell}\sigma(s)}.
\label{eq:fintermsofphi}
\end{equation}
Of course, {\it in the elastic region of the real axis},
\begin{equation}
    \Phi_\ell(s)=\frac{2q^{2\ell+1}}{\sqrt{s}}\cot\delta_\ell(s).
    \label{eq:phicot}
\end{equation}
Abusing this notation it is usual to write:
\begin{equation}
    \cot\delta_\ell(s)=\frac{\sqrt{s}}{2q^{2\ell+1}}\Phi_\ell(s)
\end{equation}
as a complex function on the whole complex-$s$ plane, although, of course, it is only the cotangent of a real angle when $s$ lies in the real axis between the elastic and any inelastic threshold.

At this point we will use a conformal expansion in order to incorporate the analyticity properties of the partial wave, taking advantage of the fact that it is analytic in the whole plane except for the cuts shown in the upper panel of Fig.~\ref{fig:anstrucpw}.
The conformal expansions we are interested in for $\pi K$ scattering are explained in detail in~\ref{app:conformal}, and they are of the type
\begin{equation}
\cot\delta_\ell(s)=\frac{\sqrt{s}}{2q^{2\ell+1}}F(s)\sum_n{B_n \omega(s)^n}.
\label{eq:generalconformal}
\end{equation}
Generically, we set $F(s)=1$. For scalar waves, which have an  Adler zero at $s_{Adler}$, we take $F(s)=1/(s-s_{Adler})$, as well as $F(s)=(s-m_r^2)$ for waves that exhibit a clear narrow resonance
and whose phase shift crosses $\pi/2$ at $m_r$.
The conformal variable is defined as:
\begin{equation}
\omega(y)\equiv \omega(y(s))=\frac{\sqrt{y(s)}-\alpha \sqrt{y_0-y(s)}}{\sqrt{y(s)}+\alpha \sqrt{y_0-y(s)}}, \quad y(s)=\left(\frac{s-\Delta_{K\pi}}{s+\Delta_{K\pi}}\right)^2, \quad y_0=y(s_0).
\label{eq:conformalvars}
\end{equation}
This change of variables, which maps the complex $s$-plane into the unit circle, 
is relatively similar to those used for $\pi\pi$ scattering in \cite{Pelaez:2004vs,Caprini:2008fc}
or $\pi K$ scattering in \cite{Cherry:2000ut}, and is explained in detail in  \ref{app:conformal}.
Suffices here to say that  such a conformal expansion ensures a rapid convergence of the series. Thus, in practice, we will find that no more than three $B_i$ coefficients are needed
for the fits to each wave in the elastic region.
Note that, for each partial wave, the $s_0$ and $\alpha$ constants are fixed, not fitted. The intuitive meaning of these two parameters is that  $s_0$ sets the maximum energy at which 
this mapping is applicable on the real axis, whereas $\alpha$ 
fixes the energy around which the expansion is centered. As explained in \ref{app:conformal}, they are chosen so that the region where data exist lies well inside the convergence region $\vert \omega\vert<1$ and with a fairly symmetric distribution of the data on the left and the right sides of the center of the expansion.

\vspace{.3cm}
{\em \underline{Partial waves in inelastic regions}}\\

First of all, an important remark is in order.
Recall that the \pik partial wave is inelastic when 
$\vert S^I_\ell(s)_{\pik}\vert<1$ or, using  Eq.~\eqref{eq:inelpw}, $\eta^I_\ell(s)<1$. 
Let us drop the $I$, $\ell$ indices for a moment. With our parameterizations, we will fit the data on the phase and the modulus of each \pik partial wave. Therefore we will have information about the inelasticity as a whole and we will not be able to discern what channels
and by how much they contribute to $\eta(s)$ being smaller than one.  However, once we use our inelastic formalism all states that contribute physically to the inelasticity are taken into account (within uncertainties). This is in contrast to the familiar ``coupled-channel models'', for which one chooses just a few coupled states (usually two-body states or approximated as such) that are the only ones to contribute to the whole inelasticity. Within these models, it is possible to separate the sources of inelasticity and include information from other channels if it exists, but the caveat is that you can also miss states if they are  not explicitly included in the model from the start.

This said, in practice, we have to choose the energy from which we allow $\eta(s)$ to be less than one.  For the $S$-wave, the inelastic region is dominated by the $K_0^*(1430)$, whose branching fractions are $\simeq93\%$ to $\pi K$ and $\simeq 8.5\%$ to $K \eta$. Thus the $K \eta$ threshold at $\simeq 1042\,$MeV  seems the natural choice to allow for inelasticity. 
In contrast, when looking at the  $P$, $D$ and $F$-waves, the  main decay channels other than $K\pi$ for the $K^*(1410)$, $K^*_2(1430)$,  $K^*(1680)$ and $K_3^*(1780)$ resonances are $ K^*(892) \pi$ and $K \rho(770) K$. The thresholds of these two states are nominally placed at $\simeq1032\,$MeV and $\simeq1266\,$MeV. One might also wonder about the $K \pi\pi$ channel but, in practice, this state is observed as one of the previous two. For a  description of the $P$-wave with a coupled channel formalism with the three $K\pi$, $K^*\pi$ and $K\rho(770)$ channels, see \cite{Moussallam:2007qc}.
Unfortunately, the $K^*(892)\pi$ and $K \rho(770)$ thresholds are not as sharply defined as the $K\eta$ threshold, due to the widths of the $K^*(892)$ and $\rho(770)$, which are $\simeq50$ and $\simeq170\,$MeV, respectively. 

Driven by our desire for simplicity we have decided to set our inelastic formalism to start at the $K\eta$ threshold, since it is sharply defined, very close to the nominal one for $K^*(892)\pi$ and well below the nominal one for $K\rho(770)$. An additional practical reason is that a well-defined threshold allows for a cusp, i.e. a discontinuity of the derivative, which facilitates the matching between the elastic and inelastic regimes. Actually, we can easily take into account the inelastic parts using a step function $\Theta_{K\eta}=\Theta(s-(m_K+m_\eta)^2)$ and ensure the continuity by using the $q_{\eta K}$ momenta, which vanishes at the $K\eta$ threshold, in the equations. Only for one wave, we will also allow for a similar treatment of the $K\eta'$ threshold. Nevertheless, we insist that, once we allow for the inelasticity to exist, all possible inelastic channels contribute to it. Our use of the $K\eta$ threshold to start our inelastic region by no means implies that we are assuming that the $K\eta$ channel dominates the inelasticity in all channels, it is just that we allow our parametrizations to acquire an inelasticity from that point. The data will then tell us how big that total inelasticity is, irrespective of what channels are producing it.

Once we have decided from where we will allow our amplitudes to become inelastic, following \cite{Pelaez:2016tgi}, the majority of our partial-wave parameterizations in the inelastic region
have been chosen to implement in a relatively simple way several resonances
 observed experimentally, while providing a  continuous match with their corresponding
parameterization in the elastic regime.
Note, however, that for  the $D^{1/2}$ and $F^{1/2}$-waves data only exist
in the inelastic region, and thus we will use an inelastic formalism
throughout the whole energy region, which reduces 
to the elastic case below $K\eta$ threshold. 
In \cite{Pelaez:2016tgi} we tried different expressions, 
including polynomial fits and splines in powers of  
the $\pi K$, $K\eta$ momenta, or the $s$ or $\sqrt{s}$ variables. However, we found
that once fitted to data,  the error bands of such parameterizations tend to be both rather small  near the elastic region and too big at larger energies. This does not reproduce well the uncertainty observed by simple inspection of the data and leads to very large correlations. 

As a matter of fact, in \cite{Pelaez:2016tgi} we found, as others before us \cite{Buettiker:2003pp},  that this region is most efficiently described 
in terms of products of exponential or rational functions, as follows:
\begin{equation}
f_\ell(s)=\frac{1}{2i\sigma(s)}\left( \prod_n{S_n}(s)-1\right).
\label{eq:inelunit}
\end{equation}
In this way it is far easier to implement resonant structures, usually overlapping, together with other background features,  while yielding more uniform uncertainty bands throughout the whole fit region. 
In particular, complex exponentials will be used to describe a non-resonant background
\begin{equation}
S_n=S_{n}^b=\exp\left[2i q_{ij}^{2\ell+1}(\phi_0+\phi_1 q_{ij}^2+...)\right],
\label{eq:inelback}
\end{equation}
with $\phi_k$ real parameters, whereas rational functions 
\begin{equation}
S_n=S_n^r=\frac{s_{rn}-s+i(P_{n}(s)-Q_{n}(s))}{s_{rn}-s-i(P_{n}(s)+Q_{n}(s))},
\label{eq:inelres}
\end{equation}
will be used to accommodate resonances and their associated poles.
Here $s_{rn}$ are real parameters and $P_{n}(s)$ and $Q_{n}(s)$ are polynomials that have the same sign over the inelastic region.  Let us remark that if these polynomials were constant, $S_n^r$ reduces to the simplest form of the familiar Breit-Wigner parameterization.
In the following subsections, we will explain in detail our choice of polynomials for different waves.

Of course, these functional forms are also chosen because they satisfy $\vert S_n\vert \leq1$ and then unitarity is satisfied trivially, even in the inelastic regime.

A continuous matching with each corresponding elastic region is achieved by fixing the $P_{n}(s)$ polynomial in $S_n$ that has the pole with the smallest $s_{rn}$.
This formalism is a modification of the parameterizations 
used by \cite{Buettiker:2003pp} in this energy region.
In order to recover the elastic case, it is enough to set
$Q_n\equiv 0$, which as explained above is of relevance for the $D^{1/2}$ and $F^{1/2}$ waves.

As already commented, near a resonance (or more precisely, its associated pole)
each of the $S_n^r$ functions bears some resemblance to the familiar Breit-Wigner functional form, which is just a simple model. Our parameterizations are much more flexible since they can incorporate  inelastic resonances even if they overlap with other resonances or analytic structures. In particular, 
when combining the $S_n$ in our complete functional form in Eq.~\eqref{eq:inelunit}, 
unitarity is satisfied exactly. This
is definitely not the case of a simple sum of Breit-Wigner amplitudes, which violates unitarity and is nevertheless often used in the literature.
Let us emphasize that the actual resonance parameters  have to be calculated with the full partial-wave expression evaluated at the pole position in the complex plane, and never from just one individual $S_n^r$ piece on the real axis.

Finally, to conclude this introduction to the generic form of our parameterizations, let us recall that we will use partial waves to describe data up to $\sim 1.8\,$GeV for \pik, and $\sim 2$ GeV for \pipikk scattering.
Beyond that energy, we will use Regge theory to describe 
the whole amplitudes $F(s,t,u)$. These parameterizations will be described last in subsection~\ref{sec:ufdregge} below. 

\subsubsection{$S$-wave}
\label{subsub:swave}

The   $S$-wave \pik scattering data were discussed in subsection \ref{subsec:Sdata}. Let us recall that the main data sets, shown in Fig.~\ref{fig:S12data}, are measured in the $f_0^{1/2}+f_0^{3/2}/2$ isospin combination. Since the $I=3/2$ partial wave has been measured independently, see Fig.~\ref{fig:s32data}, it can be used to separate the $f_0^{1/2}$ component. In practice, we will fit the two waves simultaneously. In view of the data, it is evident that there are large systematic uncertainties between different data sets and even within the same data sets. In particular, there are a few points that provide most of the $\chi^2$ of the fit and are largely incompatible with the rest of the sample.

In our first analysis of \pik \cite{Pelaez:2016klv}, we chose not to discard any data point and we followed an elaborate procedure to estimate the
uncertainties of the resulting fit. In particular, we followed
one of the techniques suggested in \cite{Perez:2014yla,Perez:2015pea}, which had been previously applied to $NN$ and $\pi\pi$ scattering. In brief, we checked several Gaussianity tests on the data with
respect to the fit and then enlarged the uncertainties of those
data points that spoilt the tests (typically those beyond 3-standard deviations). This data purge leads to a new fit upon
which the procedure is iterated until the Gaussianity test is
satisfied. We also tried other simpler approaches and we found that the uncertainty band was rather similar.

However, the method we used in \cite{Pelaez:2016klv} to estimate systematic uncertainties for this particular wave, is somewhat cumbersome and will be abandoned here.
The reason why we can do this is that, as already remarked, in this case, most of the contribution to the $\chi^2$ of any fit is due to just a few data points which are abnormally separated from the rest. Hence, we will simply discard all data points that deviated from the best fit  by more than $3.5$ standard deviations. 
As a matter of fact, and in agreement with the expectations derived from  \cite{Perez:2015pea}, given the total number of data points to be fit for the $S$ wave, deviations above $3.5$ sigmas should be extremely unlikely, but we find many more than expected. As a consequence, just by removing these dramatically inconsistent values, the $\chi^2/dof$ of the fits in all regions of the $f_0^{3/2}$ and $f_0^{1/2}+f_0^{3/2}/2$ combinations gets immediately reduced below 3. Remarkably,  the residues of the fit still follow roughly a gaussian distribution, even if we have not imposed such a feature from the onset. Finally, it is enough to re-scale the statistical uncertainties of the remaining points by a given factor to normalize  the $\chi^2/dof$. When these systematic errors are taken into account, they translate into larger uncertainties for the fit coefficients. Of course, the data outliers still remain outside the band, although in this way they do not contaminate the fit.

After removing the conflicting points we get $\chi^2_{S^{3/2}}/dof=2.6$, 
$\chi^2_{e\, S^{1/2}+S^{3/2}/2}/dof=1.4$ and 
$\chi^2_{in\, S^{1/2}+S^{3/2}/2}/dof=1.7$, where the $\chi^2_{e\, S^{1/2}+S^{3/2}/2}$ combination is the result of the fit in the elastic region, whereas the latter corresponds to the inelastic one, above the $K \eta$ threshold. One could then re-scale the uncertainties of the data by a uniform factor to get a $\chi^2/dof\sim 1$, which is roughly normally distributed.

Let us then provide the specific details of the  $S$-waves parameterizations, whose general features were previously discussed.

\vspace{.3cm} \underline{ $I=3/2$ $S$-wave}\\

For this wave, we will keep the very same simple parameterization we already introduced in \cite{Pelaez:2016klv}, although the values of the parameters will change due to our new fitting strategy. 
It is worth noting that so far there is no evidence of inelasticity up to 
$\sim$1.8 GeV and this wave can be considered  elastic.
Hence, 
 we will make use of a truncated conformal parameterization to fit the phase shift. We have checked that just three conformal parameters are enough since the addition of a fourth one does not improve the fit. Therefore, we use
\begin{equation}
\cot\delta_0^{3/2}(s)=\frac{\sqrt{s}}{2q(s_{A}-s)}\Big(B_0+B_1\omega(s)+B_2\omega(s)^2\Big).
\label{eq:cot32S}
\end{equation}
Note the Adler zero is factorized explicitly and fixed to 
its leading order within Chiral Perturbation Theory (LO ChPT): $s_{A}=\Sigma_{K\pi}\simeq (516\,$GeV$)^2$. 
For this particular wave, we choose the conformal variable in Eq.~\eqref{eq:conformalvars} with the following fixed parameters:
\begin{eqnarray}
\alpha=1.4, \quad s_0=(1.84 \, {\rm GeV})^2.
\end{eqnarray}
The parameters of the unconstrained fit to data (UFD) are shown in Table \ref{tab:S32pa}. Compared to the results using our parameters in \cite{Pelaez:2016tgi},  this UFD will produce a much better agreement with dispersion relations to start with, even though its uncertainties have been reduced. It is worth noticing that a simple fit to the whole data collection, without removing the conflicting data points as explained above, would lead to way greater deviations between the fit and the dispersive results. Moreover, its scattering length would lie further away from the sum-rule result than our new UFD.

Our resulting UFD phase is plotted as a dashed line with a light orange band in Fig.~\ref{fig:s32data}. Note that at low-energies the curve follows closely the data of \cite{Bakker:1970wg,Linglin:1973ci} whereas at high energies it is dominated by data from \cite{Estabrooks:1977xe}.  

Finally, let us remark that we also considered leaving the Adler Zero as a free parameter, instead of fixing it to the ChPT leading order value.
The motivation was that,  in subsection~\ref{subsec:subth}, we will provide the best value for this Adler zero, $\sim 0.550\,$GeV, obtained from dispersion relations using a constrained fit, whereas the LO ChPT value that we have used here is located at $\sim 0.516\,$GeV.
However, since $S$-wave data starts at energies around 0.75 GeV, roughly 200 MeV above the Adler Zero, the best fit with a free Adler zero does not improve the overall $\chi^2$ at all, whereas both the scattering lengths and the resulting Adler Zero are at odds with the dispersive values that we will obtain later. In other words, the pure unconstrained fit to data does not provide sufficient information to determine accurately the threshold and subthreshold regions, which is not unexpected considering that there are no data at threshold, and the lowest-lying data points are incompatible between themselves. Since, at this point, we are interested in a phenomenological parameterization of the data, and the subthreshold region is less relevant,  we consider that setting the Adler zero to the LO ChPT position, i.e., within $10\%$ of its actual value, is good enough for our purposes. We reiterate that the position of this Adler zero will be obtained in subsection~\ref{subsec:subth} from dispersion relations with a $\sim2\%$ accuracy.

\begin{table}[ht]
\caption{Parameters of the $S^{3/2}$-wave. In GeV$^2$ units.} 
\centering 
\begin{tabular}{c c c } 
\hline\hline  
\rule[-0.05cm]{0cm}{.35cm}Parameter & UFD & CFD \\ 
\hline 
\rule[-0.05cm]{0cm}{.35cm}$B_0$ & 2.15 $\pm$0.03  & 2.21 $\pm$0.03\\
\rule[-0.05cm]{0cm}{.35cm}$B_1$ & 3.96 $\pm$0.13  & 3.46 $\pm$0.13\\ 
\rule[-0.05cm]{0cm}{.35cm}$B_2$ & 3.15 $\pm$0.32  & 3.13 $\pm$0.32\\ 
\hline 
\end{tabular} 
\label{tab:S32pa} 
\end{table}

\vspace{.3cm} \underline{$S^{1/2}$ partial wave} \\

For the $I=1/2$ $S$-wave we will keep the very same functional form we already used in \cite{Pelaez:2016tgi}, although the values of the parameters will change slightly due to our new fitting strategy. 
Let us recall that  inelasticity in this wave has been measured above 1.3 \gev and that, as we saw in Fig.~\ref{fig:S12data}, this part of the inelastic region is dominated by the $K^*_0(1430)$ resonance, whose width is estimated in the RPP to be $\Gamma=270\pm80\,$MeV. 

Let us start then with the elastic region, for which we use the general form discussed above in terms of a truncated conformal expansion to fit the phase shift. For this particular wave we have found that just two conformal parameters are enough and therefore we use:
\begin{equation}
\cot\delta_0^{1/2}(s)=\frac{\sqrt{s}}{2q(s-s_{A})}\Big(B_0+B_1\omega(s)\Big).
\label{eq:cot12}
\end{equation}
Once again, we have factorized explicitly an Adler zero at $s_A$, whose position we take from leading order ChPT:
\begin{equation}
s_{A}=\Big(\Sigma_{K\pi}+2\sqrt{4\Delta_{K\pi}^2+m_K^2m_\pi^2}\,\Big)/5\simeq (0.486\,{\rm GeV})^2\simeq 0.236\,{\rm GeV}^2.
\label{eq:sAdler120}
\end{equation}
Actually, there are two such zeros, but here we are considering the one closer to threshold since it has the largest influence on the shape of the wave in the physical region. The other one is obtained by replacing the first plus sign with a minus \cite{GomezNicola:2007qj}. Hence, it lies inside the circular cut and thus outside the applicability region of the conformal mapping. As it happened with the $I=3/2$ $S$-wave, we fix the Adler zero instead of considering it a free parameter, because  for our fits we are interested in the physical region, where data exist, and not so much in the subthreshold region. Since the data lie quite far from threshold and thus even further away from the Adler zero, simple fits are not enough to determine it with accuracy. Later on, in subsection \ref{subsec:subth}, we will obtain the value of this Adler zero from the dispersive representation at $\simeq 0.470\,$GeV. Thus, the LO ChPT calculation we use here, which only deviates by $4\%$ from our final value, is enough for our purposes of fitting the data well, while keeping or approximating some basic features of the amplitude even outside the physical region.

In order to keep the data region reasonably centered within the conformal circle without distorting the uncertainty bands, we have found convenient
to fix the constants that define the center of the conformal variable $\omega$ to the following values (see \ref{app:conformal})
\begin{eqnarray}
\alpha=1.15, \quad s_0=(1.1 \,{\rm GeV})^2.
\end{eqnarray}

The Unconstrained Fits to Data (UFD) parameters are given in Table \ref{tab:Selparam}, and they are pretty similar to our original fit in \cite{Pelaez:2016tgi}.

\begin{table}[ht]
\caption{Parameters of the elastic $S^{1/2}$-wave. In GeV$^2$ units.} 
\centering 
\begin{tabular}{c c c } 
\hline\hline  
\rule[-0.05cm]{0cm}{.35cm}Parameter & UFD & CFD \\ 
\hline
\rule[-0.05cm]{0cm}{.35cm}$B_0$ & 0.402 $\pm$0.006        & 0.403  $\pm$0.006\\
\rule[-0.05cm]{0cm}{.35cm}$B_1$ & 0.222 $\pm$0.031         & 0.173 $\pm$0.031\\ 
\hline 
\end{tabular} 
\label{tab:Selparam} 
\end{table}

\begin{figure}[!ht]
\centering
\resizebox{0.8\textwidth}{!}{\input{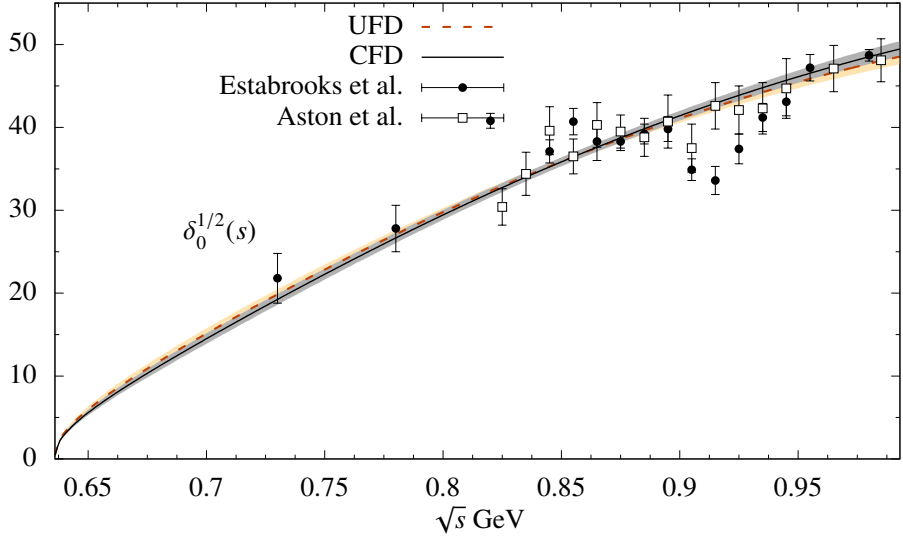}}\\
\caption{$S^{1/2}(s)$ \pik partial wave. We show the UFD fit as a dashed line, compared to the final CFD result as a continuous line. The data shown comes from \cite{Aston:1987ir} (empty squares) and \cite{Estabrooks:1977xe} (solid circles).}
\label{fig:s12elastic}
\end{figure}

The resulting phase and modulus of the $f_0^{1/2}+f_0^{3/2}/2$ combination can be seen as a dashed curve and a light-orange uncertainty band in Fig.~\ref{fig:S12data}. These are the data that we fit together with those of the $I=3/2$ $S$-wave.  Nevertheless, in this subsection we also provide in Fig.~\ref{fig:s12elastic} the $\delta^{1/2}_0$ phase-shift in the elastic region. The ``data'' in this figure are not measured directly but are extracted from the previous isospin combinations using \cite{Aston:1987ir}  and \cite{Estabrooks:1977xe}. Note that, even if there is a considerable rise in the phase, it does not reach $90^\degree$ below 1 GeV, and there is no clear sign of a resonance peak. This is the reason for the longstanding debate about the existence of a $\kap$ resonance. We will see in later sections that this shape together with dispersion relations requires the existence of a pole very deep in the complex plane that is identified with the very wide $\kap$ resonance.

Actually, our very simple UFD parameterization does have a \kap pole in the second Riemann sheet, which can be obtained using Eqs.~\eqref{eq:cot12} above with Eqs.~\eqref{eq:rescond}. It is located at: $\sqrt{s_p}=(651\pm14)-i (336\pm 5)$ MeV, which is in the ballpark of the 
precise value obtained from dispersion relations in
\cite{DescotesGenon:2006uk,Pelaez:2020uiw}, as will be explained in section \ref{sec:kapdr}. Thus, as it happened with the Adler zero, our simple UFD parameterization provides an approximation to another analytic feature of the amplitude in the non-physical region. Nevertheless, this is still a model-dependent extraction from a fit of the elastic region only which has not been constrained with dispersion relations. The rigorous and accurate model-independent extraction will be
provided in section \ref{sec:kapdr}.

Let us now turn to the parameterization of the $S^{1/2}$-wave in the inelastic region, i.e. for $s\geq (m_K+m_\eta)^2$.  
Once more we use the very same functional form  we considered  in \cite{Pelaez:2016tgi}, but with updated parameters  due to our new fitting strategy. This parameterization follows the generic inelastic form of Eq.~\eqref{eq:inelunit} discussed above,
but for this particular case, it consists of two resonant forms $S_1^r$ and $S_2^r$ and a background $S^b_0$. 
The first one will accommodate the 
$K_0^*(1430)$ that was already  discussed when we presented the $S$-wave data in subsection \ref{subsec:Sdata}.
The second one is purely phenomenological, since $\sqrt{s_{r2}}$ will come out around 1.8 GeV, well above the maximum energy where we are 
going to check and/or impose dispersion relations. Therefore we will only see its low-energy tail and for us, it is more like a background contribution. However, it will give more flexibility to mimic the low tail of resonances heavier than 1.7 GeV in this channel, like the $K^*_0(1950)$, although not their precise position or parameters. Remember that above 1.84 GeV we will be using Regge theory, not partial waves.
Thus, our $S^{1/2}$-wave parameterization reads:
\begin{equation}
f_0^{1/2}(s)=\frac{S_0^b S_1^r S_2^r-1}{2i\sigma(s)},
\end{equation} 
where
\begin{equation}
S_0^b=\exp\Big[2iq_{\eta K}(\phi_0+\phi_1 q_{\eta K}^2)\Big].
\end{equation}
For $S^r_1$ we use Eq.~\eqref{eq:inelres} with
\begin{eqnarray}
P_1(s)&=&(s_{r1}-s)\beta+e_1G_1 \frac{p_1(q_{\pi K})}{p_1(q_{\pi K}^r)}\frac{q_{\pi K}-\hat{q}_{\pi K}}{q_{\pi K}^r-\hat{q}_{\pi K}},\\
Q_1(s)&=&(1-e_1)G_1 \frac{p_1(q_{\pi K})}{p_1(q_{\pi K}^r)}\frac{q_{\eta K}}{q_{\eta K}^r}\Theta_{\eta K}(s),
\end{eqnarray}
where $p_1(x)=1+a x^2+b x^4$, $q_{i j}^r=q_{i j}(s_{r})$, $\hat{q}_{i j}=q_{i j}((m_{\eta}+m_K)^2)$
and $\Theta_{\eta K}(s)=\Theta(s-(m_K+m_\eta)^2)$ is the step function at $K\eta$ threshold.
In addition, for $S^r_2$ we use Eq.~\eqref{eq:inelres} with
\begin{eqnarray}
P_2(s)&=&e_2G_2 \frac{p_2(q_{\pi K})}{p_2(q_{\pi K}^r)}\frac{q_{\pi K}-\hat{q}_{\pi K}}{q_{\pi K}^r-\hat{q}_{\pi K}},\\
Q_2(s)&=&(1-e_2)G_2 \frac{p_2(q_{\pi K})}{p_2(q_{\pi K}^r)}\frac{q_{\eta K}}{q_{\eta K}^r}\Theta_{\eta K}(s),
\end{eqnarray}
with $p_2(x)=1+c x^2$.

Since we have chosen to  match the elastic and inelastic parameterizations at 
 $K \eta$ threshold,
we only need to demand continuity. This is ensured by defining 
$\beta\equiv 1/\cot\delta_0^{1/2}((m_K+m_\eta)^2)$, where the cotangent is now calculated from the elastic part, using \cref{eq:cot12}.

All in all, the resulting fit parameters for the inelastic region parameterization are listed in Table \ref{tab:Sinparam}.

\begin{table}[ht] 
\caption{Parameters of the $S^{1/2}$ inelastic fit.} 
\centering 
\begin{tabular}{c c c } 
\hline\hline  
\rule[-0.05cm]{0cm}{.35cm}Parameters & UFD & CFD \\ 
\hline 
\rule[-0.05cm]{0cm}{.35cm}$\phi_0$ & $-$0.08       $\pm$0.031GeV$^{-1}$   & $-$0.002       $\pm$0.031GeV$^{-1}$\\
\rule[-0.05cm]{0cm}{.35cm}$\phi_1$ & 4.64         $\pm$0.16GeV$^{-3}$  & 4.65         $\pm$0.16GeV$^{-3}$\\ 
\rule[-0.05cm]{0cm}{.35cm}$a$ & $-$5.46            $\pm$0.03GeV$^{-2}$   & $-$5.53            $\pm$0.03GeV$^{-2}$\\ 
\rule[-0.05cm]{0cm}{.35cm}$b$ &  8.1             $\pm$0.1GeV$^{-4}$  & 8.2             $\pm$0.1GeV$^{-4}$\\ 
\rule[-0.05cm]{0cm}{.35cm}$c$ & $-$1.65             $\pm$0.04GeV$^{-2}$  & $-$1.62             $\pm$0.04GeV$^{-2}$\\ 
\rule[-0.05cm]{0cm}{.35cm}$\sqrt{s_{r1}}$ & 1.401 $\pm$0.004GeV    & 1.412 $\pm$0.004GeV\\ 
\rule[-0.05cm]{0cm}{.35cm}$\sqrt{s_{r2}}$ & 1.813 $\pm$0.013GeV    & 1.800 $\pm$0.013GeV\\ 
\rule[-0.05cm]{0cm}{.35cm}$e_1$ & 1                                  & 1\\ 
\rule[-0.05cm]{0cm}{.35cm}$e_2$ & 0.179          $\pm$0.026        & 0.138           $\pm$0.026\\ 
\rule[-0.05cm]{0cm}{.35cm}$G_1$ & 0.443           $\pm$0.024GeV    & 0.439           $\pm$0.024GeV\\ 
\rule[-0.05cm]{0cm}{.35cm}$G_2$ & 0.32           $\pm$0.10GeV     & 0.25            $\pm$0.10GeV\\
\hline 
\end{tabular} 
\label{tab:Sinparam} 
\end{table} 

\begin{figure}[ht]
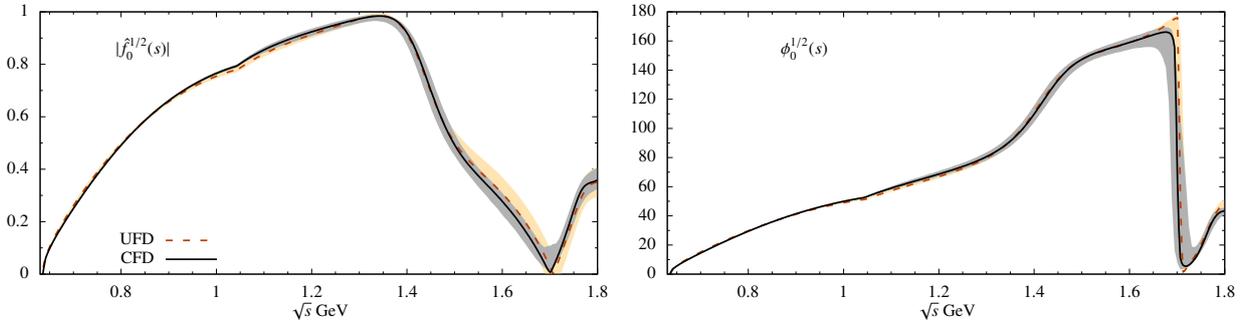

\begin{center}
\resizebox{\textwidth}{!}{\input{figures/smod12.tex} \input{figures/sphase12.tex}}
\caption{$S^{1/2}$ partial wave UFD vs CFD parameterizations for the modulus (left panel) and the phase (right panel). Notice that both fits are fairly in the whole energy region.
\label{fig:S12all} 
}
\end{center}
\end{figure}

In Fig.~\ref{fig:S12all} we show both the modulus and phase of the $\hat f^{1/2}_0=\vert \hat f_0^{1/2}\vert \exp{(i \phi_0^{1/2})}$ partial wave from threshold up to 1.7 GeV. The UFD parameterization of this wave is represented by a dashed line and the corresponding orange band for its uncertainty.
There we can see the wide structure attributed to the \kap below 1 GeV, as well as the dominant feature around 1.4 GeV due to the $K_0^*(1430)$ resonance, whose fast increase in the phase is clearly seen between 1300 and 1500 MeV. By comparing with the measured combination  $f_S\equiv f^{1/2}_0+f^{3/2}_0/2$ shown in Fig.~\ref{fig:S12data}, we see the appearance of a zero around 1.7 GeV, which is not seen in Fig.~\ref{fig:S12data} due to the presence of the $f_0^{3/2}$ component.

\subsubsection{$P$-waves}
\label{subsub:pwave}

The data for these waves were discussed in subsection \ref{subsec:Pdata}. Let us describe here the specific details of our $I=3/2$ and $I=1/2$ parameterizations, which follow the generic form described in subsection \ref{subsec:generalforms}.

\vspace{.3cm}
\underline{$P^{3/2}$ partial wave} 
\vspace{.3cm}

Since no inelasticity has been observed for this wave, we use a purely elastic formalism.
In this case, we have slightly modified our two-parameter conformal map in \cite{Pelaez:2016tgi}. The reason is that
as seen in Fig.~\ref{fig:p32wavephase}, there are no scattering data below 1 GeV. When we did the two-parameter fit, the data were described nicely, but then the scattering length becomes negative. However, this is in disagreement with the sum rule obtained from fixed-$t$ dispersion relations in \cite{Buettiker:2003pp}, where it was found $m_\pi ^3 a^{3/2}_1=(0.65\pm 0.44)\,10^{-3}$.
Moreover, as we will see in section \ref{subsec:SRth}, when calculating our own sum rules
for this report we also find positive values for $a_1^{3/2}$, which is also the case of NLO and NNLO ChPT \cite{Bernard:1990kx,Bernard:1990kw,Bijnens:2004bu}. Moreover, we have found that a negative scattering length could not be accommodated by the very accurate $P$-wave HDR near threshold.
Therefore, we have included a purely phenomenological additional 
factor, only relevant at low energies, to facilitate the phase-shift change of sign between threshold and the data region, i.e. the appearance of a zero of the amplitude so that we can also fit the  sum rule in \cite{Buettiker:2003pp}.  

Of course, we still fix the $\alpha$ and $s_0$ parameters to the values we used in \cite{Pelaez:2016tgi} and in the $S^{3/2}$ partial wave right above since they make the data to be centered within the conformal disk without producing unrealistic uncertainty bands. We take
\begin{eqnarray}
\alpha=1.4, \quad s_0=(1.84 \, {\rm GeV})^2.
\end{eqnarray}
All in all, our new parameterization reads
\begin{equation}
\cot\delta_1^{3/2}(s)=\frac{\sqrt{s}}{2q^3}\frac{s}{s-\hat s}\Big(B_0+B_1\omega(s)\Big).
\label{eq:cot32P}
\end{equation}
The $\hat s$ parameter will be the energy where this amplitude becomes zero and the phase shift changes sign.
 
 The fit parameters are given in Table~\ref{tab:Pelparam} and the resulting phase shift is shown in Fig.~\ref{fig:p32wavephase} against the data as a dashed curve whose uncertainties are covered by the orange band.
 Note that the phase shift is tiny and positive near threshold, ensuring a very small but positive scattering length, it then crosses zero around 0.75 GeV and becomes negative, describing the existing scattering data, which lies above 1 GeV. The fitted $\chi^2/dof=1.2$.

\begin{table}[ht]
\caption{Parameters of the $P^{3/2}$-wave. 
In GeV$^2$ units.} 
\centering 
\begin{tabular}{c c c } 
\hline\hline  
\rule[-0.05cm]{0cm}{.35cm}Parameter & UFD & CFD \\ 
\hline 
\rule[-0.05cm]{0cm}{.35cm}$B_0$ & $-$9.19 $\pm$2.4     & -8.39 $\pm$2.4\\
\rule[-0.05cm]{0cm}{.35cm}$B_1$ & $-$3.2   $\pm$6.6       & $-$2.4  $\pm$6.6\\ 
\rule[-0.05cm]{0cm}{.35cm}$\hat s$ & 0.57   $\pm$0.17        & 0.88    $\pm$0.17 \\
\hline 
\end{tabular} 
\label{tab:Pelparam} 
\end{table}

\vspace{.3cm}
\underline{$P^{1/2}$ partial wave}\\

In this case, we use the very same parameterization we introduced in \cite{Pelaez:2016tgi}, which follows the generic forms discussed in subsection \ref{subsec:generalforms} above. However, 
we will slightly modify the  choice of data to be fitted in the elastic region, whereas we will keep the same choice in the inelastic region.

Let us start describing the fit to the elastic region, i.e. $s\leq{(m_{\eta}+m_K)^2}$, which, as seen in Fig~\ref{fig:pwavedata}, is dominated by the $K^*(892)$. We use again the conformal parameterization of \cite{Pelaez:2016tgi}, with just three conformal parameters, namely:
\begin{equation}
\cot\delta_1^{1/2}(s)=\frac{\sqrt{s}}{2q^3}(m_r^2-s)\Big(B_0+B_1\omega(s)+B_2\omega(s)^2\Big).
\label{eq:cot121}
\end{equation}
Note we have explicitly factorized $(m_r^2-s)$
so that the phase crosses $\pi/2$ at the energy of the 
peak associated with the $K^*(892)$ resonance. At  $s=m_r^2$  the phase shift reaches 90$^\degree$.
As explained in section \ref{app:conformal}, the constants $\alpha$ and $s_0$, which define the conformal variable,
are fixed from the choice of the center of the expansion and 
the highest energy of the fit to be
\begin{eqnarray}
\alpha=1.15, \quad s_0=(1.1 \, {\rm GeV})^2.
\end{eqnarray}

There are several reasons to revisit our fitting strategy in the elastic regime.
First, although the Estabrooks et al. \cite{Estabrooks:1977xe} and Aston et al.~\cite{Aston:1987ir} data in Fig.~\ref{fig:pwavedata} may seem compatible, they are actually not, given their tiny statistical uncertainties.  
In order to estimate them, we have first followed  what we did  in \cite{Pelaez:2016tgi}. Namely, whenever two points of these data sets are incompatible, we fit their average, taking as the uncertainty the combination of their statistical  and systematic errors, the latter defined as half of their difference. 
This procedure cannot be followed for isolated points and, initially, we do not attach them any systematic uncertainty. 
We nevertheless fit the data. However, when  we find some datum severely deviated from this first fit--- typically by more than 3 $\sigma$--- we add a systematic uncertainty which is half of its deviation from the fit. Then the procedure is iterated until a reasonable $\chi^2\simeq1$ is found.
In practice, this is only needed for the two lowest-energy data from \cite{Estabrooks:1977xe} seen as clear outliers in Fig~\ref{fig:pwavedata}.
Systematic uncertainties coming from this procedure were the only ones considered in \cite{Pelaez:2016tgi}. However, isospin-violating effects are at least of the same order of statistical uncertainties. Measurements are done in the neutral channel but, due to our isospin-conserving formalism, we are considering the same mass for the neutral and charged $K^*(892)$. Hence, we will now include an explicit systematic uncertainty of $\simeq 1.8\,$\mev in the $m_r$ error bar. The rest of the data will be fitted as in \cite{Pelaez:2016tgi}. Of course,  we have checked that fitting the whole data set without including these systematic uncertainties  worsens considerably the dispersive description, yielding large deviations for the scattering lengths compared to their sum-rule values.

\begin{figure}[!ht]
\centering
\resizebox{0.8\textwidth}{!}{\input{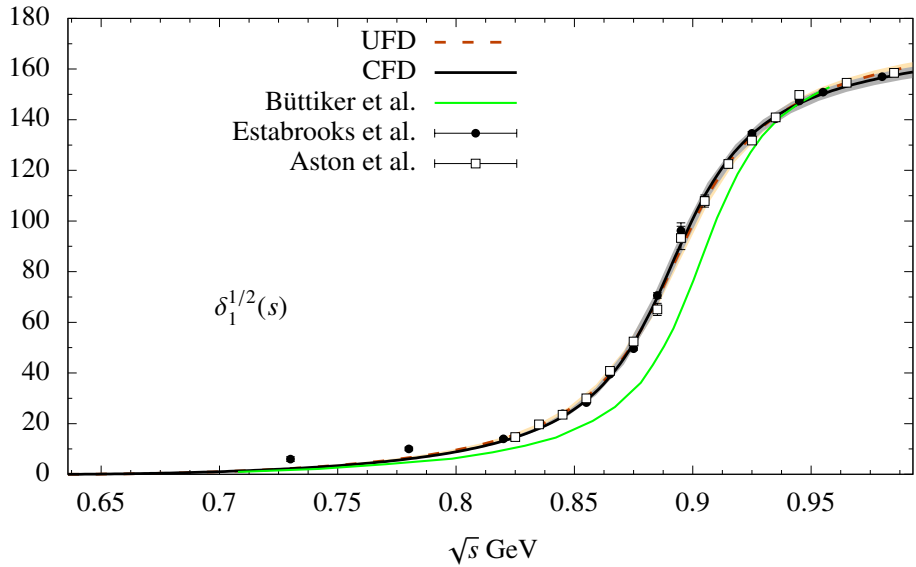}}\\
\caption{\pik scattering phase shift for the $P^{1/2}$ partial wave, we show the UFD fit as a dashed line, compared to the final CFD result as a continuous line. The data shown comes from \cite{Aston:1987ir} (empty squares) and \cite{Estabrooks:1977xe} (solid circles).
\label{fig:Pelastic}}
\end{figure}

The parameters of the unconstrained fit are listed in Table \ref{tab:Pelparam12}, and the resulting phase shift 
for the elastic region of the $P^{1/2}$ is shown in Fig.~\ref{fig:Pelastic} as a dashed line with its corresponding orange uncertainty band. Note that, as seen in Fig.~\ref{fig:p32wavephase}  the $P^{3/2}$ wave is almost compatible with zero in the elastic region, and therefore the $P^{1/2}$ phase shift shown in Fig.~\ref{fig:Pelastic} is, for all means and purposes, the same as those for  $f_P\equiv f^{1/2}_1+f^{3/2}_1/2$.

\begin{table}[ht]
\caption{Parameters of the $P^{1/2}$-wave. Note that in this case the $B_i$ are dimensionless. } 
\centering 
\begin{tabular}{c c c } 
\hline\hline  
\rule[-0.05cm]{0cm}{.35cm}Parameter & UFD & CFD \\ 
\hline 
\rule[-0.05cm]{0cm}{.35cm}$B_0$ & 0.970 $\pm$0.025      & 1.054   $\pm$0.025\\
\rule[-0.05cm]{0cm}{.35cm}$B_1$ & 0.56   $\pm$0.34        & 0.36  $\pm$0.28\\ 
\rule[-0.05cm]{0cm}{.35cm}$B_2$ & 2.66   $\pm$0.83        & 0.89   $\pm$0.83\\ 
\rule[-0.05cm]{0cm}{.35cm}$m_r$ & 0.8955 $\pm$0.0018GeV & 0.8946 $\pm$0.0018GeV\\
\hline 
\end{tabular} 
\label{tab:Pelparam12} 
\end{table}

In Fig.~\ref{fig:Pelastic} we also show, as a continuous green line, the solution of \cite{Buettiker:2003pp}, obtained {\it solving} Roy-Steiner equations, which  clearly deviates from all data. Furthermore, the mass and width associated to the $K^*(892)$ coming from this solution is at odds with all determinations listed in the RPP, by several standard deviations. It should be noticed that the \cite{Buettiker:2003pp} result for the $P$-wave is still remarkable because they do not use data in that region and it is a prediction obtained by solving Roy-Steiner equations with input from higher energies and other waves. Recall that in this report we are following the ``data-driven'' approach, i.e. using dispersion relations as  constraints on data fits, instead of ``solving'' the equations. Thus, our parameterizations describe very well the existing scattering data. Moreover, they actually contain a pole in the second Riemann sheet at: $\sqrt{s_p}=(892\pm2)-i (28.6\pm1.1)$ MeV. From this, we see that the $K^*(892)$ width that we obtain is $\Gamma\sim 57\,$MeV.

At this point, it is worth noting that in some experiments other than scattering~\cite{Link:2002ev,Link:2005ge,Bonvicini:2008jw,delAmoSanchez:2010fd,Ablikim:2015mjo,Zyla:2020zbs}, the $K^*(892)$ comes out somewhat narrower, between 45 and 50 MeV. For this reason, we have also studied an ``alternative $P$-wave'' in \ref{app:palt}, implementing a narrower $K^*(892)$, compatible with those other experiments.
The relevant observation is that, when we impose the dispersive representation on our UFD, the width becomes narrower, whereas when we impose it on the ``alternative'' $P$-wave, the width becomes wider. Thus, at the very end, the constrained fits turn out to be very similar. That is why we only comment on one solution in the main text and leave the alternative in the appendix. Finally, note that the dispersive solution in \cite{Buettiker:2003pp} (green line in Fig.~\ref{fig:Pelastic}) does not resolve this issue with the width, because it has the wrong mass.

Let us then turn to the inelastic region. 
Once more, our parameterization will be 
exactly the one we obtained in \cite{Pelaez:2016tgi}, which is of the generic form discussed in subsection \ref{subsec:generalforms}. For this wave it has three resonant shapes to accommodate the upper tail of the $K^*(892)$, which is still felt in the inelastic region, the $K^*(1410)$, which couples very little to $\pi K$,
and the $K^*(1680)$. As discussed in subsection~\ref{subsec:Pdata} the last two resonances are wide objects, and their pole positions are rather uncertain in the RPP. According to the RPP, the dominant branching ratios of the 
$K^*(1410)$ are 40\% to $K^*(892) \pi$, 6,6\% to $K\pi$ and less than 7\% to $K\rho(770)$. For the $K^*(1680)$, these are 38.7\% $K\pi$, 31.4\% $K_\rho(770)$, 30\% $K^*(892)\pi$ and only 1\% $K\eta$. 
Remember that, in our formalism, we set the inelastic region to start at the $K\eta$ threshold, but all possible inelastic channels contribute to make $\vert S^I_\ell\vert<1$, and we cannot separate individual contributions from each channel that makes $\eta(s)^I_\ell$ less than one.
We only describe the total elasticity $\eta(s)$ as given by the fit to the modulus of the $K\pi$ amplitude.
All in all, we write:

\begin{equation}
f_1^{1/2}(s)=\frac{S^r_1 S^r_2 S^r_3-1}{2i\sigma(s)},
\label{eq:pinel}
\end{equation}
where all the $S_k^r$ are of the form in Eq.~\eqref{eq:inelres}, with
\begin{eqnarray}
P_1&=&(s_{r1}-s)\beta+e_1G_1 \frac{p_1(q_{\pi K})}{p_1(q_{\pi K}^r)}\frac{q_{\pi K}^2-\hat{q}_{\pi K}^2}{(q^r_{\pi K})^2-\hat{q}_{\pi K}^2}\frac{q_{\pi K}}{q^r_{\pi K}},\nonumber\\
P_{2,3}&=&e_{2,3}G_{2,3} \frac{p_{2,3}(q_{\pi K})}{p_{2,3}(q_{\pi K}^r)}\frac{q_{\pi K}^2-\hat{q}_{\pi K}^2}{(q^r_{\pi K})^2-\hat{q}_{\pi K}^2}\frac{q_{\pi K}}{q^r_{\pi K}},\\
Q_{1,2,3}&=&(1-e_{1,2,3})G_{1,2,3} \frac{p_{1,2,3}(q_{\pi K})}{p_{1,2,3}(q_{\pi K}^r)}\left(\frac{q_{\eta K}}{q^r_{\eta K}}\right)^3\Theta_{\eta K}(s).
\nonumber
\end{eqnarray}
In addition, 
\begin{equation}
p_i (q_{\pi K})=1+a_i q_{\pi K}^2,
\end{equation}
and $\Theta_{\eta K}(s)=\Theta(s-(m_K+m_\eta)^2)$ is the step function at $K\eta$ threshold.
Again, in order to impose continuity at $K\eta$ threshold
we have defined
$\beta\equiv 1/\cot\delta_1^{1/2}((m_K+m_\eta)^2)$, where the cotangent is now calculated using the elastic parameterization in \cref{eq:cot121}.

As we saw in Fig.~\ref{fig:p32wavephase}, the $f^{3/2}_1$ amplitude is small but not entirely negligible in this region, so we always fit together the $f^{3/2}_1$ and $f^{1/2}_1+f^{3/2}_1/2$ combination.
In Table \ref{tab:Pinpa} we provide the updated parameters of the inelastic parameterization. Note that, in the inelastic region, although we fit the same parameterization and data we used in \cite{Pelaez:2016tgi}, there is a small change in the parameters since we are matching with the elastic part, for which we changed slightly our fitting strategy.

\begin{table}[ht] 
\caption{$P^{1/2}$-wave parameters in the inelastic region.} 
\centering 
\begin{tabular}{c c c } 
\hline\hline  
\rule[-0.05cm]{0cm}{.35cm}Parameters & UFD & CFD \\ 
\hline 
\rule[-0.05cm]{0cm}{.35cm}$a_1$ & $-$2.07             $\pm$0.14GeV$^{-2}$  & $-$1.60             $\pm$0.14GeV$^{-2}$  \\ 
\rule[-0.05cm]{0cm}{.35cm}$a_2$ & $-$2.11            $\pm$0.27GeV$^{-2}$   &  $-$1.79             $\pm$0.27GeV$^{-2}$  \\ 
\rule[-0.05cm]{0cm}{.35cm}$a_3$ & $-$1.34             $\pm$0.09GeV$^{-4}$   &  $-$1.37            $\pm$0.09GeV$^{-4}$\\ 
\rule[-0.05cm]{0cm}{.35cm}$\sqrt{s_{r1}}$ & 0.896  GeV (fixed)             &  0.896  GeV (fixed)    \\ 
\rule[-0.05cm]{0cm}{.35cm}$\sqrt{s_{r2}}$ & 1.344 $\pm$0.013GeV     &  1.347 $\pm$0.013GeV \\ 
\rule[-0.05cm]{0cm}{.35cm}$\sqrt{s_{r3}}$ & 1.647 $\pm$0.005GeV     &  1.657 $\pm$0.005GeV \\ 
\rule[-0.05cm]{0cm}{.35cm}$e_1$ & 1   (fixed)                                & 1 (fixed) \\ 
\rule[-0.05cm]{0cm}{.35cm}$e_2$ & 0.049           $\pm$0.008          & 0.067           $\pm$0.008 \\ 
\rule[-0.05cm]{0cm}{.35cm}$e_3$ & 0.304           $\pm$0.016          & 0.324           $\pm$0.016\\ 
\rule[-0.05cm]{0cm}{.35cm}$G_1$ & 0.034           $\pm$0.005GeV     & 0.049           $\pm$0.005GeV  \\ 
\rule[-0.05cm]{0cm}{.35cm}$G_2$ & 0.218           $\pm$0.043GeV     & 0.212           $\pm$0.043GeV \\ 
\rule[-0.05cm]{0cm}{.35cm}$G_3$ & 0.301           $\pm$0.018GeV     & 0.292           $\pm$0.018GeV  \\ 
\hline 
\end{tabular} 
\label{tab:Pinpa} 
\end{table} 

The result of our UFD for $f_P= f^{1/2}_1+f^{3/2}_1/2$  in both the elastic and inelastic regions 
can be seen in Fig.~\ref{fig:pwavedata}, as a dashed line with its corresponding orange uncertainty band.
Note the small contribution of the resonant shape around 1.4 GeV and a somewhat clearer resonant shape around 1650 MeV. The smallness of the partial-wave modulus in the 1.2 to 1.6 GeV region, translates into a rather large phase uncertainty  in that interval. The fit to data in the elastic region has  $\chi^2/dof=1.1$ and $\chi^2/dof=0.9$ in the inelastic one.

\subsubsection{$D$-waves}
\label{subsub:dwave}

For these waves,  we keep the same parameterizations and the same fitting strategy and data that we used in \cite{Pelaez:2016tgi}.

\vspace{0.3cm}
\underline{$D^{3/2}$ partial wave}\\

As we discussed in subsection~\ref{subsec:Ddata}, and seen in Figure~\ref{fig:D32phase}, only data from \cite{Estabrooks:1977xe} are available up to 1.74 GeV,
which are shown together with our fits. No inelasticity has been observed and we will thus apply our generic
elastic formalism with a conformal mapping as in Eq.~\eqref{eq:fintermsofphi}. We found that three terms are enough to get a good $\chi^2$ fit and we write
\begin{equation}
\cot\delta_2^{3/2}(s)=\frac{\sqrt{s}}{2q^5}\Big(B_0+B_1\omega(s)+B_2\omega(s)^2\Big),
\end{equation}
where the parameters $\alpha$ and $s_0$ are equal to those of the $S$ and $P$ $I=3/2$ partial waves, namely $\alpha=1.4,s_0=(1.84 \, {\rm GeV})^2$. The resulting fit
parameters are exactly the same as in  \cite{Pelaez:2016tgi}, but for the sake of completeness we repeat them here in Table \ref{tab:D32param}.
The resulting UFD curve can be seen in Fig.~\ref{fig:D32phase}, as a dashed line with an orange uncertainty band, without any remarkable feature except for its smallness. Actually, most studies neglect this wave.

\begin{table}[ht] 
\caption{Parameters of the $D^{3/2}$-wave. In GeV$^4$ units.} 
\centering 
\begin{tabular}{c c c } 
\hline\hline  
\rule[-0.05cm]{0cm}{.35cm}Parameter & UFD & CFD \\ 
\hline 
\rule[-0.05cm]{0cm}{.35cm}$B_0$ & $-$1.70  $\pm$0.12  & $-$1.78  $\pm$0.12\\
\rule[-0.05cm]{0cm}{.35cm}$B_1$ & $-$6.5  $\pm$1.7  & $-$7.88  $\pm$1.7\\ 
\rule[-0.05cm]{0cm}{.35cm}$B_2$ & $-$36.1 $\pm$8.7  & $-$56.4$\pm$8.7\\ 
\hline 
\end{tabular} 
\label{tab:D32param} 
\end{table}

\vspace{0.3cm}
\underline{$D^{1/2}$ partial wave}\\

As for the $S$ and $P$-waves, 
the $I$=1/2 $D$-wave is only measured together
with the $I$=3/2-wave in the 
$f_D \equiv f^{1/2}_2+f^{3/2}_2/2$
combination. However, we have seen that the $D^{3/2}$ is minuscule and overwhelmed by the $D^{1/2}$ contribution. The latter is dominated by  the  $K_2^*(1430)$ resonance, whose branching ratio to $\pi K$ is 
approximately 50\%, but also has smaller although significant decay branching ratios to $K^*(892)\pi$ and $K\rho(770)$. The centrifugal barrier suppression makes this wave very small until this resonance is felt. Hence,  it is most convenient to use our generic inelastic formalism, discussed in subsection~\ref{subsec:generalforms} above,  in our whole fitting range.
As we found in \cite{Pelaez:2016tgi}, it is  enough to consider a non-resonant background and a resonant-like form, to write:
\begin{equation}
f_2^{1/2}=\frac{S^b_0S^r_1-1}{2i\sigma(s)},
\label{eq:Dwave}
\end{equation}
where the background term is
\begin{equation}
S^b_0=e^{2ip(s)},
\end{equation}
with
\begin{eqnarray}
p(s)&=&\phi_0q_{\eta K}^5\Theta_{\eta K}(s)
+q_{\eta' K}^5(\phi_1+\phi_2q_{\eta' K}^2)\Theta_{\eta' K}(s),\nonumber
\end{eqnarray}
and $\Theta_{ab}=\Theta(s-(m_a+m_b)^2)$.
The resonant term for the $K_2^*(1430)$ shape, is written as
\begin{eqnarray}
S^r_1&=&\frac{s_{r1}-s+i(P_1-Q_1)}{s_{r1}-s-i(P_1+Q_1)},\\
P_1&=&e_1G_1 \frac{p_1(q_{\pi K})}{p_1(q_{\pi K}^r)}\left(\frac{q_{\pi K}}{q_{\pi K, r}}\right)^5,\nonumber\\
Q_1&=&(1-e_1)G_1 \frac{p_1(q_{\pi K})}{p_1(q_{\pi K}^r)}\left(\frac{q_{\eta K}}{q_{\eta K, r}}\right)^5\Theta_{\eta K}(s),\nonumber
\end{eqnarray}
with $p_1(q_{\pi K})=1+a q_{\pi K}^2$.
Note that here we parameterize our inelasticity allowing for two possible cusps due to the sharp two-particle thresholds of $K\eta$ and $K\eta'$.

For our unconstrained fit, we get exactly the same parameters we found in \cite{Pelaez:2016tgi}, listed here in Table \ref{tab:Dwave} for completeness.
The resulting curves can be seen as a dashed line in Fig.~\ref{fig:Dwavedata} with its corresponding orange uncertainty band attached. The shape of the $K^*_2(1430)$ is evident and well reproduced. The uncertainties are rather small except beyond 1.6 GeV where the two existing data sets separate significantly.

\begin{table}[ht] 
\caption{Parameters of the $D^{1/2}$ fit.} 
\centering 
\begin{tabular}{c c c } 
\hline\hline  
\rule[-0.05cm]{0cm}{.35cm}Parameters & UFD & CFD \\ 
\hline 
\rule[-0.05cm]{0cm}{.35cm}$\phi_0$ & 2.17         $\pm$0.26 GeV$^{-5}$     &  3.05         $\pm$0.26 GeV$^{-5}$\\ 
\rule[-0.05cm]{0cm}{.35cm}$\phi_1$ & $-$12.1       $\pm$1.7GeV$^{-5}$      & $-$16.6       $\pm$1.7GeV$^{-5}$\\ 
\rule[-0.05cm]{0cm}{.35cm}$\sqrt{s_{r1}}$ & 1.446 $\pm$0.002GeV        &  1.453  $\pm$0.002GeV\\ 
\rule[-0.05cm]{0cm}{.35cm}$e_1$ & 0.466           $\pm$0.006             & 0.455           $\pm$0.006\\ 
\rule[-0.05cm]{0cm}{.35cm}$G_1$ & 0.220           $\pm$0.010GeV        & 0.256           $\pm$0.010GeV\\ 
\rule[-0.05cm]{0cm}{.35cm}$a$ & $-$0.53             $\pm$0.16GeV$^{-2}$      &$-$0.82             $\pm$0.16GeV$^{-2}$\\
\hline 
\end{tabular} 
\label{tab:Dwave} 
\end{table}

\subsubsection{$F$-waves}
\label{subsub:fwave}

As we saw in subsection~\ref{subsec:Fdata} there are no
measurements of the $F^{3/2}$-wave below 2.5 GeV and we  will neglect it. For all means and purposes the
data in Fig.~\ref{fig:Fwavedata} is just given by the $F^{1/2}$-wave, showing a clear signal for a  $K_3^*(1780)$ resonance, whose  main branching ratios
are roughly $30\%$ to $K \rho(770)$ and around
20\% both to $\pi K$ and $K^*(892)\pi$. Note that scattering data 
have only been measured above 1.5 GeV due to the large kinematic suppression.  We will thus use, as in \cite{Pelaez:2016tgi}, an inelastic formalism dominated by that resonance. Namely
\begin{equation}
f_3^{1/2}=\frac{S^r_1-1}{2i\sigma(s)},
\label{eq:Fwave}
\end{equation}
with
\begin{eqnarray}
S^r_1&=&\frac{s_{r1}-s+i(P_1-Q_1)}{s_{r1}-s-i(P_1+Q_1)},\\
P_1&=&e_1G_1 \frac{p_1(q_{\pi K})}{p_1(q_{\pi K}^r)}\left(\frac{q_{\pi K}}{q_{\pi K, r}}\right)^{\!\!7},\nonumber\\
Q_1&=&(1-e_1)G_1 \frac{p_1(q_{\pi K})}{p_1(q_{\pi K}^r)}\left(\frac{q_{\eta K}}{q_{\eta K, r}}\right)^{\!\!7} \Theta_{\eta K}(s).
\nonumber \end{eqnarray}
Here, $p_1(q_{\pi K})=1+a q_{\pi K}^2$ and $\Theta_{\eta K}(s)=\Theta(s-(m_\eta+m_K)^2)$.

Since the data sample and the parameterization are the same, the UFD parameters for this wave are exactly those we obtained in \cite{Pelaez:2016tgi}, listed again here in Table \ref{tab:Fwave}, for completeness. The resulting curves can be seen as dashed lines and their corresponding orange uncertainty bands in Fig.~\ref{fig:Fwavedata}. The resonant shape of the $K_3^*(1780)$ is nicely reproduced. The uncertainty band is somewhat larger than for other waves because  the data error bars are also somewhat larger.

\begin{table}[ht] 
\caption{Parameters of the $F^{1/2}$-wave.} 
\centering 
\begin{tabular}{c c c } 
\hline\hline  
\rule[-0.05cm]{0cm}{.35cm}Parameters & UFD & CFD \\ 
\hline 
\rule[-0.05cm]{0cm}{.35cm}$\sqrt{s_{r1}}$ & 1.801 $\pm$0.013GeV    &   1.840         $\pm$0.013GeV\\ 
\rule[-0.05cm]{0cm}{.35cm}$e_1$ & 0.181           $\pm$0.006         &  0.173           $\pm$0.006\\ 
\rule[-0.05cm]{0cm}{.35cm}$G_1$ & 0.47           $\pm$0.05GeV    &  0.60           $\pm$0.05GeV\\ 
\rule[-0.05cm]{0cm}{.35cm}$a$ & $-0.88            \pm$0.10GeV$^{-2}$ & $-1.06\pm$0.10GeV$^{-2}$\\
\hline 
\end{tabular} 
\label{tab:Fwave} 
\end{table} 

\subsection{ $\pipikk$ Unconstrained Fits to Data}
\label{sec:UFDpipiKK}
Let us now consider the parameterization of the $t$-channel \pipikk.
Except for the $g_1^1$ partial wave, all other $\pipikk$ partial-wave parameterizations we  review here are those we introduced recently in our Roy-Steiner dispersive analysis \cite{Pelaez:2018qny}.
 Our main goal with the modifications in $g_1^1$ is to improve the description of the uncertainties associated with the data and make it flexible enough to achieve the refined level of accuracy we reach in this review.
 
 Contrary to the \pik case, which had many  resonant and non-resonant waves as well as elastic and inelastic cases, for \pipikk we only need to fit three partial waves, $g^0_0,g^1_1$ and $g^0_2$, in the inelastic regime. In general, we will describe them with phenomenological but elaborated combinations of Breit-Wigner-like forms, whose shape is clearly noticeable in the data. The only one treated differently is the $g^0_0$ whose difficulty lies in the existence of conflicting data sets that cover different energy regions. Also, the presence and parameters of some resonances there, like the $f_0(1370)$, are still somewhat controversial. Thus we have avoided resonant shapes in favor of a piece-wise but continuous and differentiable parameterization in terms of polynomials. Let us start describing this wave first.

\subsubsection{$g^0_0$ partial wave}

We saw in  Fig.~\ref{fig:g00data} that there are data in the whole region of interest 
for both the modulus $\vert g_0^0\vert$ and the phase
$\phi^0_0$. 
The data sets extend from $K\bar K$ threshold up to $t\simeq 2.4\,$GeV, but we do not fit that whole region
because from 2 GeV we will use Regge parameterizations.

For the $\phi_0^0$ phase we have already shown that, due to Watson's theorem, only one set of data makes sense. However, two sets of incompatible data
can be studied for the modulus. 
Thus, we will provide two 
alternative UFD parameterizations for $\vert g_0^0\vert$, both sharing the
same UFD parameterization for the $\phi^0_0$ phase.
Moreover, the applicability
of \pipikk dispersion relations only reaches 1.47 GeV, and this is why 
 in \cite{Pelaez:2016tgi} we decided to parameterize our
amplitudes by a two-piece function defined differently in the two regions of the inelastic regime explained in Section~\ref{subsec:S0data}. In Region I, Roy-Steiner equations will be used to test or constrain our parameterization, whereas
in Region II amplitudes are just used as input for dispersion relations.

 In order to parameterize the partial wave on each region, we will make use of Chebyshev polynomials, because they are simple and, in practice, yield very low correlations among their parameters. They are defined as:
\begin{eqnarray}
&&p_0(x)=1,\quad p_1(x)=x,     \nonumber\\
&&p_{n+1}(x)=2xp_n(x)-p_{n-1}(x).
\end{eqnarray}
Thus we first map each energy region  $i=I,II$ 
into the $x\in[-1,1]$ interval through the linear transformation 
\begin{equation}
x_i(t)=2\frac{\sqrt{t}-\sqrt{t_{min,i}}}{\sqrt{t_{max,i}}-\sqrt{t_{min,i}}}-1.
\end{equation}
Note that $p_n(1)=1$ and $p_n(-1)=(-1)^n$, for all $n$,
which will be used to ensure a smooth matching at  $\sqrt{s}=1.47\,$GeV, i.e. between the two pieces describing regions I and II, up to and including the first derivative (although only for the central value of the derivative). Between the elastic and inelastic regions, the matching at $t_K$ only ensures continuity, thus  allowing for a proper cusp behavior.

Now, for the $\phi^0_0$ phase, our parameterization is just
\begin{equation} \phi_0^0(t)=\left\{
\begin{array}{@{}rl@{}}
\sum^3_{n=0}{B_n p_n(x_I(t))}, & \text{Region I,}\\
 & \\
\sum^5_{n=0}{C_n p_n(x_{II}(t))},&\text{Region II,}
\end{array}
\right.
\end{equation}  
where we need to impose
\begin{align}
&B_0=\delta_0^{(0)}(t_K)+B_1-B_2+B_3,\\
&C_0=\phi_0^0(t_{max,I})+C_1-C_2+C_3-C_4+C_5,
\end{align}
to ensure, respectively, continuity at $K\bar K$ threshold as well as between Regions I and II.
Here, $\delta_0^{(0)}(t)$ is the \pipi phase shift that we take from 
\cite{GarciaMartin:2011cn}, leading to $\delta_0^{(0)}(t_K)=(226.5\pm1.3)^\degree$.
Continuity is enough at $t_K$ since there is a cut due to the opening of a new threshold. However, we will also impose a continuous derivative for the central value of our fit between Regions I and II, and thus the $C_1$ parameter is also fixed numerically.
The parameters of the fit are given in Table~\ref{tab:g00phase}. The total $\chi^2/dof=1.5$, which comes slightly larger than one due to the incompatibilities between data sets. 
Consequently, the
uncertainties of the parameters in Table~\ref{tab:g00phase} have been re-scaled by a factor $\sqrt{1.5}$.

\begin{table}[ht] 
\caption{Parameters of $\phi^0_0$ in the inelastic region. Note there is just one $\phi^0_0$ UFD common to the two UFD$_\text{B}$ and UFD$_\text{C}$ for $\vert g^0_0\vert$. However, once we constrain the fits there are two sets of parameters for the $\phi^0_0$ CFD$_\text{B}$ and CFD$_\text{C}$.} 
\centering 
\begin{tabular}{c c c c} 
\hline\hline  
\rule[-0.05cm]{0cm}{.35cm}Parameter & UFD & CFD$_\text{B}$ & CFD$_\text{C}$ \\ 
\hline 
\rule[-0.05cm]{0cm}{.35cm}$B_1$ & 23.6  $\pm$1.3  & 22.6 $\pm$1.3 & 24.0  $\pm$1.3\\ 
\rule[-0.05cm]{0cm}{.35cm}$B_2$ & 29.4  $\pm$1.3  & 28.1  $\pm$1.3 & 29.5  $\pm$1.3\\ 
\rule[-0.05cm]{0cm}{.35cm}$B_3$ & 0.6  $\pm$1.6  & 1.9 $\pm$1.6 &   0.7  $\pm$1.6\\ 
\rule[-0.05cm]{0cm}{.35cm}$C_1$ & 34.3932 fixed  & 29.2374 fixed & 27.6328 fixed\\ 
\rule[-0.05cm]{0cm}{.35cm}$C_2$ & 4.4  $\pm$2.6  & 4.8  $\pm$2.6 & 4.1  $\pm$2.6\\ 
\rule[-0.05cm]{0cm}{.35cm}$C_3$ &$-$32.9 $\pm$5.2  &$-$29.3 $\pm$5.2 &$-$29.1 $\pm$5.2\\ 
\rule[-0.05cm]{0cm}{.35cm}$C_4$ &$-$16.0 $\pm$2.2  &$-$12.4 $\pm$2.2 &$-$12.6 $\pm$2.2\\ 
\rule[-0.05cm]{0cm}{.35cm}$C_5$ &  7.4 $\pm$2.4  &  8.9 $\pm$2.4 &  8.5 $\pm$2.4\\ 
\hline 
\end{tabular} 
\label{tab:g00phase} 
\end{table}

In contrast to $\phi^0_0$, which only had one data set, for the modulus $\vert g^0_0\vert$
 we want to provide parameterizations for the two incompatible sets of data.
Hence, we have obtained two Unconstrained Fits to Data (UFD) in Region I:
 The one fitting Brookhaven-II data \cite{Longacre:1986fh} is called UFD$_\text{B}$, whereas
the one fitting the ``Combined'' Argonne \cite{Cohen:1980cq} and Brookhaven-I \cite{Etkin:1981sg} data is labeled UFD$_\text{C}$. 
Both of them use the same data in Region II and thus they are almost identical there, although  with slightly different parameters due to the different matching with Region I.
All in all, we use: 
\begin{equation} 
\vert g_0^0(t)\vert=\left\{
\begin{array}{@{}rl@{}}
\sum^3_{n=0}{D_n p_n(x_I(t))}, & \text{Region I,}\\
 & \\
\sum^4_{n=0}{F_n p_n(x_{II}(t))},&\text{Region II,}
\end{array}
\right.
\label{eq:ufd}
\end{equation}  
where, in order to ensure continuity between the two regions, we impose:
\begin{equation}
F_0=\vert g_0^0(t_{max,I})\vert+F_1-F_2+F_3-F_4.
\end{equation}
In addition, $F_1$ 
is fixed numerically to secure a continuous derivative for the central value.

The parameters for $\vert g_0^0\vert$ of both the UFD$_\text{B}$ and UFD$_\text{C}$  are listed in Tables~\ref{tab:g00modufd} and \ref{tab:g00modufd2}, respectively.
Let us remark that both have $\chi^2/dof\sim 1$. Their respective curves are shown as dashed lines with orange uncertainty bands in Fig.~\ref{fig:g00data}.

\begin{table}[ht] 
\begin{minipage}[t]{0.49\linewidth}
\caption{Parameters of the UFD$_\text{B}$ and CFD$_\text{B}$ fits to $\vert g^0_0\vert $.} 
\centering 
\begin{tabular}{c c c } 
\hline\hline  
\rule[-0.05cm]{0cm}{.35cm}Parameter & UFD$_\text{B}$ & CFD$_\text{B}$ \\ 
\hline 
\rule[-0.05cm]{0cm}{.35cm}$D_0$ & 0.588 $\pm$0.01  & 0.591 $\pm$0.01\\
\rule[-0.05cm]{0cm}{.35cm}$D_1$ &$-$0.380 $\pm$0.01  &$-$0.339 $\pm$0.01\\ 
\rule[-0.05cm]{0cm}{.35cm}$D_2$ & 0.12 $\pm$0.01  & 0.13 $\pm$0.01\\ 
\rule[-0.05cm]{0cm}{.35cm}$D_3$ &$-$0.09  $\pm$0.01  &$-$0.12  $\pm$0.01\\ 
\rule[-0.05cm]{0cm}{.35cm}$F_1$ &$-$0.04329  fixed  &$-$0.04312  fixed\\
\rule[-0.05cm]{0cm}{.35cm}$F_2$ &$-$0.008  $\pm$0.009  &$-$0.008  $\pm$0.009\\ 
\rule[-0.05cm]{0cm}{.35cm}$F_3$ &$-$0.028  $\pm$0.007  &$-$0.034 $\pm$0.007\\ 
\rule[-0.05cm]{0cm}{.35cm}$F_4$ & 0.026  $\pm$0.007  & 0.038 $\pm$0.007\\ 
\hline 
\end{tabular} 
\label{tab:g00modufd} 
\end{minipage}
\hfill
\begin{minipage}[t]{0.49\linewidth}
\caption{Parameters of the UFD$_\text{C}$ and CFD$_\text{C}$ fits to $\vert g^0_0\vert $.}
\centering 
\begin{tabular}{c c c } 
\hline\hline  
\rule[-0.05cm]{0cm}{.35cm}Parameter & UFD$_\text{C}$ & CFD$_\text{C}$ \\ 
\hline 
\rule[-0.05cm]{0cm}{.35cm}$D_0$ & 0.462 $\pm$0.008  & 0.446 $\pm$0.008\\
\rule[-0.05cm]{0cm}{.35cm}$D_1$ &$-$0.267 $\pm$0.013  &$-$0.236 $\pm$0.013\\ 
\rule[-0.05cm]{0cm}{.35cm}$D_2$ & 0.11 $\pm$0.01  & 0.10 $\pm$0.01\\ 
\rule[-0.05cm]{0cm}{.35cm}$D_3$ &$-$0.078  $\pm$0.009  &$-$0.087 $\pm$0.009\\ 
\rule[-0.05cm]{0cm}{.35cm}$F_1$ &$-$0.04153 fixed   &$-$0.03765 fixed \\
\rule[-0.05cm]{0cm}{.35cm}$F_2$ &$-$0.010  $\pm$0.008  &$-$0.016 $\pm$0.008\\ 
\rule[-0.05cm]{0cm}{.35cm}$F_3$ &$-$0.023  $\pm$0.007  &$-$0.023 $\pm$0.007\\ 
\rule[-0.05cm]{0cm}{.35cm}$F_4$ & 0.021  $\pm$0.006  & 0.028 $\pm$0.006\\ 
\hline 
\end{tabular} 
\label{tab:g00modufd2} 
\end{minipage}
\end{table}

\subsubsection{$g^1_1$ partial wave}

For this wave, we will keep the very same parameterization we used in \cite{Pelaez:2018qny}, which we actually took from \cite{Kuhn:1990ad, Buettiker:2003pp}.
However, we will consider some minor changes in the way we implement the fit to the  $g^1_1$ partial wave, although we will not modify the data choice, shown in Fig.~\ref{fig:g11data}. These minor changes are due to the way we added systematic uncertainties in the original fit of \cite{Pelaez:2018qny}, which we have decided to improve here.  In our  work, 
\cite{Pelaez:2018qny} we included these systematic uncertainties by multiplying the errors of all data points by an overall factor so that we got a global $\chi^2/dof\sim 1$ as is customary. However, 
most of the $\chi^2/dof$ is coming from the phase in the inelastic region, where, as seen in Fig.~\ref{fig:g11data}, the data are clearly of less quality. Thus, when multiplying all data by the same common factor we were producing an overestimated uncertainty in the other regions, which had a decent $\chi^2/dof$ from the start. Hence, here we will only multiply the uncertainties of the phase data by a factor of $\sim 1.4$ in the inelastic region, where a fit without any systematic uncertainties would yield $\chi^2/dof\sim 2$.

Apart from updating our strategy to estimate systematic uncertainties, we use the same functional form as in \cite{Kuhn:1990ad,Anderson:1999ui,Buettiker:2003pp,Pelaez:2018qny}
\begin{eqnarray}
g_1^1(t)&=&\frac{C}{\sqrt{1+r_1 q_\pi^2(t)}\sqrt{1+r_1 q_K^2(t)}} \label{ec:g11} \Big\{ \overline{BW}(t)_\rho+(\beta
+\beta_1 q_K^2(t))\overline{BW}(t)_{\rho'}+(\gamma+\gamma_1 q_K^2(t))\overline{BW}(t)_{\rho''}\Big\}. \nonumber
\end{eqnarray}
This parameterization is devised to accommodate the three vector resonances $\rho(770)$, $\rho'=\rho(1450)$, $\rho''=\rho(1700)$, 
by means of a combination of three Breit-Wigner-like shapes:
\begin{eqnarray}
\overline{BW}(t)_V&=&\frac{m_V^2}{m_V^2-t-i\Gamma_V\sqrt{t}\,\frac{2G_\pi(t)+G_K(t)}{2G_\pi(m_V^2)}},\nonumber\\
G_P(t)&=&\sqrt{t}\left(\frac{2 q_P(t)}{\sqrt{t}}\right)^3.
\end{eqnarray}
Here $m_V$ and $\Gamma_V$ stand for the ``Breit-Wigner'' mass and width of
the resonances under consideration, which are not necessarily those obtained from poles within a rigorous dispersive approach.  For our purposes they are just phenomenological parameters, whose values, together with those of the other parameters are listed in Table~\ref{tab:g11wave}.

\begin{table}[ht] 
\caption{Parameters of the $g^1_1$ wave.
Masses and widths are given in GeV whereas, $C$, $\beta_1,\gamma_1$ and $r_1$
are given in GeV$^{-2}$.}
\centering 
\begin{tabular}{c c c } 
\hline\hline  
\rule[-0.05cm]{0cm}{.35cm}Parameter & UFD & CFD \\ 
\hline 
\rule[-0.05cm]{0cm}{.35cm}$m_\rho$ & 0.7759 $\pm$0.0010 & 0.7756 $\pm$0.0010\\
\rule[-0.05cm]{0cm}{.35cm}$\Gamma_\rho$ &0.1517 $\pm$0.0016  &0.1509 $\pm$0.0016\\ 
\rule[-0.05cm]{0cm}{.35cm}$m_{\rho'}$ & 1.440$\pm$0.031 & 1.464$\pm$0.031\\
\rule[-0.05cm]{0cm}{.35cm}$\Gamma_{\rho'}$ & 0.310$\pm$0.016  & 0.339$\pm$0.016\\ 
\rule[-0.05cm]{0cm}{.35cm}$m_{\rho''}$ & 1.72 & 1.72 \\
\rule[-0.05cm]{0cm}{.35cm}$\Gamma_{\rho''}$ & 0.25  & 0.25\\ 
\rule[-0.05cm]{0cm}{.35cm}$C$ & 1.29 $\pm$0.08  & 1.35 $\pm$0.08\\ 
\rule[-0.05cm]{0cm}{.35cm}$r_1$ & 4.54 $\pm$0.64  & 3.84 $\pm$0.64\\ 
\rule[-0.05cm]{0cm}{.35cm}$\beta$ & $-$0.163 $\pm$0.004  & $-$0.167 $\pm$0.004\\ 
\rule[-0.05cm]{0cm}{.35cm}$\beta_1$ & 0.36 $\pm$0.02  & 0.35 $\pm$0.02\\ 
\rule[-0.05cm]{0cm}{.35cm}$\gamma$ & 0.09 $\pm$0.02  & 0.08 $\pm$0.02\\ 
\rule[-0.05cm]{0cm}{.35cm}$\gamma_1$ & $-$0.03 $\pm$0.03  & 0.06 $\pm$0.03\\
\hline 
\end{tabular} 
\label{tab:g11wave} 
\end{table}

The resulting curves of our unconstrained fit to data are shown in Fig.~\ref{fig:g11data} as dashed lines with an orange uncertainty band.
Note that our fit also describes the phase $\phi^1_1$ in the ``unphysical'' region below the $K \bar K$ threshold, which, given that only two-pions are observed there,  is nothing but the \pipi phase shift according to Watson's theorem. This ``unphysical'' region is completely dominated by the $\rho(770)$ resonance. Observe that the uncertainties of our fit are fairly small except for the phase above 1.4 GeV.

\subsubsection{$g^0_2$ partial wave}

For this wave, we keep the same parameterization, data sample, and fit strategy 
for the unconstrained fit that we followed in \cite{Pelaez:2018qny}. Our UFD is therefore exactly the same, but we provide it here again for completeness.
The CFD will of course change because here we will use the dispersive constraints 
simultaneously on \pik and \pipikk, instead of just the latter, as we did in \cite{Pelaez:2018qny}. In practice, our \cite{Pelaez:2018qny} parameterization had two pieces.
One piece, providing both the modulus and the phase for the ``physical region'' above $t_K$, where actual \pipikk data exist, and another piece, just for the phase below $t_K$. The latter, due to Watson's theorem, is given by the isoscalar angular momentum-2 phase shift of \pipi scattering, for which we take the parameterization obtained in the dispersive analysis of \cite{GarciaMartin:2011cn}.

We start by describing the physical region and recall that, as discussed in subsection \ref{subsec:Ddata}, the main resonances in this partial wave are the $f_2(1270)$ and $f'_2(1525)$, which are well established.  Actually, there is a clear peak for the former in the Brookhaven II data we showed in Fig.~\ref{fig:g02data} and a hint of the latter, which appears as a hunchback to the right of the main peak. However, the  $f'_2(1525)$ mass and width used in the Brookhaven II and III fits are at odds with their
present values. This will be amended in our fits. In addition, we have considered an additional $f_2(1810)$, already introduced in the Brookhaven II and III analyses \cite{Longacre:1986fh,Lindenbaum:1991tq}, contrary to what the Brookhaven collaboration did in \cite{Etkin:1982se}.
Nevertheless, the reader should keep in mind that our dispersive representations do not reach such high energies. Thus, for us, this 1.8 GeV resonance is just a convenient way of parameterizing the rise in the data on $\vert g^0_2\vert $ starting around 1.8 GeV. However, this choice  implies a particular shape for the phase at those energies, for which there are no data.

 We then use a phenomenological formula, rather similar to the one used for $g^1_1$,  given by
\begin{equation}
\hat g^0_2(t)=\frac{C\sqrt{[q_\pi(t)q_K(t)]^5}}
{\sqrt{t}\sqrt{1+r_2^2 q_\pi^4(t)}\sqrt{1+r_2^2 q_K^4(t)}}
\Big\{ e^{i\phi_1}BW(t)_1 +\beta e^{i\phi_2}BW(t)_2 
+ \gamma e^{i\phi_3}BW(t)_3 \Big\}, 
\label{eq:g02}
\end{equation}
with 
\begin{eqnarray}
BW(t)_T&=&\frac{m^2_T}{m_T^2-t-{\it i}m_T\Gamma_T(t)},\label{eq:BW}\\
\Gamma_T(t)&=&\Gamma_T\left(\frac{q_T(t)}{q_T(m_T^2)}\right)^{5}\frac{m_T}{\sqrt{t}}
\frac{D_{2}(r\, q_T(m_T^2))}{D_{2}(r\,q_T(t))}, \nonumber
\end{eqnarray}
 where $D_{2}(x)=9+3x^2+x^4$
provides the familiar Blatt-Weisskopf barrier factor for $\ell=2$, with 
a typical $r=5\,\gev^{-1}\,\simeq 1\,$fm. The index $T=1,2,3$ refers to the $f_2(1270)$, $f_2'(1525)$ and $f_2(1810)$ respectively.

Concerning the unphysical region, $t<t_K$, the relevant observation is that, since the 
contribution of the four pion state is negligible, for all means and purposes $\pi\pi$ scattering is elastic there. Then
 Watson's theorem allows us to identify $\phi_2^0=\delta_2^{(0)}$, 
where  $\delta_2^{(0)}$ is the $\pi\pi$-scattering
phase shift. At first, one could think about taking directly the 
result obtained in the \pipi dispersive analysis in \cite{GarciaMartin:2011cn}. However, we want a continuous matching with our parameterization. 
Thus, as we did in \cite{Pelaez:2016tgi}
 we just fit $\delta_2^{(0)}$ to the CFD result 
from \cite{GarciaMartin:2011cn}, using a truncated conformal expansion similar to
that in \cite{GarciaMartin:2011cn} but with one more parameter $B_2$ that is then fixed to ensure 
the continuous matching of $g^0_2$ at $K\bar K$ threshold. Namely:
\begin{eqnarray}
\cot\phi_2^{0}(t)&=&\frac{f^{1/2} }{2q_\pi^5}\big(m_{f_2(1270)}^2-t\big)m_\pi^2\left(
B_0+B_1 w(t)+B_2w(t)^2\right),
\nonumber\\
w(t)&=&\frac{\sqrt{t}-\sqrt{t_0-t}}{\sqrt{t}+\sqrt{t_0-t}},\quad
t_0^{1/2}=1.05 \, \gev, \quad
  \label{eq:D0lowparam} 
\end{eqnarray}
where
\begin{equation}
  B_2\,\omega(t_K)^2=\frac{q_\pi^5(t_K)\cot(\phi^0_2(t_K))}{m_K\big(m_{f_2(1270)}^2-t_K\big)m_\pi^2} -B_0-B_1\,\omega(t_K) ,
\end{equation}
has been fixed by continuity with the piece above $t_K$ in Eq.~\eqref{eq:g02}.

Let us note that using these parameterizations to fit the data as such, we would reproduce  with remarkable accuracy the 
description of the CFD phase-shift in \cite{GarciaMartin:2011cn}, whereas in the 
inelastic part we would obtain a $\chi^2/dof=1.4$. Looking at Fig.~\ref{fig:g02data}, this is not really due to the existence of any other physical feature but rather it seems that there is some small systematic uncertainty. Thus, as we did in \cite{Pelaez:2018qny},  we have re-scaled the data uncertainties in that region by a factor of $\sim 1.2$ and refitted. 
In Table~\ref{tab:g20wave}  we provide the parameters obtained when fitting together the inelastic part and the CFD phase-shift in \cite{GarciaMartin:2011cn}.
Once again, the latter is reproduced very nicely, but, in addition, in \cite{Pelaez:2018qny} we checked that the phase in the elastic regime is also compatible within uncertainties
with the dispersive analysis of the $\pi \pi$ $D$-wave 
using Roy and GKPY equations in \cite{Bydzovsky:2016vdx}. 
In the inelastic region, our uncertainty band covers quite conservatively the data collection.

The
resulting curves for our UFD $D$-wave parameterization are shown in Fig.~\ref{fig:g02data} as a dashed line with an orange uncertainty band. Note that for the phase we are also giving a few points obtained from \pipi scattering by Hyams et al.~\cite{Hyams:1973zf}.
Of course, the phase in the inelastic region is a pure prediction based on our assumption that it is dominated by the effect of resonances that are fitted to describe the data on the modulus. We will see that this assumption fares reasonably well against the dispersive checks. 

 Let us remark once again that, as can be seen in Fig.~\ref{fig:g02data} the original fit to data by the Brookhaven collaboration~\cite{Etkin:1982se} does not match continuously with the $\pi \pi$ $D$-wave phase shift at the $K \bar K$ threshold, thus violating Watson's theorem. On top of that, they did not include the $f_2(1810)$ resonance in their first analysis, which produces a different phase at higher energies. Our $g^0_2(t)$ does include all these features instead. 
 At this point, it is worth noticing that  the Brookhaven collaboration measured $|g^0_0-g^0_2|$ and extracted their $g^0_0$ assuming their $g^0_2$ model. It is therefore pertinent to reconsider the extraction of $g^0_0(t)$ partial wave phase. This is the reason why we give in~\ref{app:g00alt} an alternative solution for $\phi^0_0(t)$, including its dispersive constrained results. Fortunately, the effect of this wave on the other wave dispersive constraints is very small, and using our main fit or the alternative one is irrelevant. For the $g^0_0$ dispersive treatment, the differences between the main UFD fit or the alternative one appear at high energies, beyond the applicability range of our dispersion relations, which barely change their output. This is why the alternative solution is relegated to the appendix.

\begin{table}[ht] 
\caption{Parameters of the $g^0_2$ wave.} 
\centering 
\begin{tabular}{c c c } 
\hline\hline  
\rule[-0.05cm]{0cm}{.35cm}Parameter & UFD & CFD \\ 
\hline 
\rule[-0.05cm]{0cm}{.35cm}$m_{f_2(1270)}$ & 1.271 $\pm$0.0036GeV  & 1.272 $\pm$0.0036GeV\\
\rule[-0.05cm]{0cm}{.35cm}$m_{f'_2(1525)}$ & 1.522 $\pm$0.005 GeV  & 1.522 $\pm$0.005 GeV\\ 
\rule[-0.05cm]{0cm}{.35cm}$m_{f_2(1810)}$ & 1.806 $\pm$0.017 GeV  & 1.800 $\pm$0.017 GeV\\ 
\rule[-0.05cm]{0cm}{.35cm}$\Gamma_{f_2(1270)}$ & 0.187 $\pm$0.009 GeV  & 0.184 $\pm$0.009 GeV\\
\rule[-0.05cm]{0cm}{.35cm}$\Gamma_{f'_2(1525)}$ & 0.108 $\pm$0.016 GeV  & 0.116 $\pm$0.016 GeV\\
\rule[-0.05cm]{0cm}{.35cm}$\Gamma_{f_2(1810)}$ & 0.201 $\pm$0.028 GeV  & 0.180 $\pm$0.028 GeV\\
\rule[-0.05cm]{0cm}{.35cm}$\phi_{f_2(1270)}$ & $-$0.049 $\pm$0.015 & $-$0.081 $\pm$0.015 \\
\rule[-0.05cm]{0cm}{.35cm}$\phi_{f'_2(1525)}$ & 2.62 $\pm$0.16  & 2.58 $\pm$0.16\\
\rule[-0.05cm]{0cm}{.35cm}$\phi_{f_2(1810)}$ & $-$0.72$\pm$0.16  & $-$0.88$\pm$0.16  \\
\rule[-0.05cm]{0cm}{.35cm}$B_0$ & 12.5 $\pm$ 0.4 & 12.4 $\pm$ 0.4  \\
\rule[-0.05cm]{0cm}{.35cm}$B_1$ & 10.3 $\pm$ 1.0 & 13.2 $\pm$ 1.0  \\
\rule[-0.05cm]{0cm}{.35cm}$C$ & 1.82 $\pm$ 0.09 GeV$^{-2}$ & 1.80 $\pm$ 0.09  GeV$^{-2}$\\
\rule[-0.05cm]{0cm}{.35cm}$r_2^2$ & 6.68 $\pm$ 0.72 GeV$^{-4}$ & 6.85 $\pm$ 0.72 GeV$^{-4}$ \\
\rule[-0.05cm]{0cm}{.35cm}$\beta$ & 0.070 $\pm$ 0.016 & 0.083 $\pm$ 0.016  \\
\rule[-0.05cm]{0cm}{.35cm}$\gamma$ & 0.093 $\pm$ 0.02 & 0.103 $\pm$ 0.02  \\
\hline 
\end{tabular} 
\label{tab:g20wave} 
\end{table}

\subsubsection{\pipikk partial waves with $\ell>3$}
\label{sec:higherwaves}

These waves are going to be just input for our dispersion relations and their contributions will be in general small. As commented in subsection \ref{subsec:higherpw}, there are no scattering data and
thus, as we did in \cite{Pelaez:2018qny},  we will simply  use Breit-Wigner descriptions with the averaged parameters listed in the RPP.
In particular, for the $g^1_3(t)$ wave we will consider a single $\rho_3(1690)$ resonance, whereas for
the $\ell=4$ partial wave, we consider the $f_4(2050)$. Nevertheless, the latter will only be input for the $g^0_2(t)$ dispersion relation since its contribution is completely
negligible for the $g^0_0(t)$.

\subsection{ High-energy region. Regge parameterizations}
\label{sec:ufdregge}

The dispersive integrals we have seen in Section \ref{sec:DR} extend to infinite energy, although the integrands will be sufficiently suppressed at large energies to make them converge. However, we need another description, since at high energies   the partial-wave expansion is no longer valid. From the theoretical point of view this is because, strictly speaking, it is a low-energy expansion. From the practical point of view,  it is also because we only have data on a few partial waves with the lowest angular momentum and not too high energies. In particular, there are no data for $I=3/2$ \pik scattering above 1.74 GeV, although these waves present a rather monotonous behavior and it seems fine to extrapolate them a little further. For this reason, 
we have chosen  $1.84$ GeV, as done in~\cite{Pelaez:2016tgi},
as the energy beyond which we will not use \pik partial waves anymore.
Concerning  \pipikk scattering, we have chosen to stop using partial waves at 2 GeV.  It is true that data for the $g_0^0$ and $g_2^0$ waves reaches as high as 2.4 GeV, however, the $g_1^1$ scattering data ends at 1.6 GeV. Nevertheless, since the $\rho''(1700)$ is well established in the RPP and has a 250 MeV width, we think we have a fairly reasonable description of all waves up to those 2 GeV. 

The problem now is that no direct 
high-energy experimental data on $\pipikk$ nor on
$\pik$ exist. 
Notwithstanding, the high energy behavior of both processes 
can be confidently predicted from the factorization of Regge amplitudes of other hadron-scattering processes. In this regime, hadron-hadron scattering is understood as the  contribution of
the so-called Pomeron exchange (a transfer of momentum and no other quantum number, fairly well understood in terms of colorless exchanges of gluons) or 
families of resonances exchanged in the $t$-channel, called Reggeons. For pedagogical introductions we recommend \cite{Collins:1977jy,Donnachie:2002en,Gribov:2009zz}.
This is a well-established approach that we already used in \cite{Pelaez:2016tgi,Pelaez:2018qny}. In particular, for \pik, which at high energies and low $t$ is dominated by the exchange of non-strange Reggeons, we will follow our analysis in \cite{Pelaez:2016tgi} and use here the Regge-model factorization analysis presented in \cite{Pelaez:2003ky}, although later on
updated  in  \cite{GarciaMartin:2011cn,Pelaez:2016tgi}.
In contrast,  to describe 
\pipikk above 2 GeV, which involves the exchange of strange Reggeons, we will follow our analysis in \cite{Pelaez:2018qny} and use the asymptotic forms of 
the Veneziano model \cite{Veneziano:1968yb,Lovelace:1969se,Shapiro:1969km,Kawarabayashi:1969yd}, 
with the updated parameters in 
\cite{Buettiker:2003pp}, but relaxing the degeneracy with the $\rho$ trajectory assumed for simplicity in \cite{Pelaez:2003ky}. 

\vspace{.3cm}
\underline{High-energy Regge parameterizations for \pik}\\

For the \pik symmetric amplitude we have both the Pomeron $P(s,t)$ and the $f_2$-Reggeon, or 
$P'(s,t)$, exchanges:
\begin{eqnarray}
\im F^{+}_{\pi K}(s,t)=
\frac{\im F^{(I_t=0)}_{\pi K}(s,t)}{\sqrt{6}}=\frac{4\pi^2}{\sqrt{6}} f_{ K/\pi}\left[P(s,t)+rP'(s,t)\right].
\label{eq:reggef+}
\end{eqnarray}
Since we follow the notation in \cite{Pelaez:2003ky} the Regge $P$ and $P'$ amplitudes refer to \pipi scattering and for simplicity absorb the $\pi\pi-$Reggeon factor
\begin{eqnarray}
P(s,t)&=&\beta_P\psi_P(t)\alpha_P(t)\frac{1+\alpha_P(t)}{2}e^{\hat b t}\left(\frac{s}{s'}\right)^{\alpha_P(t)},\nonumber\\
P'(s,t)&=&\beta_{P'}\psi_{P'}(t)\frac{\alpha_{P'}(t)(1+\alpha_P(t))}{\alpha_{P'}(0)(1+\alpha_P(0))}e^{\hat bt}\left(\frac{s}{s'}\right)^{\alpha_{P'}(t)},\nonumber\\
\alpha_P(t)&=&1+t\alpha'_P, \qquad \psi_P=1+c_Pt,\nonumber\\
\alpha_{P'}(t)&=&\alpha_{P'}(0)+t\alpha'_{P'}, \qquad \psi_{P'}=1+c_{P'}t.
\label{eq:Pomeron}
\end{eqnarray}
Factorization allows us then to convert one 
$\pi\pi-$Reggeon vertex into a $KK-$Reggeon vertex by multiplying by the  $f_{K/\pi}$ factor, as we have done in Eq.~\eqref{eq:reggef+}. The constant $r$ is related to the branching ratio of the $f_2(1270)$ resonance to $\bar KK$ instead of $\pi\pi$.

\begin{figure}
\centering
\includegraphics[scale=1.3]{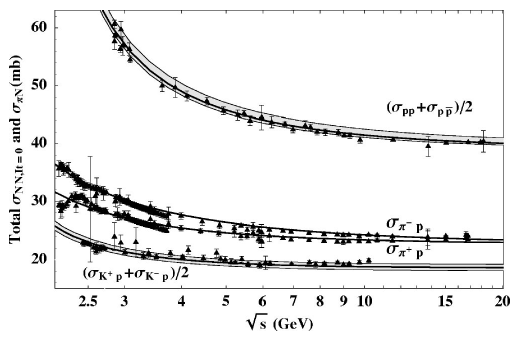}
\caption{High-energy hadron-hadron total cross-section data versus the Regge description in \cite{Pelaez:2003ky}, that we also use here although updated to the values in \cite{GarciaMartin:2011cn}. The figure is taken from \cite{Pelaez:2003ky}. }
\label{fig:totalsigma}
\end{figure}

Just for illustration, the extraction of $f_{K/\pi}$ and $r$
is particularly simple from combinations of 
data on total cross sections for proton-proton, proton-antiproton, $K^+$-proton, and $K^-$-proton,
which we show in Fig.~\ref{fig:totalsigma}. 
Then one should recall that the optical theorem tells us that the total cross section is proportional to the imaginary part of the {\it forward}, i.e. $t=0$, amplitude.  Hence, 
following \cite{Pelaez:2003ky}, for large $s$, it is possible to write:
\begin{eqnarray}
\sigma_{pp}+\sigma_{p\bar p}&\simeq& \frac{4\pi^2}{\lambda^{1/2}(s,m_p^2,m_p^2)}f_{N/\pi}^2\Big[P(s,0)+(1+\epsilon)P'(s,0)\Big],\\
\sigma_{K^+p}+\sigma_{K^-p}&\simeq&
\frac{4\pi^2}{\lambda^{1/2}(s,m_K^2,m_p^2)}f_{N/\pi}f_{K/\pi}\Big[P(s,0)+rP'(s,0)\Big],
\end{eqnarray}
where $\lambda$ is the K\"all\'en function defined in Eq~\eqref{eq:Kallenfunction}.
Fits of these two expressions to the data at sufficiently high energies, which are also shown in  Fig.~\ref{fig:totalsigma} to describe the data very well, yield 
the desired factors \cite{Pelaez:2003ky}. 

\begin{figure}[!ht]
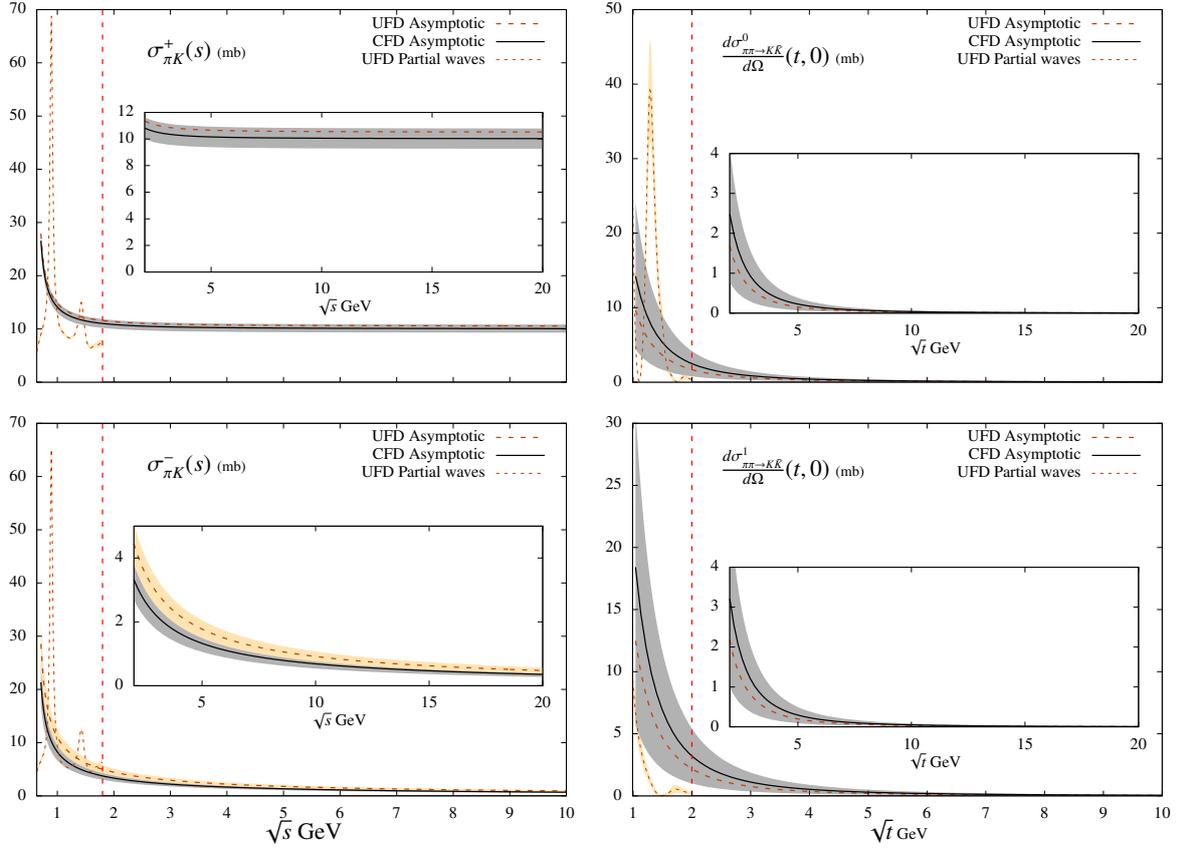

\centering
\resizebox{0.95\textwidth}{!}{\input{figures/fplusregge.tex} \input{figures/g0regge.tex}} \\
\resizebox{0.95\textwidth}{!}{\input{figures/fminusregge.tex} \input{figures/g1regge.tex}}\\
\caption{High-energy total cross sections for $\pi K$ or differential forward cross-sections $\pi\pi \to K \bar{K}$ scattering (Left and right panels respectively), in millibarns.
They are both proportional to their corresponding imaginary part of the forward amplitudes provided in the main text. Note that for the symmetric \pik scattering $\sigma^+(s)$ tends asymptotically to a constant $\simeq 10.3\,$mb. All other cross-sections of interest tend to zero asymptotically.
Regge parameterizations are used only above 1.84 GeV for \pik and 2 GeV for \pipikk (vertical lines). This is shown in detail in the insets up to 20 GeV. We show curves up to 20 GeV, since the dispersive integrals are suppressed by several factors at high energies, and their contributions from even higher energies are very small. Below 2 GeV, we use the corresponding partial wave expansion, whose central value is also shown as a dashed curve. 
Nevertheless, it can be noticed that, in that region, Regge theory is a crude average of the result obtained from partial waves. Note however that, plotting cross-sections, we cannot show the  \pipikk case below $K\bar K$ threshold, where more resonances appear. The gray and orange bands stand for the uncertainties of the CFD and UFD, respectively, although when they overlap we do not plot the uncertainty of the latter, to simplify the view.}
\label{fig:regge}
\end{figure}

Several considerations about the Pomeron we use are now in order.  When it dominates, the predicted cross-section tends to a constant value. For example,  the Pomeron $\pi K$ cross-section tends asymptotically to $\simeq 10.3\,$mb, as seen in the top left panel of Fig.~\ref{fig:regge}.
This is roughly twice the $\simeq 5\pm2.5\,$mb value used in 
\cite{Buettiker:2003pp},  inspired
by the $\pi\pi$  scattering
 asymptotic value of $6\pm 5\,$mb. However, this $\pi\pi$ value was  revisited later by members of the same group \cite{Caprini:2011ky} 
yielding $12.2\pm 0.1\,$mb for $\pi\pi$ scattering, thus supporting our larger value $\simeq10\pm3\,$mb for $\pi K$.
  The constant asymptotic value  is a simplification that works rather well up to 20-30 GeV, good enough for the accuracy we need in our amplitudes. However, it is well known now that the Pomeron cross-section, although it cannot grow asymptotically as a power of $s$, could still grow if it does not violate the Froissart bound $\sigma_P\leq \log^2s$. This slow growth of hadron total sections has been observed (see the ``Total Hadronic Cross Sections'' review at the RPP \cite{pdg}), and it would require an slightly more complicated Pomeron expression than Eq.~\eqref{eq:Pomeron}, which can also be found in \cite{Pelaez:2003ky}. Nevertheless, for our purposes here the simple Pomeron expression in Eq.~\eqref{eq:Pomeron} is accurate enough.

In addition, $r$ is expected to come out very small, and indeed it does, due to the small coupling of the $f_2(1270)$ resonance to $K\bar K$, which suppresses the $P'$ contribution. Thus one may wonder whether more Regge subleading contributions should  be considered in Eq.~\eqref{eq:Pomeron}. However, the
next Reggeon, associated with the $f_2'(1525)$ trajectory, which couples strongly to $K\bar K$ scattering, couples very little to nucleons or pions. This, together with its small intercept $\alpha_{f'_2}\simeq -0.3$, makes this Reggeon contribution very suppressed.
Therefore, the $KN$ and $\pi K$ isospin zero high-energy exchange is almost only due to the Pomeron, as assumed in Eq.~\eqref{eq:Pomeron}.

Let us now consider the antisymmetric amplitude, which is dominated by the exchange of a Reggeized-$\rho$ resonance. Then, following \cite{Pelaez:2003ky} we write:
\begin{eqnarray}
\im F^{-}_{\pi K}(s,t)=\frac{\im F^{(I_t=1)}_{\pi K}(s,t)}{2}=2\pi^2 g_{ K/\pi}\im F^{(I_t=1)}_{\pi\pi}(s,t),
\label{eq:reggef-}
\end{eqnarray}
where now $g_{ K/\pi}$ is the factorization constant that converts the $\pi\pi-$Reggeized-$\rho$
vertex into a $K\bar K-$Reggeized-$\rho$ vertex.
The Reggeized-$\rho$ contribution is:
\begin{eqnarray}
\im F^{(I_t=1)}_{\pi \pi}(s,t)&=& \beta_{\rho}\frac{1+\alpha_{\rho}(t)}{1+\alpha_{\rho}(0)}\varphi(t)e^{\hat bt}\left(\frac{s}{s'}\right)^{\alpha_{\rho}(t)},\nonumber\\
\alpha_{\rho}(t)& =&\alpha_{\rho}(0)+t\alpha'_{\rho}+\frac{1}{2}t^2\alpha''_{\rho},\nonumber\\
\varphi(t)& =&1+d_{\rho}t+e_{\rho}t^2.
\label{eq:reggerho}
\end{eqnarray}
There is not much information about the value of $g_{ K/\pi}$ although, in practice, it  can be obtained from factorization. Its extraction is somewhat more complicated than the simple case for the Pomeron trajectory that we used as an example above,
and we refer the reader to  \cite{Pelaez:2003ky}.

Let us now discuss the parameters obtained from high-energy fits and factorization. For our purposes in this review, we treat them differently depending on whether they depend on observables with strangeness or not.

In particular, all parameters in Eqs.~\eqref{eq:Pomeron} and \eqref{eq:reggerho} 
correspond to non-strange Reggeon exchanges (the Pomeron, $P'$ or $f_2$ and $\rho$) and 
have been determined in \cite{Pelaez:2003ky} and updated in \cite{GarciaMartin:2011cn}, from processes that do not involve kaons. 
Their updated values from the CFD parameterizations  are given in \cite{GarciaMartin:2011cn} are listed in Table~\ref{tab:regge}.
Hence, as we did in \cite{Pelaez:2016tgi}, we will keep their values fixed both for our unconstrained and constrained fits here.

\begin{table}
\begin{minipage}[t]{0.49\linewidth}
\centering 
\caption{Values of Regge parameters obtained in \cite{Pelaez:2004vs,GarciaMartin:2011cn}. Since these could be fixed using reactions other than $\pi K$ or \pipikk scattering,
they will be fixed both in our UFD and CFD parameterizations.
\label{tab:regge} } 
\begin{tabular}{c c} 
\hline\hline  
\rule[-0.15cm]{0cm}{.55cm} Regge & Used both for \\ 
\rule[-0.15cm]{0cm}{.55cm} Parameters & UFD and CFD \\ 
\hline 
$s'$ & 1 GeV$^{2}$                                  \\
$\hat b$ & 2.4                 $\pm$0.5 GeV$^{-2}$       \\ 
$\alpha'_{P}$ & 0.2       $\pm$0.1 GeV$^{-2}$       \\
$\alpha'_{P'}$ & 0.9 GeV$^{-2}$                     \\ 
$c_{P}$ & 0.6             $\pm$1 GeV$^{-2}$         \\
$c_{P'}$ & $-$0.38          $\pm$0.4 GeV$^{-2}$       \\ 
$\beta_{P}$ & 2.50        $\pm$0.04                 \\
$c_{P}(0)$ & 0            $\pm$0.04                 \\
$\beta_{P'}$ & 0.80       $\pm$0.05                 \\
$c_{P'}(0)$ & $-$0.4        $\pm$0.4                  \\
$\alpha_{P'}(0)$ & 0.53   $\pm$0.02                 \\
$\alpha_{\rho}(0)$ & 0.53 $\pm$0.02                 \\ 
$\alpha'_{\rho}$ & 0.9 GeV$^{-2}$                   \\
$\alpha''_{\rho}$ & $-$0.3 GeV$^{-4}$                 \\ 
$d_{\rho}$ & 2.4          $\pm$0.5 GeV$^{-2}$       \\
$e_{\rho}$ & 2.7          $\pm$2.5                  \\
$\beta_{\rho}$ & 1.47     $\pm$0.14                 \\
\hline 
\end{tabular}
\end{minipage}
  \hfill
  \begin{minipage}[t]{0.49\linewidth}
  \centering
\caption{Values of Regge parameters 
involving strangeness. They are all allowed to vary from our UFD to our CFD sets except for $\alpha_{K^*}$ and $\alpha'_{K^*}$ since they are both determined from linear Regge trajectory fits to strange resonances. 
\label{tab:reggepiK} } 
\begin{tabular}{c c c} 
\hline\hline  
\rule[-0.15cm]{0cm}{.55cm} Regge & UFD & CFD\\ 
\hline           
$f_{K/\pi}$  &  0.67 $\pm$0.02     &  0.64$\pm$0.02    \\
$g_{K/\pi}$  &  0.70 $\pm$0.09     &  0.52$\pm$0.09    \\
$r$          &  0.050$\pm$0.010     &  0.056$\pm$0.010     \\ 
\hline           
$\alpha_{K^*}$     &  0.352  &  0.352                  \\
$\alpha'_{K^*}$    &  0.882 GeV$^{-2}$  &  0.882 GeV$^{-2}$      \\ 
$\lambda$    &  11.0$\pm$4.0          &  13.4$\pm$4.0      \\
\hline 
\end{tabular} 
\end{minipage}
\end{table} 

In contrast, the determination of the
$f_{K/\pi}$, $r$ and $g_{K/\pi}$ factors requires input from kaon interactions.
In principle, both of them were 
determined in \cite{Pelaez:2003ky} from $KN$ factorization
and we take the $f_{K/\pi}$ and $r$ values from that reference. 
The values and uncertainties of $f_{K/\pi}$ and $r$ are rather robust.
However, for
$g_{K/\pi}$ we take our UFD updated value from the forward dispersion 
relation study of $\pi K$ scattering in 
\cite{Pelaez:2016tgi}, which is further constrained.
We provide their values in Table~\ref{tab:regge}. We will also add in quadrature a systematic uncertainty associated with the errors coming from $\pi \pi$ that we keep fixed.
Since their determination 
involves kaon interactions, we will allow them to vary when constraining
our fits with dispersion relations, i.e. from the UFD to the CFD sets.
However, in the tables it is seen that the change is always within 2 sigmas of the UFD values.

Once the  parameters needed for the high energy description of \pik have been determined, we obtain the curves plotted in the left panels of  Fig.~\ref{fig:regge}. Note
that, to ease the comparison with the other hadron total cross-sections shown in Fig.~\ref{fig:totalsigma}, we have also plotted cross sections for the symmetric and antisymmetric isospin combinations in Fig.~\ref{fig:regge}.

At this point, some technical remarks about Regge parameterizations are in order.
It is often said that Reggeons and resonances are dual representations of hadronic physics ( see \cite{Collins:1971ff,Gribov:2009zz} for textbook introductions) and which one is better suited to describe some energy region depends on whether it can be better approximated with just a few Reggeons or a few resonances, instead of the full collection of them. However, this duality is semi-local,
 which, for our purposes, means that Regge theory at low energies does not necessarily yield the correct value of an amplitude at every value of the energy, but instead it yields the correct value {\it on the average} over a sufficiently large energy region. Of course, since we use Regge theory only at high energies and only inside integrals, this is good enough. To illustrate this ``on the average'' description, in Fig.~\ref{fig:regge} we have extrapolated the Regge bands below the region where we use them, i.e. below 1.84 GeV (vertical dashed lines). As expected, they do not  describe the resonance peaks, but they seem to average them.

Still, one might wonder how well the detailed partial-wave reconstruction 
agrees with the Regge description around 1.84 GeV,  where we shift from one description to another inside our integrals for \pik scattering. In Fig.~\ref{fig:regge} we can check that they are fairly consistent within uncertainties,  except for the  symmetric amplitude 
(upper left panel) for which they do not match by several standard deviations. 
We insist this is not incorrect, because one description is local and the other is an averaged description. 
 Furthermore, this mismatch does not produce any noticeable effects in the Roy-Steiner equations below 1 GeV. However, it does produce a significant effect when studying the forward dispersion relations close to this point. The dispersive output of the FDR, coming from and integral, will change smoothly between the values of the partial-wave reconstruction below 1.84 GeV and the Regge values above. As a result, the partial-wave input and the dispersive output for $F^+(s)$ are not compatible around 1.8 GeV, as will be shown in Fig.~\ref{fig:fdrchecks}.
 Actually, this is one of the main reasons why we cannot claim to have a data description consistent with the $F^+$ FDR up to 1.84 GeV, but only up to $\sim$1.7 GeV. 
 Of course, we have looked at whether it was possible to start using our Regge description from an energy  where it matches the partial-wave reconstruction. Unfortunately, the closest point available below 2 GeV is  around 1.4-1.5 GeV, too low to rely on Regge theory for \pik scattering,  whereas the closest point above is at around 2.2 GeV. Even though this last point may seem appealing, there are no data for the $I=3/2$ partial waves above  1.74 GeV, and as  already explained in~\cite{Pelaez:2016tgi} and in the introduction to this section, the partial-wave truncated series are not reliable much beyond that. In addition, we do not want to use the Regge description at too-low energies.
 All in all, we thus consider that a value around 1.84 GeV is well suited to shift from the partial-wave to the Regge description, and we then will not constrain the amplitudes above 1.7 GeV.

\vspace{.3cm}
\underline{High-energy Regge parameterizations for \pipikk}\\

To describe the high-energy region of  $\pi\pi\rightarrow K\bar K$, we need to consider the exchange of strange Reggeons, which are much worse known than the Pomeron, $P'$, and $\rho$ trajectories. Following \cite{Pelaez:2018qny} we will consider that  the two dominant trajectories, which are those of the $K^*_1(892)$ and $K_2^*(1430)$, are degenerate. Hence we will describe them with a common trajectory 
$\alpha_{K^*}(s)=\alpha_{K^*}+\alpha'_{K^*}s$,
whose parameters are listed in Table~\ref{tab:regge}. They are obtained from their linear Regge trajectories and will thus be kept fixed for both our UFD and CFD sets.
In practice, all these features are incorporated in the
 dual-resonance Veneziano-Lovelace 
model \cite{Veneziano:1968yb,Martin:1976mb,Lovelace:1969se,Shapiro:1969km,Kawarabayashi:1969yd},
which was already used before us in the Roy-Steiner context for $\pi K$ scattering
\cite{Ananthanarayan:2001uy, Buettiker:2003pp}. 

Nevertheless, let us recall that, as explained in Section~\ref{sec:DR}, for the dispersive analysis of \pipikk we will make use of hyperbolic dispersion relations defined along the curve $(s-a)(u-a)=b$. Then, for a given $t$, $s_b$ is the value of $s$ that lies in the previous hyperbola, which, in view of Eq.~\eqref{eq:sboft}, for large $t$ behaves as $s_b\rightarrow a$. Therefore, inside the integrals for the high-energy region of $t$, we need:
\begin{align}
&\frac{\im G^0(t,s_b)}{\sqrt{6}}=\frac{\im G^1(t,s_b)}{2}=\frac{\pi\lambda(\alpha'_{K^*}t)^{\alpha_{K^*}+a \alpha'_{K^*}}}{\Gamma(\alpha_{K^*}+a \alpha'_{K^*})}
\Big[1+\frac{\alpha'_{K^*} b}{t}(\psi(\alpha_{K^*}+a \alpha'_{K^*})-\log(\alpha'_{K^*} t)\Big)\Big],
\label{eq:vene}
\end{align}
where $\psi$ is the polygamma function.  
Note that, although numerically small, we have kept an $O(b/t)$ term, 
since it allows us to recover, as a check, the expressions in \cite{Buettiker:2003pp}, where $a=0$.

The last parameter $\lambda$ was estimated in \cite{Pelaez:2018qny}
assuming, for its central value, exact degeneracy between the $\rho$ and $K^*$ trajectories. Therefore, we take the $\rho$ contribution in   Eq.~\eqref{eq:reggef-} at 2 GeV and match it with
the expression from the degenerate Veneziano model with
its original parameter $\alpha_\rho^V=0.475$.
This leads to
\begin{equation}
\lambda\simeq 
\frac{2\pi\Gamma(\alpha_\rho^V)}{\alpha'^{\alpha_{\rho}^V}_{K^*}}
 4^{\alpha_\rho-\alpha_\rho^V}\simeq 10.6\pm2.5.
\label{eq:lambdaestimate}
\end{equation}
This value is compatible with the one estimated in \cite{Buettiker:2003pp}, $\lambda=14\pm 4$. 
Conservatively we add in quadrature a 30\% uncertainty as a crude estimate of the breaking of degeneracy and, rounding up, we write 
\begin{equation}
\lambda \simeq 11\pm4,
\label{eq:lambda}
\end{equation}
also listed in Table~\ref{tab:reggepiK}. 
Once again, given that this trajectory involves strangeness and also because it is a crude estimate, we will allow this value to vary from exact degeneracy when constraining our fits to obtain the CFD sets.
As we will see in Section \ref{sec:CFD}, after 
imposing the dispersive constraints we will obtain
$\lambda=13.4\pm4.0$, which, if we  use degeneracy between the $\rho$ and $K^*$ families, 
suggests $g_{K/ \pi}\sim 0.6\pm0.2$, in fair agreement with the CFD
value used here that comes from the dispersive $\pik$ study.

Once these Regge parameters have been fixed, our Regge parameterizations of the $\pi K$ total cross-sections  and the imaginary part of the forward $\pipikk$ amplitudes are shown in Fig.~\ref{fig:regge}.

\begin{figure}[!ht]
\centering
\resizebox{0.9\textwidth}{!}{\input{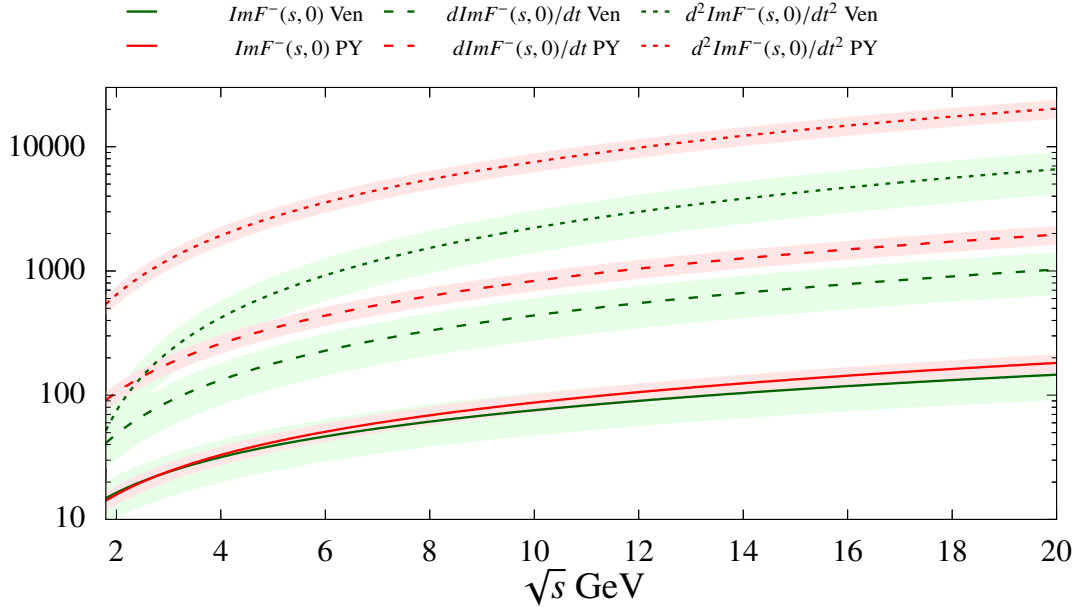}}
\caption{Comparison between the Regge model of~\cite{Pelaez:2003ky} (red lines) and the Veneziano-Lovelace model~\cite{Veneziano:1968yb,Martin:1976mb,Lovelace:1969se,Shapiro:1969km,Kawarabayashi:1969yd} (green lines). Note that although their results for Im$F^-(s,0)$ (continuous lines) are very similar and overlap within uncertainties, their first and second  derivatives (dashed and dotted lines, respectively) can differ substantially from one another to the point of being clearly incompatible. The fixed-$t$ sum rules for higher $\ell$ partial waves are dominated by such contributions, so that for these dispersion relations we have added a systematic uncertainty for the derivatives to cover both models, as detailed in section~\ref{sec:applications}.}
\label{fig:reggecomp}
\end{figure}

To end this section, two technical remarks are in order. 

First, the Regge models we have shown, either based on \cite{Pelaez:2003ky}
or the asymptotic forms of the Veneziano model, are particularly robust for 
the high-$s$ dependence and small $t$. As shown  in
Fig.~\ref{fig:reggecomp}, for $\im F^-(s,t)$ these two models produce consistent descriptions of the close-to-forward region in the $s$ variable. As a result, it is irrelevant whether we use one or the other inside our FDR and $S$, $P$ partial-wave projected dispersion relations. However, the $\ell\geq2$ fixed-$t$ sum rules of section~\ref{sec:sumrules} receive substantial contributions from this asymptotic region from terms with derivatives on the $t$-variable, in particular second and third-order derivatives. 
Unfortunately,  the second or third derivatives with respect to $t$ 
of these models are no longer compatible. This is illustrated in Fig.~\ref{fig:reggecomp} already for the first and second derivatives. Thus, just for the $D$-wave sum rules, we have considered the average of the two models with a combined uncertainty coming from the statistical ones and 
a systematic error estimated as the difference between the two models. Fortunately, this contribution is negligible for the sum rules obtained from hyperbolic dispersion relations, which are therefore much more precise and reliable.

Second, let us recall that for \pipikk the only dispersive representation we have in the physical region comes from partial-wave HDR relations
obtained in Section~\ref{sec:DR} above. In such case, we take integrals over $b$ for a family of $(s-a)(u-a)=b$ hyperbola,
while keeping the {\it negative} value $a=-10.9 M_\pi^2$ fixed (see \ref{app:Applicability}). 
This means that the exponent $\alpha_{K^*}+a \alpha'_{K^*}<\alpha_{K^*}$
and thus the Regge contribution to the $\pi\pi\rightarrow\bar KK$ dispersion relations, given the same number of subtractions,
is suppressed when compared to its size in \cite{Buettiker:2003pp}, where $a=0$.
This allows us to consider 
fewer subtractions without Regge contributions becoming too large. Of course, this same suppression occurs for the high-energy $\pipikk$ contributions to the \pik partial-wave dispersion relations obtained from HDR, where now $a=-10 M_\pi^2$.

\subsection{ Tests of the dispersive representation.}
\label{sec:DRTests}

Once again, we recall the reader that we aim at providing a simple set of parameterizations that are consistent with basic requirements as
analyticity, unitarity, and crossing. In the previous subsections, we have obtained such a relatively simple description of  data, paying particular attention to both  statistical and systematic uncertainties. Our parameterizations even contain some basic features known to exist in these two-meson amplitudes (cuts, poles, resonance poles, Adler zeros...). However, in this section, we will show that they are still not good enough. Actually, we are going to show now that, by itself, the previous Unconstrained Fit to Data (UFD)  description fails to satisfy to different degrees several of the dispersion relations we have derived in Section \ref{sec:DR}. 
In passing, we will also show what is in practice the weight of different contributions to each dispersion relation.
The present section will then justify the need for a Constrained Fit to Data (CFD)  that we will present in Section \ref{sec:CFD} below.

Let us then list the dispersive equations detailed in Section~\ref{sec:DR} that we will study here: the forward dispersion relations (FDRs) in Eqs.~\eqref{eq:FDRTan} and \eqref{eq:FDRTsym}, 
 the partial-wave relations obtained either  from fixed-$t$ dispersion relations (FTPWDR) as in Eq.~\eqref{eq:sftpwdr} or from Hyperbolic dispersion relations (HPWDR) as in Eqs.~\eqref{eq:shdrfm},\eqref{eq:shdrfmi}, \eqref{eq:pwhdr} and \eqref{eq:pwhdr1}. Let us recall that we use two versions of the HDR for the $F^-$ amplitude, with a different number of subtractions,
 so that both  $f_\ell^-$ and $g^1_1$ will have two versions of their HPWDR. The main reason to consider a different number of subtractions is that while the unsubtracted $F^-$ relation is adequate to obtain a sum rule at threshold, thus constraining efficiently the low-energy region, its Regge contribution is not negligible. Moreover, its crossed channel contribution, in particular, that from $g^1_1$, plays an important role at low energies, which entails the correlated fulfillment of dispersion relations for both  \pik and \pipikk. Hence, in order to get a different constraint, keeping the Regge asymptotic region almost negligible, so that the partial waves that build $F^-$ are dominated by their own input, we will consider also partial-wave relations from the once-subtracted $F^-$ HDR. Note, however, that this is not needed for the FTPWDR, since in this case both the high-energy region and the crossed channel contributions are largely suppressed.

How do we test then these integral relations?
We follow a similar approach to what we did in \cite{Pelaez:2016tgi,Pelaez:2018qny}. Namely, we first introduce a $\chi^2/dof$-like function measuring the deviation between the input coming from the data fits and the output of a given dispersion relation
\begin{equation}
\hat d^{2}=\frac{1}{N} \sum_{i=1}^{N}\left(\frac{d_{i}}{\Delta d_{i}}\right)^{2}.
\label{eq:dhatdef}
\end{equation}
By ``output'' we mean the integral part of the dispersion relation and by ``input'' the non-integral part that is calculated directly from the data fit.
The $d_i$ above, with $i=1,...N$, are the differences between input and output at a set of discrete energies, and $\Delta d_{i}$ are the uncertainties of the difference, obtained from the errors of the initial parameters.
Each dispersion relation will then have an associated distance which provides a fair estimate of its fulfillment. In particular, we will use $N=50$ points equally spaced in $\sqrt{s_i}$ between the $\pi K$ threshold and 1.7 GeV for the two FDRs. We will then use $N=30$ points, again equally spaced in $\sqrt{s_i}$, between the $\pi K$ threshold and 0.98 GeV for both the $\pik$ FTPWDR and HPWDR. Finally, we define the $\pipikk$ HPWDR distances by using again 30 points but now spaced between the $K \bar K$ threshold and 1.47 GeV. Furthermore, we will use as input our fitted partial waves below $s'\leq 1.84$ GeV$^2$ for \pik scattering, and below $t'\leq 2$ GeV$^2$ for \pipikk, using above those energies the Regge asymptotic contributions described in section~\ref{sec:ufdregge} above.

We consider that a given dispersive equation is satisfied within uncertainties if this $\chi^2/dof$-like function is 1 or less. Of course, it is not enough that this $\hat d^2$ should be one in the whole energy region, we have to make sure that this happens uniformly and there are no  regions where the fulfillment is bad but they are compensated by other regions where the agreement is very good. For this reason, we will present both global numbers, but also plots of the difference between input and output and their corresponding uncertainty.

\begin{figure}[!ht]
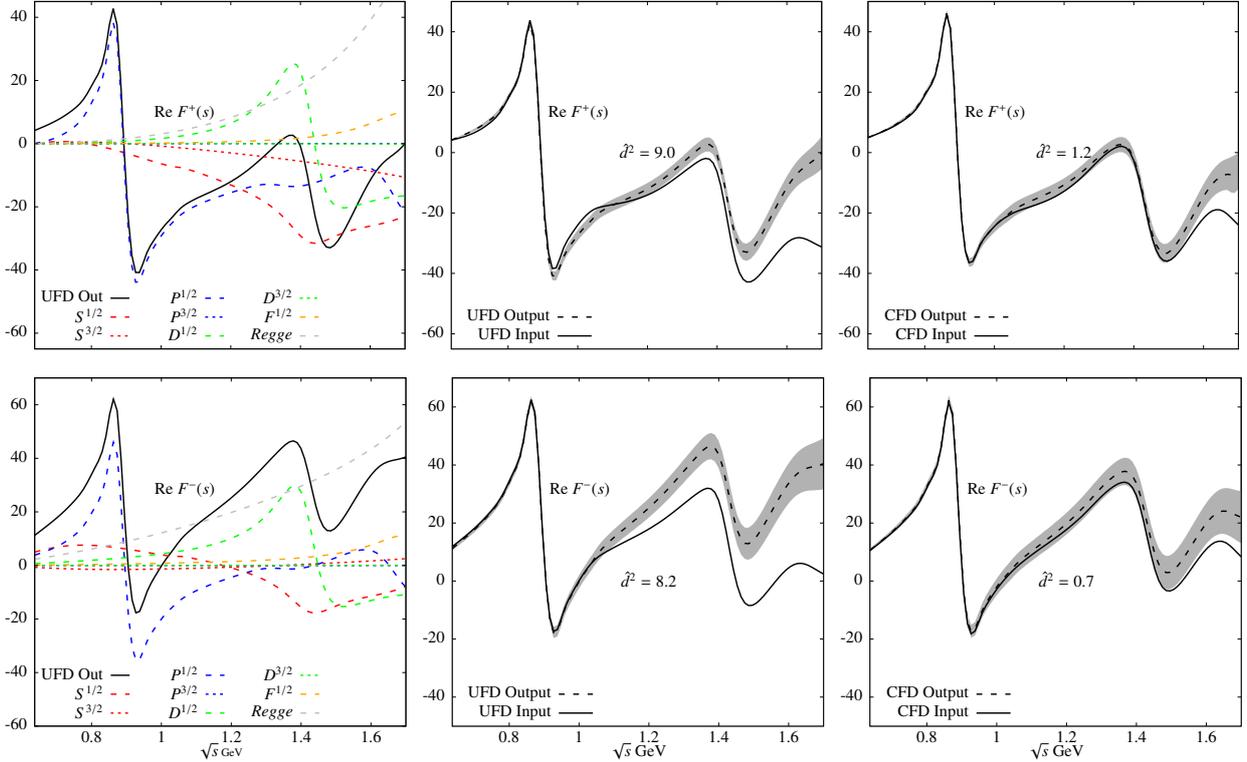

\centering
\resizebox{\textwidth}{!}{\input{figures/fplusparts.tex} \input{figures/fplus.tex} \input{figures/fpluscfd.tex} }\\
\resizebox{\textwidth}{!}{\input{figures/fminusparts.tex} \input{figures/fminus.tex} \input{figures/fminuscfd.tex}}\\
\caption{Checks of the $F^{+}(s)$ and $F^{-}(s)$ forward dispersion relations.  The gray bands are the uncertainties in the difference between input and output, although we attach them to the output to ease the comparison. Note the large inconsistencies between input and output for the UFD and the nice consistency of the CFD, except in the region above 1.6 GeV for $F^+$, which we cannot fix while keeping a decent description of data there. In the first column, we show the size of different contributions for each dispersion relation.}
\label{fig:fdrchecks}
\end{figure}

\subsubsection{Dispersive tests for \pik}

\vspace{.3cm}
\underline{Tests of forward dispersion relations}\\

Starting then with the FDRs for \pik,  our results here are very similar to those we obtained in \cite{Pelaez:2016tgi}, but updated with the new parameterizations and uncertainties. Thus, we show in the central panels of Fig.~\ref{fig:fdrchecks} the input and output of the two FDRs for $F^+$, i.e. Eq.~\eqref{eq:FDRTsym}, and $F^-$, Eq.~\eqref{eq:FDRTan} (upper and lower panels respectively) as a function of $\sqrt{s}$. We show one error band, which is the uncertainty in the difference between the two, attached to the ``output'' central value. If a dispersion relation is to be satisfied, both ``input'' and ``output'' curves should overlap within this uncertainty band. The inconsistency of both FDRs when using the UFD as input is obvious.  On average, the input and output lie more than 3 standard deviations away from each other when considering most of the energy region from $\pi K$ threshold up to 1.74 GeV, although the deviation is larger above 1 GeV.

In addition, in the left panels, we provide the 
decomposition of the integral ``output'' for each relation.
It is worth noticing that up to 1.2 GeV both amplitudes are dominated by the $P^{1/2}$-wave, i.e. the $K^*(892)$ resonance.
The shape around 1.4 GeV is dominated by the $D^{1/2}$, which corresponds to the $K^*_2(1430)$, and from that point there is no clear dominance. Above 1.6 GeV  Regge contributions become dominant, although they do not produce any distinct shape but just a monotonous rise. Beyond 1.8 \gev, the UFD violates dispersion relations by more than 5 sigmas and we do not plot it anymore. 

Compared to the previous analysis we carried out in \cite{Pelaez:2016tgi}, the low-energy region is better described by the updated  $S$ and $P$ fits. However, at higher energies, the  observed inconsistencies are similar, if not worse as now our uncertainties have shrunk slightly with the update. Actually, we already saw in \cite{Pelaez:2016tgi}, and have found again with the updated parameterizations, that, even imposing the FDRs as constraints, it is not possible to have a simultaneous fair description of data and FDRs beyond 1.7 GeV for $F^-$ or 1.6 GeV for $F^+$. This is partially due to the non-smooth match shown in Fig.~\ref{fig:regge}, but also caused by the fact that there is no centrifugal suppression anymore so that many partial waves contribute to the full amplitudes with similar strength.

Hence, as we concluded in \cite{Pelaez:2016tgi} and we have checked again here with updated fits, the UFD fails to satisfy the FDRs, particularly at high energies, and asks for improvement. This will be achieved in Section \ref{sec:CFD} by constraining the data fits with dispersion relations, including these FDRs.

\vspace{.3cm}
\underline{Tests for partial waves from fixed-$t$ dispersion relations }\\

The previous FDR test the whole amplitudes $F^{\pm}(s,0)$, which in the region of applicability are built as sums of partial waves. However, in Section~\ref{sec:DR} we also derived dispersion relations for individual partial waves. 
For \pik scattering, we will test the $S$ and $P$ partial waves, i.e. $\ell=0,1$, both in the symmetric and antisymmetric isospin combinations. Higher partial waves are input for these.
Remember that these partial-wave dispersion relations can be obtained either from fixed-$t$, i.e. FTPWDR, or HDR dispersion relations.

\begin{figure}
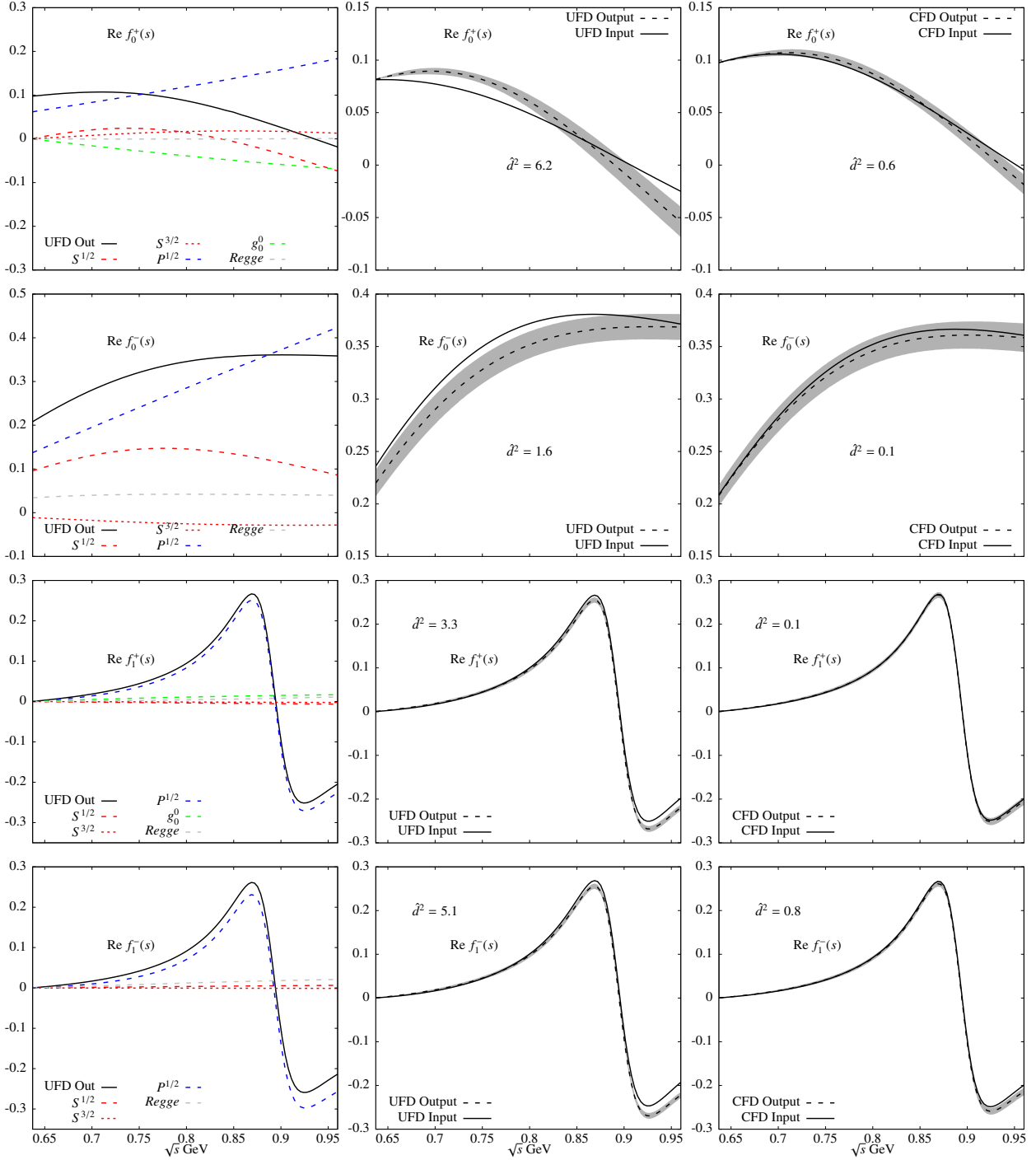

\centering
\resizebox{\textwidth}{!}{\input{figures/ff0plusparts.tex} \input{figures/ff0plus.tex} \input{figures/ff0pluscfd.tex} }\\
\resizebox{\textwidth}{!}{\input{figures/ff0minusparts.tex} \input{figures/ff0minus.tex} \input{figures/ff0minuscfd.tex} }\\
\resizebox{\textwidth}{!}{\input{figures/ff1plusparts.tex} \input{figures/ff1plus.tex} \input{figures/ff1pluscfd.tex} }\\
\resizebox{\textwidth}{!}{\input{figures/ff1minusparts.tex} \input{figures/ff1minus.tex} \input{figures/ff1minuscfd.tex} }\\
\caption{Tests of the $f^{+}_\ell(s), f^{-}_\ell(s)$ fixed-$t$ partial-wave dispersion relations.  The gray bands are the uncertainties in the difference between input and output, although we attach them to the output to ease the comparison. Note the large deviations between input and output in the UFD and the remarkable agreement within uncertainties of the CFD partial waves. We show in the first column the size of different contributions to each dispersion relation.}
\label{fig:ftchecks}
\end{figure}

Thus, in the central panels of Fig.~\ref{fig:ftchecks} we show the output versus input for the four FTPWDR, obtained from Eq.~\eqref{eq:sftpwdr}, for the $f^\pm_\ell$ partial waves, with $\ell=0,1$. 
As a technical remark, note that in Fig.~\ref{fig:ftchecks} we only show our curves up to 0.98 GeV, although these FTPWDR are applicable up to $s\sim 1.1 \gev^2$. However,  for simplicity and the sake of comparison, we will only impose all partial-wave dispersion relations up to $\sqrt{s}=0.98\, \gev$, which is the highest allowed value for the HPWDR.

Now, as can be seen in Fig.~\ref{fig:ftchecks}, the deviations between input and output are not as large as for FDRs.
However, there is a clear deviation in the $f^+_0$ case (Top center), not only because the overall $\hat d^2=6.2$ but because the shapes of input and output do not look very similar. The 
best one is $f^-_0$ which, on the average, deviates by less than 1.3 standard deviations. The input and output for both vector waves $f^+_1$ and $f^-_1$ seem to lie very close, but the uncertainties are very small and on average they deviate by more than 1.8 and 2.2 sigmas, respectively.

On the left panels of Fig.~\ref{fig:ftchecks} we show the size of different contributions to these FTPWDR.
It is worth mentioning that the $P^{1/2}$-wave dominates completely the vector dispersion relations, the other contributions being practically zero. Thus, the inconsistency of these relations will require some honing of the $P^{1/2}$ and in particular our $K^*(892)$ description. For scalar waves, the most salient feature is that the Regge contribution is completely negligible and the observed shape is a cancellation between the dominant $P^{1/2}$ contribution and the subdominant $S$-waves and, for the $f^+_0$, even the $g^0_0$ wave of the crossed channel. It is also worth mentioning that the contribution from the \pipikk crossed channel is small. The largest one comes from $g^0_0$ and we can see that is much smaller than the dominant ones. For this reason, it does not make any difference to use as input our UFD$_B$ or UFD$_C$ (which we actually used for these plots).

Let us remark that this check would be much worse in case we had used the original fits of  \cite{Pelaez:2016tgi} instead of the new ones. The main effect comes from the UFD $S^{3/2}$ partial wave, which is now much closer to the final constrained result. Therefore this comparison supports our updated parameterization and disfavors substantially the lowest-lying data points of~\cite{Estabrooks:1977xe}.

In any case, although somewhat better satisfied than the FDRs, the partial-wave tests obtained from fixed-$t$ dispersion relations also suggest revisiting the data fits using them as constraints, as we will do in Section~\ref{sec:CFD}.

\begin{figure}[!b]
\centering
\resizebox{\textwidth}{!}{\input{figures/f0minusparts} \input{figures/f0minus.tex} \input{figures/f0minuscfd.tex}}\\
\resizebox{\textwidth}{!}{\input{figures/f1minusparts.tex} \input{figures/f1minus.tex} \input{figures/f1minuscfd.tex} }\\
\caption{Checks of the $f^{-}_\ell(s)$ non-subtracted hyperbolic partial-wave dispersion relations. The gray bands are the uncertainties in the difference between input and output, although we attach them to the output to ease the comparison. Note the large deviations between input and output in the UFD and the remarkable agreement of the three CFD partial waves.  In the first column, we show the size of different contributions to each dispersion relation.}
\label{fig:hdrchecks}
\end{figure}

\vspace{.3cm}
\underline{Tests for partial-waves from hyperbolic dispersion relations}\\

First of all, recall that there are HPWDR for both the \pik channel, i.e. Eqs.~\eqref{eq:shdrfm} and \eqref{eq:shdrfmi}, as well as for the \pipikk channel, i.e. Eqs.~\eqref{eq:pwhdr} and \eqref{eq:pwhdr1}. For the former, we only test the $S$ and $P$-waves in the isospin symmetric or antisymmetric combinations, whereas for the latter we test $g^0_0, g^1_1$ and $g^0_2$. Also, recall that we have two versions, depending on whether we subtract the $F^-$ dispersion relation or not. In addition, we have two alternative fits for $g_0^0$, i.e. UFD$_B$ and UFD$_C$, but we will see that their contribution to dispersion relations other than their own are very small, if any, and taking one or the other yields the same result. Unless otherwise stated, we use UFD$_C$ when $g^0_0$ is needed.

Thus, in the central column of Fig.~\ref{fig:hdrchecks} we  show the tests
for the $f^-_\ell$ \pik partial waves, with the  $F^-$ HDR unsubtracted. The large inconsistency of the scalar wave in the top central panel is striking: the output lies uniformly well below the input, by more than 3.7 standard deviations on the average. Moreover, the inconsistency for the vector channel in the bottom central panel is as large as for the scalar one, i.e.
an average of more than 3.6 standard deviations, although since the uncertainty band is much smaller, the central values of the curves look very close in the plot.
These are the largest inconsistencies we will find when using the UFD.

On the left panels of Fig.~\ref{fig:hdrchecks} we show the size of the different contributions. In both cases, the dominant ones are the $P^{1/2}$-wave and  the crossed-channel $g^1_1$ 
contributions.  
In the scalar case
they suffer a large cancellation among themselves,  and the curve shape is given by the scalar $S^{1/2}$-wave.
The Regge part is negligible in both cases.
 The strong violations we report are also related to the deviation from the dispersive $F^-$ sum rule we found in \cite{Pelaez:2016tgi}. Actually, note that, for the scalar case, the low-energy region is dominated by the $g^1_1$ partial wave. Even more specifically, the part that dominates the $g^1_1$ wave is the pseudo-physical region, included as input between $4 m_\pi^2$ and $4 m_K^2$. This may seem surprising, since the phase in that region comes from the \pipi $P$-wave, taken from the dispersive study of \cite{GarciaMartin:2011cn}, but it is worth remembering this is produced by the peak of the $\rho$ resonance, as it appears in \pipikk, which cannot be measured.
 Actually, the UFD modulus (Fig.~\ref{fig:gdispmod}), obtained employing an unsubtracted Muskhelishvili-Omn\`es dispersion relation as explained in section \ref{sec:DR} is small around the $\rho$ resonance. This adds to the fact that the phase-shift crosses $90^\degree$ there, producing a large imaginary part contributing to the dispersion relations. We will see that this modulus will suffer a remarkable change when imposing the dispersive constraints on $g^1_1$.

\begin{figure}
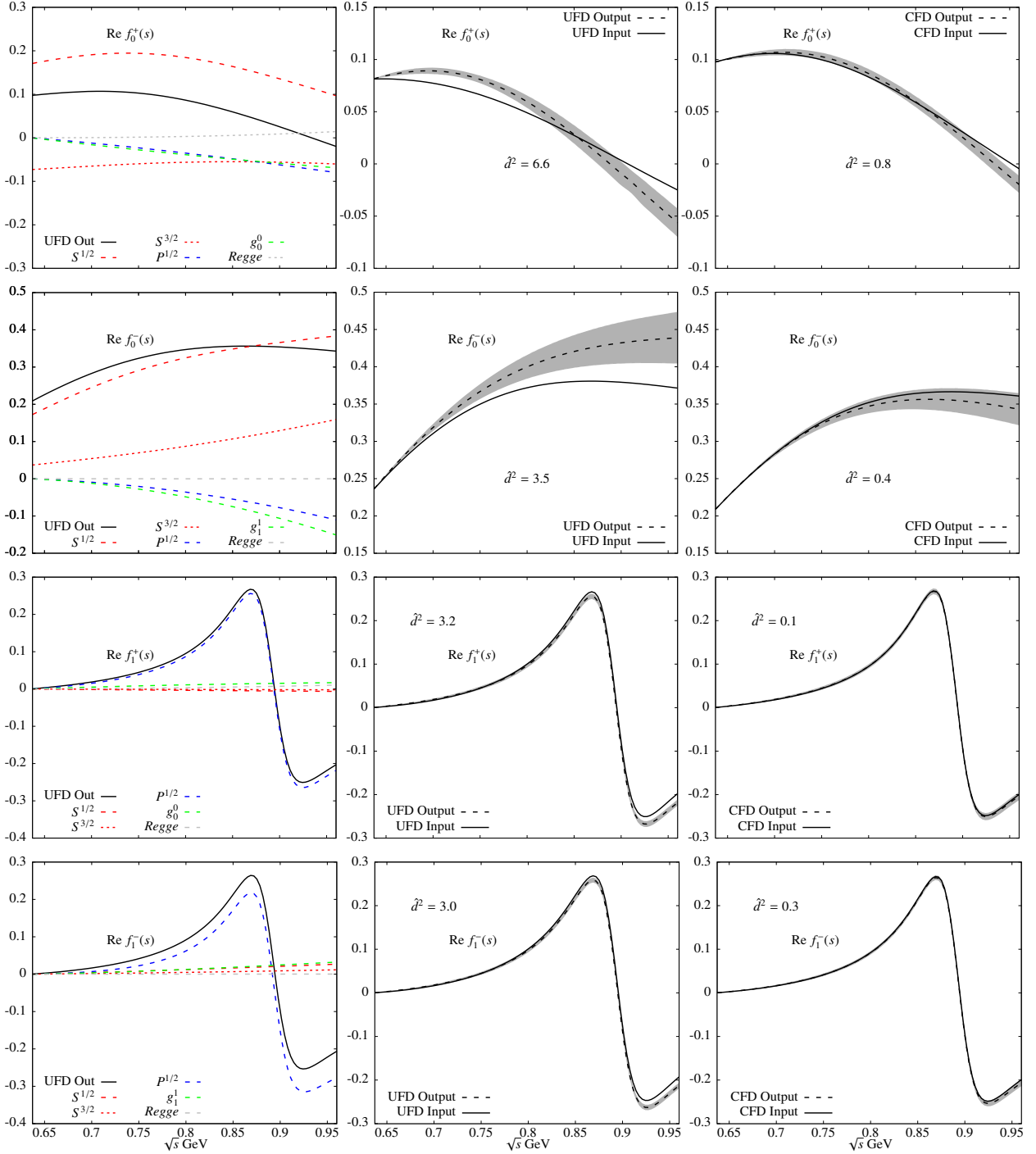

\centering
\resizebox{\textwidth}{!}{\input{figures/f0plusparts.tex} \input{figures/f0plussub.tex} \input{figures/f0plussubcfd.tex} }\\
\resizebox{\textwidth}{!}{\input{figures/f0minussubparts.tex} \input{figures/f0minussub.tex} \input{figures/f0minussubcfd.tex} }\\
\resizebox{\textwidth}{!}{\input{figures/f1plusparts.tex} \input{figures/f1plussub.tex} \input{figures/f1plussubcfd.tex} }\\
\resizebox{\textwidth}{!}{\input{figures/f1minussubparts.tex} \input{figures/f1minussub.tex} \input{figures/f1minussubcfd.tex}}\\
\caption{Checks of the $f^{+}_\ell(s)$ and $f^{-}_\ell(s)$ once-subtracted hyperbolic partial-wave dispersion relations.
 The gray bands are the uncertainties in the difference between input and output, although we attach them to the output to ease the comparison. Note the large deviations between input and output in the UFD and the remarkable agreement of the four CFD partial waves. In the first column, we show the size of different contributions to each dispersion relation.}
\label{fig:hdrcheckssub}
\end{figure}

When looking at the $f^+_\ell$ and $f^-_\ell$ once-subtracted HPWDR tests in Fig.~\ref{fig:hdrcheckssub}, we find again a rather large disagreement. Somewhat larger indeed for the scalar waves than the vector ones. The deviations between input and output are slightly worse but relatively similar in size to those we already observed in Fig.~\ref{fig:ftchecks} in the partial-wave tests coming from FTPWDR. 
Comparing the panels in the left column of Fig.~\ref{fig:hdrcheckssub} versus their counterparts in
Fig.~\ref{fig:ftchecks}, we see that, for  vector waves, the $K^*(892)$ dominates both kinds of dispersion relations, as should be expected. 
However, for all waves, the respective sizes of the other contributions for the two kinds of dispersion relations are rather different.  As a consequence, FTPWDR and HPWDR provide independent tests, and both show a clear inconsistency in the UFD parameterization.
Let us also remark that the $g^0_0$ contribution is very small and therefore it is irrelevant whether we use our CFD$_B$ or CFD$_C$ parameterization as input (we used the latter for these plots).

 Compared to the HPWDR unsubtracted case in Fig.~\ref{fig:hdrchecks}, we find a smaller disagreement for $f^-_0$ but  note that this time the output comes {\it above} the input. 
Thus, the disagreement between the unsubtracted and subtracted $f^-_0$ outputs, which in principle should be equal, is dramatic.
These tests would have been worse if the original fits of \cite{Pelaez:2016tgi} had been used instead of the ones presented here. The $g^1_1$ partial wave is now less dominant, albeit big, and several cancellations produce a way more stable result, almost dominated by its own input. In particular, the Regge contributions have been suppressed. However, the $a^-_0$ scattering length is much more relevant now.

In summary, we find that the UFD no matter how nicely it describes the data, fails to satisfy well almost all the dispersion relations under consideration: FDRs and partial-wave dispersion relations either coming from fixed-$t$ or HDR. We will dedicate the whole Section \ref{sec:CFD} to find a constrained fit that will describe fairly well and simultaneously all dispersion relations and the data within uncertainties, not only for \pik but also for \pipikk, whose UFD we will see next that also fails to pass the dispersive tests.

\subsubsection{Dispersive tests for \pipikk and prediction for the unphysical region}

Let us now recall that only partial-wave dispersion relations from HDR can be used to test or constrain this process, and only up to 1.47 GeV  (see \ref{app:Applicability}). These tests were already studied by us in \cite{Pelaez:2018qny}, but here we have updated them to the new parameterizations, finding relatively similar results. In addition, the dispersive representation yields a prediction for the modulus of the partial waves in the unphysical region, where data on the modulus do not exist, but the phase can be obtained from \pipi scattering. This region is relevant because it is input for other dispersion relations in this and the crossed channel. 

As a technical remark, we have slightly changed the value of the matching point of the $g^1_1(t)$ Muskhelishvili-Omn\`es  from $\sqrt{t_m}=1.2\, \gev$ used in \cite{Pelaez:2018qny} to $\sqrt{t_m}=1\,$GeV. The reason is that this new point makes the pseudo-physical region more sensitive to changes occurring in the low-energy physical one, thus allowing us to have more room for improvement when using the dispersion relations as constraints. At the same time, it reduces our uncertainties and uses the fit to the data on a larger input region, as explained in section~\ref{sec:MO}.

\begin{figure}[!ht]
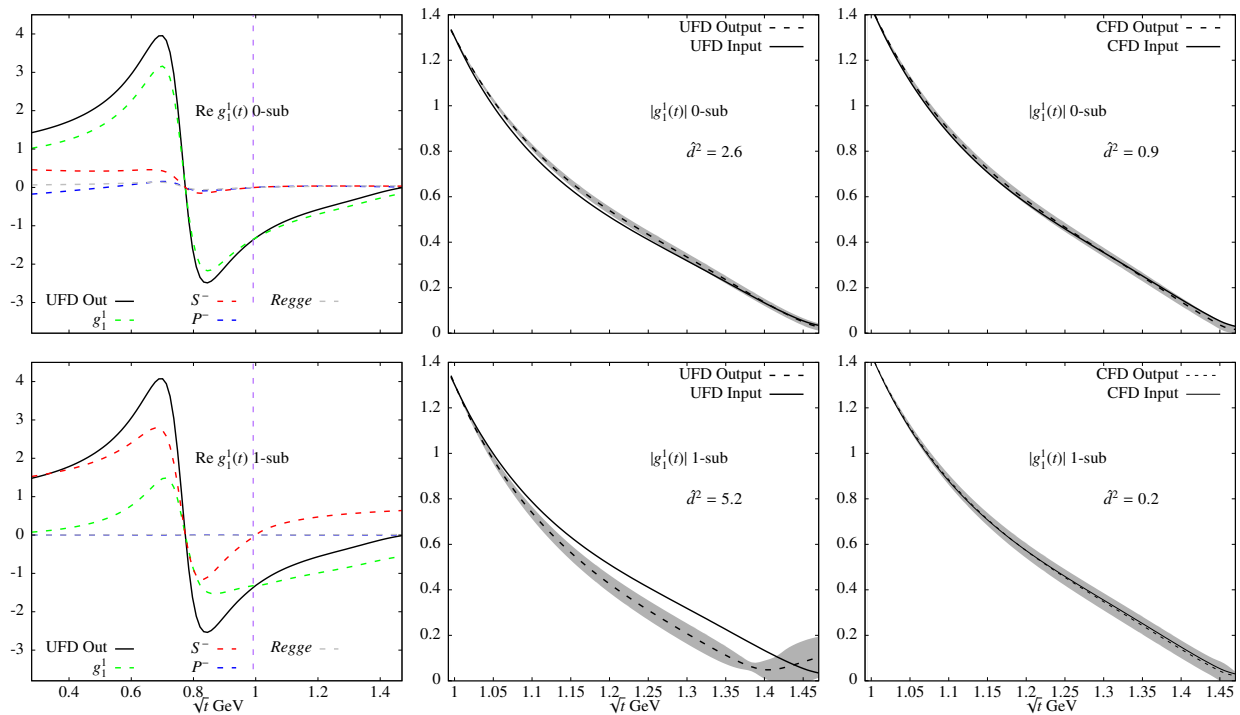

\centering
\resizebox{\textwidth}{!}{\input{figures/g11parts.tex} \input{figures/g11.tex} \input{figures/g11cfd.tex} }\\
\resizebox{\textwidth}{!}{\input{figures/g11subparts.tex} \input{figures/g11sub.tex} \input{figures/g11subcfd.tex} }\\
\caption{Checks of the $g^1_1(t)$ partial waves  with none (Top) and  one subtraction (Bottom) for the $F^-$ amplitude.  The gray bands are the uncertainties in the difference between input and output, although we attach them to the output to ease the comparison. Note the inconsistency for the UFD case (central column), particularly for the once-subtracted case, which disappears for the CFD (right column). On the left column, we show the size of different contributions to the dispersive output.
The vertical dashed lines on the left panels signal the $K\bar K$ threshold.}
\label{fig:g11checks}
\end{figure}

Thus, in Fig.~\ref{fig:g11checks}, we show the results for both the unsubtracted and subtracted $g^1_1$ dispersion relations (top and bottom panels, respectively).
Note that we can only compare input and output for the modulus in the physical region since the unphysical one is just output. 
Both waves show some inconsistency with the dispersive output, milder for the unsubtracted case.
By comparing the upper and lower left panels, we can see how  the weight of different contributions changes considerably by subtracting or not the dispersion relation. While the unsubtracted output is largely dominated by the $g^1_1$ wave itself, the largest contribution to the subtracted one  comes from $S$-waves, particularly in the region below the $\rho$ mass. 

\begin{figure}[!ht]
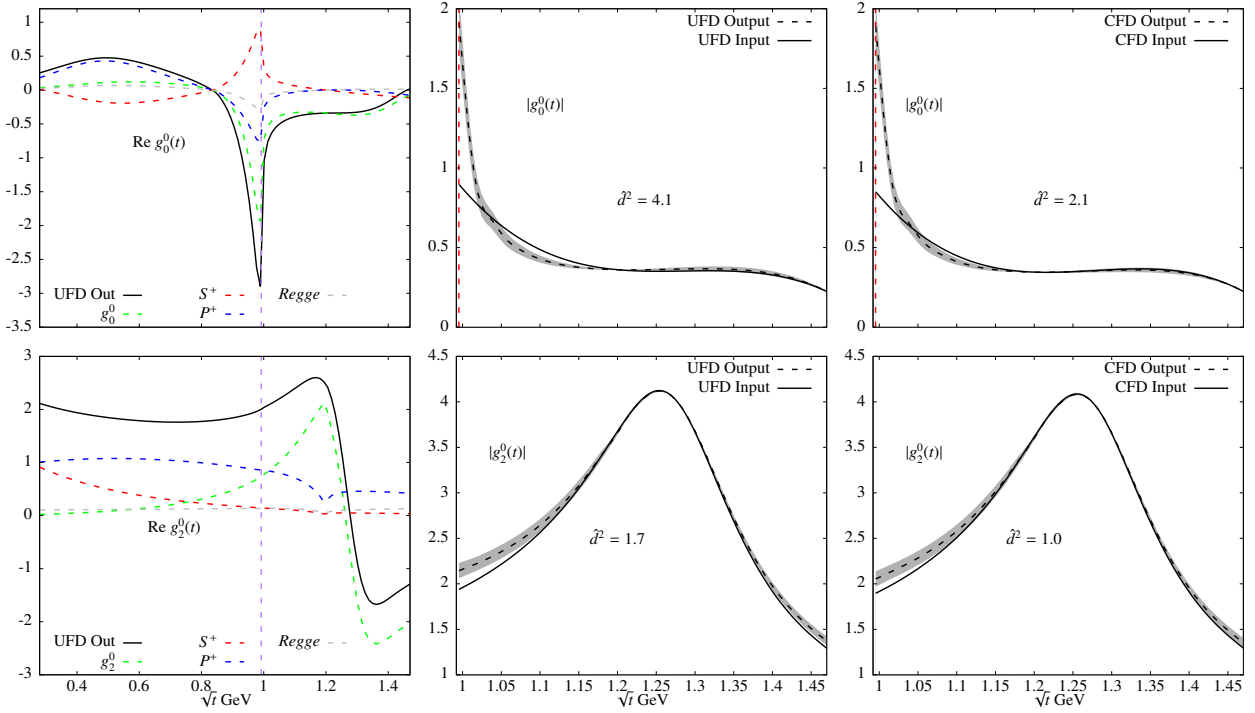

\centering
\resizebox{\textwidth}{!}{\input{figures/g00parts.tex} \input{figures/g00.tex} \input{figures/g00cfd.tex} }\\
\resizebox{\textwidth}{!}{\input{figures/g02parts.tex} \input{figures/g02.tex} \input{figures/g02cfd.tex} }\\
\caption{Checks of the $g^0_0(t)$ and $g^0_2(t)$ partial waves  with one subtraction for the $F^+$ amplitude.  The gray bands are the uncertainties in the difference between input and output, although we attach them to the output to ease the comparison. Note the consistency improvement from the UFD to the CFD results (central and right columns, respectively). For the $g^0_0$ we show as UFD and CFD the UFD$_C$ and CFD$_C$, since the UFD$_B$ and CFD$_B$ are of similar qualitative behavior. Nevertheless, even for the CFD, the region close to the $K \bar K$ threshold (vertical dashed line)  cannot be well described with our isospin-symmetric formalism at this level of precision. This is particularly evident for the $g^0_0$ since that region is enhanced by the presence of the $f_0(980)$ resonance and its isospin violating mixing with the $a_0(980)$. On the left column, we show the size of different contributions to the dispersive output. }
\label{fig:g00checks}
\end{figure}

Concerning $g^0_0$, the dispersive test is shown in
the top central panel of Fig.~\ref{fig:g00checks} for the UFD$_C$. Once again, we find an overall disagreement between input and dispersive output, although it is concentrated below 1.2 GeV and more intensely in the $\sim$ 20 MeV region right above $K\bar K$ threshold. 
Beyond this near-threshold region the agreement is much better. We already found this behavior in \cite{Pelaez:2018qny}. Actually, this was the most likely place for our isospin-conserving approximation to fail, since there are indeed two thresholds, one for $K^0 \bar K^0$ and another for $K^+ K^-$, separated by  $\simeq 8\,$MeV, but our isospin-symmetric formalism only has one.  
In other waves this effect is less relevant, but 
in the $S$-wave it is enhanced by the isospin-violating mixing of the nearby $f_0(980)$ and $a_0(980)$ resonances \cite{Achasov:1979xc,Hanhart:2007bd}. 

In the corresponding left panel, we also see that the dispersive output for the $g^0_0$ wave is dominated below threshold by the \pik $P$-wave, whereas $g^0_0$ itself dominates above $K\bar K$ threshold. However, right at threshold there is a cancellation between three significant contributions, i.e. those from the $g^0_0$, $P$, and $S$-waves. The Regge contribution is rather small and well under control. Note that we $\pi\pi$ phase contribution is always subsumed inside other terms through the Omn\`{e}s function and that is why we do not consider it separately.

In addition, in the lower panels of Fig.~\ref{fig:g00checks}, we show the dispersive tests for the tensor wave $g^0_2$. In this case, the agreement is quite acceptable, save the very near threshold. Once again the isospin-violating
splitting of the 
$K\bar K$ threshold commented above may play a role in this mismatch, although it is milder than for the $g^0_0$ wave since
for $g^0_2$ no resonances are sitting right at $t_K$. Well above $K \bar K$ threshold, the wave is largely dominated by the $f_2(1270)$ resonance and by the \pik $P$-wave in the unphysical region.

\begin{figure}[!ht]
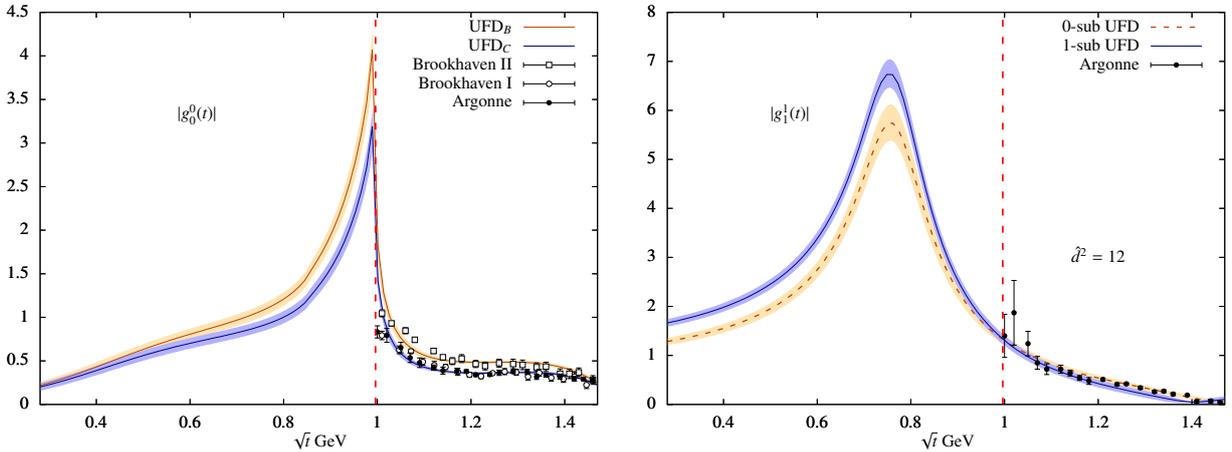

\centering
\resizebox{\textwidth}{!}{\input{figures/g00disp.tex} \input{figures/g11disp.tex}}\\
\caption{Dispersive result for the modulus of the UFD $g^0_0(t)$ and $g^1_1(t)$ partial waves. Left: We show the two alternative $g^0_0$ fits UFD$_B$ and UFD$_C$. Since they differ in the physical region, they also differ below. Right: Notice the discrepancy between the  unsubtracted and once-subtracted dispersion relations for $g^1_1(t)$, even if using the same UFD input in the physical region. We have calculated as an illustration the average $\hat{d}$ distance between the two $g^1_1$ dispersion relations in the pseudo-physical region between $2 m_\pi$ and $2 m_K$, divided by the relative uncertainty between them. They are clearly incompatible by more than 3 standard deviations.}
\label{fig:gdispmod}
\end{figure}

Finally, in Fig.~\ref{fig:gdispmod} we show the dispersive 
predictions for the modulus of the $g^0_0$, both for UFD$_B$ and UFD$_C$, as well as  the modulus of $g^1_1$, but this time down to the unphysical region and versus data in the physical region.
Recall there are no  data for these moduli in the unphysical region. Our only input there is the \pipi phase shift with the corresponding quantum numbers. These moduli are calculated using the Muskhelishvili-Omn\`{e}s formalism explained in Section \ref{sec:DR}.
Obviously, the predictions for the unphysical region of $g^0_0$ differ when using the UFD$_B$ or UFD$_C$, since they provide different inputs.  In both cases we clearly see a peak corresponding to the $f_0(980)$ resonances and a lower energy ``bump'' that is associated with the $\sigma/f_0(500)$  resonance, which is well-known to be very wide and not a Breit-Wigner-like shape \cite{Pelaez:2015qba}.
In general, since the modulus of the UFD$_B$ is larger than that of the UFD$_C$ in the physical region, it is also larger in the unphysical one.
Fortunately, we have already seen that the contributions of this wave to other dispersion relations are not significant, and thus they barely change when using one input or the other. 

In contrast, we have seen that  $g^1_1$ plays a relevant role, if not dominant, for several dispersion relations, even for the crossed channel and in particular for the \pik scalar waves (where the controversial scalar resonance \kap lies). Remarkably, as seen in the right panel of Fig.~\ref{fig:gdispmod},  {\it for the same UFD input}, we get two incompatible predictions for the unphysical region, depending on whether they come from the unsubtracted or once-subtracted dispersion relation. In principle, they should be the same. As we have seen, this is in part responsible for the dramatic difference in the  dispersive output for the \pik $S$-wave. This will severely affect the extraction of the \kap pole that we will discuss in section \ref{sec:kappa}.

Thus, the conclusion we reach from this first test is that fairly good-looking fits to the data do not fulfill the dispersive representation. 
And this happens in different energy regions and affects different contributions in both the $s$ and $t$ channel processes. Fortunately, it will be possible to amend this in the next section.

\section{Constrained Fits to Data}
\label{sec:CFD}

\subsection{Fulfillment of the dispersive representation.}
\label{sec:FulfillDR}

In the previous sections, we have first
obtained, within the isospin-symmetric approximation, a set of 16 dispersion relations, which are a consequence of first principles like causality and crossing symmetry, that the scattering amplitudes should satisfy. 
Next, we have obtained
a relatively simple description of the existing scattering data on both \pik and $\pipikk$ scattering, which also gives a fair representation of the existing statistical and systematic uncertainties. These are our Unconstrained Fits to Data (UFD). 
 However, as we already found in \cite{Pelaez:2016tgi,Pelaez:2018qny}, we have just seen that using our UFD as input for the dispersion relations, the output is inconsistent with the input, in some cases not by much, but in others by a rather large difference.

To address these inconsistencies, in \cite{Pelaez:2016tgi} we constrained the \pik amplitudes to satisfy forward dispersion relations (FDR), whereas in \cite{Pelaez:2018qny} we constrained the \pipikk amplitudes with partial-wave hyperbolic dispersion relations (HPWDR), i.e. hyperbolic Roy-Steiner equations,  but  keeping the \pik input from \cite{Pelaez:2016tgi} fixed.
We should also recall that there is also a \pik scattering dispersive analysis \cite{Buettiker:2003pp} {\it solving}, in the elastic region, partial-wave dispersion relations coming from fixed-$t$ dispersion relations and keeping the \pipikk input fixed from phenomenological models not  constrained dispersively. This ``solving'' approach was also followed for other processes in \cite{Colangelo:2001df,Ananthanarayan:2000ht,Buettiker:2003pp,Ditsche:2012fv,Hoferichter:2015hva}.

However, no matter whether one solves the dispersion relations or uses them as constraints, one might wonder about correlations between the crossed \pik and \pipikk channels. Namely, whether solving or constraining one channel would require the modification of the other to ensure consistency.
This kind of coupled analysis between the $s$ and $t$ channels, including both physical regions, has never been performed for the \pik system and is the main original result of this review, from which all other original results derive.

How we impose the dispersion relations in our fit is similar to what we have done in our previous works. We will allow the parameters of our fits to vary to improve the consistency with dispersion relations, but keeping their uncertainties fixed, whose corrections we consider second-order effects. Then we make use of the averaged $\hat\chi^2$-like distances $\hat d_i^2$, already defined in Eq.~\eqref{eq:dhatdef}, as penalty functions, by minimizing them together with the $\chi^2$ of the data fits.  In this work, each penalty function $\hat d^2_i$ is assigned the weight $W_i$ given in Table \ref{tab:DRweigths}. In order to arrive at those values of $W_i$, we first considered each  $W_i$ to be roughly the apparent number of degrees of freedom of the respective curve, and then we tune it until, after minimization, we get a $d^2_i$ reasonably close to one in the whole range of energies where the dispersion relation is applied. 
The dependence of the minimization procedure on  these weights is smooth and we have tried different choices, leading to very similar final results consistent within uncertainties. The one we provide in the tables ensures a quite uniform consistency of the dispersion relations throughout their applicability region.
We will see that some dispersion relations yield a very small $\hat d^2_i$ once the other ones are satisfied.

In addition, we  define another distance, $\Delta_p$, between the initial and final parameters of the high-energy asymptotic formulas described in Section \ref{sec:ufdregge}, divided by their uncertainties. The weight of these parameters will be simply their total number.  Finally, we will add these distances to the $\chi^2$ of the data, although the data $\chi^2$ for each partial wave has also been weighted by its apparent degrees of freedom $W_{pw\, i}$, that we list in Table \ref{tab:pwweigths}. All in all, the final formula to be minimized reads
\begin{equation}
    \sum_i \left( W_{pw\,i}\chi^2_i+W_i \hat d^2_i\right)+\Delta^2_{p}+W_{FDR}\, \hat  d^2_{FDR}+W_{p'}\Delta^2_{p'}.
    \label{eq:tominimize}
\end{equation}

The result of minimizing this quantity is our Constrained Fit to Data (CFD), which  satisfies all our dispersion relations within uncertainties, while still providing a fairly good description of data.
The parameters of this CFD have already been provided in the tables of section \ref{sec:UFD} together with the UFD ones, but here we will comment on its features. 
A technical remark is in order first, though. After minimization, most of the parameters vary by 1 and 2 $\sigma$ from their UFD to their CFD values. However, in practice, there are a few highly correlated parameters, for which their CFD result could lie far from the UFD one. Nevertheless, there is always another solution, with a negligibly larger $\chi^2$,
where these few parameters remain closer to their UFD counterparts. Since we have decided to keep the CFD uncertainties fixed to their UDF values, those latter values are preferred. For this reason, we have also added once again the $\Delta_{p_i}^2$ of every parameter that deviates by more than 3$\sigma$ from its UFD value.
We denoted by $\Delta_{p'}^2$ the sum of all these contributions, which is summed at the end of \cref{eq:tominimize} above. Given that any contribution to this sum starts at 9, and the other contributions to the total function to be minimized are expected to be of order one, we have found that $W_{p'}\sim 0.1$ is enough to 
ensure that only a handful of our parameters are deviated by a bit more than 3 $\sigma$ from their UFD values, and very rarely beyond that. We have also checked that the dispersive results and the data description do not vary substantially when adding this penalty, it just helps produce a CFD set of parameters without problematic outliers.
\begin{table}
\caption{List of weights used to determine our final penalty functions that constrain the data fits.} 
\begin{minipage}[t]{0.33\linewidth}
\vspace{0.3cm}
\centering 
\begin{tabular}{|l|c|c|} 
\hline
Weight/DR & $F^+$ & $F^-$ \\ 
\hline 
\rule[-0.05cm]{0cm}{.35cm}FDR & 8 &  8\\
\hline 
\hline
Weight/DR & $f^+_0$ & $f^+_1$ \\
\hline 
\rule[-0.05cm]{0cm}{.35cm}Fixed-$t$ & 3 &  3\\
\rule[-0.05cm]{0cm}{.35cm}HDR & 3 &  3 \\
\hline
\end{tabular} 
\label{tab:DRweigths} 
\end{minipage}
\begin{minipage}[t]{0.33\linewidth}
\vspace{0.3cm}
\centering 
\begin{tabular}{|l|c|c|} 
\hline
Weight/DR & $f^-_0$ & $f^-_1$ \\
\hline 
\rule[-0.05cm]{0cm}{.35cm}Fixed-$t$ & 3 &  3\\
\rule[-0.05cm]{0cm}{.35cm}HDR 0-sub & 3 &  3 \\
\rule[-0.05cm]{0cm}{.35cm}HDR 1-sub & 3 &  3 \\
\hline
\end{tabular} 
\end{minipage}
\begin{minipage}[t]{0.33\linewidth}
\vspace{0.3cm}
\centering 
\begin{tabular}{|l|c|c|c|} 
\hline
Weight/DR & $g^0_0$ & $g^1_1$ & $g^0_2$ \\
\hline 
\rule[-0.05cm]{0cm}{.35cm}HDR 0-sub &   & 3  &  \\
\rule[-0.05cm]{0cm}{.35cm}HDR 1-sub & 5 & 3 & 4 \\
\hline
\end{tabular} 
\end{minipage}
\end{table}

\begin{table}
\caption{List of weights used to determine our final $\chi^2_{data}$ for $\pik$ and $\pi \pi \to K \bar K$ scattering} 
\hspace*{1cm}
\begin{minipage}[t]{0.62\linewidth}
\centering 
\begin{tabular}{|l|c|c|c|c|c|c|c|c|} 
\hline
\rule[-0.05cm]{0cm}{.35cm}  & $S^{1/2}$ & $S^{3/2}$ & $P^{1/2}$ & $P^{3/2}$ & $D^{1/2}$ & $D^{3/2}$ & $F^{1/2}$ & $G^{1/2}$  \\ 
\hline 
\rule[-0.05cm]{0cm}{.35cm}$W_{pw\,i}$ & 12 & 3 & 12 & 3 & 5 & 2 & 3 & 2\\
\hline 
\end{tabular} 
\label{tab:pwweigths} 
\end{minipage}
\begin{minipage}[t]{0.28\linewidth}
\centering 
\begin{tabular}{|l|c|c|c|} 
\hline
\rule[-0.05cm]{0cm}{.35cm}  & $g^0_0$ & $g^1_1$ & $g^0_2$  \\ 
\hline 
\rule[-0.05cm]{0cm}{.35cm}$W_{pw\,i}$ & 10 & 10 & 8\\
\hline 
\end{tabular}
\end{minipage}
\end{table}

Let us then show the improvement in the fulfillment of dispersion relations by comparing the UFD versus CFD $\hat d_i^2$ for all our dispersion relations. 

Thus, in Table \ref{tab:DRtests} we have collected all the UFD $\hat d_i^2$ that we obtained for \pik dispersion relations in Section \ref{sec:DRTests} and we have displayed them together with the $\hat d_i^2$ resulting from the CFD. The improvement is dramatic: except for the $F^+$ FDR, all the CFD averaged distances between input and output are one or less, showing a remarkable fulfillment of dispersion relations. 
The $\hat d^2$ for $F^+$ is 1.2, but in the top right panel of Fig.~\ref{fig:fdrchecks} we see that almost all the deviation comes from the region around and above 1.6 GeV. We already found this problem in \cite{Pelaez:2016tgi} and it simply means that we are not able to obtain a consistent description of this FDR above 1.6 GeV if we still want to describe the data there. As discussed in section \ref{sec:ufdregge},  the deviation is due to the non-continuous matching between the data and the Regge asymptotic formulae, which are an ``average'' description,  which occurs at 1.84 GeV, as shown in Fig.~\ref{fig:regge} (top left panel). Nevertheless, below 1.6 GeV the consistency is  remarkable. 
The agreement is particularly impressive for the FDRs and the $\pik$ unsubtracted HPWDR, taking into account the huge deviations they had in the UFD case.

\begin{table}
\caption{Average $\hat d^2$ distances for the various $\pik$ dispersion relations using the UFD or the CFD parameterizations. The improvement in the CFD case is remarkable.} 
\begin{minipage}[t]{0.49\linewidth}
\vspace{0.3cm}
\centering 
\begin{tabular}{|l|c|c|} 
\hline
\rule[-0.05cm]{0cm}{.35cm}FDR & UFD & CFD \\ 
\hline 
\rule[-0.05cm]{0cm}{.35cm}$F^+(s)$ & 9.0 &  1.2\\
\rule[-0.05cm]{0cm}{.35cm}$F^-(s)$ & 8.2 &  0.7\\
\hline 
\hline
\rule[-0.05cm]{0cm}{.35cm}FTPWDR & UFD & CFD \\
\hline
\rule[-0.05cm]{0cm}{.35cm}$f^+_0(s)$ & 6.2 &  0.6\\
\rule[-0.05cm]{0cm}{.35cm}$f^-_0(s)$ & 1.6 &  0.1\\
\rule[-0.05cm]{0cm}{.35cm}$f^+_1(s)$ & 3.3 &  0.1\\
\rule[-0.05cm]{0cm}{.35cm}$f^-_1(s)$ & 5.1 &  0.8\\
\hline
\end{tabular} 
\label{tab:DRtests} 
\end{minipage}
\begin{minipage}[t]{0.49\linewidth}
\vspace{0.3cm}
\centering 
\begin{tabular}{|l|c|c|} 
\hline
\rule[-0.05cm]{0cm}{.35cm}HPWDR 0-sub & UFD & CFD \\ 
\hline 
\rule[-0.05cm]{0cm}{.35cm}$f^-_0(s)$ & 14.1 &  0.7\\
\rule[-0.05cm]{0cm}{.35cm}$f^-_1(s)$ & 13.3 &  0.7\\
\hline 
\hline
\rule[-0.05cm]{0cm}{.35cm}  HPWDR 1-sub & UFD & CFD \\
\hline
\rule[-0.05cm]{0cm}{.35cm}$f^+_0(s)$ & 6.6 &  0.8\\
\rule[-0.05cm]{0cm}{.35cm}$f^-_0(s)$ & 3.5 &  0.4\\
\rule[-0.05cm]{0cm}{.35cm}$f^+_1(s)$ & 3.2 &  0.1\\
\rule[-0.05cm]{0cm}{.35cm}$f^-_1(s)$ & 3.0 &  0.3\\
\hline
\end{tabular} 
\end{minipage}
\end{table}

A substantial improvement is also reached when using the CFD in the HPWDR for the \pipikk channel, as shown in Table \ref{tab:DRcrosstests}. 
We show results for the CFD$_C$ set, but the CFD$_B$ set is very similar. All these dispersion relations are well satisfied, with a $\hat d^2$ smaller than one, except for the  $g^0_0$, whose $\hat d^2=2.1$, which is nevertheless half of what was found with the UFD.
Notwithstanding, as shown in the top right panel of Fig.~\ref{fig:g00checks}, almost the whole contribution to $\hat d^2$ for this dispersion relation comes from the region very near the $K\bar K$ threshold, which we are afraid cannot be described consistently within the isospin symmetric formalism. Beyond that threshold region, the agreement is once again remarkable and the averaged $\hat d^2$ would be roughly one. 

\begin{table}
\caption{Average $\hat d^2$ distances for the various $\pi \pi \to K \bar{K}$ dispersion relations. 
The improvement in the consistency of the CFD with respect to the UFD is remarkable. For the CFD, the only $\hat d^2>1$ is that of $g^0_0(t)$, but the largest contribution comes from the 10-20 MeV region right above the $K \bar K$ threshold and is mostly due to isospin breaking effects that cannot be accommodated within our isospin symmetric formalism. Otherwise, its CFD $\hat d^2$ is less than one.}
\vspace{0.3cm}
\centering 
\begin{tabular}{|l|c|c|} 
\hline
\rule[-0.05cm]{0cm}{.35cm}HPWDR & UFD & CFD \\ 
\hline 
\rule[-0.05cm]{0cm}{.35cm}$g^0_0(t)$ & 4.1 &  2.1\\
\rule[-0.05cm]{0cm}{.35cm}$g^0_2(t)$ & 1.7 &  1.0\\
\rule[-0.05cm]{0cm}{.35cm}$g^1_1(t)$ 0-sub & 2.6 &  0.9\\
\rule[-0.05cm]{0cm}{.35cm}$g^1_1(t)$ 1-sub & 5.2 &  0.2\\
\hline
\end{tabular} 
\label{tab:DRcrosstests} 
\end{table}

This said it is not enough to check that the averaged distances $\hat d^2_i$ are close to one or less.
It is also important that this global fulfillment of dispersion relations is uniform in their applicability domain. This is indeed the general case with the only two exceptions already noted. We illustrate this rather uniform consistency throughout the energy regions of interest in the right panels of Figs.~\ref{fig:fdrchecks} to \ref{fig:g00checks}. 

In particular, in the right column of Fig.~\ref{fig:fdrchecks} we show the output of the two FDRs versus the input when using the CFD. The remarkable improvement in the whole energy region can be seen by comparing these figures to their UFD counterpart in the central column of the same figure.
For both $F^\pm$ FDRs, the CFD input and output now overlap within errors, except in the region around 1.6 GeV. However, for the $F^-$ FDR (lower right panel), the deviation in that region is only very slightly outside the uncertainty band, whereas for the $F^+$ (upper right panel) the deviation between input and output is rather large and we must conclude, as we already commented above and already found in \cite{Pelaez:2016tgi} that, whereas the $F^-$ FDR consistency is very good even up to 1.7 GeV,  we cannot find a consistent description of the $F^+$ beyond $\sim1.6$ GeV unless we force a large deviation from data.

Let us now discuss the improvement in the consistency of the FTPWDR. Thus, in the right panels of Fig.~\ref{fig:ftchecks} we show their input versus output when using the CFD, which has to be compared with their UFD counterparts in the central column of the same figure.
Only for $f^+_0$ and $f^-_1$ we can see that the input lies very slightly outside the uncertainty band in the region around 0.95 GeV. Thus, the consistency between CFD input and output for all partial-wave dispersion relations from fixed-$t$ is rather uniform.

Something similar happens with the $\pi K$ HPWDR.
We show first the CFD case for the $f^-_\ell$ obtained with the unsubtracted $F^-$ in the right panels of Fig.~\ref{fig:hdrchecks}. The improvement with respect to the UFD case, shown in the central panel, is the most striking of them all.
The huge inconsistencies of the UFD case have disappeared, and only beyond 0.9 GeV we can observe a very slight deviation outside the uncertainty band. Next, we show the once-subtracted CFD case in the right column of Fig.~\ref{fig:hdrcheckssub}. Once again we find a very nice and quite uniform consistency within uncertainties in the whole energy region, correcting all the rather large discrepancies found in the UFD case (central panels).

The situation with the \pipikk HPWDR is illustrated in the right columns of Figs.~\ref{fig:g11checks} and ~\ref{fig:g00checks}.
Once again, the inconsistencies we found for the UFD set (central panels) disappear when using the CFD as input. The only exception, as already commented, is the region close to $K\bar K$ threshold, particularly for the $g_0^0$ wave, which is enhanced by the presence of the $f_0(980)$ resonance. We attribute this discrepancy to using an isospin-symmetric formalism, which is not well suited, at this level of precision, to the actual existence of two different thresholds for $K^0\bar K^0$ and $K^+K^-$.  As a matter of fact, it is well known that isospin violation is enhanced around these thresholds due to the simultaneous presence of the $f_0(980)$ and $a_0(980)$ and their  mixing  \cite{Achasov:1979xc,Hanhart:2007bd}. But beyond, say, 10 or 20 MeV above $t_K$ the agreement between input and output is very nice again.

One might wonder if our dispersion relations become strongly correlated among themselves and then we overweight their combined constraint. Of course, some correlation is unavoidable, because they share the same input.  Also, having one more subtraction 
we introduce a dependence on a combination of threshold parameters through the additional subtraction constant, which is not really an independent parameter from the parameterizations. However, we have seen in the previous section 
that, by using different subtractions or fixed-$t$ instead of hyperbolic dispersion relations, the input is weighted very differently, to the point that different dispersion relations have different dominant contributions, yielding different central values and uncertainties. In an ideal world with perfect data without uncertainties, they should be the same, but when we have data and uncertainties weighted differently on each dispersion relation, it is not trivial that they are all satisfied for all values of $s$. In particular, the $W_i$ are chosen so that  the final $\hat d^2$ is uniformly close or smaller than one in the whole applicability region. This said, the existence of correlations is seen because forcing the dispersion relations with worse unconstrained $\hat d^2$ (larger than 5 or 10) to be satisfied uniformly, produces a general improvement in the others, which, starting from not so large $\hat d^2$ end up with values much smaller than 1. This can be due to their correlations or to the different sizes of the uncertainties.  Nevertheless, in practice, the penalty function of the dispersion relations that start with the smaller distances has little real weight, and its contribution to the constrain is minimal.

Of course, now that we have shown the remarkable consistency of the CFD, one might wonder if the constraints in the fits have spoiled badly the data description. We have already advanced that the CFD is still a fairly good description of data, although some deviations from the UFD to the CFD are worth noticing and we will comment on them next in detail. 

When discussing the quality of the fits we should  first recall that we have included systematic uncertainties in the data and that this makes the UFD have $\chi^2/dof\sim$ 1 on the average. After imposing the whole set of 16 dispersion relations on the fits, and using the same procedures to calculate systematic uncertainties, we find that the CFD set has $\chi^2/dof\sim 1.6$. Thus, the price to pay to satisfy the dispersion relations is that our fits central values lie on the average about 1.25 standard deviation from data. This means that we will not find large deviations from CFD to UFD, except maybe in some particular waves and regions. However, the $\hat d^2$ per dispersion relations decreases from 5.5 for the UFD to 0.6 for the CFD.

\subsection{$\pik$ CFD}
\label{sec:CFDpik}

Let us start discussing how each \pik partial wave is modified in detail. Remember these are used only up to 1.84 GeV. Beyond that, we use Regge parameterizations.

\subsubsection{CFD $S$-waves}
As it can be seen in Fig.~\ref{fig:S12data}, the whole  CFD  $S$-wave  barely changes
with respect to its UFD in the whole energy region from \pik threshold up to 1.8 GeV. The very small deviations between the CFD and UFD central values are well within their uncertainties. However, that  is for the $f_0^{1/2}+f_0^{3/2}/2$ combination and we will see next that whereas the $S^{1/2}$ barely changes, the $S^{3/2}$ suffers a somewhat larger variation.

\vspace*{.3cm} \underline{$S^{3/2}$ partial wave} \\

The CFD is shown in Fig.~\ref{fig:s32data} and it mostly changes with respect to the UFD curve below 1 GeV, becoming less negative and moving away from the Estabrooks et al. data \cite{Estabrooks:1977xe}, which is clearly disfavored by the dispersive constraint. This is the main cause why the $\chi^2/dof$ goes from 2.6 to a bit above 3. As a consequence, we will see in subsection \ref{subsec:SRth} below that the CFD scattering length is also less negative and then remarkably consistent with its sum-rule value, contrary to the UFD case. In general, this CFD variation makes the direct calculation of most threshold parameters and their combinations to agree much better with the more robust results from sum rules, as we  will see in subsection \ref{subsec:SRth}.
        
The CFD parameters have already been provided in Table~\ref{tab:S32pa}. There we can see that  they are rather similar to their UFD counterparts, except for $B_2$, which lies above 3 standard deviations away.

\begin{figure}[!ht]
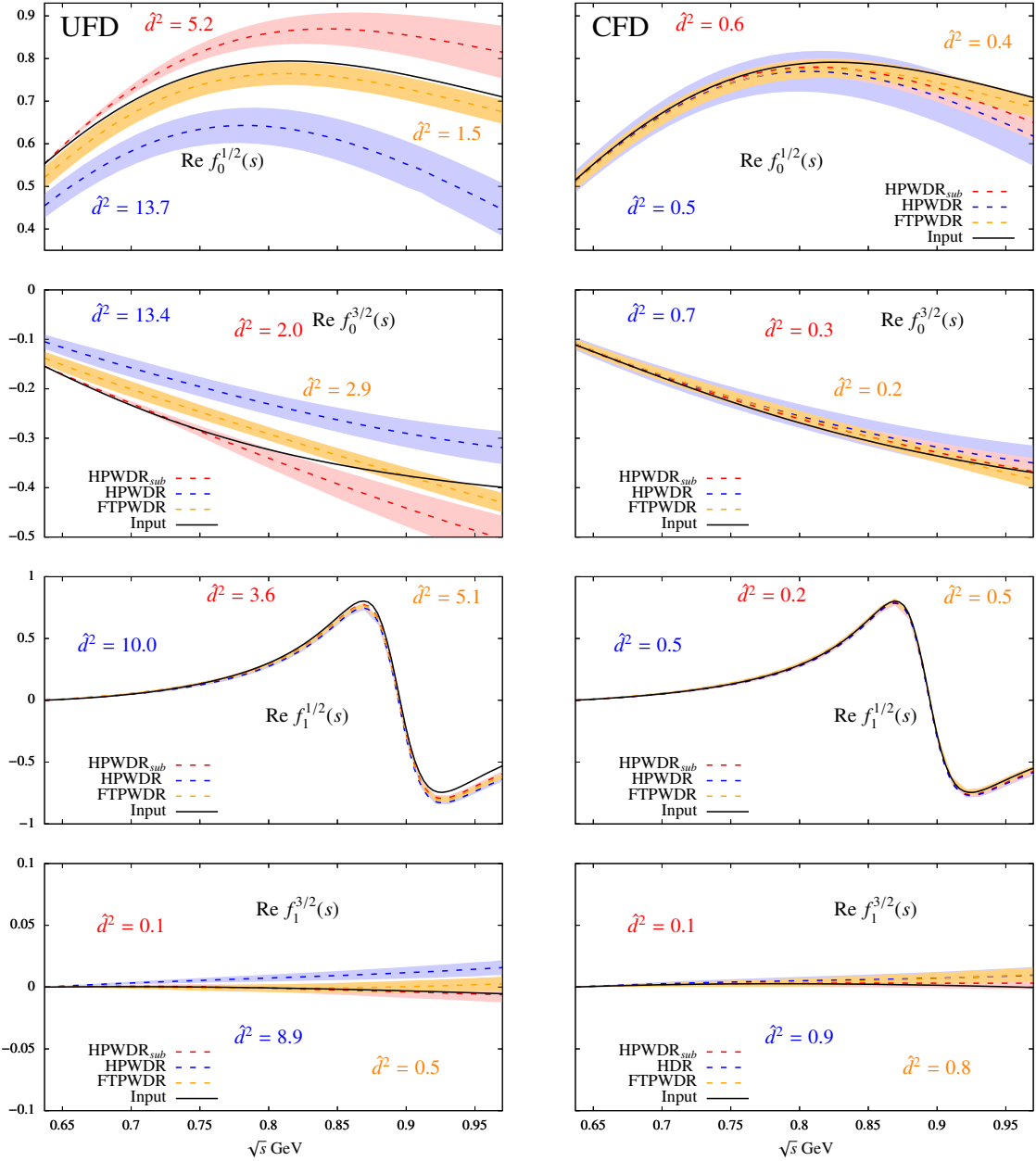

\centering
\resizebox{0.9\textwidth}{!}{\input{figures/kappa12ufd.tex} \hspace{0.3cm} \input{figures/kappa12cfd.tex}}\\
\resizebox{0.9\textwidth}{!}{\input{figures/kappa32ufd.tex} \hspace{0.3cm} \input{figures/kappa32cfd.tex}}\\
\resizebox{0.9\textwidth}{!}{\input{figures/kappa121ufd.tex} \hspace{0.3cm} \input{figures/kappa121cfd.tex}}\\
\resizebox{0.9\textwidth}{!}{\input{figures/kappa321ufd.tex} \hspace{0.3cm} \input{figures/kappa321cfd.tex}}\\
\caption{Dispersive and direct results for UFD (left) versus CFD (right) real parts of the $f^{1/2}_0(s),f^{3/2}_0(s),f^{1/2}_1(s)$ and $f^{3/2}_1(s)$ partial waves, in respective order from top to bottom. The uncertainty bands correspond to the uncertainty in the difference between the input and the respective dispersive representation, although we have attached them to their dispersive central value for simplicity. In the HPWDR and HPWDR$_{sub}$ cases the $F^-$ HDR has been used unsubtracted and once-subtracted, respectively.}
\label{fig:alltogether}
\end{figure}

Up to here, we have been discussing our parameterization of the $S^{3/2}$-wave. However, one of the goals of this report is to provide  robust and model-independent descriptions of \pik partial waves, by using the dispersive representation. We may wonder if we have achieved our goal, but so far we have shown the consistency of the dispersive representation for symmetric and antisymmetric $f^\pm_\ell$ waves. Consistency is then also expected  for the  $f^I_\ell$ in the isospin basis. Thus in Fig.~\ref{fig:alltogether} we show the three dispersive calculations of the real part of $f^I_\ell$, namely, those coming from FTPWDR (orange) as well as unsubtracted or subtracted HPWDR (blue and red, respectively). The input from the CFD parameterization is represented by a continuous black line and the error bands represent the uncertainty on the difference between this input and the corresponding dispersive representation.

In particular, in the second row of Fig.~\ref{fig:alltogether} we see the dispersive results for $f^{3/2}_0$ in the elastic region. The UFD results (left) are largely inconsistent since the different dispersion relations for this partial wave do not agree with each other. However, we can see that this inconsistency disappears for the CFD (right), and all the dispersive representations overlap and agree with our direct use of the CFD parameterizations, which is therefore very robust.

\vspace{.3cm} \underline{$S^{1/2}$ partial wave} \\  

In the elastic region, shown in Fig.~\ref{fig:s12elastic}, the change from UFD to CFD is almost imperceptible. Actually, the $\chi^2/dof$ goes from 1.4 to around 1.8,  and the UFD and CFD scattering lengths are compatible. In addition, we already commented that our simple parameterization of the elastic region does have a  \kap pole in the second Riemann sheet, which for the CFD now lies at $\sqrt{s_p}=(673\pm13)-i (331\pm 5)$ MeV.
It is relatively close to the UFD pole we gave in subsection~\ref{subsub:swave}. However, 
it is interesting to note that, despite the CFD and UFD curves being almost indistinguishable in the real axis, their naive continuation to the complex plane displaces the real part of this pole position by almost two-standard deviations. This is an illustration of the instability of naive extrapolations of models to the complex plane, which cannot be trusted for precision studies, particularly for wide resonances. This is why a dispersive approach is needed to determine rigorously the \kap pole position and residue, as we will review in section \ref{sec:kappa} below.

The CFD versus UFD curves are shown in Fig.~\ref{fig:S12all} in the complete energy range of the study. There we also see that in the inelastic regime  they are also consistent within uncertainties. Once more we find almost the same $\chi^2$.

The fact that we find fewer deviations between UFD and CFD than in our previous analysis in ~\cite{Pelaez:2016tgi}, is  due to the improvement in the $S^{3/2}$-wave, needed to separate the $S^{1/2}$ component, and to our new fitting strategy, since when we constrain the fit now we also fit the data and not the parameters of the UFD fit as done in the past.

The CFD parameters of this wave were already given in Tables \ref{tab:Selparam}  and \ref{tab:Sinparam}. Many of them barely move. However, there are several that change by around two standard deviations. Nevertheless, correlations among them leave the curves within one standard deviation.

So far, we have been discussing our parameterization of the $S^{1/2}$-wave. Concerning the dispersive outputs,  in the first line of Fig.~\ref{fig:alltogether} we have gathered all the dispersive results for $f^{1/2}_0$ and compared them with the direct use of the CFD.
On the left, we see that, although the results coming from FTPWDR are roughly consistent with the input, the two hyperbolic determinations are at odds with the UFD input and even more inconsistent among themselves. This is very important, for instance, for the determination of the \kap pole, since, strictly speaking, that pole is only within the reach of HPWDRs, not the projected fixed-$t$. Therefore, the same UFD input would yield incompatible \kap poles depending on what HPWDR is used. We will see this in detail in section \ref{sec:kappa}.

Nevertheless, all these problems disappear for the CFD parameterization (top right), since the three dispersive representations are consistent among themselves and with the CFD parameterization. We thus consider that the latter is a very robust and consistent description of this controversial wave.
    
\subsubsection{CFD $P$-waves}
    
The CFD result for the $P$-wave, obtained in the $f_P\equiv f^{1/2}_1+f^{3/2}_1/2$ combination  was already shown in Fig.~\ref{fig:pwavedata}, versus data and the UFD. There, it can be noticed that the CFD and UFD are pretty similar in the elastic region but there is some visible deviation in the inelastic regime. Of course, since in the inelastic region the uncertainties are rather large, the deviation  between CFD and UFD is within or only slightly off the uncertainty, except maybe beyond 1.6 GeV, where our dispersive representation does not reach. We, therefore, conclude that the CFD still provides a fairly reasonable description of data. In terms of the separated isospin contributions the situation is as follows:

\vspace{.3cm} \underline{$P^{3/2}$ partial wave} \\ 

As it can be seen in Fig.~\ref{fig:p32wavephase}, the $P^{3/2}$ phase barely changes from the UFD (orange) to the CFD (black continuous curve with the grey band). Recall that, for all means and purposes, this wave is elastic up to 1.8 GeV, and thus is completely determined by its phase. The  $\chi^2/dof$ of the data, with the same systematic uncertainty estimations used for the UFD fit, changes from  1.2 for the UFD to around 2 for the CFD. 

However, it can be noticed that the central value of the CFD phase stays positive from threshold up to $\sqrt{s}\simeq 1\,$GeV. As we will see in \ref{sec:sumrules}, this makes the scattering length larger and somewhat closer to the sum rule and ChPT results. Actually, all our dispersive sum rules in Table~\ref{tab:leparameters} in section \ref{sec:sumrules}, the predictions coming from the ones of B\"uttiker et al.~\cite{Buettiker:2003wj} and the ChPT determinations~\cite{Bernard:1990kw,Bijnens:2004bu} point to a slightly positive scattering length, even though  the  partial wave far from threshold has to be negative according to the existing data, which lie above 1 GeV.

The CFD parameters of this wave are listed in Table~\ref{tab:Pelparam}. They lie  within less than one-third of the uncertainties of their UFD counterpart, except for $\hat s$, which signals the energy where the phase changes sign, which has moved  by two standard deviations up to $(0.88\pm0.17)\,$GeV$^2$.

The consistency of this CFD parameterization with its dispersive representations can be seen in the bottom right panel of in Fig.~\ref{fig:alltogether}, whereas for  the UFD case (bottom left) there was a considerable disagreement with the HPWDR when the $F^-$ is used unsubtracted.

\vspace{.3cm} \underline{$P^{1/2}$ partial wave} \\ 

Given that the $P^{3/2}$-wave is so small, the plot in 
Fig.~\ref{fig:pwavedata} already tells us that the 
change from UFD to CFD for the $S^{1/2}$ is very mild.

In particular, using the same systematic uncertainty estimates, the  $\chi^2/dof$ changes from 1.1 to around 2.1 in the elastic region, which is plotted in Fig.~\ref{fig:Pelastic}. Nevertheless, the data error bars are so small that the CFD and UFD difference is almost imperceptible to the eye, except in the region above the $K^*(892)$ nominal mass, which is mostly responsible for the increase in $\chi^2$. This produces a non-negligible shift in the value of the phase at the  matching point used by the Paris group~\cite{Buettiker:2003pp}. As already explained in~\cite{Pelaez:2016tgi}, we think this might contribute to the deviation from the data of the dispersive solution in \cite{Buettiker:2003pp}, which is shown as a green line in Fig.~\ref{fig:Pelastic}.

The CFD parameters of the elastic part were given in Table~\ref{tab:Pelparam12} and although $B_1$ and the peak mass of the resonance $m_r$ are consistent with their UFD counterparts, $B_0$ and $B_2$ lie more than two standard deviations apart. As we will see in section \ref{sec:sumrules}, this has important consequences for the slope threshold parameter, which for the CFD becomes consistent with the sum rule and the ChPT estimate, whereas the UFD result was not.

The elastic parameterization of this wave also has a pole in the second Riemann sheet associated with the $K^*(892)$ resonance. For the CFD case this is found at $\sqrt{s_p}=(891\pm2)-i (27\pm 1)$ MeV, which overlaps within uncertainties with the UFD pole we gave in subsection~\ref{subsub:pwave}. Although this is a narrow resonance and its pole parameters are relatively stable when extracted from simple parameterizations of the data around its peak, we will also provide a rigorous dispersive value in section \ref{sec:kappa}.

Once again we use Fig.~\ref{fig:alltogether} to illustrate  the consistency and robustness of the CFD parameterization of this wave, displayed in the third row of panels.
Despite the small uncertainty bands, it can be noticed that using the UFD (left), the input did not agree with the three dispersive representations. However, they all agree well within uncertainties for the CFD. We, therefore, consider that the description of the scattering data in the elastic region is very robust.

As already commented in section~\ref{sec:Data} the elastic phase shift could also be obtained from other sources. In \ref{app:palt} we have studied this possibility, but the resulting CFD is almost indistinguishable from the CFD presented here.

In the inelastic region, the phase changes by a fair margin, better visible in Fig.~\ref{fig:pwavedata}, since the uncertainties are larger and the $\chi^2$ increases from 0.9 up to somewhat more than 2. It seems that the $K^*(1410)$ interference with the other two resonances should be slightly different than that obtained from unconstrained fits.

The parameters of the inelastic CFD for this wave were already given in Table~\ref{tab:Pinpa}. There we can see that they are very similar to their UFD counterparts, but given the size of the uncertainties, the variations reach three standard deviations in a few instances.

\subsubsection{CFD $D$-waves}

Contrary to the previous cases, we have not used their partial-wave dispersion relations as constraints.
Nevertheless, they are still modified because they are input for the forward dispersion relations and for the other partial-wave dispersion relations.

\vspace{.3cm} \underline{$D^{3/2}$ partial wave} \\ 

Let us recall that this wave is tiny up to 1.8 GeV, barely reaching $-2^\degree$ at most. The CFD result can be found in Fig.~\ref{fig:D32phase}. It can be seen that it is fairly consistent with the UFD up to $\simeq1.2\,$GeV, but beyond that energy, the CFD prefers an even less negative phase, closer to $-1^\degree$. As a result the data  $\chi^2/dof$ increases from around 1.1 to around 1.6, which we consider still fairly reasonable. Most of this $\chi^2$ comes from two outliers that were also inconsistent with the UFD and the rest of the data and, in contrast, the last data point which was an UFD outlier overlaps with the CFD within uncertainties.

The CFD parameters of this wave were given in Table~\ref{tab:D32param} and they are all consistent with their UFD counterparts, except $B_2$, which lies roughly $2.5$ deviations away.

In any case, the effect of this wave is always very small and is even smaller for the CFD.

\vspace{.3cm} \underline{$D^{1/2}$ partial wave} \\ 

Given the fact that the $f^{3/2}_2$ wave is so small, the $f_D \equiv f^{1/2}_2+f^{3/2}_2/2$ data and curves shown in Fig.~\ref{fig:Dwavedata} illustrate in practice the 
$f^{1/2}_2$ changes from CFD to UFD.
The modulus of this wave barely changes around the $K^*_2(1430)$ peak region. However, 
there is a visible deviation in the modulus far from the resonance and in the phase 
right after the peak. 
This should not be surprising since the data far from the resonance are not very reliable or consistent with each other. All in all, the $\chi^2/dof$ increases from 1.2 to roughly 3. Most of this deviation comes from the last three points above 1.6 GeV of the modulus measured by Aston et al. \cite{Aston:1987ir} as well as a few points which are outliers both for the UFD and CFD and quite distant from the other data. In general, it seems like the modulus clearly prefers the solution by Estabrooks et al.~\cite{Estabrooks:1977xe} over the one by Aston et al.~\cite{Aston:1987ir}, particularly at higher energies.

The CFD parameters of this wave have already been listed in Table~\ref{tab:Dwave} and they all change by about 2 deviations. Once more, finding somewhat larger changes in this wave should not be surprising given the poor quality and/or consistency of data outside the resonance peak, as shown in Fig.~\ref{fig:Dwavedata}.
    
    \subsubsection{CFD $F$-wave}
For this wave, seen in Fig.~\ref{fig:Fwavedata}, there are only data above 1.5 GeV, and very few for the phase, all of them dominated by the $K^*_3(1780)$. 
Both the UFD and CFD yield a fairly good description of the resonance although the $\chi^2/dof$ increases from around 1 to 1.5. The CFD parameters are given in Table~\ref{tab:Fwave} and their change is sizeable, often beyond 2 standard deviations,  although due to correlations the CFD and UFD overlap within uncertainties, as shown in Fig.~\ref{fig:Fwavedata}.

\subsection{$\pipikk$ CFD}
\label{sec:CFDpipiKK}

After discussing \pik,  we turn our attention to \pipikk partial waves to show how they changed from the UFD to the CFD set. Remember these are used only up to 2 GeV. Beyond that, we use Regge parameterizations.

\subsubsection{CFD I=0 $S$-wave}

Let us recall that here we had two sets of incompatible data for the modulus and their corresponding UFD$_B$ and UFD$_C$, which differ substantially below 1.47 GeV. They both shared the same UFD fit for the phase.
Their respective CFD 
were already shown in Fig.~\ref{fig:g00data}.

It can be noticed that in Region II, i.e. above 1.47 GeV, the new CFD$_B$ and CFD$_C$ are once again perfectly compatible within uncertainties both with one another and with the UFD.  The same can be said about the phase, although now in the whole energy region.

However, there are sizable changes in the modulus of both UFDs to their respective CFDs in ``Region I'', i.e. below 1.47. Actually, in both cases, the CFD modulus becomes smaller than its UFD counterpart below 1.25 GeV
and larger between 1.3 and 1.47 GeV.
This leads to a more pronounced local minimum around 1.2 GeV and a maximum around 1.35 GeV than in their respective UFDs. This double structure is flatter for the CFD$_C$.  
All in all the $\chi^2/dof$ increases from roughly 1 up to 1.4 for CFD$_C$, and up to 2 for the CFD$_B$.
Overall the CFD seems more stable compared to our previous determination in~\cite{Pelaez:2018qny} due to our constraining simultaneously  $\pik$ and $\pipikk$.
   
The parameters of both CFD fits for the phase were already given in Table~\ref{tab:g00phase}. Note that although we only had one UFD phase, we now have two phases CFD$_B$ and CFD$_C$. This is because each one is constrained together with their respective moduli, which are different. Nevertheless, the two CFD$_B$ and CFD$_C$ phases are almost identical and therefore indistinguishable in  Fig.~\ref{fig:g00data}, both overlapping with the UFD. Consequently, their parameters only differ slightly and are always compatible well within uncertainties, both among themselves and with the UFD.

The parameters for the CFD$_B$ and CFD$_C$ moduli are given in Tables~\ref{tab:g00modufd} and \ref{tab:g00modufd2}. In this case, the $D_i$ parameters controlling the ``Region I''  clearly differ from their UFD counterparts, whereas the $F_i$ are perfectly compatible with their respective UFD values.

Concerning the dispersive solutions, which we showed in the physical region in Fig.~\ref{fig:g00checks} for the UFD$_C$ (Top center) and CFD$_C$ (Top Right), the decrease in the modulus below 1.25 that we have just commented on makes the CFD consistent with the dispersive result. This is the main reason for the improvement in $\hat d^2$. Something similar happens for the CFD$_B$.
However, as we have repeatedly noted, we cannot accommodate well the 20 MeV near threshold, which we think is due to our dealing with an isospin symmetric model. Nevertheless, apart from that region, our CFD result is very robust. As it happened in our previous work \cite{Pelaez:2018qny}, there is no clear-cut preference for the CFD$_B$ or CFD$_C$ description in terms of the $\chi^2$ and $d^2$. Fortunately, we have already shown that using one or the other as input for our other dispersion relations is irrelevant. In order to illustrate the results of these other dispersion relations, in this work we have chosen the UFD$_C$ and CFD$_C$.

With \pipikk and \pik data and dispersion relations alone we cannot exclude any of the two incompatible data sets. As it happened in \cite{Pelaez:2018qny},  the ``dip'' solution of the elasticity, favored by the $\pi\pi\rightarrow\pi\pi$  dispersive analysis in \cite{GarciaMartin:2011cn}, is consistent with the CFD$_B$  assuming just the two coupled states $\pi\pi$ and $K \bar{K}$, whereas the CFD$_C$ would need a non-negligible contribution from other states like, possibly, $4\pi$.  Nonetheless, as pointed out by one of the authors of~\cite{Longacre:1986fh} in~\cite{Morgan:1993td} there could be some normalization ambiguity affecting the CFD$_B$ solution. Indeed, the CFD$_B$ deviates somewhat more from its UFD$_B$ than  the CFD$_C$ from its UFD$_C$. In any case, as we concluded in \cite{Pelaez:2018qny}, we have found that both data sets can be reasonably well described with consistent CFD parameterizations and none of them should be discarded a priori. 

\begin{figure}[!ht]
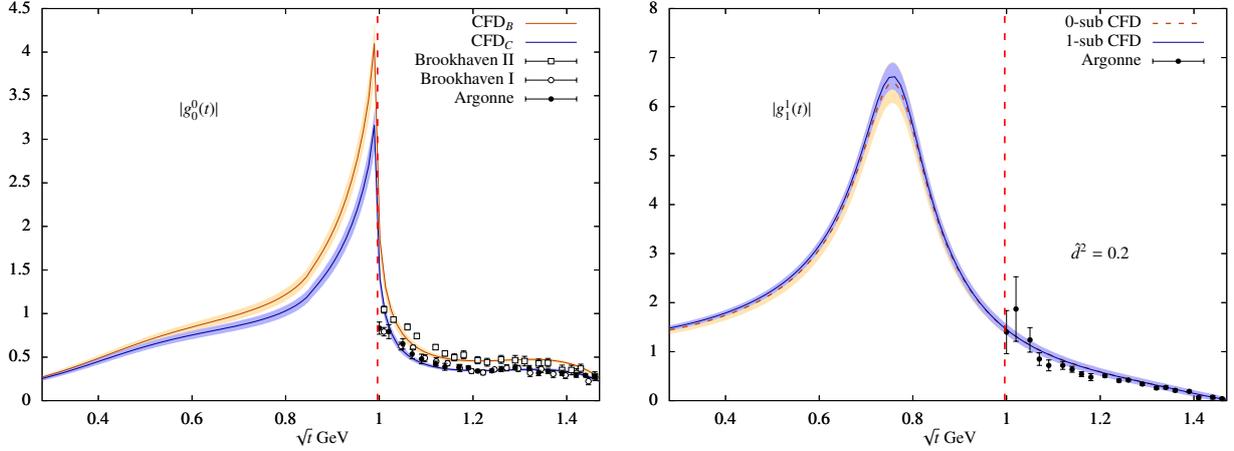

\centering
\resizebox{\textwidth}{!}{\input{figures/g00dispcfd.tex} \input{figures/g11dispcfd.tex}}\\
 \caption{Dispersive results for the modulus of the $g^0_0(t)$ and $g^1_1(t)$ partial waves using CFD as input. Left: The existence of two acceptable $g^0_0(t)$ CFD descriptions above $K\bar K$ threshold leads to two different solutions in the unphysical region. Right: Notice that the non subtracted and the once-subtracted dispersion relations for $g^1_1(t)$ are perfectly compatible even well below the physical region. We have calculated as an illustration the average $\hat{d}$ distance between the two dispersion relations, divided by the relative uncertainty.
 The vertical dashed line signals the $K\bar K$ threshold.}
\label{fig:pseudocfd}
\end{figure}

The CFD results in the unphysical region are shown in the left panel of Fig.~\ref{fig:pseudocfd}, both for the CFD$_B$ and CFD$_C$. Of course, there are no data in this region, but they can be compared to their UFD counterparts
that we showed in the left panel of Fig.~\ref{fig:gdispmod}. Neither the CFD$_C$ nor the CFD$_B$ reveal significant changes.

Finally, let us recall that the phase of the tensor wave
of the Brookhaven collaboration  \cite{Etkin:1982se} is just a model that violates Watson's theorem near $K\bar  K$ threshold and does not include an $f_2(1810)$, listed in the RPP, as we have done for that wave. Their phase is therefore different from ours, as shown in the right panel of  Fig.~\ref{fig:g02data}. Unfortunately, this wave was used to extract the data on the $I=0$ $S$-wave.
We have discarded this phase near threshold in our fits, which satisfy then Watson's theorem, but one might wonder what would happen if we used our UFD tensor wave to extract the $I=0$ $S$-wave
phase instead of the Brookhaven model. Thus, we have also considered in \ref{app:g00alt} an alternative $g_0^0$ UFD and CFD. Fortunately, they only differ significantly from the UFD and CFD presented here in the phase, but not the modulus,
and only above 1.6 GeV, i.e. outside the applicability range of our $g_0^0$ dispersive analysis, and in a region whose contribution is insignificant for other dispersion relations.

\subsubsection{CFD I=1 $P$-wave}

As can be seen in Fig.~\ref{fig:g11data}, in the physical region, $g^1_1(s)$ barely changes from UFD to CFD. The only small changes, still consistent within uncertainties, occur in the modulus near the $K \bar K$ threshold and in the phase and modulus above 1.5 or 1.6 GeV, where scattering data cease to exist. As a consequence, the CFD parameters, already listed in Table ~\ref{tab:g11wave}, barely change with respect to their UFD values or vary within uncertainties. The only exceptions are $\gamma_1$, with a different sign and three standard deviations away from its UFD counterpart and $\Gamma_{\rho'}$, two sigmas away. This is not surprising since the $\rho''(1700)$ and $\rho'(1450)$ parameters are not very robust.

The consistency of this wave either with its unsubtracted or subtracted hyperbolic dispersion relation is remarkable, as already seen in the right panels of Fig.~\ref{fig:g11checks}. Note in the left panels that the relative size of the contributions to these two dispersion relations is rather different, which makes us even more confident in the correct determination of this wave in the physical region.

Moreover, this new CFD parameterization also solves the inconsistency in the {\it unphysical} region that we showed in the right panel of Fig.~\ref{fig:gdispmod}. Recall that {\it using the same } UFD input, we obtained two incompatible predictions for the modulus of $g_1^1$ below 
$K \bar K$. Notice that there are no data in that region, however, it yields an important contribution to other dispersion relations. However, when we look at the CFD modulus, displayed in the right panel of Fig.~\ref{fig:pseudocfd}, we see that there is an impressive agreement between the unsubtracted and subtracted results. 

We, therefore, consider that this wave is very robust from the $\pi\pi$ threshold up the 1.47 GeV, i.e. {\it both in the unphysical and physical regions}. This is very important because this unphysical region is a relevant contribution to some dispersion relations, in particular for the \pik scalar waves and the precise determination of the \kap pole that we will provide in section~\ref{sec:kappa}.
    
\subsubsection{CFD $I=0$ $D$-wave}

  The CFD $g^0_2$ wave is almost identical to the UFD, as seen in Fig.~\ref{fig:g02data}. Actually, for the modulus, it may seem that there is only one curve with its uncertainty, but this is because it almost perfectly overlaps with the UFD. For the CFD phase, the central values deviate a little bit above 1.5 GeV, but are always inside the uncertainty band, in a region where there are no data at all on the phase. Let us also remark that the CFD we found here is also similar to the one we obtained in~\cite{Pelaez:2018qny}.
  
  Given this little variation, it is not a surprise that the CFD parameters, listed in Table~\ref{tab:g20wave}, change so little with respect to their UFD values. Many parameters do not change at all, some in the last digit, and most within one standard deviation. Only two of them lie above 2 standard deviations away.
  
  Let us also recall that our solution does not suffer from the violation of Watson's theorem in the model used by \cite{Etkin:1981sg} for the phase. In addition, we have also included an $f_2(1810)$ resonance, to describe the small bump in the modulus in that region whose shape is clearly visible in the phase, which therefore deviates from the Brookhaven-I model above 1.5 GeV. Since Brookhaven-I used this tensor wave to extract the scalar isoscalar wave, the use of our solution would have led to different $g^0_0$ data. This alternative solution is studied in \ref{app:g00alt}. Fortunately, using it does not change our results in the region of applicability of our dispersive analysis.

\subsection{High energy Region. CFD Regge parameterizations}

Finally, our CFD Regge parameterization also does not change much.
This can be seen in Fig.~\ref{fig:regge}. All UFD central curves lie within the uncertainties of their CFD counterparts. The only exception is the reggeized  $\rho$ resonance contribution to \pik scattering (bottom left panel), whose contribution clearly lies below the UFD curve. We already found this decrease when imposing only FDRs to \pik  \cite{Pelaez:2016tgi}. Actually, it is responsible for a substantial reduction in the $\hat d_{F^-}^2$ FDR. It is not very surprising that this contribution, governed by the factorization constant $g_{K/\pi}$, suffers a sizable change, since as already commented when we introduced it, there is little information on it, and its extraction 
from factorization is somewhat more complicated than for other resonances, see \cite{Pelaez:2003ky}. 

The Regge parameters for the CFD are listed in Tables~\ref{tab:regge}  and \ref{tab:reggepiK}. Recall that those parameters that can be fixed using reactions other than $\pi K$  or \pipikk scattering  are kept fixed for both UFD and CFD and are given in Table~\ref{tab:regge}.
In contrast, those parameters involving strangeness have been allowed to vary in the fits and are given in Table~\ref{tab:reggepiK}.
Note that those related to the reggeized-$K^*$ exchange barely change from UFD to CFD.
In contrast, the factorization constant $g_{K/\pi}$ decreases by slightly more than two standard deviations, consistently with the observed decrease in the reggeized-$\rho$ contribution commented above.
The Pomeron and $f_2$ factorization constants, $f_{K/\pi}$ and $r$, also change, but they compensate each other and the effect of this change is small as shown in the top left panel of Fig.~\ref{fig:regge}.

\section{Strange resonances and the $\kap$}
\label{sec:kappa}

\subsection{State of the art}
\label{sec:kapreview}

Despite the fact that Quantum Chromodynamics (QCD) was formulated almost half a century ago the existence and/or precise properties of some of its lowest-lying states are still under debate, in particular in the strangeness sector. A reliable determination of these strange resonances is crucial for both their own classification in multiplets, as well as for our understanding of intermediate-energy QCD interactions and the low-energy chiral perturbation theory regime. In addition, these strange resonances appear as a result of  $\pi K$ scattering or re-scattering, and contribute as final states to many of the hadronic processes with net strangeness. Hence, the shape of heavier decays, or their Dalitz plot, depends upon the precise determination of these low-energy interactions. In particular, this is the case of the many hadronic decays of the heavy $B$ and $D$ mesons, whose description nowadays relies on resonance-exchange models. Nonetheless, many of the modern experimental efforts are being directed towards these heavier processes, in particular CP violation or new physics searches. Thus a precise knowledge of these strange resonances is mandatory for model-independent studies of  heavier sectors.

\begin{figure}[!ht]
\centering
\includegraphics[width=0.42\linewidth]{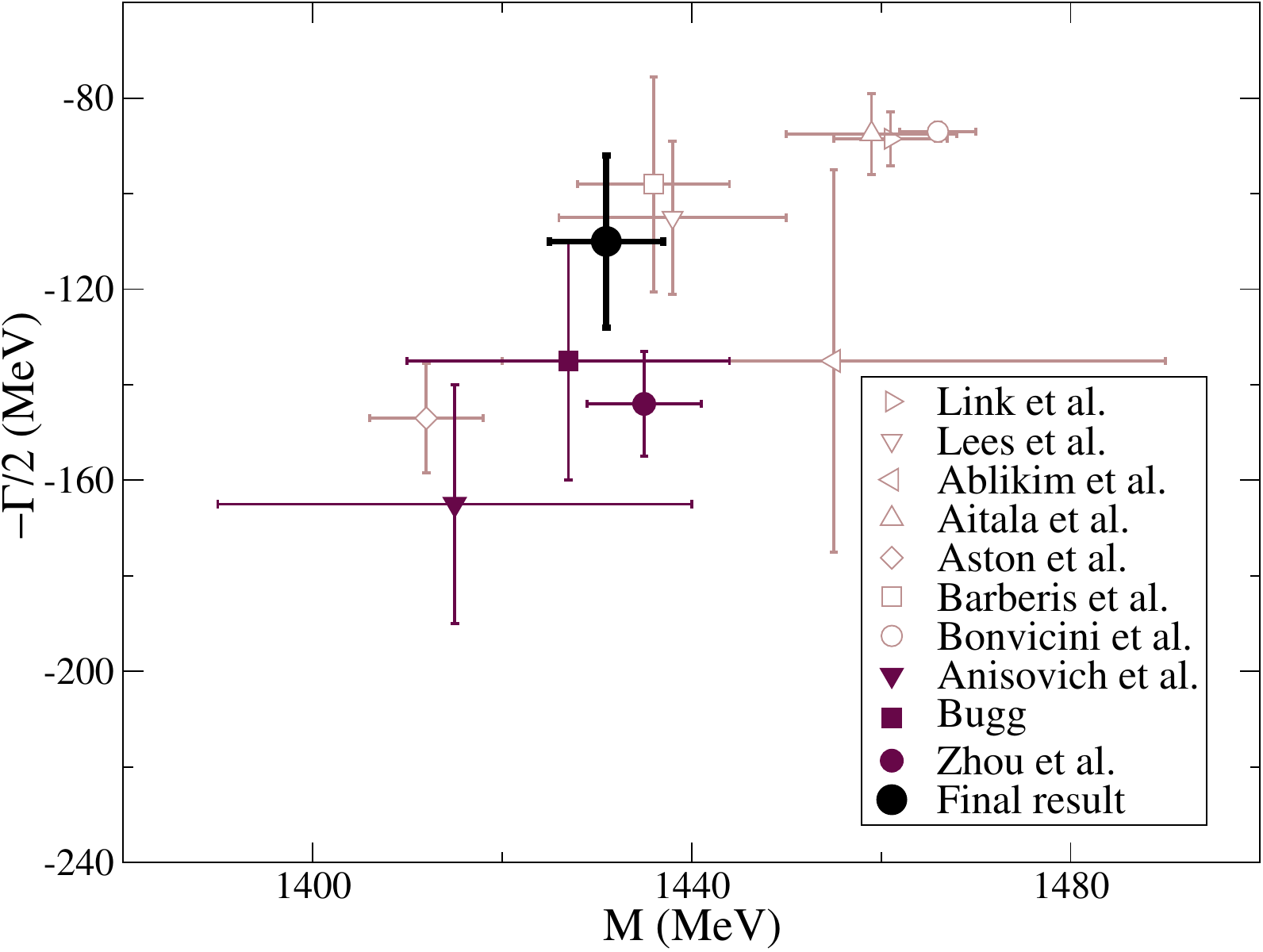} \hspace{1.cm} \includegraphics[width=0.42\linewidth]{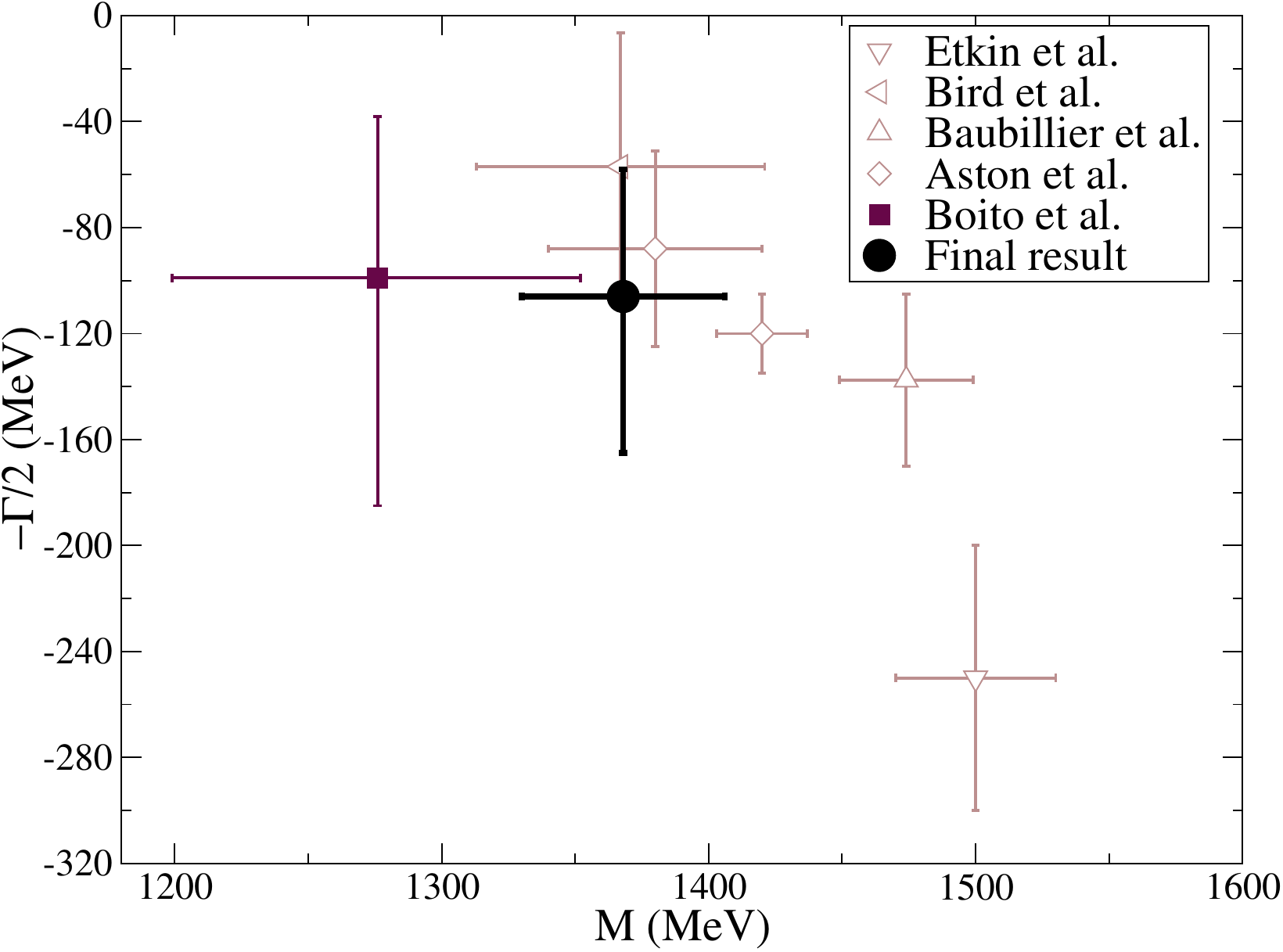}\\
\includegraphics[width=0.42\linewidth]{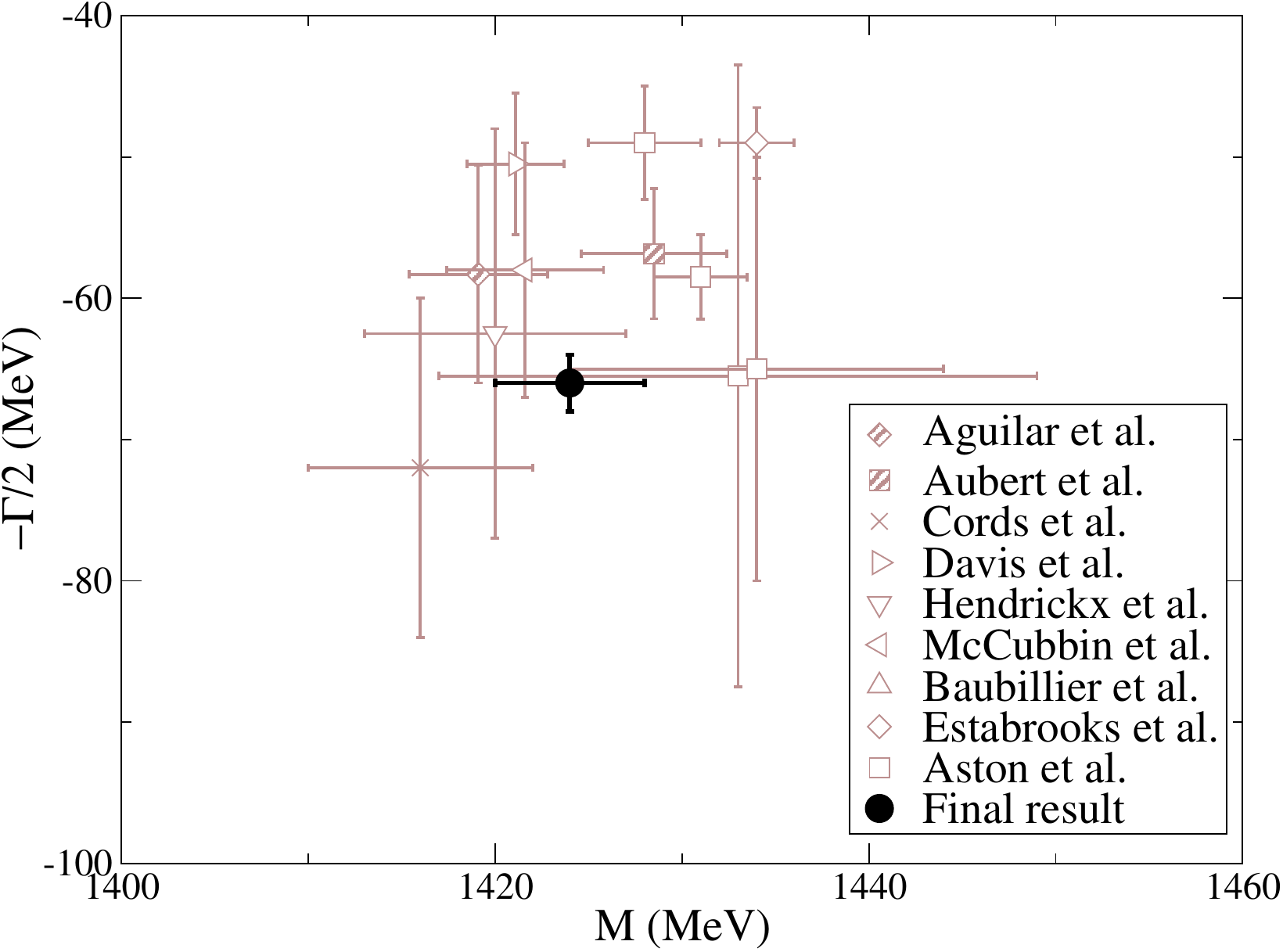} \hspace{1.cm} \includegraphics[width=0.42\linewidth]{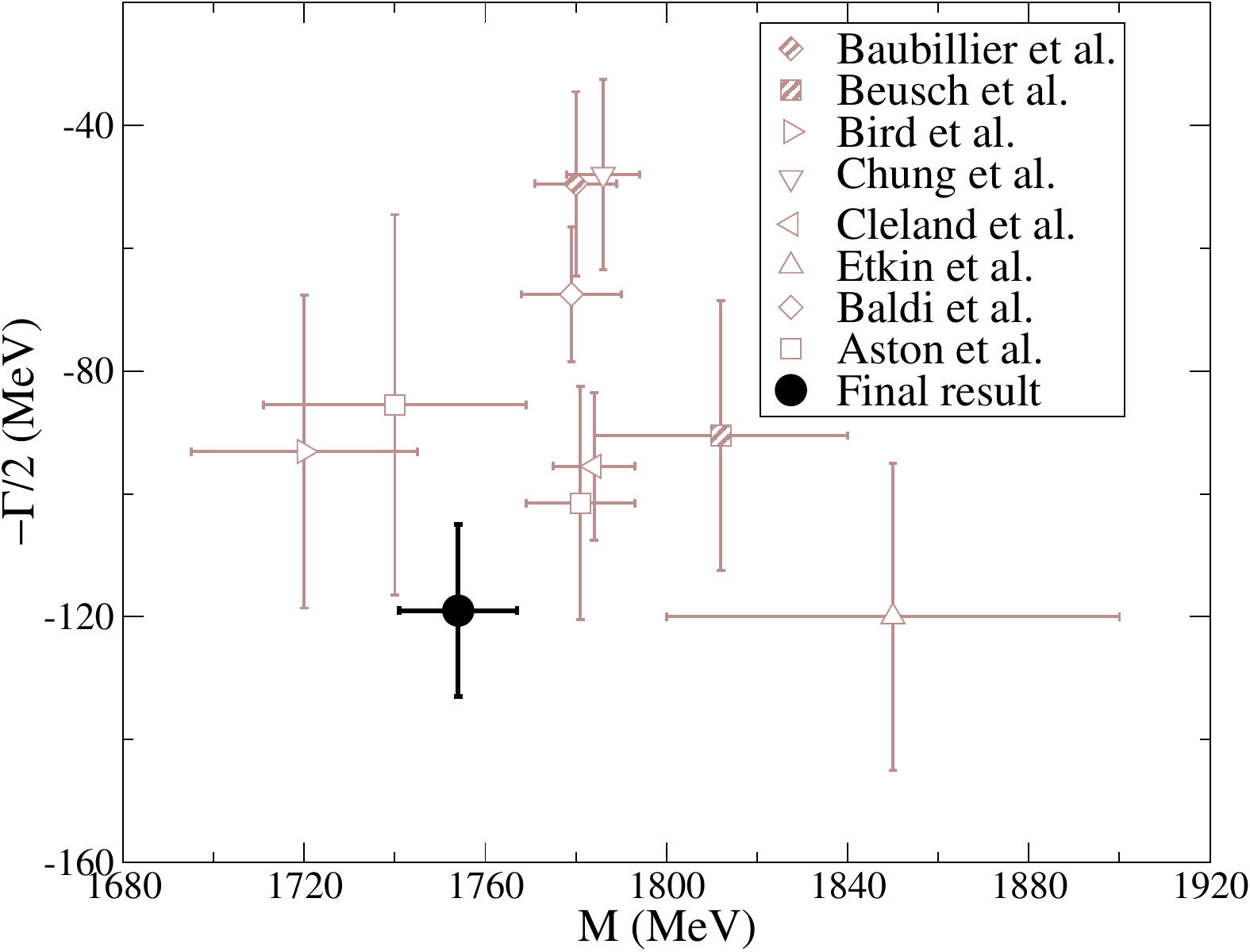}
\caption{\rm \label{fig:kpades} Figures taken from \cite{Pelaez:2016klv}.
Determination of the $K_0^*(1430)$ (top-left), $K^*(1410)$ (top-right), $K_2^*(1430)$ (bottom-left) and $K_3^*(1780)$ (bottom-right) poles obtained using the Pad\'e sequence method (``Final Result'' \cite{Pelaez:2016klv}). Other values correspond to those
listed in the RPP compilation as main results \cite{Zyla:2020zbs}, Zhou et al. \cite{Zhou:2006wm}, D.Bugg \cite{Bugg:2003kj}, Anisovich et al. \cite{Anisovich:1997qp}, Bonvicini et al. \cite{Bonvicini:2008jw}, Barberis et al. \cite{Barberis:1998tv}, Aitala et al. \cite{Aitala:2002kr}, Ablikim et al. \cite{Ablikim:2005kp}, Lees et al. \cite{Lees:2014iua}, Link et al. 
\cite{Link:2002ev}, Boito et al. \cite{Boito:2008fq}, Baubillier et al. \cite{Baubillier:1982eb}, Bird et al. \cite{Bird:1988qp}, Etkin et al. \cite{Etkin:1980me}, Estabrooks et al. \cite{Estabrooks:1977xe}, Mccubbin et al. \cite{McCubbin:1974yy}, Hendrickx et al. \cite{Hendrickx:1976ha}, Davis et al. \cite{Davis:1969sf}, Cords et al. \cite{Cords:1972hh}, Aubert et al. \cite{Aubert:2007ur}, Aguilar et al. \cite{AguilarBenitez:1972bw}, Baldi et al. \cite{Baldi:1976ua}, Cleland et al. \cite{Cleland:1982td}, Chung et al. \cite{Chung:1977ji}, Beusch et al. \cite{Konigs:1978at}
.}
\end{figure}

Besides their interest as ingredients to describe other processes, the properties and nature of many of these light resonances have been controversial for decades. This is the case of the famous \kap resonance, which ``still needs confirmation'' according to the RPP~\cite{Zyla:2020zbs}. The confirmation of the existence of this resonance is crucial for completing the lightest scalar nonet of mesons and, given its clear similarities with the \sig resonance, it would rule out the \sig glueball interpretation, already quite disfavoured in the literature (see \cite{Pelaez:2015qba} for a review). Additionally, there are six other strange resonances that dominate \pik scattering amplitudes below roughly 1.8 GeV. These are: the scalar $K_0^*(1430)$, the vectors $K^*(892)$, $K^*(1410)$ and $K^*(1680)$, the tensor $K_2^*(1430)$ and the $\ell=3$ resonance $K_3^*(1780)$. Most  experimental studies on these resonances are based either on $\pi K$ scattering experiments studying reactions like $K N\to \pi K N'$~\cite{Davis:1969sf,Mercer:1971kn,AguilarBenitez:1972bw,Cords:1972hh,McCubbin:1974yy,Hendrickx:1976ha,Baldi:1976ua,Estabrooks:1977xe,Chung:1977ji,Cleland:1982td,Aston:1987ir} or $\pi K$ rescattering from heavy-meson decays~\cite{Aitala:2002kr,Ablikim:2005kp,Bonvicini:2008jw,Lees:2014iua}. We have gathered in Fig.~\ref{fig:kpades} the pole determinations of the $K^*_0(1430)$, $K^*(1410)$,$K^*_2(1430)$ and $K^*_3(1780)$, as compiled in the Review of Particle Physics \cite{pdg}. The large spread of values for each resonance and the existence of multiple incompatible determinations are evident. The main problem is that almost all of those values suffer from systematic deviations and crude model extractions. Their biggest source of systematic uncertainty is that these resonance poles have not been extracted from a sound analytic continuation of the $T$-matrix to the complex  $s_p=M-i\, \Gamma/2$ plane, but, for the most cases,  their listed parameters come out from some sort of Breit-Wigner parameterization. This is not a model-independent definition of a resonance, but a narrow-width approximation for isolated resonances and it has been applied to these strange resonances even though they are not particularly narrow,  overlap often with other nearby resonances, or lie close to relevant thresholds.
Moreover, even the definition of the Breit-Wigner mass and width
has a sizable dependence on the particular choice of Breit-Wigner parameterization, as well as on the background present in the process that is being fitted. In the following sections we will describe and summarize how analytic techniques can improve our amplitude analyses and produce robust and yet simple ways of determining resonance parameters.

Regarding lattice QCD calculations on the strangeness sector, the first dynamical QCD determinations of resonances have only been published in the recent past. There exist several analysis regarding the prominent $K^*(892)$ resonance using $N_f=2+1$ staggered quarks~\cite{Fu:2012tj} and $N_f=2$ Wilson quarks~\cite{Lang:2012sv,Prelovsek:2013ela,Bali:2015gji}. The $K^*(892)$ is a rather narrow resonance and the use of a naive Breit-Wigner shape may be justified given the available precision, making its extraction from lattice QCD easier. Nonetheless, a precise determination of the parameters of this resonance is crucial from a lattice QCD perspective, as it allows to compare different approaches and to extrapolate these results to physical pion masses, where this channel is very well determined.

More challenging is the analysis and determination of the \kap resonance, for which  a very elaborated use of the Lüscher formalism is required~\cite{Luscher:1990ux,Rummukainen:1995vs,Bedaque:2004kc,Kim:2005gf,Fu:2011xz,Leskovec:2012gb,Gockeler:2012yj,He:2005ey,Lage:2009zv,Bernard:2010fp,Doring:2011vk,Doring:2011nd,Agadjanov:2013kja,Doring:2012eu,Hansen:2012tf,Briceno:2012yi,Guo:2012hv,Briceno:2017max}. Recent lattice QCD analyses, including $N_f=2+1$ Wilson quarks~\cite{Wilson:2014cna,Dudek:2014qha,Brett:2018jqw,Wilson:2019wfr} have been able to extract both the $S$ and $P$-waves at various quark masses. We already illustrated their phase shifts in the introduction in Fig.~\ref{fig:hadspeckap}.  
For nonphysically heavy pions \cite{Dudek:2014qha,Brett:2018jqw},  the \kap was found as a virtual bound state, compatible with unitarized NLO  ChPT predictions~\cite{Nebreda:2010wv,Guo:2018zss}.
Nevertheless, it seems  that once the pion mass starts approaching the physical value the extraction
of the \kap pole by analytic continuations becomes more unreliable. However, at higher pion masses the \kap resonance appears robustly and stably as a result of fitting the partial wave.  Other exploratory studies of the \kap and $K^*(892)$ resonances obtained from lattice QCD can be found in ~\cite{Prelovsek:2010kg,Fu:2011wc,Fu:2011xb,Fu:2011xw,Alexandrou:2012rm,Lang:2012sv}. 
Heavier strange resonances become a great challenge for lattice QCD at close to physical pion masses, as they can in principle couple to several multi-body hadron channels. However, some of these aforementioned works have been able to determine their parameters at higher pion masses.

\subsection{Analytic techniques for inelastic resonances}
\label{sec:pades}

Let us first discuss the inelastic strange resonances. As pointed out throughout the previous sections, a dispersion relation is the most rigorous tool for performing the analytic continuation of a given amplitude. Nevertheless, we do not know a simple set of partial-wave dispersion relations that could be applied up to arbitrarily high energies. We saw in section~\ref{sec:DR} that $\pi K$ partial-wave dispersion relations are only applicable up to roughly the beginning of the inelastic region~\cite{Hite:1973pm,Johannesson:1976qp}. More complicated dispersion relations which improve the convergence region exist for $\pi\pi$ scattering~\cite{Auberson:1974ai,Auberson:1974xu,Auberson:1975yx,Mahoux:1974ej,Roy:1978tu,Roy:1978mj}, although they have never been applied to $\pi\pi$  experimental data. No equivalent set of equations has been derived so far for $\pi K$ scattering.

Moreover, even if we do have a dispersive representation of a partial wave in a given area, it is formulated in the first Riemann sheet, whereas resonance poles lie in the second or other Riemann sheets. In section \ref{sec:Notation}, and only for the elastic case, we provided a simple formula, Eq.~\eqref{ec:firsttosecondsheet}, to access this second Riemann sheet and to obtain the pole position using Eqs.~\eqref{ec:resonancecondition} and \eqref{eq:rescond}. However, in the inelastic case, since more channels are open, giving access to more sheets, there is no straightforward relation between the first and the contiguous or second sheet.
Only within certain approximations, there are simple
equations to write the amplitude in different sheets, but they not only require the knowledge of the amplitude in the first,
but also the amplitudes to the other accessible channels.
This is relatively simple to implement in coupled-channel models restricted to two-body scattering in just a few different channels, particularly if all the complications of left and circular cuts can be neglected, which is a fair approximation in the inelastic regime (but not in the elastic one).
Nevertheless, remember that we aim to avoid model dependencies.

For the reasons just explained, there is a growing interest in exploiting other analytic techniques to extract resonance poles, which could eliminate or produce smaller systematic uncertainties than those of simple model-based extractions. We list here a few of the most successful approaches: The conformal mapping expansion~\cite{Cherry:2000ut,Yndurain:2007qm,Caprini:2008fc}, which includes the dynamical cuts into a conformal mapping variable, thus producing, as a result, a simple fit that could partially mimic some of the analytic requirements. These mappings have been used in our UFD and CFD fits as explained in section~\ref{sec:UFDpik}. Other successful techniques are the Laurent~\cite{Oller:2004xm,Guo:2015daa} and the Laurent-Pietarinen~\cite{Svarc:2013laa,Svarc:2014zja,Svarc:2014aga,Svarc:2014sqa,Svarc:2015usk} methods mostly applied to baryon resonances, which disentangle the resonance pole (Laurent expansion), from a conformal map (Pietarinen expansion) over the partial waves. Two other methods have been used to extract low-energy mesonic resonances. The first one makes use of the Schlessinger continued-fraction method~\cite{Schlessinger:1968} and has been successfully implemented in several different problems~\cite{Tripolt:2016cya,Tripolt:2018xeo,Binosi:2019ecz}. Finally, there is the  method of sequences of Pad\'e approximants~\cite{Masjuan:2013jha,Masjuan:2014psa,Caprini:2016uxy,Pelaez:2016klv}, which will be the central topic of this subsection. They are all different approaches to determine resonance poles without assuming any model relating their coupling to their width parameters. Thus they are much less model-dependent and robust than the usual Breit-Wigner approach used so far in most determinations, very often not even meeting the Breit-Wigner applicability conditions.

All these analytic methods require the knowledge of the scattering amplitude, employing data or a previously fitted parameterization. However, as explained in detail in the introductory sections, determining the experimental data on meson-meson experiments is a complicated task, plagued with systematic uncertainties. As a result, fitting a model to the original data samples may not be robust enough for some of these resonances, on top of which none of these models implement  analyticity, crossing, and unitarity at the same time so that the original fits would be partially lacking first-principle constraints.
Most often they only consider the scattering of  two-body states.

A satisfactory solution follows from applying these analyticity techniques over a parameterization previously  constrained  to satisfy dispersion relations, as the CFDs obtained either for $\pi\pi$ \cite{GarciaMartin:2011cn} or $\pi K$ \cite{Pelaez:2016tgi} scattering. This approach was successfully carried out in \cite{Masjuan:2014psa,Caprini:2016uxy} and \cite{Pelaez:2016klv}, respectively. Thus, right below, we summarize the Pad\'e-sequence method and results \cite{Pelaez:2016klv} when applied to $\pi K$.

The Pad\'e approximant of order $[N/M]$ 
of a function $F(s)$ is defined as a rational function that satisfies
\begin{equation}
P^N_M(s,s_0)= \frac{Q_N(s,s_0)}{R_M(s,s_0)}\simeq F(s)+O\Big((s-s_0)^{M+N+1}\Big),
\end{equation}
where $Q_N(s,s_0)$ and $R_M(s,s_0)$ are polynomials in $s$ of $N^{th}$ and $M^{th}$ degree, respectively, and the expansion is centered around a given point $s_0$.

According to the Montessus de Ballore theorem~\cite{Montessus:1902} this sequence of Pad\'e approximants 
can be used to reach the next continuous Riemann sheet of an amplitude,
which is where resonance poles exist
\cite{Masjuan:2013jha,Masjuan:2014psa,Caprini:2016uxy,Pelaez:2016klv}. Notice again that these Padés do not assume any model relationship between the pole position and residue, so that there is way less model dependence when extracting resonances with this method.

For simplicity, in this section we focus on the extraction of isolated resonances, although more resonances can be included in a straightforward way~\cite{Pelaez:2016klv}. For isolated poles, the denominator $R_M(s,s_0)$ should be of first degree, and thus the Padé approximant is defined as:

\begin{equation}
P^N_1(s,s_0)=\sum^{N-1}_{k=0}{a_k(s-s_0)^k+\frac{a_N(s-s_0)^N}{1-\frac{a_{N+1}}{a_N}(s-s_0)}}.
\end{equation}

If one expands the amplitude as a Taylor expansion in the real axis around $s_0$ then the constants $a_n=\frac{1}{n!}F^{(n)}(s_0)$ are given by the $n^{th}$ derivative of the amplitude. Therefore the pole and residue read
\begin{equation}
s_p^{N}=s_0+\frac{a_N}{a_{N+1}},~Z^{N}=-\frac{(a_N)^{N+2}}{(a_{N+1})^{N+1}}.
\label{eq:poleresidue}
\end{equation}

Note that, customarily, the coupling of a resonance to an amplitude is defined as follows:
\begin{equation}
\vert g\vert^2=\frac{16\pi(2\ell+1)|Z|}{[2q(s_p)]^{2\ell}},
\label{eq:coupling}
\end{equation}
where $q(s)$
is the center-of-mass momentum of the scattering system and
$\ell$ the angular momentum of the partial wave.

If the resonance is not isolated, i.e., there are several overlapping resonances or nearby thresholds, extra poles can be introduced into the Padé by increasing the degree of $R_M(s,s_0)$. 
These new poles will take account of the other resonances or they will mimic the other analytic structures present in the amplitude. In principle, nearby thresholds are associated with cuts in the complex plane, which can be approximated up to the desired accuracy by including a sufficiently large number of extra poles.

The methodology implemented in~\cite{Pelaez:2016klv} to extract strange resonances from the CFD fit in \cite{Pelaez:2016tgi} can be summarized as follows:
\begin{itemize}
    \item We define the variation in the pole position when changing our calculation from $N^{th}$ to $(N+1)^{th}$ degree. Thus, for each $N$, starting from $N=1$, we define $\Delta \sqrt{s_p^N}_{sys}\equiv\left|\sqrt{s_p^N}-\sqrt{s_p^{N-1}}\right|$, which is calculated for a grid of $s_0$ in a big region surrounding the resonance. We consider our best $s_0$ as the one which minimizes these uncertainties, thus improving the convergence of the series.
    \item The statistical uncertainty is thus added by employing a Montecarlo re-sampling over the fits.
    \item We truncate the Pad\'e sequence at a given $N$ when $\Delta \sqrt{s_p^N}_{sys}$ is considerably smaller than the statistical uncertainty.
\end{itemize}

This procedure makes use of the smallest possible number of derivatives, which usually offers better stability and smaller deviations for the statistical uncertainties. Unfortunately, different parameterizations compatible within uncertainties with the CFD could yield somewhat different derivatives. This might be the only source of some relatively mild model or parameterization dependence, which we will consider as an additional systematic error. Thus, in order to estimate this systematic uncertainty, we  extract the resonance pole implementing the whole Pad\'e sequence method but using different functional forms fitted to describe the CFD within uncertainties. We consider this approach to produce a robust, precise determination of resonance parameters and their uncertainties but avoiding or at least reducing considerably
any model dependence.

Our recent results applying this Pad\'e sequence method are shown in Fig.~\ref{fig:kpades} for  the poles of strange resonances appearing in the $\pi K$ inelastic region below 1.8 GeV.
They can be compared to the values listed in the RPP~\cite{Zyla:2020zbs}. This includes the scalar $K_0^*(1430)$, the vector $K^*(1410)$, the tensor $K_2^*(1430)$ and the $\ell=3$ resonance $K_3^*(1780)$. The $K^*(892)$ and \kap have also been extracted as shown in Table~\ref{tab:kpades}. However, these two are not discussed here as they will be discussed in more detail in the next section, since elastic resonances can be calculated directly from partial-wave dispersion relations. 
From Fig.~\ref{fig:kpades} it can be noticed that, even with the attached systematic uncertainty, our analytic determinations, which do not assume a specific model parameterization, are competitive in precision with the other existing values.

\begin{table}[htb]
\setlength{\tabcolsep}{3pt}
\caption{Poles obtained from the Pad\'e sequence method \cite{Pelaez:2016klv} using as input the CFD results obtained in \cite{Pelaez:2016tgi}.
The uncertainty for $\sqrt{s_p}$ and $\vert g\vert$ include statistical and theoretical (systematic) errors. } 
\centering 
\begin{tabular}{c c c} 
\hline\hline  
\rule[-0.15cm]{0cm}{.15cm}  & $\sqrt{s_p}$(MeV) & $\vert g\vert$ \\ \hline
\rule[-0.15cm]{0cm}{.15cm} $\kap$ & $(670\pm18)-i(295\pm28)$ & $(4.47\pm 0.40)$ GeV \\
\rule[-0.15cm]{0cm}{.15cm} $K_0^*(1430)$ & $(1431\pm6)-i(110\pm19)$ & $(3.82\pm 0.74)$ GeV \\
\rule[-0.15cm]{0cm}{.15cm} $K^*(892)$ & $(892\pm1)-i(29\pm1)$ & $(6.1\pm 0.1)$ \\
\rule[-0.15cm]{0cm}{.15cm} $K^*(1410)$ & $(1368\pm38)-i(106^{+48}_{-59})$ & $(1.89^{+1.77}_{-1.34})$ \\
\rule[-0.15cm]{0cm}{.15cm} $K_2^*(1430)$ & $(1424\pm4)-i(66\pm2)$ & $(3.23\pm0.22)$ GeV$^{-1}$ \\
\rule[-0.15cm]{0cm}{.15cm} $K_3^*(1780)$ & $(1754\pm13)-i(119\pm0.14)$ & $(1.28\pm0.14)$ GeV$^{-2}$\\
\hline
\end{tabular} 
\label{tab:kpades} 
\end{table}

We want to remark that the region occupied by different determinations of each resonance pole  in Fig.~\ref{fig:kpades} is much larger than the RPP estimated average. One of the reasons is that only the values with solid symbols are $T$-matrix poles and used to obtain such an estimate. The rest are just Breit-Wigner parameters, which as discussed above are not a model-independent definition of a resonance, but just a narrow-width approximation valid for isolated resonances. Let us remark that most of these extractions rely on some, but not always the same, Breit-Wigner-like parameterization since it is often modified with further crude model assumptions. In particular, often these parameterizations include barrier factors, or fits with summations of Breit-Wigner forms which do not fulfill unitarity, etc... thus producing big systematic spreads. Notice also how the determination of the width for all these resonances is generally in worse shape than their mass (note that the mass and width scale is different). This is not surprising since the apparent width can become easily process-dependent when not extracted from a rigorous analytic continuation to the pole. 
Our analytic techniques avoid all these caveats and therefore yield a robust determination of the parameters, whose accuracy is competitive too, if not better than, present estimates.

Finally, this formalism can be adapted in a straightforward way to study other inelastic channels involving meson-meson scattering processes. In a possible future application, we are planning to apply the method to the $f_0(1370)$, $f_0(1500)$ and $f_0(1710)$ resonance poles using as input the dispersively constrained parameterizations of $\pi \pi \to \pi \pi$~\cite{Pelaez:2019rwt} and the CFD of $\pi \pi\to K\bar K$ that we have just provided here in section \ref{sec:CFD}.


\subsection{Dispersive determination of the \kap resonance from data}
\label{sec:kapdr}

\begin{flushright} {\it 
``The weight of the evidence should be proportioned to the strangeness of the facts''\\
{\footnotesize Principle of Laplace, as restated by T. Flournoy in ``India to the Planet Mars'' (1900),\\ when commenting Laplace's ``Essai philosophique sur la probabilit\'e'' (1825).}}
\end{flushright}
\begin{flushright} {\it     ``We are beginning to think that $\kappa$ should
be classified along with flying saucers,\\ the
Loch Ness Monster, and the Abominable Snowman.''\\
{\footnotesize A.H.Rosenfeld et al.``Data on Particles and Resonant States'' review (1967)  \cite{ROSENFELD:1967zz}.}}
\end{flushright}

\begin{figure}[!ht]
\centerline{\includegraphics[width=0.6\textwidth]{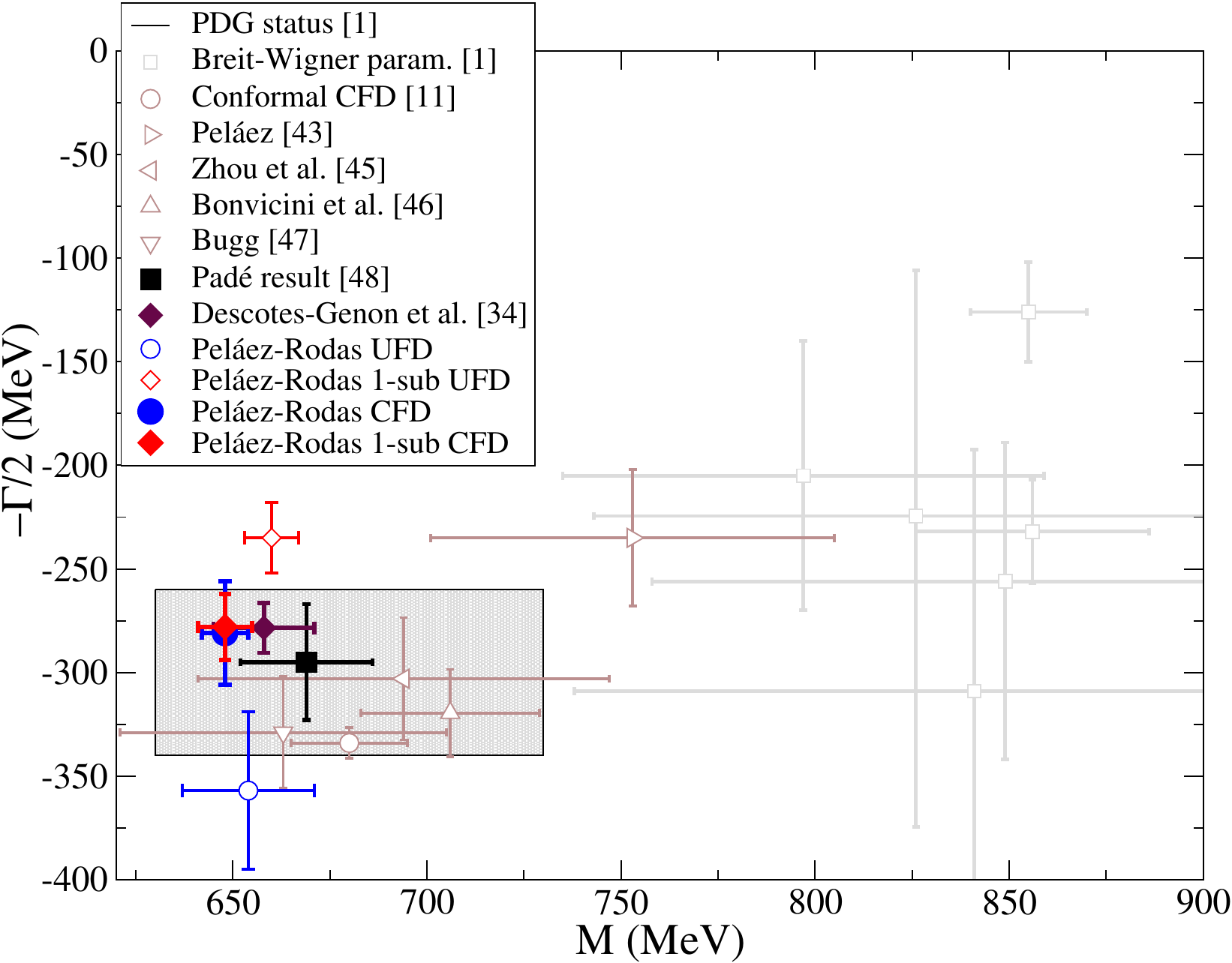}}
 \caption{\rm \label{fig:polekappa} Determinations of the $\kap$ pole in the complex plane as illustrated in \cite{Pelaez:2020uiw}.  Of course, the numbering  of the references in \cite{Pelaez:2020uiw} is not the same as in here, which we provide next. In particular,
 Breit-Wigner parameterizations are taken from the RPP compilation \cite{Zyla:2020zbs} (called PDG in \cite{Pelaez:2020uiw}), which also includes: Descotes-Genon et al. \cite{DescotesGenon:2006uk}, Bonvicini et al. \cite{Bonvicini:2008jw}, Bugg \cite{Bugg:2009uk}, Pel\'aez \cite{Pelaez:2004xp}, Zhou et al. \cite{Zhou:2006wm} and the ``Pad\'e Result'' \cite{Pelaez:2016klv}. The conformal CFD is the simple analytic extrapolation of our parameterization in \cite{Pelaez:2016tgi}. 
 The ``PDG status'' shaded rectangle covers the mass and width estimate in the RPP.
 Our dispersive extractions of the \kap pole are also included, using as input for the PWHDR either the UFD or CFD parameterizations.
 Red and blue points use for the $F^-$ amplitude a once-subtracted or an unsubtracted dispersion relation, respectively. Notice that even when using dispersive approaches the extraction is not fully stable, and the UFD values are inconsistent with one another and deviated by around 2 $\sigma$ from the CFD ones. Only once Roy-Steiner equations are imposed as a constraint for the CFD parameterizations, do both pole determinations fall on top of each other, thus producing a negligible systematic uncertainty, as explained in the text. Figure taken from~\cite{Pelaez:2020uiw}.}
\end{figure}

As explained above the situation regarding the \kap resonance has been debated for several decades, partially because of how unstable its determination is. First, there are several different extractions relying on relatively simple models  \cite{Jaffe:1975fd,Jaffe:1976ig,vanBeveren:1986ea,Bugg:1996ki,Ishida:1997wn,Oller:1997ng,Oller:1997ti,Oller:1998hw,Oller:1998zr,Black:1998zc,Black:1998wt,Close:2002zu,Kelkar:2003iv,Maiani:2004uc,Pelaez:2004xp,Jaffe:2004ph,Amsler:2004ps,vanBeveren:2005ha,Giacosa:2006tf,Jaffe:2007id,Fariborz:2015bff}. Of course, as repeatedly explained, the use of conflicting data sets, which do not fulfill the first principle's requirements, together with the large model dependencies produce a huge spread of results as shown in Fig.~\ref{fig:polekappa}. This is once again the situation explained in the previous section, and in Fig.~\ref{fig:kpades}, where simple models can produce huge systematic spreads. 

Furthermore, the \kap pole is one of 
the widest resonances and therefore its pole lies deep in the complex plane.
Consequently, as illustrated in the top panel of Fig.~\ref{fig:anstrucpw}, the 
region of its nominal mass is as close to the pole as it is to the $\pi K$ threshold, the Adler zero, or even the left and circular cuts. Hence a precise and reliable extraction cannot be fully understood without a model-independent way of performing the necessary analytic continuation into the complex plane (for recent reviews we refer to Ref.~\cite{Pelaez:2015qba,Yao:2020bxx}), making sure that all the required analytic structures are accounted for. Of course, those models that include some of these basic features, as well as unitarity---which is essential for a resonant behavior--- and some reasonable input from data, generically produce a pole that is not too far from its actual position. Note that it is important to check that the pole lies within the applicability region of the partial-wave expansion in the complex plane (see \ref{app:Applicability}). At any rate, models cannot be used for a precise description of this resonance.

All pole positions listed in Fig.~\ref{fig:polekappa} are taken from the RPP~\cite{Zyla:2020zbs}, although there are many other values, like those quoted just above. Most of these determinations (light gray) are simply Breit-Wigner parameters obtained as a result of several fits to different processes. However, not only this is not a model-independent definition of a resonance pole, but it is actually a wrong approximation. Recall that the Breit-Wigner formula is devised for narrow resonances well isolated from other analytic features. From Fig.~\ref{fig:anstrucpw}, it is evident that it is not suitable for a resonance so deep in the complex plane, so close to a threshold and other singularities, and so close to the Adler zero imposed by chiral symmetry. The spread of the results in Fig.~\ref{fig:polekappa}  using a Breit-Wigner formalism speaks for itself about this inadequacy. The rest of the results shown in the figure for the \kap resonance include at least some basic features arising from QCD~\cite{Bugg:2003kj,Bugg:2005ni,Bugg:2005xx,Bugg:2006sz,Zhou:2006wm,Bugg:2009uk,Bonvicini:2008jw,Bugg:2009uk,Pelaez:2016klv}. Some others resort to the unitarization of ChPT either by using the $N/D$ or Inverse Amplitude Method (IAM)~\cite{Oller:1997ng,Oller:1998hw,Oller:1998zr,Pelaez:2004xp,Nebreda:2010wv} (For recent reviews of ChPT unitarization see \cite{Pelaez:2015qba,Oller:2019rej,Yao:2020bxx,Pelaez:2021dak}). Generically, these approximate the left-hand cut contributions at a given order in ChPT, and make use of dispersion relations to unitarize the ChPT amplitude of a given order. An alternative way of implementing dispersion relations over ChPT in this channel is the one of the Beijing group~\cite{Zheng:2003rw,Zhou:2006wm}, which produces a rather stable result for the pole position. Finally, a very sound determination comes from the dispersive study of the Paris group~\cite{DescotesGenon:2006uk}. Here the authors make use of a solution to FTPWDR in~\cite{Buettiker:2003pp} obtained without using data in that energy region for the S and P waves. Then they use that solution as input into HPWDR to extract the \kap resonance. It is also very relevant to remark that in their analysis they showed that the \kap pole lies within the applicability range of their choice of HPWDR. 

On the lattice QCD front, this resonance has also been tackled recently, following developments in the calculation of meson-meson scattering phase shifts, which we illustrated in Fig.~\ref{fig:hadspeckap}. Unfortunately, the pole is not a direct observable, but once again it has been extracted using models.
Nevertheless, as can be seen in~\cite{Dudek:2014qha,Brett:2018jqw}, at  $m_\pi\simeq 400\,$MeV the \kap appears as a virtual bound state, compatible with what we know from unitarized NLO  ChPT~\cite{Nebreda:2010wv,Guo:2018zss}. However, using lighter pion masses between 200 and 400 MeV the pole extraction becomes rather unstable \cite{Wilson:2019wfr}, even though the  basic features of the partial wave are well described, as seen in Fig.~\ref{fig:hadspeckap}. In the aforementioned work the Lüscher formalism \cite{Luscher:1991cf} was extensively used to produce many different energy levels for each pion mass, which produces a very constrained result on the real axis. The authors of this study, when referring to lighter $m_\pi$,   stated on the scalar wave that ``Even with precise information about the amplitude for real energies, the analytic continuation required to reach any pole is sufficiently large that a unique result is not found'' (\cite{Wilson:2019wfr}). This supports our idea that a more elaborated, dispersive analysis is needed at lower $m_\pi$  to extract accurate information when performing analytic continuations. This same behavior can be seen for the \sig, where for heavier $m_\pi$ the extraction is pretty stable~\cite{Prelovsek:2010kg}, but becomes rather unstable for pion masses near the physical one~\cite{Briceno:2016mjc,Briceno:2017qmb,Guo:2018zss}.

\begin{figure}[!ht]
\centerline{\includegraphics[width=0.49\textwidth]{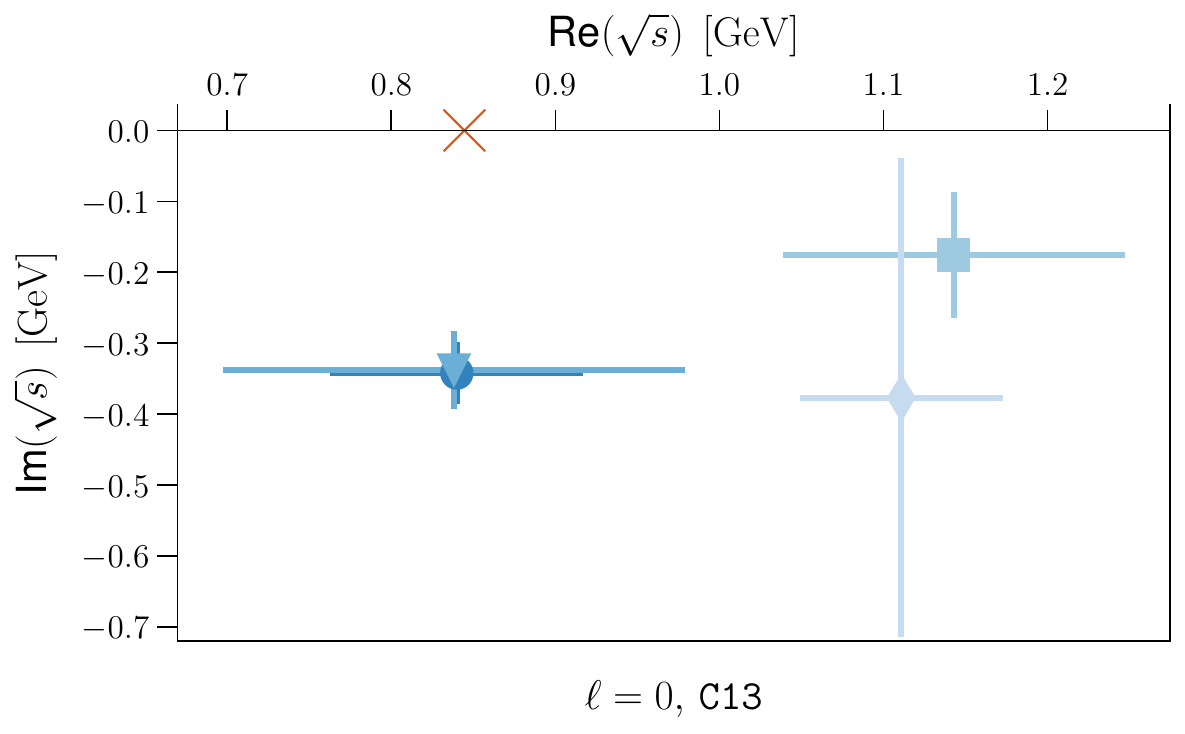} \includegraphics[width=0.49\textwidth]{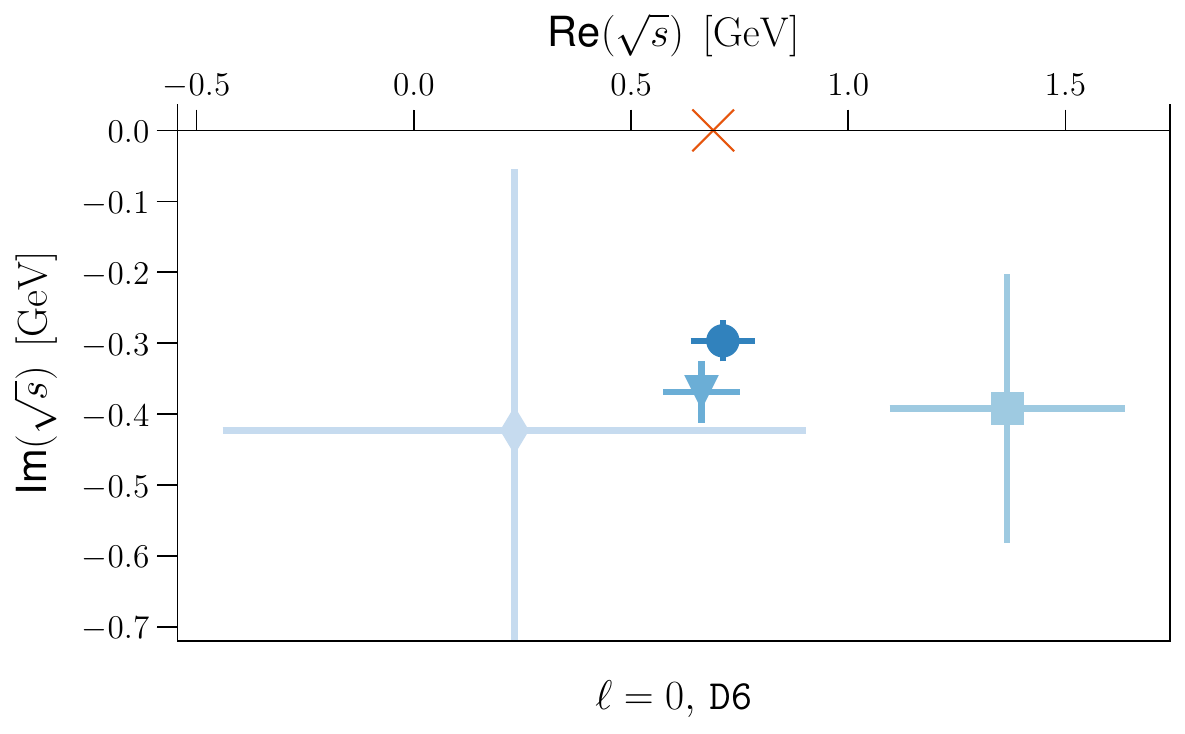}}
 \caption{\rm \label{fig:polekappalattice}
$\kap$ poles from lattice calculations~\cite{Rendon:2020rtw} with $m_\pi\simeq 317\,$MeV (left) and $m_\pi\simeq176\,$MeV (right).  Note how the parameterizations including the Adler zero (darker markers) have a smaller spread when the pion mass is lighter, and they produce little systematic spread between themselves. Figure taken from~\cite{Rendon:2020rtw}.}
\end{figure}

It is not so clear how large  the effect of the Adler zero is when the pion mass is really heavy, and these resonances appear as bound or virtual bound states. Nonetheless, the main problem when dealing with lighter extractions is that the subthreshold features and left-hand cuts start playing a non-negligible role in the extraction. Thus one needs to include as much information as possible to get a reliable determination of the resonance pole position. There exists a very recent lattice QCD determination of the \kap resonance at close-to-physical pion mass~\cite{Rendon:2020rtw}. We have included their figure as Fig.~\ref{fig:polekappalattice} to illustrate our previous explanation. In the left panel, the heavier pion-mass results ($m_\pi\simeq$ 317 MeV) are depicted, whereas the lighter pion-mass results are  shown on the right panel ($m_\pi\simeq$ 176 MeV). Of the four different parameterizations used by the authors, the darker ones are compatible among themselves, and their uncertainties shrink when the pion mass decreases. These two include the Adler zero at LO in ChPT, which produces a significant improvement when extracting the resonance, as the lighter $m_\pi$, the closer to the resonance lies the Adler zero. The other two parameterizations are built with $K$-matrix and effective range formalisms. This picture may illustrate just how challenging the extraction of broad resonances is, and why incorporating first principle features from QCD into our amplitude analyses is paramount to the robust determination of the QCD spectra. Lattice QCD collaborations will be able to access close-to-physical pion mass in the near future and thus, to extract their resonance poles, an approach based on constraining the amplitudes and the data, like ours, seems to be a must.

Nevertheless, despite all this growing support for its existence and properties, the \kap still carries the ``Needs Confirmation'' label in the RPP \cite{pdg}. 
Thus, following the quotation at the beginning of this subsection, we have very recently proportioned further weight of evidence in support of the existence and precise parameter determination of this strangeness carrying resonance.

Let us first remember that in \cref{eq:sftpwdr,eq:shdrfm,eq:shdrfmi}  we have provided several partial-wave projected fixed$-t$  and hyperbolic dispersion relations (FTPWDR and HPWDR, respectively)  to constrain the partial waves. 
In principle, one could think about using FTPWDR rather than HPWDR, because their dependence on \pipikk is much smaller. However by comparing their respective applicability domains in Fig.~\ref{fig:rvpik} versus Fig.~\ref{fig:rvpikhdr}
we see that, even though their applicability on the real axis is rather similar and overlaps widely with the region of the \kap nominal mass, the domain of validity 
in the complex plane of the FTPWDR is much smaller and, unfortunately, it cannot reach the position of the \kap pole. This is why the Paris group, as so do we, have to resort to HPWDR. Note that
the choice of the $a$ parameter defining our hyperbolae is paramount. As explained in section~\ref{sec:DR} and \ref{app:Applicability}, we could think of changing $a$ to increase the applicability region of the HDR on the real axis, as we have actually done for \pipikk. However, this would narrow down the applicability region in the complex plane, as depicted in Fig.~\ref{fig:rvpikhdr}, leaving out the \kap pole. Thus a compromise between the best $a$ on the real axis and the complex plane must be found, which we consider satisfactory for $a=-10 m_\pi^2$. Of course, right outside their applicability domains, we do not expect an abrupt disruption, but just a smoothly increasing disagreement  between both sides of the dispersion relations, as the discrepancies are produced by the $\pi K$ box diagrams shown in Fig.~\ref{fig:boxdiagrams}. Nevertheless, the rigorous determination of the pole with precision demands that we remain inside the applicability domain.

Thus, in \cite{Pelaez:2020uiw} we have used our data parametrizations inside the HPWDR and we have found that there is indeed a pole
in the expected region in the scalar channel with isospin 1/2.
The values of the pole position and the modulus of the coupling are shown in Table~\ref{tab:poleskappa}, compared with other existing determinations in Fig.~\ref{fig:polekappa} above.

\begin{table}[!hbt] 
\caption{Table taken from~\cite{Pelaez:2020uiw}. Comparison between various poles and residues of the \kap resonance. The last two lines are our dispersive outputs using the CFD as input. We consider the last line as our best result. Here we also provide the phase (in radians) of the coupling.}
\centering 
\begin{tabular}{l  c  c  c } 
\hline
\hline
 & \hspace{0.2cm} $\sqrt{s_{pole}}$\,\, (MeV) & \hspace{0.2cm} $|g|$\,\, (GeV) & $\phi_g$\,\, (rad) \\
\hline
\rule[-0.2cm]{-0.1cm}{.55cm} $K^*_0(700)$ \cite{DescotesGenon:2006uk}& \hspace{0.2cm} $(658\pm 13)-i (279\pm 12)$ & \hspace{0.2cm} --- & --- \\
\rule[-0.2cm]{-0.1cm}{.55cm} $K^*_0(700)$ \cite{Pelaez:2016klv}& \hspace{0.2cm} $(670\pm 18)-i (295\pm 28)$ & \hspace{0.2cm} $4.4\pm0.4$ & --- \\
\rule[-0.2cm]{-0.1cm}{.55cm} $K^*_0(700)$ 0-sub & \hspace{0.2cm} $(648\pm 6)-i (283\pm 26)$ & \hspace{0.2cm} $3.80\pm0.17$ & $-1.50\pm0.05$\\
\rule[-0.2cm]{-0.1cm}{.55cm} {\bf $\bf{K^*_0(700)}$ 1-sub}& \hspace{0.2cm}  $\bf{(648\pm 7) -i (280\pm 16)}$ & \hspace{0.2cm} $\bf{3.81 \pm 0.09}$ & $-1.49\pm0.07$\\
\hline
\hline
\end{tabular} 
\label{tab:poleskappa} 
\end{table}

Note that we have calculated the poles using both the HPWDR obtained with one or no subtractions for the antisymmetric $F^-$ amplitude (red and blue symbols in Fig.~\ref{fig:polekappa}). In addition, we have used  as input both the UFD (hollow symbols) and CFD (solid symbols).

The first striking feature when we observe our dispersive results in that figure is that the results using the UFD as input with one or no subtractions are incompatible with each other.
Actually, the two poles, which should be the same, differ by around 3-4 $\sigma$ even when they are obtained from the same input.
 This outcome comes from the fact that, as already seen in  section~\ref{sec:FulfillDR} and shown in Fig.~\ref{fig:alltogether}, the UFD is inconsistent with the dispersive representation. This result is very important because it illustrates how {\it even when using the same input in the real axis} 
 the extrapolation to the complex plane can be rather unstable depending on the method used unless one makes sure that the
 input satisfies a whole set of dispersion relations. Moreover, if this happens even using dispersion relations to make the analytic continuation, one can imagine that using any model the instability could be even larger.

 In contrast, when we use the CFD, the poles move 
 away from their UFD values by around 2 deviations and the  results from the unsubtracted and once-subtracted HPWDR  agree perfectly well, as seen in the last two lines of Table~\ref{tab:poleskappa}.
 Actually, they agree so well that they are hard to distinguish in Fig.~\ref{fig:polekappa}.

 Although both CFD determinations agree remarkably well, we have highlighted in Table~\ref{tab:poleskappa} the one obtained using the subtracted antisymmetric amplitude $F^-$ because we have found that it is the most stable  under variations of the partial waves and  modifications of the $a$ parameter. This is not surprising considering that the unsubtracted $F^-$ is dominated mostly by the $\pi \pi \to K \bar K$ $g^1_1(t)$ partial wave, in particular by its pseudo threshold region, as shown in Figs.~\ref{fig:hdrchecks} and~\ref{fig:pseudocfd}. We thus consider that the result coming from the once subtracted case is more robust, and favor it as our final result.

In the second line of Table~\ref{tab:poleskappa} we have also provided the 
value of the \kap pole that we obtained by applying the method of Pad\'e sequences reviewed in the previous section. It can  be seen that, although  it is much less precise, it perfectly overlaps within uncertainties with our dispersive determinations, thus validating the method. In addition, in the first line of the same table we supply the previous dispersive result of the Paris Group \cite{DescotesGenon:2006uk}, with which we agree well within uncertainties. We want to emphasize once again that they did not use data on the S and P waves in the elastic region, but solutions of the FTPWDR equations with input from higher energy data and other partial waves. In contrast, ours is obtained from a data fit using the data in that region. They are therefore independent determinations with  different input and rather different extractions. 
For instance, this shows that the fact that their P-wave prediction in \cite{DescotesGenon:2006uk} does not describe data, as seen in Fig.~\ref{fig:Pelastic}, does not affect dramatically the \kap determination.

In conclusion, we are confident that this is the kind of confirmation required to settle the existence of this state. Moreover, the pole parameters and couplings are now known with great precision, about an order of magnitude less than currently evaluated in the RPP.

Finally, following the same methods, but applied to the $P$-partial wave,  it is also possible to determine the pole and residue of the vector $K^*(892)$ resonance. Actually, as we saw in the third row of Fig.~\ref{fig:alltogether}, the CFD agreement for the three $f^{1/2}_1(s)$ dispersion relations is remarkable. Note that, since this resonance is much narrower, its associated pole lies very near the real axis and all three partial-wave dispersion relations (four, taking into account the two different subtractions for $F^-$) can be used for its determination. The difference between using one or another is in the 1 MeV range. We have thus averaged them to find the pole that was already advanced in  \cite{Pelaez:2020uiw}: 
$$\sqrt{s_{K^*(892)}}=(890\pm 2) -i (25.6\pm 1.2)\,\mev, $$ 
and its dimensionless residue has modulus
$\vert g\vert= (5.69\pm0.12)$ and phase $\phi_g=-(0.076\pm0.008)$ (in radians).

\subsection{On the nature of the $\kap$}
\label{subsec:nature}

There is growing evidence for the existence of hadrons whose composition falls beyond the ordinary quark-antiquark classification of mesons or the three-quark classification of baryons. Most modern investigations of the dynamical models that form exotic resonances focus on individual angular momenta. This provides a framework to extract the properties of a given resonance at complex-energy values. However, in this section we will briefly summarize our efforts in combining dispersion relations and the analytic properties of amplitudes for complex angular momenta. This helps us to connect resonances of different spins under their Regge trajectories in the squared mass vs. spin, $(s=M^2,J)$, plane (see \cite{Collins:1971ff} for a textbook introduction to Regge Theory). A well-known feature of ordinary hadrons is that they can be classified into real and linear Regge trajectories with an almost universal slope of around 0.9 GeV$^{-2}$. In the case of mesons, it can be naively interpreted in terms of the tension of the rotating flux tube between the quark and antiquark. As a result, strong deviations from this behavior suggest a different microscopic nature.

We are particularly interested in the investigation of the internal structure of the \kap resonance, debated for many decades. Many different models have determined its structure to be different than that of $q \bar q$ mesons. In particular, many past works show some predominant meson-meson (also called molecular) or tetraquark nature for this state~\cite{Jaffe:1976ig,vanBeveren:1986ea,Oller:1998hw,Oller:1998zr,Black:1998wt,Black:1998zc,Close:2002zu,Pelaez:2003dy,Pelaez:2004xp,vanBeveren:2006ua,Giacosa:2006tf,Wolkanowski:2015jtc,Agaev:2018fvz}. Moreover, many of these works find striking similarities between the \kap and \sig resonances, pointing to a similar internal composition. This last property would rule out the idea of the \sig being a glueball, as the \kap, due to its strangeness, cannot be one.

As explained above, Regge Theory emerges as an application of analytic constraints in the complex angular momentum plane. Even though no such thing as an analytic formula for hadronic states exists, we know $q\bar q$ meson states should be classified unambiguously by a straight Regge trajectory, with a universal slope. However, both \kap and \sig resonances are at odds with this classification~\cite{Anisovich:2000kxa,Masjuan:2012gc}, which illustrates that it is not possible to accommodate them into Regge families of resonances, or find suitable ``Regge partners''. Moreover, we shall review below how from dispersion theory it can be shown that the \kap Regge trajectory does not follow the ordinary behavior and has  non-ordinary Regge parameters. Its trajectory is 
not dominated by its real part, which is not linear, and its slope is 5 times smaller than the universal one for ordinary hadrons. On top of that, the behavior of this trajectory at low energies is similar to that of the \sig, as will be seen in Fig.~\ref{Fig:yukawacomp}. This is yet another piece of evidence supporting the idea that the \kap and \sig resonances are non-ordinary hadrons.

In the following, we will briefly summarize the elastic dispersive formalism that allows us to calculate the Regge trajectory of a resonance just from its pole parameters. There are no fits to other resonances involved. More details can be found for mesons in the following Refs.~\cite{Epstein:1968vaa,Chu:1969ga,Collins:1977jy,Fiore:2000fp,Gribov:2003nw,Londergan:2013dza,Carrasco:2015fva,Pelaez:2017sit} or in Refs.~\cite{Fiore:2004xb,Fernandez-Ramirez:2015fbq,Silva-Castro:2018hup} for baryons. This formalism will be applied here to the $K^*(892)$, $K_0^*(1430)$, and \kap resonances. 
The $K^*_0(1430)$ is not completely elastic, but the formalism is still applicable since it predominantly decays to $\pi K$ with a branching ratio of around $(93\pm 10)\%$. The other two are purely elastic. The first two are considered ordinary $q\bar q$ mesons, and their trajectories come out as such. In contrast, the \kap is believed to be a non-ordinary resonance and,  as we will see next, this is supported by the fact that its trajectory does not follow the expected $q \bar q$ behavior.

We first assume that the partial wave can be approximated by the nearest resonance that dominates the line shape as
\begin{equation}
   t_l(s)  = \frac{\,\beta(s)}{l-\alpha(s)\,} + f_{background}(l,s),
   \label{eq:reggepw}
\end{equation}
where $\alpha(s)$ is the Regge trajectory and $\beta(s)$ the residue of the resonance pole. 
Then, from elastic unitarity 
\begin{equation}
    \im \alpha(s)=\sigma_{\pi K}(s) \beta(s),
\end{equation}
where  $\sigma_{\pi K}(s)$ is the two-particle phase space, already defined in Eq.~\eqref{eq:phasespace}. Finally by making use now of the Schwarz reflection symmetry one gets $\alpha^*(s)=\alpha(s^*)$ and $\beta^*(s)=\beta(s^*)$. Note that $\beta(s)$ is real in the real axis above threshold.

Since we are dealing with meson-meson scattering, 
we already saw in Fig.~\ref{fig:anstrucpw} that partial waves have three main analytic structures in the first Riemann sheet. First and foremost, the right-hand cut is created by $s$-channel unitarity. Second, the left-hand cut is produced by crossed channel interactions, and lastly, the circular-cut is produced when the two mesons have unequal masses. However, it can be proven  \cite{Collins:1977jy} that the Regge trajectory $\alpha(s)$ and the residue $\beta(s)$ are only related to the right-hand cut, starting at the first two-particle threshold.  It is, therefore, possible to write the customary dispersion relation for the Regge trajectory as
\begin{equation}
    \re \alpha(s)=\alpha(0)+\frac{s}{\pi}\int_{s_{th}}^{\infty}\frac{\im \alpha(s')}{s'(s'-s)}ds' ,
\end{equation}
where we have included one subtraction to ensure convergence, and we have factorized out explicitly the intercept of the trajectory. More subtractions can be included straightforwardly, as can be seen in~\cite{Carrasco:2015fva}. It is now clear from these two equations above that obtaining a closed dispersive description for both $\alpha(s)$ and $\beta(s)$ demands solving a system of coupled integral equations. In the case of the $\beta(s)$ function, one first defines the reduced residue as $\gamma(s)=\beta(s)\hat{s}^{-\alpha(s)} \Gamma(\alpha(s)+3/2)$, 
 with $\hat s= 4p^2/s_0$. In order to have the right dimensions,
we have introduced a scale $s_0$, conveniently set
to $s_0 = 1 \gev^2$ without losing generality. This reduced residue is an analytic function except for a cut on the real axis above threshold, where its phase is known and hence can be obtained from a Muskhelishvili-Omn\`es function, in a relatively similar approach to the one we followed in section \ref{sec:MO}. For more details on the derivation, we refer the reader to Refs.~\cite{Londergan:2013dza,Carrasco:2015fva,Pelaez:2017sit}.

The final closed system of equations, for the $\pi K$ case, reads

\begin{align}
\mbox{Re} \,\alpha(s) & =   \alpha_0 + \alpha' s +  \frac{s}{\pi} PV \int_{(m_K+m_\pi)^2}^\infty ds' \frac{ \mbox{Im}\alpha(s')}{s' (s' -s)}, \label{iteration1}\\
\mbox{Im}\,\alpha(s)&=  \frac{ \sigma_{\pi K}(s)  b_0 \hat s^{\alpha_0 + \alpha' s} }{|\Gamma(\alpha(s) + \frac{3}{2})|}
 \exp\Bigg( - \alpha' s[1-\log(\alpha' s_0)] 
+  \!\frac{s}{\pi} PV\!\!\!\int_{(m_K+m_\pi)^2}^\infty\!\!\!\!\!\!\!ds' \frac{ \mbox{Im}\alpha(s') \log\frac{\hat s}{\hat s'} + \mbox{arg }\Gamma\left(\alpha(s')+\frac{3}{2}\right)}{s' (s' - s)} \Bigg), 
\label{iteration2}\\
 \beta(s) &=    \frac{ b_0\hat s^{\alpha_0 + \alpha' s}}{\Gamma(\alpha(s) + \frac{3}{2})} 
 \exp\Bigg( -\alpha' s[1-\log(\alpha' s_0)] 
+  \frac{s}{\pi} \int_{(m_K+m_\pi)^2}^\infty \!\!\!\!\!\!\!ds' \frac{  \mbox{Im}\alpha(s') \log\frac{\hat s}{\hat s'}  + \mbox{arg }\Gamma\left(\alpha(s')+\frac{3}{2}\right)}{s' (s' - s)} \Bigg),
 \label{betafromalpha}
 \end{align}
where $PV$ denotes the principal value.

\begin{figure}[!ht]
\begin{center}
\resizebox{.9\textwidth}{!}{%
\includegraphics[height=3.3cm]{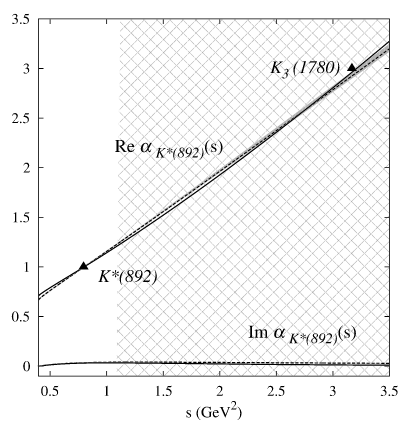} \hspace{0.1cm}\raisebox{0.05\height}{\includegraphics[height=3.cm]{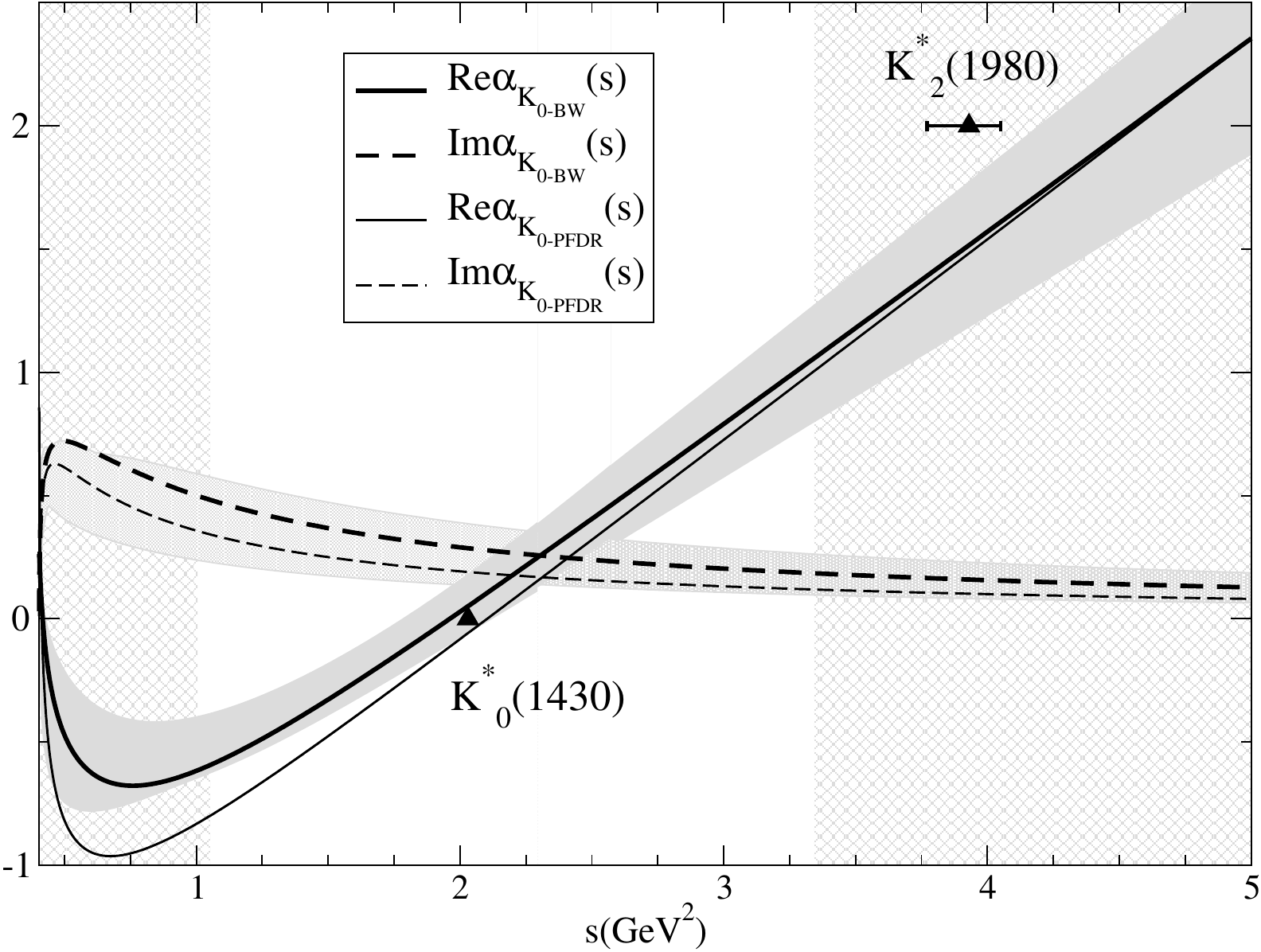}}}\\
\resizebox{.9\textwidth}{!}{%
\includegraphics[height=3.3cm]{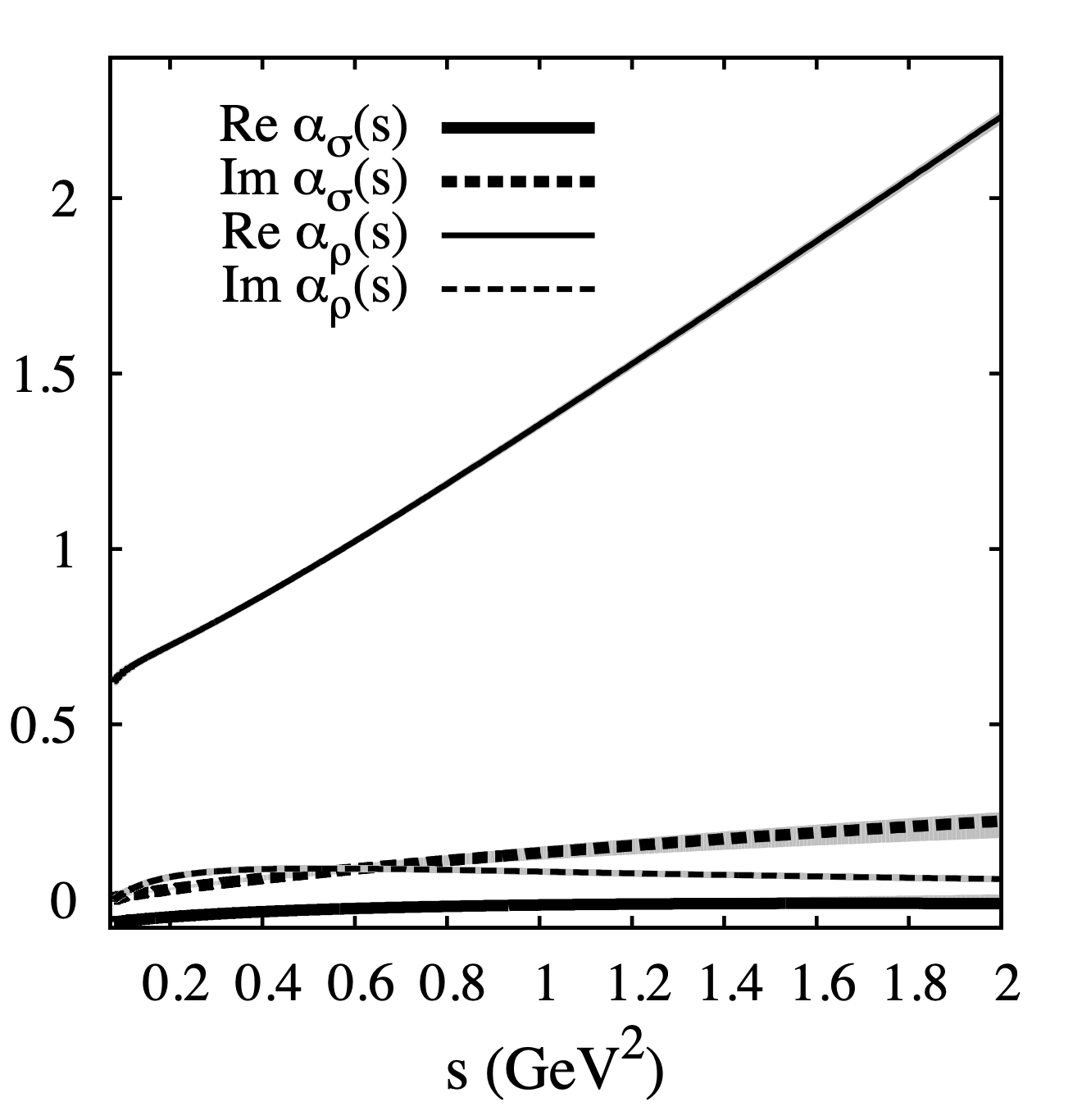} \raisebox{0.1\height}{\includegraphics[height=2.9cm]{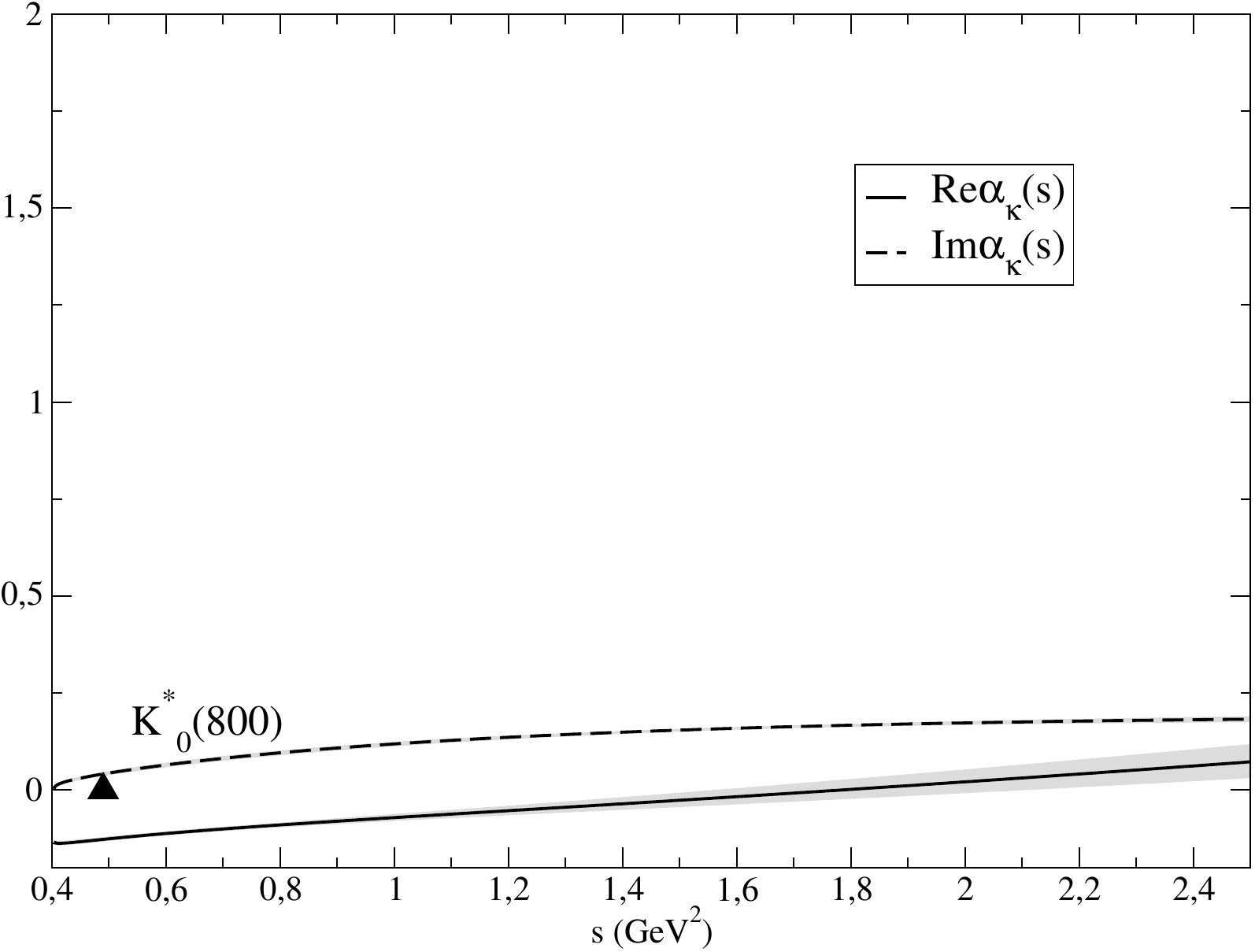}}}
\caption{  Ordinary vs. non-ordinary  Regge trajectories
obtained from a dispersive representation with only their lightest pole as input. Let us recall that for ordinary Regge trajectories the imaginary part is negligible and the real part should be a rising straight line. Notice the ordinary linear Regge trajectories of the  $K^*(892)$ (Top left), $K_0^*(1430)$ (Top right, we provide two extractions, either from its Breit-Wigner shape or with Padé sequences from data constrained with FDRs) and $\rho(770)$ (Bottom left).  
In contrast, the \sig (Also bottom left) and \kap (Bottom right)  have non-ordinary trajectories. Figures are taken from \cite{Londergan:2013dza,Carrasco:2015fva,Pelaez:2017sit}, when the $K^*_0(700)$ was still called $K^*_0(800)$.}
\label{fig:kregges}
\end{center}
\end{figure}

As can be easily seen this coupled system of equations depends on three free parameters, which are determined once we impose that the partial wave of \cref{eq:reggepw} should contain a pole at a given position with a given residue. In particular, in \cite{Pelaez:2017sit}, we used the pole positions extracted in~\cite{Pelaez:2016klv} employing the Pad\'e sequence method, just explained in section \ref{sec:pades} above, as the inputs for the \kap and $K_0^*(1430)$ resonances. One minor technical remark is in order: for the scalar wave, we also imposed on the integral equations the existence of an Adler zero at the LO ChPT position.

The real and imaginary parts of the resulting trajectories are shown in Fig.~\ref{fig:kregges}, compared to similar calculations for other resonances \cite{Londergan:2013dza,Carrasco:2015fva}.
All their resulting parameters are listed in Table~\ref{tab:kappap32}.  The method successfully yields the required Regge parameters and at the same time, the resulting amplitude approximates reasonably well the energy region surrounding each resonance (see the details in ~\cite{Londergan:2013dza,Carrasco:2015fva,Pelaez:2017sit}), consistently with being the dominant contribution there. Notice that the purpose is not to reproduce the partial waves within uncertainties in the whole elastic region, as that could only be achieved by introducing backgrounds that are neglected here.

\begin{table}[!htb]
\caption{Dispersively calculated Regge trajectories \cite{Londergan:2013dza,Carrasco:2015fva,Pelaez:2017sit}.}
\centering 
\begin{tabular}{c c c c} 
\hline\hline  
\rule[-0.15cm]{0cm}{.15cm}  & $\alpha_0$ & $\alpha'$ GeV$^{-2}$ & $b_0$ \\ \hline
\rule[-0.15cm]{0cm}{.15cm} $\rho(770)$ & $0.52\pm0.02$ & $0.902\pm0.004$ & \\
\rule[-0.15cm]{0cm}{.15cm} $K_0^*(1430)$ & $-1.28^{+0.01}_{-0.17}$ & $0.81^{+0.01}_{-0.04}$ & $2.5^{+1.1}_{-0.4}$\\
\rule[-0.15cm]{0cm}{.15cm} $K^*(892)$ & $0.32\pm0.01$ & $0.83\pm0.01$ & $0.48\pm0.03$ \\
\hline
\rule[-0.15cm]{0cm}{.15cm} $\sig$ & $-0.090^{+0.004}_{-0.012}$ & $0.002^{+0.050}_{-0.001}$ & \\
\rule[-0.15cm]{0cm}{.15cm} $\kap$ & $-0.27\pm0.03$ & $0.11\pm 0.09$ & $0.45^{+0.11}_{-0.08}$ \\
\hline
\end{tabular} 
\label{tab:kappap32} 
\end{table}

Let us now briefly discuss these results. The solutions for the $\rho(770),K^*(892)$ and $K_0^*(1430)$ yield ordinary, straight Regge trajectories, at least near and beyond the resonance under consideration, whose slopes are consistent with the universal one. Moreover, when checking higher spin partners, these fall within these trajectories or very close to them, even though the parameters of the ``straight line'' were not fitted, but obtained from the dispersive representation using only the pole parameters of the lighter particle. On top of that, the real part of the trajectory dominates over the imaginary part above the resonance mass. In contrast,  the \kap and \sig trajectories do not fit this universal pattern. These trajectories are highly nonlinear, and the slopes are many times smaller than the ordinary ones. This indicates that the scale of the dynamics that govern the binding of these resonances is larger than typical quark-antiquark dynamics. Furthermore, the imaginary part of the trajectory is larger than its real part, which is the opposite of what is expected for ordinary mesons. This indicates a strong non-ordinary behavior for these two states.

Finally, we also show in Fig.~\ref{Fig:yukawacomp} the behavior of both the \sig and \kap Regge trajectories in the $( \re \alpha, \im \alpha )$ plane. Let us remark that the low-energy part of these trajectories resembles those of non-relativistic Yukawa potentials~\cite{Lovelace:1962yza,Barut:1962fpy,Ahmadzadeh:1963ith}. In particular below 2 GeV both trajectories can be approximated by a given potential $V(r)= Ga \exp(-r/a)/r$, which they match from $s=-\infty$ to the two-particle thresholds. At the two-particle thresholds, the imaginary part becomes greater than zero, with a very step rising, which once again can be described by the Yukawa potentials. Up to $s=2\,\gev^2$ the $\sig$ trajectory is almost equal to that of a $G=2$ Yukawa potential, while the line with $G=1.4$ is quite similar to the $\kap$ trajectory. We can thus
estimate the effective ranges of the closest Yukawa potential in the $\sig$ case, namely \cite{Londergan:2013dza}
$a_{\pi\pi}=0.5\,\gev^{-1}\simeq0.1\,$fm,
as well as in the
$\kap$ case: $a_{\pi K}=0.36\,\gev^{-1}\simeq 0.07\,$fm.
As explained in~\cite{Pelaez:2017sit}, 
the ratio between them is numerically very near the inverse of the ratio
between the reduced masses of the respective scattering systems.
Thus, it seems that the pion and kaon masses
set the scale for the binding dynamics and therefore support again a possible ``molecular'' nature. Let us nevertheless remark that the effective range of the Yukawa potentials needed to mimic these trajectories at low energies is rather small for a meson \cite{Pelaez:2017sit},  of the same order of the scalar radius found in~\cite{Albaladejo:2012te} for the \sig.

\begin{figure}[!ht]
\begin{center}
\includegraphics[width=0.6\textwidth]{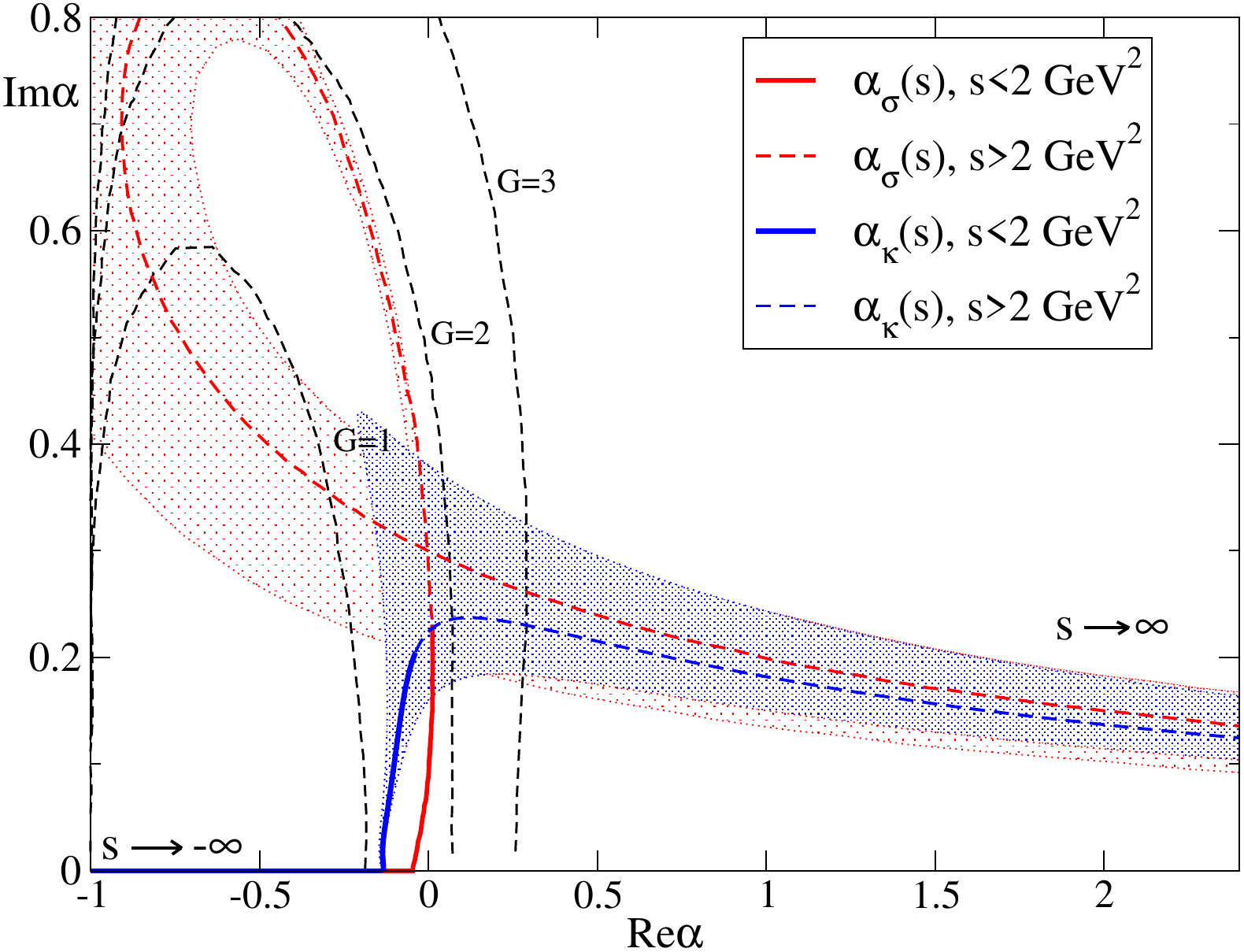}
 \caption{\rm \label{Fig:yukawacomp} 
 Trajectories of the $\sig$ (red line and area covering uncertainties) and $\kap$ (blue line and area covering uncertainties) in the $(\re \alpha,\im \alpha)$ plane as functions of $s$. The lines are continuous and thick for $s<2\, \gev^2$, which is where, at most, the method could be justified. Above that energy, we just show the extrapolation of the results.
 We also show, for comparison, three black dashed lines, labeled with different values of $G$, which correspond to the trajectories of the corresponding Yukawa potentials  $V(r)= Ga \exp(-r/a)/r$
 \cite{Lovelace:1962,Barut:1962fpy,Ahmadzadeh:1963ith,Burke:1969re} with $a$'s given in the text. This is a colored version of our figure in~\cite{Pelaez:2017sit}.}
\end{center}
\end{figure}

Altogether, our results support a predominantly non-ordinary microscopic nature for the \kap, whose behavior is very similar to that of the \sig. These results may suggest that their properties are predominantly controlled by meson-meson dynamics. 

\section{Other applications}
\label{sec:applications}

\subsection{\pik threshold and subthreshold parameters from sum rules}
\label{sec:sumrules}

As we have already explained in the introduction, the values of the threshold parameters,  defined in Eqs.~\eqref{eq:scldef} and \eqref{eq:lowex}, are of interest for our understanding of low-energy hadron physics, since, in principle,  this is the most suitable regime for low-energy effective field theory. Namely, with the $\pik$  threshold parameters, we are testing SU(3) Chiral Perturbation Theory (ChPT) and its convergence when the strange quark is taken into account since its mass is not as small as those of non-strange quarks. We emphasize again that we are working in the isospin conserving approximation and for studies of isospin violation in \pik we refer to ~\cite{Nehme:2001wf,Nehme:2001wa,Kubis:2001bx,Kubis:2001ij}

The problem to extract these threshold parameters is that the kinematic suppression makes it very hard to obtain precise data, or even data at all, close to threshold. For instance, see all the data plots in Section~\ref{sec:piKData}
for \pik partial waves and note that the lowest energy data are at least 100 MeV above threshold, and even those points are rather isolated and usually with large uncertainties. Therefore, extracting threshold parameters directly from scattering data depends strongly on the precise parameterization or model used for the extrapolation down to threshold. It is advisable to avoid such a strong model dependence. As we will see below, some determinations from $\pi K$ atom lifetimes provide information on scattering lengths, although their uncertainties are rather large.

 Sum rules are obtained from dispersion relations evaluated at particular points or limits and this makes their prediction much more robust than direct extractions using a parameterization of the data. The reason lies in the integral nature of the sum rule, which makes rather irrelevant the details of a particular parameterization around threshold, thus suppressing  very strongly, if not completely, any model or parameterization dependence. In addition, an integral determination typically yields a much smaller uncertainty. Therefore, generically, sum rules provide the most accurate and robust method to determine  threshold parameters.

Moreover, sum rules for threshold parameters provide additional consistency tests for data parameterizations.
In this section we will first discuss the scalar scattering lengths, since they have attracted quite some attention recently, discussing also the Adler zeros in these waves. Next, we will present values for both scattering lengths and slopes for all waves up to angular momentum 2 and finally discuss the subthreshold parameters.

\subsubsection{Scalar scattering lengths}
\label{subsec:scl0}

At present, there is a great deal of interest in the values of the $S$-wave scattering lengths from ChPT, dispersion theory, and lattice gauge theory communities, since some tension exists between the determinations using these different techniques. This was illustrated in Fig.~\ref{fig:scl} in the introduction, where we can see in the $(a_0^{1/2},a_0^{3/2})$ plane that, for both scattering lengths, lattice QCD results (in red) tend to produce lower values  than dispersive results (in brown and green). The tension is somewhat more evident for the $a_0^{1/2}$ channel, which, as we have seen, is probably the most controversial one and includes the \kap meson. Concerning ChPT, the LO receives sizable NLO corrections that take the values closer to lattice results, but in order to reach the dispersive results the NNLO corrections should be even bigger, casting doubts about the good convergence of $SU(3)$ ChPT. Actually, 
in the extensive ChPT review in \cite{Bijnens:2014lea} it is shown that the $\pi K$ scattering lengths have the worst convergence of all the observables under consideration.
The precise values of all these calculations are shown in  Table.~\ref{tab:scatteringlengths}.

\begin{table}
\small
\begin{tabular}{l|l|l|l}
\hline
Reference &$m_\pi a_0^{1/2}$&$m_\pi a_0^{3/2}$& Description\\
\hline
B\"uttiker et al. (2004) \cite{Buettiker:2003pp} & $0.224\pm0.022$ & $-0.0448\pm0.0077$& Dispersive Roy-Steiner\\
Pel\'aez-Rodas (2016) \cite{Pelaez:2016tgi} & $0.220\pm0.010$ & $-0.0540^{+0.010}_{-0.014}$& Fit constrained with FDR\\
\hline
Bijnens-Ecker (2014) \cite{Bijnens:2014lea} & $0.142$ & $-0.071$&ChPT LO\\
Bijnens-Ecker (2014) \cite{Bijnens:2014lea} & $0.173(0.169)$ & $-0.064(-0.066)$&ChPT NLO fit 14(free fit)\\
Bijnens-Ecker (2014) \cite{Bijnens:2014lea} & $0.224(0.226)$ & $-0.048(-0.047)$&ChPT NNLO fit 14(free fit)\\
\hline
Miao et al.(2004) \cite{Miao:2004gy}& $-$ & $-0.056\pm0.023$& lattice, improved Wilson quenched\\
NPLQCD (2006) \cite{Beane:2006gj}& $0.1725\pm0.0017^{+0.0023}_{-0.0156}$ & $-0.0574\pm0.016^{+0.0024}_{-0.0058}$& lattice. Domain-wall valence\\
Flynn-Nieves (2007) \cite{Flynn:2007ki}& $0.175\pm0.017$& $-$& lattice+Omn\`{e}s Dispersion Relation\\
Fu (2012) \cite{Fu:2011wc}
& $0.1819\pm0.0035$& $-0.0512\pm0.0018$& lattice, staggered, moving wall source\\
PACS-CS (2014) \cite{Sasaki:2013vxa}& $0.182\pm0.053$& $-0.060\pm0.006$& lattice, improved Wilson\\
ETM (2018) \cite{Helmes:2018nug}& $-$ & $-0.059\pm0.002$& lattice, twisted mass. \\
\hline
\end{tabular}
\caption{Previous determinations of $\pi K$ scalar scattering lengths from various approaches. Note that lattice results tend to yield lower values than dispersive results for both scattering lengths.
\label{tab:scatteringlengths} }
\end{table}

Experimentally the best determination, avoiding model dependencies, is the combination
\begin{equation}
 a_0^-=\frac{1}{3}(a_0^{1/2}-a_0^{3/2})=0.11^{+0.09}_{-0.04} m_\pi^{-1}\quad  {\rm (DIRAC)},
\end{equation}
obtained in 2017 by the DIRAC Collaboration \cite{Adeva:2014xtx} by studying the lifetime of $\pi K$ atoms at CERN. The relation between the decay width of these atoms and the scattering lengths including isospin violation and QED corrections can be found in \cite{Schweizer:2004ir,Schweizer:2004qe}.
The observation of these atoms and their decay lifetime is a remarkable experimental achievement, but the uncertainty of the result, which is represented by a beige band in Fig.~\ref{fig:scl}, labeled DIRAC 17, is not enough to discern between present lattice QCD or dispersive results.

So far, the best dispersive determinations come from the 2004 Roy-Steiner analysis of \cite{Buettiker:2003pp}
(Brown ellipse in Fig.~\ref{fig:scl})
or our 2016 fit to data constrained with forward dispersion relations \cite{Pelaez:2016tgi} (Green result labeled ``FDR-CFD-old'' in Fig.~\ref{fig:scl}). Note, however, 
that their uncertainties are rather large and, as explained several times, that the  authors  of \cite{Buettiker:2003pp} {\it solve} the Roy-Steiner equations without using data in the \pik elastic region. Also, our result in \cite{Pelaez:2016tgi} does not have separate dispersive constraints for each partial wave, but for the whole isospin amplitudes. Actually we only had a sum rule for $a_0^-$; namely, Eq.~\eqref{eq:scla0fdr} below. In what follows we will provide new independent values obtained from sum rules coming from Roy-Steiner equations for each partial wave. Let us emphasize  that no convergent sum rule for $a_0^+$ can be obtained using the dispersion relations provided in this work. Therefore, whenever we quote the dispersive $a_0^I$ value in the isospin basis, its component orthogonal to $a_0^-$ comes directly from the CFD parameterization described in sections~\ref{sec:UFDpik} and~\ref{sec:CFDpik}.

In addition, we will provide different sum rules for the most controversial threshold parameters. In later subsections, we will use them to obtain precise values of the scattering lengths and slopes. Having different sum rules for the same observables, and therefore weighting differently the input, will also provide strong consistency tests of our data parameterizations. 

At this point, a technical remark is in order. Threshold parameters correspond to $s=m_+^2$, but setting $s$ directly to this value, which coincides with the lower integration limit, would not allow calculating the principal value needed for real $s$ in the dispersion relations.
One should then exert caution to define these sum rules,  as the $s\rightarrow m^2_+$ limit has to be taken after the dispersion relation has been evaluated with its principal value. For example, a sum rule for the $a^-_0$ scattering length can be obtained from the $F^-$ FDR, \eqref{eq:FDRTan},  as:
\begin{equation}
a^-_0=\frac{ m_\pi m_K}{2 \pi^2 m_+}\lim_{s\rightarrow m^2_+}PV\!\!\int^{\infty}_{m_+^2}{\frac{\im F^-(s')}{(s'-m_-^ 2)(s'-s)}ds'}.
\end{equation}
Fortunately, 
the imaginary part of the amplitude $\im F^-(s')\propto \sqrt{s-m^2_+}$ suppresses the singularity at threshold, and hence the principal value can be removed when taking the limit, to yield
\begin{equation}
a^-_0=\frac{ m_\pi m_K}{2 \pi^2 m_+}\!\!\int^{\infty}_{m_+^2}{\frac{\im F^-(s')}{(s'-m_-^ 2)(s'-m_+^2)}ds'}.
\label{eq:scla0fdr}
\end{equation}
Thus, in this case, the combined appearance of the $s\rightarrow m_+^2$ limit and the principal value allows us to get rid of both in the final expression.

However, obtaining a simple algebraic expression for the $S$-wave $b^I_{\ell}$ parameters requires a slightly more elaborated, but rather standard, manipulation \cite{Nebreda:2012ve,GarciaMartin:2011cn,Kaminski:2008fu} to deal with the divergent contribution arising from the  $\im f^I_0(s)/(s'-s)$ behavior close to threshold. Namely, since this divergence is proportional to $(a^I_0)^2/(s'-m_+^2)^{3/2}$ we can add to our dispersive expression 
this factor multiplied by
\begin{equation}
 PV\!\!\int_{m_+^2}^{\infty}\frac{ds'}{(s'-s)\sqrt{s'-m_+^2}}=0.
\end{equation}
When this is done  the cancellation of the divergence is directly seen inside the integral, which allows us to get rid of the principal value and the limit outside. The price to pay is the appearance of a new term in the integrand.

In particular, under these manipulations, we can write sum rules for scattering lengths and slopes coming from $F^+$ \cite{Pelaez:2016tgi}\footnote{We fixed an errata with respect to our published expression in  \cite{Pelaez:2016tgi}.} as follows:
\begin{eqnarray}
{SR}_1&\equiv& b^{1/2}_0+3a^{1/2}_1+\frac{a^{1/2}_0m_+}{2 m_\pi m_K} 
\label{eq:sr1} \\
&=&
\frac{m_+}{8 \pi^2 m_\pi m_K}
 \!\!\int^{\infty}_{m_+^2}\!\!\!\!ds'\left[\frac{\im F^+(s')+2\im F^-(s')}{(s'-m_+^2)^2}-8m_+\pi\sqrt{m_\pi m_K}\frac{(a^{1/2}_0)^2}{(s'-m_+^2)^{3/2}}- \frac{\im F^+(s')-2\im F^-(s')}{(s'+m_+^2-2\Sigma_{\pi K})^2}\right],\nonumber
\end{eqnarray}
and
\begin{eqnarray}
SR_2&\equiv&b^{3/2}_0+3a^{3/2}_1+\frac{a^{3/2}_0m_+}{2 m_\pi m_K}
\label{eq:sr2}\\
&=& \frac{m_+}{8 \pi^2 m_\pi m_K}
  \!\!\int^{\infty}_{m_+^2}\!\!\!\!ds'\left[\frac{\im F^+(s')-\im F^-(s')}{(s'-m_+^2)^2}-8m_+\pi\sqrt{m_\pi m_K}\frac{(a^{3/2}_0)^2}{(s'-m_+^2)^{3/2}}- \frac{\im F^+(s')+\im F^-(s')}{(s'+m_+^2-2\Sigma_{\pi K})^2}\right].\nonumber
\end{eqnarray}
Note that in \cite{Pelaez:2016tgi}, as long as we only  considered FDRs, which are not projected in partial waves, we took these sum rules as additional constraints for our partial-wave fits. Since here we are also considering partial-wave dispersion relations, these are no longer needed as constraints and we will see they are still well satisfied.

If  we used FTPWDR
we would obtain for $a_0^-$ the same sum rule already obtained from FDR in Eq.~\eqref{eq:scla0fdr}.
However, using the HDR at $t=0$ and $b=\Delta^2+a^2-2 \Delta a$ the dispersion relation for $F^-/G^1$ can be related to  $a^-_0$, leading to the following sum rule \cite{Karabarbounis:1980bk,Buettiker:2003pp}:
\begin{align}
\frac{8\pi m_+a^-_0}{m^2_+-m^2_-}=\frac{1}{2\pi}\!\!\int^{\infty}_{4m_{\pi}^2}\frac{dt'}{t'}\im\frac{ G^1(t',s'_{b})}{\sqrt{(t'-4m^2_\pi)(t'-4m^2_K)}} 
+\frac{1}{\pi}\int^{\infty}_{m_+^2}ds'\frac{\im F^-(s',t'_{b})}{\lambda_{s'}}.
\label{eq:scla0hdr}
\end{align}
This sum rule is dominated by the \pipikk amplitude and thus, for all means and purposes, it is independent of the value obtained with  Eq.~\eqref{eq:scla0fdr}. In practice, we will calculate this sum rule with $a=-10.9m_\pi^2$, which maximizes the Roy-Steiner applicability in the \pipikk channel. Note that we only obtain a sum rule for $a_0^-$ from the HDR without subtractions for $F^-$, since $a_0^-$ is input in our subtracted $F^-$ HDR case.

At this point, we can use the UFD and CFD parameterizations to determine the \pik threshold parameters, which can be done either directly from the parameterizations, or from the sum rules, which, being obtained from an integral would suppress the \pik parameterization dependence. 
Thus, in Table \ref{tab:lowpara} we have collected the scattering lengths obtained directly from the UFD and CFD parameterizations. 
Of course, we think the constrained results are better because they are consistent with dispersion relations. 
We can see that there is a change from UFD to CFD, by roughly 1.5 standard deviations, which brings our CFD results very close to those of the dispersive calculation of the Bonn-Paris group \cite{Buettiker:2003pp}. We are thus providing an independent dispersive confirmation of those older dispersive  results,  this time using data together with the dispersive constraints, whereas in \cite{Buettiker:2003pp} they were obtained from Roy-Steiner solutions.

\begin{table}[!ht] 
\caption{$S$-wave scattering lengths ($m_\pi$ units).  }
\vspace{0.3cm}
\centering 
\begin{tabular}{l  c  c  c } 
\hline &
 \hspace{0.2cm} Ref. \cite{Buettiker:2003pp} & \hspace{0.2cm} UFD & \hspace{0.2cm} \bf{CFD}  \\
\hline\hline  
\rule[-0.2cm]{-0.1cm}{.55cm} $a^{1/2}_0$ & \hspace{0.2cm} $0.224\pm0.022$ & \hspace{0.2cm} $0.241\pm0.012$ & \hspace{0.2cm} $\bf{0.224\pm0.010}$ \\
\rule[-0.2cm]{-0.1cm}{.55cm} $a^{3/2}_0$ & \hspace{0.2cm} $-0.0448\pm0.0077$ & \hspace{0.2cm} $-0.067\pm0.012$ & \hspace{0.2cm} $\bf{-0.048\pm0.006}$  \\
\hline
\end{tabular} 
\label{tab:lowpara} 
\end{table}

\begin{figure}[!ht]
    \centering
    \resizebox{0.8\textwidth}{!}{\input{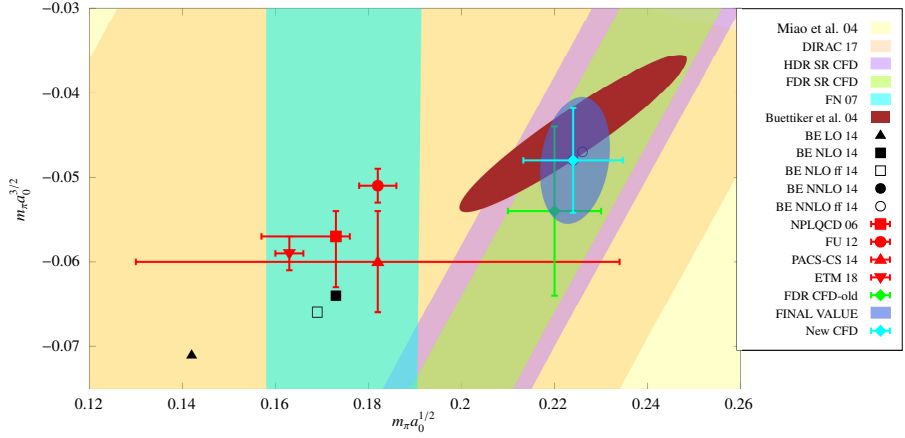}}
    \caption{Comparison between various sum rules and determinations coming from different Roy-Steiner equations, compared with both lattice QCD predictions and ChPT calculations. The references in the legend are as in Fig.~\ref{fig:scl}, except that we are now providing as a cyan diamond our new result obtained directly from our dispersively constrained fit to data (CFD), listed in boldface in Table~\ref{tab:lowpara}, as well as our best or ``Final value'', shown as a blue ellipse and
    listed in Table~\ref{tab:isospinscl}. The latter is obtained by combining the dispersive sum rules explained in section \ref{sec:sumrules}. Note that the two axes are plotted with different scales to maximize the visibility of the region of interest.}
    \label{fig:sclcfd}
\end{figure}

We are showing our results obtained directly from our CFD parameterization as a cyan diamond and cross in Fig.~\ref{fig:sclcfd}, which overlaps nicely with the brown ellipse from \cite{Buettiker:2003pp}. Note that it also overlaps with our old result \cite{Pelaez:2016tgi}, where we only used FDRs as constraints. The use of those FDRs together with the Roy-Steiner equations in this review has moved the central value within the old error bars, 
and has reduced considerably our uncertainties. 
Note that the errors in Table~\ref{tab:lowpara}, plotted as a cross in Fig.~\ref{fig:sclcfd}, are given as uncorrelated since they come from the direct CFD parameterization, where we treat all input parameters as uncorrelated and thus each partial wave as independent. This is a conservative error estimate, but
we will discuss below the correlation between scattering lengths once we use the $a_0^-$ numerical values from sum rules.

The nice consistency of the CFD versus the inconsistency of the UFD is illustrated in Table~\ref{tab:sumrule}. There we see an almost perfect agreement between the $a_0^-$ obtained directly from the CFD parameterization, and the sum rules in Eqs.~\eqref{eq:scla0fdr} and \eqref{eq:scla0hdr} using CFD as input. In contrast, for the UFD, the inconsistencies reach the three standard-deviation levels.

\begin{table}[ht]
\caption{\label{tab:sumrule}  Comparison of the value $a^-_0$  obtained directly from the parameterization versus the value obtained from different sum rules, coming from forward dispersion relations (FDR) Eq.~\eqref{eq:scla0fdr} or from hyperbolic dispersion relations (HDR), Eq.~\eqref{eq:scla0hdr}.}
\vspace{0.3cm}
\centering 
\begin{tabular}{c c c } 
\hline
 & UFD & CFD  \\
\hline\hline
\rule[-0.2cm]{-0.1cm}{.55cm} $3 m_\pi a^-_{0,\,direct}$ & 0.309$\pm$0.016 & 0.273$\pm$0.012 \\
\rule[-0.2cm]{-0.1cm}{.55cm} $3 m_\pi a^-_{0,\,FDR}$  & 0.290$\pm$0.010 & { 0.275$\pm$0.010} \\
\rule[-0.2cm]{-0.1cm}{.55cm} $3 m_\pi a^-_{0,\,HDR}$ & 0.253$\pm$0.015 & 0.274$\pm$0.016 \\
\hline
\end{tabular} 
\end{table}

 Therefore, sum rules not only provide stringent constraints but sometimes also the most reliable and precise results for threshold parameters, suppressing model dependencies too.
Thus, once we have shown that these two sum-rule determinations with CFD input are consistent, they provide a very stringent and robust result for $a_0^-$ which we have included in Fig.~\ref{fig:sclcfd} as  light green and purple bands. Let us recall that they are rather independent since the sum rule from FDR, i.e. Eq.~\eqref{eq:scla0fdr} (the ``FDR SR CFD'' green band), is dominated by the scalar \pik scattering wave, whereas the sum rule from HDR, i.e Eq.~\eqref{eq:scla0hdr} (the ``HDR SR CFD'' purple band) is dominated by the \pipikk scattering contribution. Incidentally, had we used as input the UFD parameterization, which is not consistent with dispersion relations, these two bands would not even overlap.

A relatively similar pattern is observed in Table \ref{tab:sumrules12}, where we compare the results of the sum rules $SR_1$ and $SR_2$ in Eqs.~\eqref{eq:sr1} and \eqref{eq:sr2} using as input either the UFD or CFD parameterizations with their respective values obtained directly from the fits. Once again the CFD results are wonderfully consistent,
whereas the UFD  shows some tension at the 2 standard deviation level.

\begin{table}[ht]
\caption{\label{tab:sumrules12}  Comparison between the sum rules of Eqs. \eqref{eq:sr1} and \eqref{eq:sr2} calculated using the UFD or CFD parameterizations as input, versus the values obtained directly from each parameterization.}
\vspace{0.3cm}
\centering 
\begin{tabular}{c c c } 
\hline
 & UFD & CFD  \\
\hline\hline
\rule[-0.2cm]{-0.1cm}{.55cm} $SR_1$ & $0.187\pm0.006$ & {\bf 0.182$\pm$0.006} \\
\rule[-0.2cm]{-0.1cm}{.55cm} ${\rm Direct}_{SR_1}$ & 0.163$\pm$0.010 & 0.176$\pm$0.013 \\
\rule[-0.2cm]{-0.1cm}{.55cm} $SR_2$  & $-0.042\pm0.004$ & ${\bf -0.039\pm0.003}$ \\
\rule[-0.2cm]{-0.1cm}{.55cm} ${\rm Direct}_{SR_2}$ & $-0.052\pm0.005$ &$-0.037\pm0.006$ \\
\hline
\end{tabular} 
\end{table}

So far we have not discussed partial-wave dispersion relations to obtain sum rules, but they can also be expanded around threshold to obtain further integral expressions for threshold parameters. 
At first, it may seem we have three more sum rules since we can expand the partial-wave fixed-$t$ dispersion relations (FTPWDR) as well as the unsubtracted and once-subtracted partial-wave hyperbolic dispersion relations (HPWDR).
However, for the scalar scattering lengths, the HPWDR sum rule with one subtraction coincides with the direct CFD calculation, whereas the unsubtracted HPWDR is almost identical to~\eqref{eq:scla0hdr}, the only difference is that $a=0$ for the latter. Finally, the FTPWDR sum rule is identical to~\eqref{eq:scla0fdr}.

Hence, in practice we just have two independent $a_0^-$ sum rules, i.e., Eqs.~\eqref{eq:scla0fdr} and
\eqref{eq:scla0hdr}. We can now use them to get our final and more reliable values of the scalar scattering lengths in the isospin basis, which are more robust than those in Table~\ref{tab:lowpara}, obtained directly from the CFD. Of course, for the orthogonal combination $a_0^{1/2}+a_0^{3/2}$ (or for $a_0^+$ when needed), which does not have a sum rule, we still take the direct CFD parameterization.
Thus, the numerical values 
for the scalar scattering lengths in the isospin basis are listed
in Table~\ref{tab:isospinscl}.
The first two columns are obtained by using values from each one of our $a_0^-$ sum-rules: FDR Eq.~\eqref{eq:scla0fdr} and HDR Eq.~\eqref{eq:scla0hdr}.   
Nevertheless, given that the two $a_0^-$ sum rules have rather different inputs, we have calculated their weighted average and statistical error, to which we have added
a systematic uncertainty\footnote{Calculated as the sum in quadrature of their separation from the mean, divided by two. }.
This constitutes our ``Final Value'', provided in the third column, for which we quote correlated uncertainties in the isospin basis, whose
correlation matrix is provided in the next column.
The reason is that, contrary to CFD direct values in Table~\ref{tab:lowpara}, the  calculation for our Final Values is not done from independent partial waves in the isospin basis. Instead, we use the  weighted average of sum rules for $a_0^-$. These correlated uncertainties are
represented by an ellipse in Fig.~\ref{fig:sclcfd}.
It can be seen that, after rounding up numbers, the central Final Values are almost identical to the CFD result, and the uncertainties are very similar. It is also nicely compatible with previous dispersive or ChPT determinations,  provided in the last two columns of Table~\ref{tab:isospinscl}. Note that our estimated uncertainty is somewhat smaller than in \cite{Buettiker:2003pp}, in particular, due to the small uncertainty from the FDR sum rule.

In summary, our calculations of the scalar $\pi K$ scattering lengths show that simple unconstrained fits to data (UFD) yield inconsistent values between the ``direct'' input and the sum rule output and even between different sum rules. This is amended with the constrained fits to data (CFD), which show a remarkable consistency. In this review, we have thus updated our CFD values for the scattering lengths and we have also provided robust
and model-independent
results from sum rules for $a_0^-$,
which lead to our ``Final Values'' in the isospin basis.
Both the CFD and ``Final Values'' are remarkably compatible among themselves and with previous dispersive determinations \cite{Buettiker:2003pp}, including our own in \cite{Pelaez:2016tgi}, although our uncertainties are now much smaller than then. In addition, the consistency between our different sum rules, which are dominated by very different inputs, provides very robust and parameterization-independent constraints. We can therefore conclude that the two standard deviation tension between dispersive analyses and the bulk of lattice results still remains.

\begin{center}
\begin{table}[!hb]
\centering
\footnotesize
\begin{tabular}{l|c|c|c|c|c|c}
\hline
&\multicolumn{3}{c|} {This work sum rules with CFD input}  & \multicolumn{1}{c|}{Correlation} & Sum rules \cite{Buettiker:2003pp}&  NNLO ChPT  \\
 & FDR & HDR & \textbf{Final Value} & matrix &  Fixed-$t$ &  \cite{Bijnens:2004bu} and 
\cite{Bijnens:2014lea}$^*$\\
\hline
\hline
\rule[-0.175cm]{-0.1cm}{.5cm} $m_\pi a_0^{1/2}$ & 0.226$\pm$ 0.010 &  0.225$\pm$ 0.012 & \textbf{0.225$\pm$0.008} & \multirow{2}{*}{ $\begin{pmatrix} 1 & 0.04 \\ 0.04 & 1 \end{pmatrix}$ } & 0.224$\pm$0.022 & 0.224$^*$  \\
\rule[-0.175cm]{-0.1cm}{.5cm} $m_\pi a_0^{3/2}$  $\times$ 10&  $-$0.489$\pm$0.052 & $-$0.485$\pm$0.066 & \textbf{$-$0.480$\pm$0.067}& & $-$0.448$\pm$0.077 & $-$0.471$^*$ \\
\hline
\end{tabular}
\caption{Determination of the $\pi K$ scalar scattering lengths using the sum rule for $a_0^-$ with our CFD as input. Let us remind that the combination orthogonal to $a_0^-$  comes directly from the CFD parameterization and not from a sum rule. 
This is why we provide the correlation matrix in the isospin basis.  }
\label{tab:isospinscl}
\end{table}
\end{center}

In this subsection, we concentrated on the particular case of scalar scattering lengths.
Let us now consider other partial waves and higher orders in the threshold expansion.

\subsubsection{Sum rules for other threshold parameters of partial waves up to $\ell=2$.}
\label{subsec:SRth}

We provide in this subsection the sum rules for the $\ell=1,2$ scattering lengths $a^I_{\ell}$ as well as the $\ell=0,1,2$ slopes $b^I_{\ell}$. These are calculated from the low-momentum expansion around threshold of the partial-wave dispersion relations obtained either from ``fixed-$t$'' dispersion relations (FTPWDR) Eq.~\eqref{eq:sftpwdr} or from hyperbolic dispersion relations in Eqs.~\eqref{eq:shdrfm} and \eqref{eq:shdrfmi} (HPWDR). For this, we first expand their respective kernels (given in \ref{app:Kernels}) in a $q^2$ power series that we match to the expansion in Eq.~\eqref{eq:lowex}. We have truncated the expansion at second order on each wave and up to $\ell=2$ included. Not only do they provide tests for our amplitudes, but also, since all of them have been previously calculated either from sum rules \cite{Buettiker:2003pp} or within ChPT   \cite{Bijnens:2004bu}, we will be able to  compare with those predictions.

The analytic expressions are rather long and we will only provide here the most relevant ones, namely the  $P$-wave scattering lengths and the $S$-wave slopes for the $+$ and $-$ amplitude isospin combinations. Analytic expressions become unmanageable as the order of the power of $q^2$ grows, but can be obtained following the method we just described. We will nevertheless provide numerical values for all scattering lengths and slopes up to $\ell=2$ at the end of the subsection.

Let us then provide first the sum rules for the $a_1^I$ scattering lengths,  which  we organize according to the type of dispersion relation they come from:

\begin{itemize}
\item Scattering length sum rules from FTPWDR for $\ell=1$:

\begin{align}
a^+_1&=\frac{1}{\pi}\!\!\int_{m_+^2}^{\infty}ds'\left( \frac{-2}{3(s'-m_-^2)^2}\im f^+_0(s') +
\frac{2s'^2+4m_+^2(s'-m_-^2)-2m_+^4}{(s'-m_-^2)^2(s'-m_+^2)^2}\im f^+_1(s')\right.  \nonumber \\ 
&\left.+ \frac{-10}{3(s'-m_-^2)^2} \im f^+_2(s')\right) +\frac{1}{\pi}\int_{4 m_\pi^2}^{\infty}dt' \frac{2}{3\sqrt{3}t'^2} \im g^0_0(t') + \frac{5 \lambda(t')}{24\sqrt{3}t'^2}\im g^0_2(t')+d_{a^+_1}, \nonumber \\
a^-_1&=\frac{1}{\pi}\!\!\int_{m_+^2}^{\infty}ds' \left(   \frac{2}{3(s'-m_-^2)^2}\im f^-_0(s') + \frac{s'^2-2s'm_-^2+m_+^4}{(s'-m_-^2)^2(s'-m_+^2)^2}\im f^-_1(s')\right. \nonumber \\
&\left.+ \frac{s'^2-2s'(3m_-^2-2m_+^2)+m_+^4}{3(s'-m_-^2)^2(s'-m_+^2)^2}\im f^-_2(s')\right)+d_{a^-_1}, \label{eq:asfromft}
\end{align}
where $d_{a^+_1},d_{a^-_1}$ stand for the contributions from higher partial waves, which are numerically very small but were included in our numerical calculations.
\item Scattering length sum rules from HPWDR  for $\ell=1$:
\begin{align}
a^+_1&=\frac{1}{\pi}\!\!\int_{m_+^2}^{\infty}ds'\left( \frac{-2}{3(s'-m_-^2)^2}\im f^+_0(s') +
\frac{2 m_+^4 \left(s'-a\right)-2 s' \left(2 a m_-^2-3 a s'+s'^2\right)+4 m_+^2 s'
   \left(m_-^2-s'\right)}{\left(a-s'\right) \left(s'-m_-^2\right)^2 \left(s'-m_+^2\right)^2}\im f^+_1(s')\right.  \nonumber \\ 
&\left.+ \frac{10 \left(\frac{6 a s' \left(-a m_-^2+m_+^2 \left(m_-^2-a\right)+2 a s'-s'^2\right)}{\left(a-s'\right)^2
   \left(s'-m_+^2\right)^2}-1\right)}{3 \left(s'-m_-^2\right)^2} \im f^+_2(s')\right) +\frac{1}{\pi}\int_{4 m_\pi^2}^{\infty}dt' \left( \frac{2}{3\sqrt{3}t'^2} \im g^0_0(t') + \frac{5 (\lambda(t')+6 a t')}{24\sqrt{3}t'^2}\im g^0_2(t')\right)+d_{a^+_1}, \nonumber \\
a^-_{1}&=\frac{1}{\pi}\!\!\int_{m_+^2}^{\infty}ds'\left( \frac{2}{3 \left(s'-m_-^2\right)^2}\im f^-_0(s') +\frac{2 m_+^4 \left(a-s'\right)+2 s' \left(s' \left(a+s'\right)-2 a m_-^2\right)+4 m_+^2 s'
   \left(m_-^2-s'\right)}{\left(a-s'\right) \left(s'-m_-^2\right){}^2 \left(s'-m_+^2\right){}^2}\im f^-_1(s') \right. \nonumber \\
&\left.+\frac{10}{3 \left(a-s'\right)^2 \left(s'-m_-^2\right){}^2 \left(s'-m_+^2\right)^2}\im f^-_2(s')\times\right. \nonumber \\
& \left. \Bigg(m_+^4 \left(a^2-2 a s'-5 s'^2\right)-2 m_+^2 s' \left(-2 a^2-3 m_-^2 \left(a+s'\right)+a
   s'+s'^2\right)+s' \left(s' \left(a^2+4 a s'+s'^2\right)-6 a m_-^2
   \left(a+s'\right)\right)\Bigg)  \rule{0mm}{8mm} \right) \nonumber  \\ 
  & +\frac{1}{\pi}\int_{4 m_\pi^2}^{\infty}dt' \frac{t'+4m_\pi m_K}{2\sqrt{2}t'^2} \im g^1_1(t')+d_{a^-_1}, \nonumber \\
a^-_{1, sub}&=\frac{1}{\pi}\!\!\int_{m_+^2}^{\infty}ds'\left( \frac{2 \left(m_--m_+\right) \left(m_-+m_+\right)}{3 \left(s'-m_-^2\right){}^2 \left(s'-m_+^2\right)}\im f^-_0(s') -\frac{2 \left(m_--m_+\right) \left(m_-+m_+\right) \left(m_+^2+s'\right)}{\left(s'-m_-^2\right){}^2
   \left(s'-m_+^2\right){}^2}\im f^-_1(s') \right. \nonumber \\
&\left.-\frac{10 \left(m_--m_+\right) \left(m_-+m_+\right) \left(m_+^2 \left(a-s'\right)+s' \left(5 a+s'\right)\right)}{3
   \left(a-s'\right) \left(s'-m_-^2\right){}^2 \left(s'-m_+^2\right)^2}\im f^-_2(s') \right) +\frac{1}{\pi}\int_{4 m_\pi^2}^{\infty}dt' \frac{\sqrt{2}m_\pi m_K}{t'^2} \im g^1_1(t')+d_{a^+_{1,sub}},
\end{align}
\end{itemize}
where the last one is obtained from an HDR with one subtraction for $F^-$. Once again the $d_i$ terms are due to higher partial waves and they are very small, although we computed them in our calculations.
Note that $a$ is one of the parameters that define the hyperbolae, which we have chosen as $a=-10 m_\pi^2$ to impose the HDR in our CFD parameterization.

Let us now provide the analytic expressions for the sum rules of the $b_0^\pm$ parameters. Remember that, to perform the double limit between the principal value and $s\rightarrow m_+^2$, the singular part of the integrands must be removed, as explained above.

\begin{itemize}
    \item Scattering slopes sum rules from FTPWDR  for $\ell=0$:
\end{itemize}
\begin{align}
b^+_0&=\frac{1}{\pi}\!\!\int_{m_+^2}^{\infty}ds'\left( \frac{2s'^2+4m_+^2(s'-m_-^2)-2m_+^4}{(s'-m_-^2)^2(s'-m_+^2)^2} \im f^+_0(s')-\frac{m_+^2}{m_\pi m_K}\frac{(a^+_0)^2}{(s'-m_+^2)^{3/2}}\right. \nonumber \\
&\left.+ \frac{-6}{(s'-m_-^2)^2}\im f^+_1(s') + \frac{10(s'^2+2m_+^2(s'-m_-^2)-m_+^4)}{(s'-m_-^2)^2(s'-m_+^2)^2} \im f^+_2(s')\right) \nonumber \\ 
&+\frac{1}{\pi}\int_{4 m_\pi^2}^{\infty}dt'\left( \frac{-2}{\sqrt{3} t'^2}\im g^0_0(t') + \frac{5\lambda(t')}{8\sqrt{3}t'^2}\im g^0_2(t') \right)  +d_{b^+_0}, \nonumber \\
b^-_0&=\frac{1}{\pi}\!\!\int_{m_+^2}^{\infty}ds'\left( (\frac{\Sigma}{m_\pi m_K (s'-m_-^2)}+\frac{m_+^2}{m_\pi m_K (s'-m_+^2)}) \im f^-_0(s')-\frac{m_+^2}{m_\pi m_K}\frac{(a^-_0)^2}{(s'-m_+^2)^{3/2}}\right. \nonumber \\
&\left.+ \frac{(3/2m_+^4+3m_-^4+3/2m_+^2m_-^2-6 s'(\Sigma+m_\pi m_K))}{m_\pi m_K(s'-m_+^2)(s'-m_-^2)^2}\im f^-_1(s')\right. \nonumber \\
&\left.+\frac{m_+^2(5/2m_+^4+5m_-^4+5/2m_+^2m_-^2)+20 s' (m_+^4+2/3m_-^4-2/3m_+^2m_-^2)+10 s'^2 (\Sigma+m_\pi m_K)}{m_\pi m_K(s'-m_+^2)(s'-m_-^2)^2}\im f^-_2(s')\right) +d_{b^-_0}, 
\end{align}
\begin{itemize}

\item Scattering slopes sum rules from  HPWDR for $\ell=0$:
\end{itemize}
\begin{align}
b^+_0&=\frac{1}{\pi}\!\!\int_{m_+^2}^{\infty}ds'\left( \frac{2s'^2+4m_+^2(s'-m_-^2)-2m_+^4}{(s'-m_-^2)^2(s'-m_+^2)^2} \im f^+_0(s')-\frac{m_+^2}{m_\pi m_K}\frac{(a^+_0)^2}{(s'-m_+^2)^{3/2}}\right. \nonumber \\
&\left.-\frac{6 \left(\frac{2 a \left(s'-m_-^2\right)}{\left(a-s'\right) \left(s'-m_+^2\right)}+1\right)}{\left(s'-m_-^2\right){}^2}\im f^+_1(s')\right. \nonumber \\
&\left.-\frac{10}{\left(a-s'\right)^2 \left(s'-m_-^2\right)^2
   \left(s'-m_+^2\right){}^2} \im f^+_2(s')\times\right. \nonumber \\
&\left. \left(2 m_+^2 \left(m_-^2 \left(a^2+a s'+s'^2\right)-s' \left(4 a^2-2 a
   s'+s'^2\right)\right)-s' \left(a^2 \left(6 m_-^2-11 s'\right)+4 a
   s'^2+s'^3\right)+m_+^4 \left(a-s'\right)^2\right)\right) \nonumber \\ 
&+\frac{1}{\pi}\int_{4 m_\pi^2}^{\infty}dt'\left( \frac{-2}{\sqrt{3} t'^2}\im g^0_0(t') + \frac{5(\lambda(t')+6 a t')}{8\sqrt{3}t'^2}\im g^0_2(t') \right) +d_{b^+_0}, \nonumber \\
b^-_{0}&=\frac{1}{\pi}\!\!\int_{m_+^2}^{\infty}ds'\left( -\frac{2 \left(m_+^4 \left(m_-^2-2 s'\right)+m_+^2 \left(-6 m_-^2 s'+2 m_-^4+3 s'^2\right)+m_-^2
   s'^2+m_+^6\right)}{\left(m_--m_+\right) \left(m_-+m_+\right) \left(s'-m_-^2\right)^2
   \left(s'-m_+^2\right){}^2} \im f^-_0(s') \right. \nonumber \\
  &\left. -\frac{6 \left(m_-^2 s' \left(a+s'\right)+m_+^4 \left(s'-a\right)+m_+^2 \left(3 s' \left(a+s'\right)-m_-^2 \left(a+5
   s'\right)\right)-2 a m_-^4\right)}{\left(m_--m_+\right) \left(m_-+m_+\right) \left(a-s'\right) \left(s'-m_-^2\right)^2
   \left(s'-m_+^2\right)}\im f^-_1(s') \right. \nonumber \\
&\left. -\frac{10}{\left(m_--m_+\right) \left(m_-+m_+\right) \left(a-s'\right)^2 \left(s'-m_-^2\right)^2
   \left(s'-m_+^2\right)^2}\im f^-_2(s')\times \right. \nonumber \\
 &\left. \left(m_+^6 \left(a^2-2 a s'-5 s'^2\right)+m_+^4 \left(m_-^2 \left(a^2+16 a s'+13
   s'^2\right)-2 s' \left(4 a^2+7 a s'+s'^2\right)\right) \right. \right. \nonumber \\
&\left.~\left. +m_+^2 \left(-6 m_-^2 s' \left(-a^2+2 a
   s'+s'^2\right)+3 s'^2 \left(a^2+4 a s'+s'^2\right)+2 m_-^4 \left(a-s'\right)
   \left(a+2 s'\right)\right)\right. \right. \nonumber \\
&\left.~\left.   +m_-^2 s' \left(s' \left(a^2+4 a s'+s'^2\right)-6 a m_-^2
   \left(a+s'\right)\right)\right)\right) \nonumber \\ 
   & +\frac{1}{\pi}\int_{4 m_\pi^2}^{\infty}dt' \frac{3(t'((m_\pi+m_K)^2-m_\pi m_K)- 4(m_\pi m_K)^2)}{2\sqrt{2}m_\pi m_Kt'^2} \im g^1_1(t') +d_{b^-_0}, \nonumber \\
b^-_{0,sub}&=\frac{1}{\pi}\!\!\int_{m_+^2}^{\infty}ds'\left( -\frac{2 \left(m_--m_+\right) \left(m_-+m_+\right) \left(m_+^2+s'\right)}{\left(s'-m_-^2\right){}^2
   \left(s'-m_+^2\right){}^2} \im f^-_0(s')+\frac{6 \left(m_--m_+\right) \left(m_-+m_+\right)}{\left(s'-m_-^2\right){}^2 \left(s'-m_+^2\right)}\im f^-_1(s')\right. \nonumber \\
&\left. -\frac{10 \left(m_+^2-m_-^2\right) \left(a \left(5 s'-m_+^2\right)+s' \left(m_+^2+s'\right)\right)}{\left(a-s'\right)
   \left(s'-m_-^2\right){}^2 \left(s'-m_+^2\right){}^2}\im f^-_2(s') \right) \nonumber \\
   & -\frac{1}{\pi}\int_{4 m_\pi^2}^{\infty}dt' \frac{3\sqrt{2}m_\pi m_K}{t'^2} \im g^1_1(t')+d_{b^-_{0,sub}},
\end{align}
where the last one is obtained from an HDR with one subtraction for $F^-$.
Recall that we have set $a=-10m_\pi^2$.
As before, the $d_i$ terms are due to higher partial waves, which we computed in our calculations, although they are numerically very small.

Now, our CFD satisfies the dispersive representation and we have already seen that it yields consistent results for the scalar scattering-length sum rules, whereas the UFD does not necessarily do so. Thus, we  provide in Table~\ref{tab:leparameters} the evaluation of all previous sum rules, as well as the scattering lengths for the $D$-waves and the slopes for the $P$ and $D$-waves--- for which we have not provided analytic expressions--- using as input our CFD parameterizations obtained in section \ref{sec:CFD}. 
The first five columns are the results of this work. 
The first three correspond to  sum-rule results from FTPWDR and from HPWDR with one or no subtractions for $F^-$. The fourth column contains what we consider our ``Final Value'' for each parameter, which we have obtained assuming that each one of the three sum-rule values, $x_i\pm\delta x_i,\, i=1...3$, for a given observable $x$ is an independent quantity. This is, of course, an approximation,
although we have already shown how different the size of the contributions from different input are for each kind of dispersion relation. We have then used as weights their inverse squared uncertainties $\omega_i=1/(\delta x_i)^2$ to define:
\begin{equation}
\bar x\pm \delta \bar x\equiv\Big(\sum_{i=1}^{3} \omega_i x_i\Big)/\Big(\sum_{i=1}^{3} \omega_i \Big)\pm\Big(\sum_{i=1}^{3} \omega_i \Big)^{-1/2},\quad  \Delta\bar x\equiv \delta\bar x+d_Sx,
\label{eq:combineerrors}
\end{equation}
where $\bar x\pm \delta \bar x$ are the weighted average and the statistical uncertainty (see the RPP Introduction section \cite{pdg}).
However, as we have repeatedly emphasized a substantial part of our uncertainties has a systematic nature. Thus, to stay on the conservative side, we have also added linearly  a systematic uncertainty $d_S x$, defined as half the difference between the maximum and minimum central values of the individual sum rules for that observable. Therefore what we are showing in the ``Final Value'' column in Table~\ref{tab:leparameters} is $\bar x\pm\Delta \bar x$.

In the table we also list for comparison the ``direct'' calculations from the CFD parameterization, as well as, in the column before the last, the results of the sum rules obtained from ``fixed-$t$'' dispersion relations by the Bonn-Paris group \cite{Buettiker:2003pp}. Finally, in the last column, we provide the
NNLO SU(3) ChPT values from \cite{Bijnens:2004bu}, except for the  scalar threshold parameters for which we use the ``main numbers'' in  the review \cite{,Bijnens:2014lea}, in which the values of the scalar scattering lengths from  \cite{Buettiker:2003pp} were included in the fit and therefore they are very consistent with the previous column.

\begin{center}
\begin{table}[!hb]
\centering
\footnotesize
\begin{tabular}{l|c|c|c|c|c|c|c}
\hline
&\multicolumn{4}{c|} {This work sum rules with CFD input}  & \multicolumn{1}{c|}{This work} & Sum rules \cite{Buettiker:2003pp}&  NNLO ChPT  \\
 & FTPWDR & HPWDR & HPWDR$_{sub}$ & \textbf{Final Value} & direct CFD &  Fixed-$t$ &  \cite{Bijnens:2004bu} and 
\cite{Bijnens:2014lea}$^*$\\
\hline
\hline
\rule[-0.175cm]{-0.1cm}{.5cm} $m_\pi^3 b_0^{1/2}$ $\times$ 10 & 1.05$\pm$ 0.04 & 1.05$\pm$0.07 & 1.15$\pm$ 0.04 & \textbf{1.09$\pm$0.07}& 0.95$\pm$0.04& 0.85$\pm$0.04 & 1.278    \\
\hline
\rule[-0.175cm]{-0.1cm}{.5cm} $m_\pi^3 b_0^{3/2}$ $\times$ 10& $-$0.43$\pm$0.02 & $-$0.41$\pm$0.03  &$-$0.45$\pm$0.02 & \textbf{$-$0.43$\pm$0.03} & $-$0.36$\pm$0.04 & $-$0.37$\pm$0.03 & $-$0.326    \\
\hline
\rule[-0.175cm]{-0.1cm}{.5cm} $m_\pi^3 a_1^{1/2}$ $\times$ 10 & 0.228$\pm$0.010  & 0.218$\pm$0.008  & 0.222$\pm$0.006 & \textbf{0.222$\pm$0.009} & 0.20$\pm$0.04 & 0.19$\pm$0.01  & 0.152       \\
\hline
\rule[-0.175cm]{-0.1cm}{.5cm} $m_\pi^5 b_1^{1/2}$ $\times$ 10$^{2}$ & 0.58$\pm$0.03 & 0.59$\pm$0.03  & 0.60$\pm$0.03 & \textbf{0.59$\pm$0.02}&   0.5$\pm$0.2& 0.18$\pm$0.02   & 0.032      \\
\hline
\rule[-0.175cm]{-0.1cm}{.5cm} $m_\pi^3 a_1^{3/2}$ $\times$ 10$^{2}$ & 0.15$\pm$0.05 & 0.19$\pm$0.05  & 0.17$\pm$0.04 & \textbf{0.17$\pm$0.05} &  0.15$\pm$0.11& 0.065$\pm$0.044   & 0.293         \\
\hline
\rule[-0.175cm]{-0.1cm}{.5cm} $m_\pi^5 b_1^{3/2}$ $\times$ 10$^{3}$ & $-$0.94$\pm$0.09 & $-$0.97$\pm$0.08 & $-$1.03$\pm$0.07 &  \textbf{$-$0.99$\pm$0.09}&  $-$1.04$\pm$0.8& $-$0.92$\pm$0.17  & 0.544         \\
\hline
\rule[-0.175cm]{-0.1cm}{.5cm} $m_\pi^5 a_2^{1/2}$ $\times$ 10$^{3}$ & 0.60$\pm$0.13  & 0.54$\pm$0.03   & 0.55$\pm$0.02  & \textbf{0.55$\pm$0.05} & 0.53$\pm$0.05 & 0.47$\pm$0.03   & 0.142        \\
\hline
\rule[-0.175cm]{-0.1cm}{.5cm} $m_\pi^7 b_2^{1/2}$ $\times$ 10$^{4}$ & $-$0.89$\pm$0.10  & $-$0.96$\pm$0.09   & $-$0.95$\pm$0.09  & \textbf{$-$0.94$\pm$0.09} & 0.20$\pm$0.02 & $-$1.4$\pm$0.3   & $-$1.98       \\
\hline
\rule[-0.175cm]{-0.1cm}{.5cm} $m_\pi^5 a_2^{3/2}$ $\times$ 10$^{4}$ & $-$0.05$\pm$0.60  & $-$0.12$\pm$0.16  & $-$0.19$\pm$0.15 & \textbf{$-$0.15$\pm$0.18} & $-$0.09$\pm$0.03 & $-$0.11$\pm$0.27   & $-$0.45         \\
\hline
\rule[-0.175cm]{-0.1cm}{.5cm} $m_\pi^7 b_2^{3/2}$ $\times$ 10$^{4}$ & $-$1.12$\pm$0.10  & $-$1.14$\pm$0.09   & $-$1.14$\pm$0.09 & \textbf{$-$1.13$\pm$0.06} & $-$0.03$\pm$0.01& $-$0.96$\pm$0.26   & 0.61   \\
\hline
\end{tabular}
\caption{Determination of the $\pi K$ threshold parameters using our CFD  as input for the sum rules listed throughout this section calculated from the partial-wave dispersion relations  FTPWDR, HPWDR, or HPWDR$_{sub}$. The latter refers to the HPWDR sum rules obtained when one subtraction is used in the $F^-$ HDR. We also provide for comparison the direct values of the CFD parameterizations as well as previous results using sum rules from Roy-Steiner equations obtained from fixed-$t$ dispersion relations by the Bonn-Paris Group\cite{Buettiker:2003pp}. The last column lists NNLO ChPT results 
from \cite{Bijnens:2004bu} except for the scalar scattering lengths for which we used the update in the main fit of \cite{Bijnens:2014lea}. Recall that we are only imposing the $S$ and $P$ Roy-Steiner equations, so that our CFD $D$ waves are not devised to describe the threshold region precisely, but just the region where they are not completely negligible far from threshold. The same happens for the $I=3/2$, $\ell=1$ wave. We consider our sum rule results to be more reliable than the direct CFD values, particularly for the three last waves, for which no scattering data exist below 1 GeV.
Our ``Final Value'' for each threshold parameter, obtained by combining the three sum-rule results as explained in the text, is listed in boldface in the central column.}
\label{tab:leparameters}
\end{table}
\end{center}

Note that we have imposed Roy-Steiner equations only for our CFD $S$ and $P$-waves. 
These are the ones for which we expect the best agreement, first  between the different values of the sum rules and also with the direct CFD result. Actually, in the previous subsection we have already seen that the agreement is remarkable for the scalar scattering lengths.
For the scalar slopes, there is a few-percent deviation which translates into a two-sigma level average deviation.  We consider this quite acceptable because one has to keep in mind that there are no precise data or simply no data at all for these waves until roughly 100 MeV above threshold. Note that, as shown in Table \ref{tab:lepcontri}, the largest contribution 
for each of the scalar sum rules is a different one. Therefore, the fact that all our sum rules are fairly compatible among themselves is highly non-trivial. In addition, we are quite consistent with the dispersive results of \cite{Buettiker:2003pp}, except for the slope $b_0^{1/2}$, which is lower than ours, but not too far from the direct CFD value, although the NNLO ChPT estimate seems to prefer our sum-rule values.

Concerning the vector threshold parameters for the $I=1/2$ wave, we find a remarkable agreement between our  direct CFD results and all our sum rules, partly due to the larger uncertainty of the CFD.  Both the scattering length and slope sum-rule results of \cite{Buettiker:2003pp}, are significantly lower than all our sum rules, particularly the slope. This is most likely because their description of the $K^*(892)$ resonance yields a larger mass than observed in $\pi K$ scattering. Nonetheless, they are  fairly compatible with our direct CFD value,  once again due to the large uncertainty of the latter.

When discussing the 
$I=3/2$, $\ell=1$  wave, we have to keep in mind that it is so small that it is usually neglected in the literature and it has no data below 1 GeV, very far from threshold (see Fig.~\ref{fig:p32wavephase}). For this reason, we did not expect our direct CFD result to be very reliable. However, the fact that the sum rules for this wave are dominated by other waves allows us to obtain very accurate values for its threshold parameters.
Remarkably, all three sum rules are perfectly consistent among themselves at the one-sigma level, even when their uncertainties are much smaller than that of the CFD direct result, which is nevertheless very consistent with them.
The sum-rule results of \cite{Buettiker:2003pp} are perfectly compatible with our direct CFD calculation and consistent at roughly the 2-sigma level with our sum rules.

\begin{table}[!htb]
\centering 
\footnotesize
\begin{tabular}{l|c|c|c|c|c|c|c|c|c|c|c|r}
\hline 
\rule[-0.2cm]{-0.1cm}{.55cm}&\multicolumn{12}{c} {Largest contribution to the sum rules}\\
\rule[-0.2cm]{-0.1cm}{.55cm} & $a_0^{1/2}$ & $b_0^{1/2}$ & $a_0^{3/2}$ & $b_0^{3/2}$ & $a_1^{1/2}$ & $b_1^{1/2}$ & $a_1^{3/2}$ & $b_1^{3/2}$ & $a_2^{1/2}$ & $b_2^{1/2}$ & $a_2^{3/2}$ & $b_2^{3/2}$\\
\hline
\hline
\rule[-0.2cm]{-0.1cm}{.55cm} FTPWDR & $f^{1/2}_0(s)$ & $f^{1/2}_0(s)$ & $a^+_0$ & $f^{1/2}_1(s)$ & $f^{1/2}_1(s)$ & $f^{1/2}_1(s)$ & $g^{0}_0(s)$ & $g^{0}_0(s)$ &  Regge$_{\pi K}$ & $g^{0}_0(s)$ & Regge$_{\pi K}$ & $g^{0}_0(s)$  \\
\hline
\rule[-0.2cm]{-0.1cm}{.55cm} HPWDR & $g^{1}_1(s)$ & $g^{1}_1(s)$ & $g^{1}_1(s)$ & $g^{1}_1(s)$ & $g^{1}_1(s)$ & $f^{1/2}_1(s)$ & $f^{1/2}_1(s)$ & $g^{0}_0(s)$ &  $g^{1}_1(s)$ & $g^{0}_0(s)$ & $g^{0}_0(s)$ & $g^{0}_0(s)$  \\
\hline
\rule[-0.2cm]{-0.1cm}{.55cm} HPWDR$_{sub}$ & $a^{1/2}_0$ & $a^{-}_0$ & $a^{3/2}_0$ & $a^{-}_0$ & $f^{1/2}_1(s)$ & $f^{1/2}_1(s)$ & $a^-_0$ & $g^{0}_0(s)$ &  $g^{1}_1(s)$ & $g^{0}_0(s)$ & $g^{0}_0(s)$ & $g^{0}_0(s)$  \\
\hline
\end{tabular}
\caption{Largest single contribution to each one of the sum rules listed in table \ref{tab:leparameters}. Notice that for higher angular momentum the FTPWDR are dominated by asymptotic physics. Thus we have included a systematic uncertainty associated with the different Regge models as described in the text. Recall that for the scalar scattering lengths $a_0^I$ HPWDR$_{sub}$ is the same as the direct CFD calculation as it includes subtractions for both isospins.}
\label{tab:lepcontri}
\end{table}

 The $D$-waves are the first for which we do not impose their own Roy-Steiner equations. They are modified from UFD to CFD indirectly because they enter as input in the Roy-Steiner equations of the $S$ and $P$-waves, and also in the forward dispersion relations.
 Remember we have chosen existing Breit-Wigner-like  parameterizations to describe the data around the resonance and that there are no data below roughly 1 GeV, where they are almost negligible. Hence, the direct parameterizations carry little, if any, information about threshold dynamics. Thus, in principle, our CFD results are not expected to be very reliable there. The situation may be even worse for the $I=3/2$ case because it is almost negligible everywhere (see Fig.~\ref{fig:D32phase}). 
As a matter of fact, we were only interested in the bulk of their contribution to dispersion relations above 1 GeV. Surprisingly, the  scattering lengths from sum rules are perfectly consistent with each other for both the $I=1/2$ and $I=3/2$ partial waves, and even with their CFD parameterization as well as with the results in \cite{Buettiker:2003wj}.
  The slopes of the $D$-wave CFD are meaningless, but our sum rules are fairly consistent among themselves and with \cite{Buettiker:2003wj}.
  Note that we confirm the negative sign for $b_2^{3/2}$, at odds with the positive NNLO ChPT estimate. Of course, the direct CFD slope parameters for the $D$-waves are less reliable than those obtained from sum rules, as there is a lack of data, we chose very simple parameterizations and we have not imposed partial-wave dispersion relations on these waves.
  
As a technical remark, let us note that, as shown in Table \ref{tab:lepcontri},
while the asymptotic region for $\ell=2$ is almost negligible for sum rules obtained from HPWDR, it dominates the FTPWDR sum rules for $D$-wave scattering lengths, and also plays a significant role for the slopes. Unfortunately, 
these contributions present a sizable, although not dominant, dependence on second or even third-order derivatives on the $t$-variable of the Regge model. 
As explained in section \ref{sec:ufdregge} we use two different asymptotic models: On the one hand, the one for the $\pi K$ channel \cite{Pelaez:2003ky}, which comes from the $\pi \pi$ Regge model through factorization, and, on the other hand, the Veneziano model \cite{Veneziano:1968yb} for $\pi \pi \to K\bar{K}$. As shown in section \ref{sec:ufdregge} these two models produce fairly consistent descriptions of the close-to-forward region on the $s$ variable. As a result, it is irrelevant whether we use one or the other inside our $S$ and $P$-waves dispersion relations. But, unfortunately,  their second and third derivatives with respect to $t$ 
are incompatible and this is precisely a relevant contribution to the $D$-wave sum rules ( Fig.~\ref{fig:reggecomp} illustrates that this incompatibility already occurs for the first and second derivatives). Thus for our $D$-wave sum rules,  we provide in Table \ref{tab:leparameters} the average of the two models
with a combined uncertainty coming from the statistical error and 
a systematic error estimated as the difference between the two.

All in all, we therefore consider that our CFD and its uncertainty estimates for the $S$ and $P$ waves give a very consistent description of the threshold parameters. The $D$ waves, which are needed as input and were devised to reproduce data above 1 GeV and particularly their dominant resonance region, still provide a decent description at threshold of the scattering lengths, but not the slopes, although we can obtain reliable values for them from our sum rules. 

To summarize, our most reliable and accurate ``Final Values'' for threshold parameters are those coming from sum-rule determinations
and are listed in tables \ref{tab:isospinscl} and \ref{tab:leparameters}. This is one of the main novelties in this report and one of the main applications of our constrained parameterizations.

\subsubsection{Adler zeros and subthreshold parameters}
\label{subsec:subth}

There are certain subthreshold quantities of interest.
First of all, Adler zeros, which are zeros of scalar partial waves that appear due to the spontaneous chiral symmetry breaking pattern of QCD. 
In particular, Goldstone bosons should couple derivatively among themselves and therefore their scattering amplitude should have a zero at $s=0$. However, pions and kaons are not pure Goldstone bosons and have a small mass (not so small for kaons). At low energies and sufficiently far and below threshold the amplitude basically behaves as a polynomial so we still expect a zero not exactly at $s=0$ but displaced by a magnitude of $O(M_P^2)$
where $M_P$ is the mass of the pseudoscalar meson. These zeroes are a dynamical feature of chiral symmetry, which are not seen in other waves where there are always zeros right at threshold due to the $q^{2\ell}$ behavior.
These zeros were first found by S.L. Adler \cite{Adler:1964um} using current algebra, which is equivalent to LO ChPT. We have already used their LO ChPT values inside our low-energy $S^{1/2}$ and $S^{3/2}$ parameterizations in Eqs.~\eqref{eq:cot12} and \eqref{eq:cot32S}. These are $\sqrt{s_{A, LO}}=\sqrt{\Sigma_{\pi K}}\simeq 0.516\,$ GeV for $I=3/2$ and $\simeq0.486\,$GeV for isospin 1/2, as obtained from Eq.~\eqref{eq:sAdler120}.
Their position in the complex $s$ plane has been represented in Fig.~\ref{fig:anstrucpw}.

Of course, their LO value does not have to be the true position, we just use it as a reasonable value to define our parameterization. But we have calculated their actual values by looking for zeros of
partial-wave dispersion relations in the real axis below threshold, using either the UFD or the CFD parameterizations as input. The results are listed in Table~\ref{tab:adlerzeros}. There we see that the three kinds of dispersion relations yield very consistent values, which makes our result very robust. The UFD results are closer to the LO ChPT result. 
However, the CFD values are somewhat different from those of UFD and roughly 3-sigmas away from the LO calculation, which could be clear evidence that NLO ChPT corrections are needed at this level of precision.

\begin{table}[ht] 
\caption{ Adler zero positions $\sqrt{s_A}$ (GeV), for the $I=1/2$ and $I=3/2$ $S$-waves  from dispersion relations using as input either the UFD or CFD parameterizations. Recall that the LO ChPT result is $\simeq$0.486 GeV for I=1/2 and $\simeq$0.516 GeV for I=3/2. }
\vspace{0.3cm}
\centering 
\begin{tabular}{l c c c c} 
\hline
 & \hspace{0.2cm} UFD $I=1/2$ & \hspace{0.2cm} \bf{CFD $I=1/2$} & UFD $I=3/2$ & \hspace{0.2cm} \bf{CFD $I=3/2$}\\
\hline\hline  
\rule[-0.2cm]{-0.1cm}{.55cm} $\sqrt{s_{A_{FTPWDR}}}$ & $0.477^{+0.0010}_{-0.007}$ & $0.466^{+0.006}_{-0.005}$ & $0.530^{+0.013}_{-0.016}$ & $0.549^{+0.008}_{-0.0010}$\\
\rule[-0.2cm]{-0.1cm}{.55cm} $\sqrt{s_{A_{HPWDR}}}$ & $0.473^{+0.011}_{-0.009}$ & $0.466^{+0.007}_{-0.005}$ & $0.537^{+0.016}_{-0.019}$ & $0.551^{+0.009}_{-0.0010}$\\
\rule[-0.2cm]{-0.1cm}{.55cm} $\sqrt{s_{A_{HPWDR_{sub}}}}$ & $0.481^{+0.008}_{-0.008}$ & $0.470^{+0.010}_{-0.005}$ & $0.532^{+0.013}_{-0.016}$ & $0.552^{+0.008}_{-0.010}$\\
\hline
\end{tabular} 
\label{tab:adlerzeros} 
\end{table}

As a technical remark, one may wonder if we could change our value of the Adler zero in our parameterizations to this CFD result. However, we should recall that our parameterizations are only chosen to describe data in the physical region. We have indeed adopted a particular analytic form to be able to imitate some basic features (like Adler zeros, or some resonant given shape), but our parametrizations are not model-independent, just a reasonable fit to data to be used as input for the dispersion relations. We do not need them to be accurate outside the physical region. That is what dispersion integrals are for.
In addition, we chose our low-energy partial wave parameterizations as truncated conformal expansions avoiding the circular-cut and the Adler zero is very close to that cut, or even inside as the $I=1/2$ dispersive CFD case.
It makes no sense to try to be accurate with the truncated conformal expansion at the border or outside its applicability limit.
Thus, the robust results are the dispersive ones and, as it happened with the $\kap$ pole, the UFD and CFD parameterizations can only be expected to  provide a fair but  model-dependent  approximation of features in the complex plane outside the physical real axis.

Other interesting quantities are the coefficients of the so-called subthreshold expansion, obtained around the $t\rightarrow 0$ and $\nu\rightarrow 0$ limit. In the $\pi K$ scattering case this means $s=u=\Sigma_{\pi K}$. The ChPT calculations are expected to converge better in this region because there is no threshold singularity. Fortunately, our set of dispersion relations can be continued below the physical region, so that we can make use of these sets to determine the amplitudes and their derivatives with high accuracy. We adopt the standard definition:

\begin{equation}
    F^+(s,t)=\sum_{i\,j} C^+_{i j}\left(\frac{t}{m_{\pi^+}^2}\right)^i\left(\frac{\nu}{4 m_{\pi^+} m_K}\right)^{2j}, 
    \quad
    F^-(s,t)=\frac{\nu}{4 m_{\pi^+} m_K} \sum_{i\,j} C^-_{i j}\left(\frac{t}{m_{\pi^+}^2}\right)^i\left(\frac{\nu}{4 m_{\pi^+} m_K}\right)^{2j},
\end{equation}
where the parameters $C_{i j}$ are dimensionless and $\nu=s-u$. We have obtained their algebraic expressions by expanding
the Kernels in \ref{app:Kernels} in terms of $t,\nu$.
These expansions are then introduced in~\cref{eq:sftpwdr,eq:shdrfm,eq:shdrfmi}  to obtain the corresponding sum rules. However, these are lengthy expressions and we will just list the numerical results in Table~\ref{tab:stparameters}. Note we provide values 
using our CFD parameterization as input for our three kinds of dispersion relations (without projecting on partial waves) obtained either from Fixed-$t$ or HDR (either with no or one subtraction for $F^-$). We also provide the symmetric amplitude evaluated at the Cheng-Dashen point $\nu=0, t=2m_\pi^2$, which is of interest for the next subsection. Previous calculations obtained within dispersive and ChPT approaches are also listed for comparison.

\begin{table}[!ht]
\begin{center}
\small
\begin{tabular}{l|l|l|l|l|l|l}
\hline
&\multicolumn{3}{c|} {This work sum rules with CFD input}   & Sum rules &  NNLO ChPT & Sum rules \\
 & Fixed-$t$ & HDR &  HDR$_{sub}$ & B\"uttiker et al. \cite{Buettiker:2003pp}& Bijnens et al. \cite{Bijnens:2004bu} & Lang et al. \cite{Lang:1979zr}  \\
\hline
\hline
\rule[-0.175cm]{-0.1cm}{.5cm} $C^+_{0 0}$ & 1.52$\pm$0.56 & like fixed-$t$ &  & 2.01$\pm1.10$ & 0.278 & $-$0.52$\pm$2.03 \\
\hline
\rule[-0.175cm]{-0.1cm}{.5cm} $C^+_{1 0}$  & 0.96$\pm$0.11 & 1.04$\pm$0.11 &  & 0.87$\pm0.08$ & 0.898 & 0.55$\pm$0.07   \\
\hline
\rule[-0.175cm]{-0.1cm}{.5cm} $C^+_{0 1}$ & 2.34$\pm$0.05 & like fixed-$t$ &  & 2.07$\pm0.10$ & 3.8 & 2.06$\pm$0.22 \\
\hline
\rule[-0.175cm]{-0.1cm}{.5cm} $C^+_{1 1}$ & $-$0.047$\pm$0.006 & $-$0.050$\pm$0.006 &  & $-$0.066$\pm0.010$ & $-$0.10 & $-$0.04$\pm$0.02 \\
\hline
\rule[-0.175cm]{-0.1cm}{.5cm} $C^-_{0 0}$ & 9.11$\pm$0.35 & 9.54$\pm$0.38 & 9.04$\pm$0.39 & 8.92$\pm0.38$ & 8.99 & 7.31$\pm$0.90      \\
\hline
\rule[-0.175cm]{-0.1cm}{.5cm} $C^-_{1 0}$ & 0.45$\pm$0.05 & 0.38$\pm$0.02 & 0.39$\pm$0.02 & 0.31$\pm0.01$ & 0.088 & 0.21$\pm$0.04    \\
\hline
\rule[-0.175cm]{-0.1cm}{.5cm} $C^-_{0 1}$  & 0.68$\pm$0.02 & 0.66$\pm$0.02 & 0.68$\pm$0.02  & 0.62$\pm0.06$ & 0.71 & 0.51$\pm$0.10    \\
\hline   
\rule[-0.175cm]{-0.1cm}{.5cm} $F^+_{CD}$  & 3.55$\pm$0.64 & 3.71$\pm$0.64 &  & 3.90$\pm1.50$ & 2.11 &     \\
\hline    
\end{tabular}
\caption{Determinations of the coefficients of the $\pi K$ subthreshold expansion from various approaches. Although we use 3 different dispersive families, for $C^+_{00}$ and $C^+_{01}$ 
the HDR is exactly the same as the fixed-$t$ one. Note that the ``+'' cases do not have two different HDR values because their HDR is always once-subtracted}
\label{tab:stparameters}
\end{center}
\end{table}

Note that our  sum-rule determinations are very consistent among themselves, all of them within one standard deviation of the others. Our results, although equal in sign and relatively similar in size to those obtained in \cite{Buettiker:2003pp} are sometimes significantly different in terms of standard deviations. Note also that our results
have in most cases somewhat smaller uncertainties.

\subsection{$\pi K$ $\sigma$-term}
\label{sec:sigterm}

Another interesting application of our results is in the calculation 
of the so-called $\pi K$ $\sigma$-term.
This requires an extrapolation of the $\pik$
amplitude to an unphysical point, which can be achieved in a robust and model-independent way thanks to our dispersive formalism.

The relevance of $\sigma$-terms  is due to their
relation to $\langle H\vert m_q q \bar q \vert H\rangle$, which tests the QCD mass terms inside a hadron $H$. The $\sigma$-term is nothing but the scalar form factor of the $H$ hadron evaluated at zero momentum transfer, which intuitively yields a contribution to the hadron mass and also sets the normalization of the scalar form factor itself. Unfortunately, no single scalar hadronic probe is accessible experimentally at low-energies, and therefore it has to be obtained indirectly through the scattering of two hadrons, one of which is the hadron of interest, which appears also in the final state. The simplest possibility is, therefore, the scattering process $\Phi(q) H(p)\rightarrow \Phi(q') H(p')$, where $\Phi$ is a pseudo-Nambu-Goldstone Boson i.e. $\pi, K ,\eta$, most frequently, a pion.  Note that in the $t$-channel of this process it is possible to exchange two Goldtone bosons to form a scalar current that would therefore couple to the $H \bar H$ system. The use of Goldstone bosons leads to a low-energy theorem \cite{Cheng:1970mx,Brown:1971pn}, to be introduced below, that relates the scattering amplitude to the scalar form factor.

The archetypal example is  the $\pi N$ $\sigma$-term, because it probes the role of quark masses in the total mass of the nucleon (see \cite{Alarcon:2021dlz} for an up-to-date pedagogical and historical introduction).  This is actually the  system for which the low-energy theorem was first introduced in \cite{Cheng:1970mx}, although it was later corrected in \cite{Brown:1971pn}. There are many  phenomenological determinations~\cite{Koch:1982pu,Gasser:1988jt,Gasser:1990ce,Gasser:1990ap,Fettes:2000xg,Pavan:2001wz} and effective-field-theory-based calculations~\cite{Fettes:2000xg,Alarcon:2011zs,Alarcon:2012nr,Ditsche:2012fv,Hoferichter:2015dsa,Hoferichter:2016ocj,RuizdeElvira:2017stg,Fernando:2018jrz}. Generically, modern phenomenological data-driven determinations cluster around a value of $\simeq 60$ MeV. However, there is some tension with modern lattice-QCD calculations~\cite{Durr:2015dna,Yang:2015uis,Abdel-Rehim:2016won,Bali:2016lvx,Borsanyi:2020bpd}, which favor a value closer to $\simeq 40$ MeV \footnote{While this manuscript was under referee's revision, a new Lattice-QCD determination was published~\cite{Gupta:2021ahb}, which value is compatible with the dispersive determinations.}.
For our purposes, the most interesting study is the dispersive analysis of $\pi N$ scattering in \cite{Ditsche:2012fv,Hoferichter:2015dsa} since the authors used as their framework a system of hyperbolic dispersion relations similar to the ones described in this work, hence allowing for robust extractions of the
$\pi K$  $\sigma$-term for different low energy inputs~\cite{Hoferichter:2015dsa,Hoferichter:2016ocj,RuizdeElvira:2017stg}. 

Since our dispersive formalism is very similar, it is  straightforward to apply it to calculate the $\pi K$ $\sigma$-term~\cite{Gasser:2000wv,Frink:2002ht}, setting $H=K$.  Let us first define the scalar form-factor of the kaon, $\Gamma_K(t)$, as:
\begin{equation}\Gamma_{K}(t)=\left\langle K^{0}\left(p^{\prime}\right)|\hat{m}(\bar{u} u+\bar{d} d)| K^{0}(p)\right\rangle, \quad \hat{m}=\frac{1}{2}\left(m_{u}+m_{d}\right), \quad t=(p'-p)^2,
\end{equation}
Then, the $\pi K$ $\sigma$-term is defined as:
\begin{equation}
    \sigma_{\pi K}\equiv \frac{\Gamma_K(0)}{2m_\pi}.
\end{equation}
Now, as commented above, the scalar form factor is not directly measurable but, thanks to chiral symmetry, is related to the $\pi K$ symmetric amplitude through the following low-energy theorem, in which a ``soft-pion'' is probing the kaon \cite{Brown:1971pn}:
 \begin{equation}
 F_\pi^{2} F^+(t, \nu)=\Gamma_{K}(t)+q^{\prime \mu} q^{\nu} r_{\mu \nu},
 \end{equation}
where $F_\pi$ is the pion decay constant and $q^{\prime \mu} q^{\nu} r_{\mu \nu}$ is the so-called remainder, which is not determined by chiral symmetry, although it must have the same analytic structure as the scattering amplitude. This remainder is suppressed by evaluating the expression at the  Cheng-Dashen point \cite{Cheng:1970mx}, $t=2m_\pi^2$, $\nu=(s-u)=0$, where:
\begin{equation}
F_\pi^{2} F_{CD}^{+}=\Gamma_{K}\left(2 m_{\pi}^{2}\right)+\Delta^{CD}_{\pi K}.
\end{equation}
For brevity $F_{CD}^{+}, \Delta^{CD}_{\pi K}$ stand, respectively, for the scattering amplitude and remainder evaluated at the Cheng-Dashen point.
The final step to obtain $\sigma_{\pi K}$ is to evaluate the difference 
\begin{equation}
\Delta_{\sigma}=\Gamma_{K}\left(2 m_{\pi}^{2}\right)-\Gamma_{K}(0).
\end{equation}
All in all, we can recast $\sigma_{\pi K}$ as follows:
\begin{equation}
2 m_{\pi} \sigma_{\pi K}=F_\pi^{2} F_{CD}^{+}-\Delta_{\sigma}-\Delta^{CD}_{\pi K}.
\end{equation}

Of course, the very $\sigma_{\pi K}$ and all these quantities can be evaluated using ChPT \cite{Frink:2002ht,Bijnens:2004bu}. However, in the spirit of this report, we aim at a data-driven determination. Remember that we expect the remainder to be small at the Cheng-Dashen point. We also expect  $\Delta_\sigma$ to be relatively small since it is the difference between not too distant values of $t$ (and the scalar form factor only has the right cut starting at twice that distance). These expectations have been explicitly checked within NLO ChPT \cite{Frink:2002ht}. Therefore the dominant term should be  the one containing $F_{CD}^{+}$,
which is the part that we have indeed calculated from data using our CFD as input for  the sum rule.

Actually, in the previous section, we presented our  evaluation of two subthreshold sum-rules for $F^+_{CD}$: one obtained from fixed-$t$ dispersion relations and another one from HDR (note that for the symmetric case we always have one-subtraction). The numerical results are displayed in the final row of Table~\ref{tab:stparameters} and they come out perfectly compatible with one another and with similar uncertainties. In this case, we do not combine the two results as we did for the threshold parameters in \cref{eq:combineerrors}, because now the two values are strongly correlated by their subtraction constant.
Hence, we have just taken their average, and for the uncertainty we have added half of their difference to their statistical error, which is the same for both. On the whole:
\begin{equation}
F^+_{CD}=3.6\pm0.7,
\end{equation}
which is the main result of this section. In particular, 
this number is compatible with the  estimate obtained by the Bonn-Paris group \cite{Buettiker:2003pp}, although our result is slightly lower and our uncertainties are twice smaller. In contrast, it is in a $\sim 2\sigma$  tension with the  value  estimated from NNLO ChPT  in \cite{Bijnens:2004bu} (no errors), which is already about 50\% larger than the range of values from 1.2 to 1.4 estimated to NLO in \cite{Frink:2002ht}. 
Note that with our $F^+_{CD}$ value, the relative sizes of the remainder and $\Delta_\sigma$ terms become even smaller.

As a future prospect, in order to extract completely the $\sigma_{\pi K}$-term
from data in a robust and precise way, one would need~\cite{Gasser:2000wv}:
\begin{itemize}
    \item A data-driven determination of $F_{CD}^{+}$, like the one we have just obtained here from a dispersive sum rule using our CFD as input.
    \item A determination of the remainder $\Delta^{CD}_{\pi K}$. Remember this has been made small on purpose, by choosing the Cheng-Dashen point. The best calculation is that of \cite{Frink:2002ht}, based on NLO ChPT,
    that yields: $$
   \Delta^{CD}_{\pi K}= [0.013... 0.021]\,m_\pi^2.$$
   
    Thus, the remainder estimate\footnotetext{We thank B. Kubis for explaining to us that, although in \cite{Frink:2002ht} other ranges were obtained when using $F^2=F_K^2$ or $F_\pi F_K$, which might be legitimate from the point of view of the chiral expansion, being a soft {\it pion} SU(2) theorem the ``correct'' choice is $F^2=F_\pi^2$.} is about two orders of magnitude less than our value $F_\pi^2F^+_{CD}=(1.6\pm0.64) m_\pi^2$. This is a consistency check that the use of the Cheng-Dashen point truly suppresses the remainder.
     In that work, it has also been shown that, even though corrections can appear on each separated term, they cancel in the difference. This should keep on happening even for higher-order corrections, so we think that the theoretical estimation of the smallness of this difference might be fairly reliable, even when we have seen that one of the terms in the difference is much larger than its NLO value.

    \item Calculate the $\Delta_{\sigma}$ difference dispersively. Remember this only concerns the scalar form factor evaluated at two different momentum transfers below threshold, where no further analytic structures exist and the dependence should be smooth.  We think it could also be obtained from data by using the coupled-channel MO formalism for the $\pipikk$ scalar form factor  developed in  \cite{Hoferichter:2012wf}. Maybe it could even be possible to update some of the input used there with our CFD. 
\end{itemize}

Unfortunately, we have not been able to find any lattice calculations of the kaon sigma term. There are calculations of the kaon mass dependence on quark masses (see for instance the review~\cite{MILC:2009mpl} and references therein) and in principle an extraction of this sigma term seems feasible. However, such a calculation is far beyond our scope and capabilities, since one has to deal with the usual lattice technicalities of finite-volume effects, continuum extrapolation, physical mass extrapolation, different actions, scale setting, etc... (see \cite{Fodor:2012gf} and references therein for a review on light hadron masses from lattice). There are indirect extractions of the sigma term, or more precisely the scalar form factor of the kaon from lattice data, but using ChPT or a resummed version of it (see \cite{Daub:2012mu,Bernard:2012ci,Daub:2015xja}. These result in an estimate of $\Gamma_K(0)$ in the range 0.4 to 0.6. We nevertheless hope that our result, and the persistent tension with ChPT calculations, or at least the apparent need for large ChPT corrections, can raise interest in the lattice community to calculate this number.

In summary, using our CFD as input for two sum rules for the $F^+$ symmetric amplitude, we have been able to provide a robust and precise determination, from a dispersive analysis of data, of the dominant contribution to the kaon $\sigma$-term. As it happened with other quantities of interest, our value is consistent  with previous dispersive results, although our smaller uncertainties reveal, once more, some tension with the perturbative calculations within $SU(3)$ ChPT. 

\subsection{$\pi \pi \to K \bar{K}$ and  $(g-2)_\mu$}
\label{sec:g-2}

Another topic of interest, where dispersion relations provide a solid framework to work with, is the hadronic contribution to the $(g-2)_\mu$. Explaining in detail the different approaches towards the $(g-2)_\mu$ determination is out of our scope and we refer the reader to the recent review on Ref.~\cite{Aoyama:2020ynm}. Nevertheless, we
comment very briefly on the motivation for its study. The anomalous magnetic moment of the muon can be determined both theoretically, either data-driven or from lattice QCD, and experimentally (see for instance \cite{Bennett:2002jb,Bennett:2004pv,Bennett:2006fi}). Of all the Standard Model contributions to $(g-2)_\mu$ both the QED and electroweak contributions have been obtained from perturbation theory to a high degree of accuracy. Nowadays the largest  factor to the final uncertainty comes from  hadronic contributions, which have to be calculated in a non-perturbative way. These are classified according to their diagram topology into the so-called Hadron Vacuum Polarization ($HVP$) and Hadronic Light by Light ($HLbL$) terms. In the past these contributions were determined using various models, hence producing large systematic uncertainties.  
However, in the last few years, considerable improvements have been made to this topic, led by dispersive formalisms applied to existing data analyses including mesonic final states. A theoretical study group, the \emph{Muon $g-2$ Theory Initiative}~\cite{Aoyama:2020ynm} was recently formed, to lead the analyses as well as weighting and averaging the theoretical evaluations regarding the various $(g-2)_\mu$ contributions.
Their recent final result \cite{Aoyama:2020ynm} reads
\begin{equation}
a_{\mu}^{\mathrm{SM}}=116591810(43) \times 10^{-11}.
\end{equation}

The experimental determination made almost 20 years ago by the E821 collaboration  at Brookhaven~\cite{Bennett:2002jb,Bennett:2004pv} had at the time of the publication smaller uncertainties than both the $HVP$ and $HLbL$  theoretical predictions. At present,
two major experiments are underway to refine the experimental uncertainties with more data and higher statistics. The first one is the Fermilab Muon $g-2$ collaboration (FNAL) \cite{Grange:2015fou}, the second one is the proposed g-2/EDM experiment J-PARC initiative, which will perform a new experimental design \cite{Saito:2012zz}. Both of them expect to reduce the $(g-2)_\mu$ error by a factor of 4.
During the referee revision of this manuscript, the FNAL collaboration released its first result \cite{Muong-2:2021ojo}, which still has not achieved the precision goal. It is in excellent agreement with the previous  E821 measurement and, when both are combined, yield the present experimental average
\begin{equation}
   a_\mu^{exp} =116592061(41)\times 10^{-11}\quad (0.35 {\rm ppm}).
\end{equation}

This theoretical determination lies below the experimental value by $4.2\,\sigma$. Nonetheless, a recent lattice QCD determination of the leading-order hadronic vacuum polarization \cite{Borsanyi:2020mff}, with similar precision to dispersive studies, yields a larger value than those previous theoretical determinations. This result, together with the rest of the dominant contributions would eliminate the discrepancy with data.

Further studies will be needed to make the dispersive, lattice, and experimental numbers agree or to confirm the discrepancy. 

\subsubsection{Hadronic-vacuum polarization}

As explained above, both the $HVP$ and $HLbL$ contribute to similar order to the muonic $(g-2)_\mu$ term. Thus a precise and robust calculation of these quantities is paramount to extract the most rigorous value and compare properly between the various approaches involved in the determination of this observable.

The hadronic vacuum polarization can be best extracted from $e^+e^-\to {\rm hadrons}$.  The biggest contribution, by far, comes from the $\pi \pi$ channel, due to its low threshold. Recent dispersive analyses have been applied to this $e^+e^-\to \pi^+ \pi^-$ channel \cite{Colangelo:2018mtw}, which includes modern high statistic data samples. This work is a dispersive implementation based on the well-known Roy equations \cite{Roy:1971tc}, and is an update over a previous solution on $\pi \pi$ scattering \cite{Caprini:2011ky}, with a slight modification of the $P$-wave. After taking into account the various subtleties required to analyze the new experimental data, the most robust determination of the two-pion contribution to the Hadronic-vacuum polarization reads
\begin{equation}
\left.a_{\mu}^{\pi \pi}\right|_{\leq 1 \mathrm{GeV}}=495.0(1.5)(2.1) \times 10^{-10}.
\end{equation}

\begin{figure}[!ht]
\begin{center}
\centerline{\includegraphics[width=0.37\textwidth]{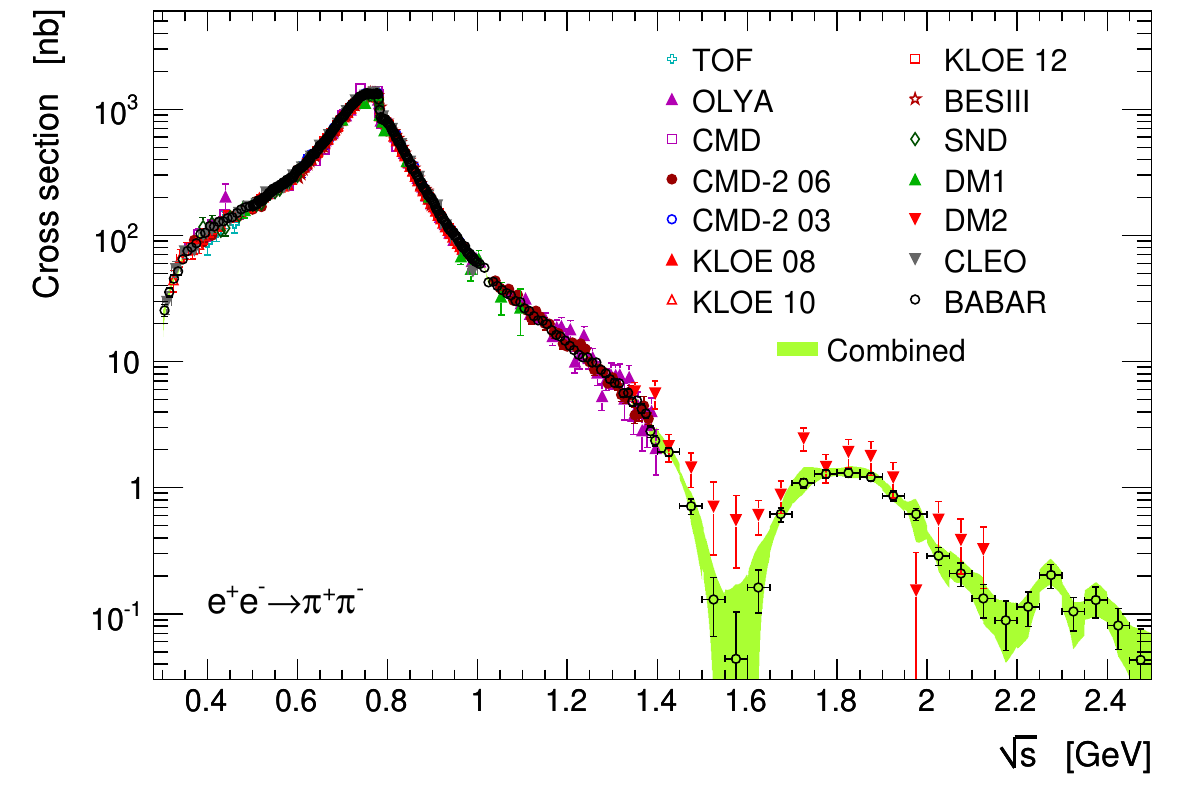}
\includegraphics[width=0.62\textwidth]{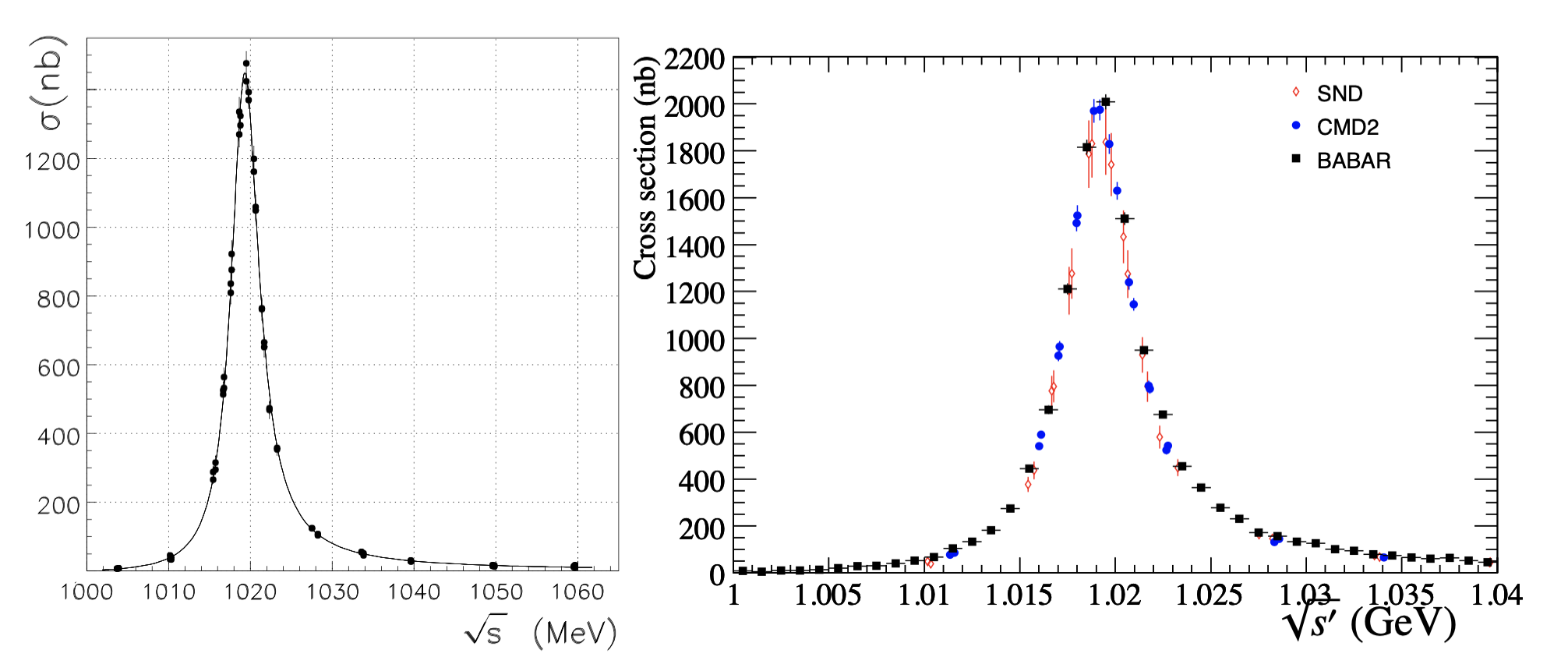} }
 \caption{\rm \label{fig:eetohadrons} 
Figures taken from \cite{Davier:2019can} (left),~\cite{Achasov:2000am} (center) and~\cite{Lees:2013gzt} (right). Cross sections of the reactions $e^+e^-\to \pi^+ \pi^-$ (left panel),  $e^+e^-\to K_S K_L$ (center panel) and $e^+e^-\to K^+ K^-$ (right panel). Notice how the main contribution to the $HVP$, which is the region around the $\rho-\omega$ resonances mixing, is purely elastic.}
\end{center}
\end{figure}

One may wonder if including the $K\bar K$ inelasticity could potentially modify this result, however as seen in Table VI of Ref.~\cite{Aoyama:2020ynm} the contribution saturates very close to the $\rho-\omega$ peaks, which are shown in Fig.~\ref{fig:eetohadrons} (left panel). If one takes into account that the $K\bar K$ threshold of a $P$-wave does not produce any noticeable cusp effect then the result quoted above should not vary appreciably. However the uncertainties are dominated by statistics, and as seen above they are pretty low, so including a larger region and the inelasticity could produce in principle a non-negligible effect.

The second and third biggest contributions to the $HVP$ of the muon are $\pi^+\pi^-\pi^0$ and $K^+K^-/K_S K_L$ (center and right panels of Fig.~\ref{fig:eetohadrons}) respectively. These are already one order of magnitude smaller both in the magnitude and uncertainty than the $\pi \pi$ contribution. Of the two, the latter is dominated by the $\phi(1020)$ resonance decay into two kaons, which cannot be determined from our dispersive approach. The next low energy contribution would be the $4\pi$ channel, which is of the same order as the $3\pi$ and $KK$ channels. Finally, only a few higher-energy channels add relatively small contributions compared to the dominant $\pi\pi$ final state.

\subsubsection{$HLbL$}

One of the largest sources of uncertainty to the calculation of the $(g-2)_\mu$ is the $HLbL$ scattering. Unfortunately, this contribution cannot be calculated perturbatively, and thus must be determined from data analyses or lattice QCD, as it is done for the $HVP$.

However, the $HLbL$ contribution is much more challenging in structure, as it is created by four-point functions. As can be seen in Fig.~1 of~\cite{Nyffeler:2017ohp}, at low energies the mesonic contribution is dominated by several resonances, in particular the $f_0(980)$ and $a_0(980)$, as well as by charged pion and kaon loops and the pion/kaon box. For a recent and  detailed review, we refer the reader again to Ref.~\cite{Aoyama:2020ynm}.

Modern approaches consist of implementing dispersive formulations to relate the off-shell contributions to the $(g-2)_\mu$ with on-shell, measurable quantities. Then, each of these contributions has a physical unambiguous definition. These projects aim to determine with high accuracy separated contributions to the $(g-2)_\mu$, and elucidate the experimental uncertainties required to improve the knowledge on the anomalous magnetic moment.

One of the first dispersive formalisms is explained in detail in Ref.~\cite{Pauk:2014rfa}, where they define a dispersion relation for the Pauli form factor. Albeit successful in reproducing the pion-pole contribution of a $VMD$ model to the $(g-2)_\mu$, no explicit formulas have been derived so far for generic contributions.

The other main approach~\cite{Colangelo:2014dfa,Colangelo:2015ama} consists of describing in a dispersive way the polarization tensor for off-shell photon-photon scattering, for which the analytic structures of all the components must be first derived. The contribution to the anomalous magnetic moment of the muon can thus be determined from projections over this four-point function. This would be the equivalent of the dispersive description of the $HVP$. However, the fact that we are now dealing with four-point functions makes the task highly complicated.

Several works have been recently published making use of this formalism to determine different pion contributions to the $(g-2)_\mu$ like the pion pole~\cite{Hoferichter:2018kwz,Hoferichter:2018dmo}, two pion contributions~\cite{Colangelo:2014pva,Colangelo:2017fiz} or the pion box~\cite{Colangelo:2017qdm}. These are considered to be part of the dominant $HLbL$ contributions to the $(g-2)_\mu$.

\begin{figure}[!ht]
\begin{center}
\centerline{\includegraphics[width=0.48\textwidth]{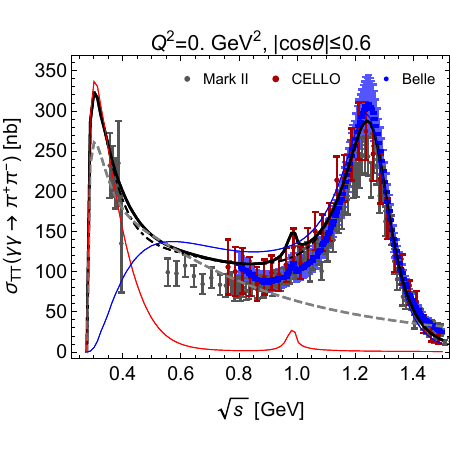}\includegraphics[width=0.48\textwidth]{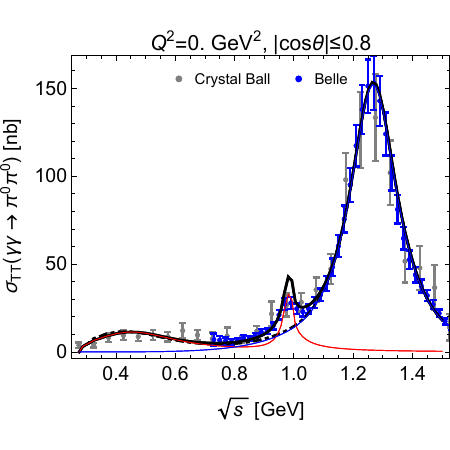}}
 \caption{\rm \label{fig:ggtopipi} 
The figure is taken from \cite{Danilkin:2018qfn}. Cross-sections of the reactions $\gamma\gamma\to \pi^+ \pi^-$ (left panel) and $\gamma\gamma\to \pi^0 \pi^0$ (right panel). Notice the enhancement at around 1 GeV as a result of including the $K\bar K$ channel, which gives rise to the $f_0(980)$ resonance contribution, depicted as the continuous red line in both figures.}
\end{center}
\end{figure}

Extensions to the kaon box seem to be straightforward~\cite{Aoyama:2020ynm}. However, as seen from Dyson-Schwinger equations, the kaon-box contribution is very suppressed compared to the leading pion one~\cite{Eichmann:2019bqf}, once again due to the larger kaon mass. Nevertheless, as explained in section 4.6.1. of Ref.~\cite{Aoyama:2020ynm} two-kaon contributions could produce larger effects. This channel not only produces some contribution to the rescattering effects inside the pion/kaon loops but also gives rise to a better constraint over the $f_0(980),a_0(980)$ and $f_2(1270),f'_2(1525)$ resonances contributing to the $(g-2)_\mu$.

Finally, several recent dispersive analyses have been published studying the one or doubly virtual photon scattering into hadrons $\gamma^{(*)}\gamma^*\to\pi\pi\,(K\bar K)$~\cite{Danilkin:2017lyn,Danilkin:2018qfn,Danilkin:2019opj,Deineka:2019bey,Hoferichter:2019nlq}, most of them including both $\pi \pi$ and $K\bar{K}$ final states, and their contributions to the anomalous magnetic moment of the muon. In some of these works, a previous dispersive determination carried out by our group~\cite{Pelaez:2018qny} was already implemented, as shown in Fig.~\ref{fig:ggtopipi}, which entails the relevance of this reaction for the precise extraction of the $(g-2)_\mu$.  For more details on the topic, we refer the reader to Ref.~\cite{Danilkin:2019mhd}.
We hope that our final dispersive results presented in this report will help these groups to obtain a more reliable determination of the $HLbL$ contribution to the $(g-2)_\mu$ following their various approaches.

Note: While this manuscript was under referee's revision, an analysis of the $f_0(980)$ contribution to  ${HLbL}$ was completed \cite{Danilkin:2021icn}. It studies $\gamma\gamma\rightarrow\pi\pi(K\bar{K})$ with a Muskhelishvili-Omnès method, whose functions are taken from an $N/D$ analysis \cite{Danilkin:2020pak} that fits the  $\pi\pi\rightarrow\pi\pi$ Madrid-Krakow dispersive analysis \cite{GarciaMartin:2011cn} and the $\pi\pi\rightarrow K\bar{K}$ analysis presented here. The result, well below $10^{-11}$, is fairly similar to previous narrow widths estimates, although it illustrates how to go beyond this simple approximation and include rescattering effects using as an ingredient our dispersive results.

\section{Summary \& Conclusions}
\label{sec:Summary}

Let us come to an end by  summarizing the main items we have reviewed and our main new results. For the sake of clarity and brevity, we omit in this summary any citation since we have provided aplenty in the main text.

Studies of \pik and \pipikk scattering have been the subject of considerable interest since they were first measured around four decades ago. These interactions are not only relevant by themselves, but also because they appear in the final states of numerous hadronic processes. In addition, they are key to hadron spectroscopy, providing
one of the main sources of information on the existence and properties of strange resonances --- and therefore on the classification of mesons in symmetry multiplets as well as their inner structure. Moreover, in their low-energy regime, these interactions constitute a test-ground for Chiral Perturbation Theory (ChPT) as well as the most recent lattice QCD developments.

Unfortunately, for many decades, the study of \pik and \pipikk scattering data has been hindered by the conflict within and between different data sets and affected by large model dependencies.
Moreover, we have reviewed here how these data are also in severe conflict with several dispersion relations. However, the wealth of data on hadronic processes  collected over the last years, the unprecedented statistics obtained in present and planned experiments, together with the recent theoretical advances in lattice QCD and ChPT, call for a consistent, precise, model-independent, and easy to implement description of \pik and \pipikk amplitudes.

In this review, we have shown how this demand can be met by constraining the data description with dispersion relations. As a matter of fact, the two problems commented above can be overcome with the use of dispersion theory, which is a direct consequence of the strong analyticity constraints derived from causality and relativistic crossing symmetry.

Hence, the main result of this review is:
\begin{itemize}
    \item 
a dispersively constrained and precise, but still rather simple, simultaneous description of \pik and \pipikk scattering data and its uncertainties, which is consistent with an ample sample of dispersion relations.
This is what we have called the ``Constrained Fit to Data'' (CFD) parameterization set. 
\end{itemize}
We have then illustrated several applications of using this CFD as input to obtain other results of interest: 
\begin{itemize}
    \item a rigorous dispersive determination of the existence and parameters of the controversial \kap scalar meson, 
    \item model-independent determinations of other strange resonances below 2 GeV, reducing their model dependencies employing analytic techniques,
    \item precise and model-independent values of the threshold and subthreshold parameters, using sum rules derived from different dispersive representations.
\end{itemize}
In addition, we have reviewed other dispersive applications
that use or may profit from some of the previous results, namely, the dispersive study of the non-ordinary \kap Regge trajectory, the $\pi K$ sigma term, as well as the contribution of $\pipikk$ to some hadronic corrections  in the calculation of $(g-2)_\mu$.

Hence, after stating our motivation and goals in section \ref{sec:intro}, we have reviewed in section \ref{sec:Data} the existing data, explaining the 
conflicts both within some given data sets and between different experiments.

In section \ref{sec:DR} we have presented first a brief pedagogical introduction to dispersion relations, reviewing also seminal works and the state of the art for \pik and \pipikk dispersive approaches.
Next, we have derived the different kinds of dispersion relations of relevance for our purposes, although the detailed and lengthy expressions of their integral kernels are provided in \ref{app:Kernels}. The reason to use several types of relations is twofold. On the one hand, they have different applicability regions, both in the real axis and on the complex plane. On the other hand, even for the same observables, they may have different inputs or they may weigh them in non-equivalent ways. The discussion of the applicability ranges, which we have shown how to maximize, is probably the most technical one and has been relegated to the long and thorough \ref{app:Applicability}. In particular, we use forward dispersion relations (FDR) because, apart from their simplicity, they only involve $\pi K$ amplitudes, and constrain them up to 1.6 GeV. However, in order to study $\pi K$ partial waves individually, we need to project them out of either fixed-$t$ (FTPWDR) or hyperbolic dispersion relations (HPWDR).  Both of them are limited to $\sim$1 GeV for real values of the energy. The FTPWDRs depend little on \pipikk, but cannot reach the \kap pole in the complex plane, whereas the HPWDRs reach this pole region, but have sizable contributions from this crossed channel. At this point, it is therefore relevant to consider also  FTPWDR and HPWDR for the crossed channel \pipikk, which, as a technical difficulty, need input from the pseudo-physical regime below the $K \bar K$ threshold, where no data exist and the amplitude has to be treated with the so-called Muskhelishvili-Omn\`{e}s method, explained in section \ref{sec:MO}.

Thus, after the various dispersive representations have been introduced, we have presented in section \ref{sec:UFD} a set of Unconstrained Fits to Data (UFD), paying particular attention to the evaluation of systematic uncertainties. For this we have used very simple parameterizations, but flexible enough to accommodate later the dispersive constraints. Our aim with these simple choices is that they should be easy to implement in future studies, either of phenomenological or experimental nature. 
However, at the end of section \ref{sec:UFD}, we have reviewed how,  this UFD, which is a very nice looking fit to data, was shown in some of our previous works to be inconsistent with FDR for \pik and HPWDR for \pipikk. Moreover, we have shown here that the UFD also fails to satisfy the \pik FTPWDR 
and reach the largest inconsistencies for the \pik HPWDR.
It is therefore evident that simple fits to data, including fits using particular models, are not adequate to produce a precise and reliable description of \pik and \pipikk.

Consequently, in section \ref{sec:CFD}, we have presented our main result: the Constrained Fit to Data (CFD). We have explicitly shown that it satisfies our collection of dispersion relations while still describing fairly well the data. Of course, some deviations from the best possible fit to experiment occur, but these are in general relatively mild and are needed to ensure the fulfillment of the dispersive representations. It is also worth noticing that this is the first time that both the \pik and \pipikk channels are constrained simultaneously, which completes our previous dispersive analyses.

Once the CFD is available, we use it in section \ref{sec:kappa} to study the strange resonances that appear in \pik scattering. First, it is used as input to build sequences of Pad\'e approximants out of successive derivatives of the amplitude. 
This is relevant because dispersion relations are formulated in the first Riemann sheet, whereas the poles associated with resonances lie on the contiguous one. Remarkably, these Pad\'e series are shown to converge to the analytic continuation of the amplitude to the next continuous Riemann sheet.
The advantage of the method is that the existence of poles and their parameters can thus be determined by avoiding model-dependent assumptions. The only caveats are the
sources of uncertainty, which, apart from those of the CFD, come from the numerical calculation of higher derivatives and the truncation of the series. Nevertheless, the results are very robust and competitive with model-based determinations.

In the elastic region it is even possible to get rid of those additional caveats and uncertainties since there is a direct relation between the first and second sheets and resonances can be 
studied with a fully model-independent and precise dispersive formalism.  
Actually, in section \ref{sec:kappa} we have reviewed our recent work where, using our dispersive analysis of data,  we confirmed the existence of the \kap that the Review of Particle Physics is asking for. In addition, we supply a high precision determination of its parameters, which come in fair agreement with a previous dispersive determination using the FTPWDR prediction inside HPWDR by the Paris group.
We concluded this section with a dispersive determination of the \kap Regge trajectory, which strongly supports its non-ordinary nature.

We started section \ref{sec:applications}  using the CFD as input for a precise and model-independent determination
of $\pi K$ threshold and subthreshold parameters.
We provide the expressions of these sum-rules, derived from dispersive integrals. 
Threshold and subthreshold parameters are of relevance to test Chiral Perturbation Theory or to determine its low-energy constants. In this context, the most significant result is a new dispersive and precise determination of the scalar \pik scattering lengths. Our values are consistent with the dispersive solution of the Paris group, although our result is obtained from dispersively constrained data fits rather than a solution to dispersion relations. Therefore, the tension between dispersive values and existing lattice results lingers on.

Finally, also in section \ref{sec:applications} we address other possible applications of the dispersive constraints, which require further input besides our scattering parameterization. This is the case of the  $\sigma$-term determination, as well as the \pipikk  possible contribution  to the $(g-2)_\mu$.

In conclusion, dispersion relations are a very powerful tool for hadron physics. In this report we have not only reviewed but also provided new results concerning their application to \pik and \pipikk scattering data up to $\sim 1.6$ and $\sim1.5$ GeV, respectively. Our main new outcome is a consistent and precise constrained description of these data in terms of simple parameterizations easy to implement for further studies. We have shown here how the use of this approach allows us to determine the existence and/or properties of strange resonances, to provide precise determinations of low-energy observables, as well as several other applications. We hope these results can be of use for further studies both for the phenomenology and experimental communities working on hadron physics.

 \section*{Acknowledgements}
 
The authors would like to thank J. Ruiz de Elvira for many useful discussions. The authors would also like to thank B. Kubis, L. Leskovec, and A. Pilloni for useful comments and conversations. We also thank I. Danilkin for several corrections. This project has received funding from the Spanish Ministerio de Ciencia e Innovación grant PID2019-106080GB-C21 and the European Union’s Horizon 2020 research and innovation program under grant agreement No 824093 (STRONG2020). AR acknowledges the financial support of the U.S. Department of Energy contract DE-SC0018416 at the College of William \& Mary, and contract DE-AC05-06OR23177, under which Jefferson Science Associates, LLC, manages and operates Jefferson Lab.
 
\appendix



\section{Conformal expansion for elastic waves}
\label{app:conformal}

Let us briefly describe here the kind of parameterizations that we have been using for our fits in the elastic region, commenting on the features that make them particularly well suited for that case. Similar conformal expansions have been used for a long time in hadron physics \cite{Ciulli:1969fw,Ciulli:1969gr,Cutkosky:1970pd,Nogova:1973fq,Nogova:1973hk,Ciulli:1977ub,Caprini:1977uc} and have been  recently revived for the particular cases of $\pi\pi$ scattering in \cite{Pelaez:2004vs,Yndurain:2007qm,Caprini:2008fc}
or $\pi K$ scattering 
in \cite{Cherry:2000ut,Rendon:2020rtw}. The specific parameterization described here was introduced by us in \cite{Pelaez:2016tgi}.

\begin{figure}
\begin{center}
\includegraphics[width=0.8\textwidth]{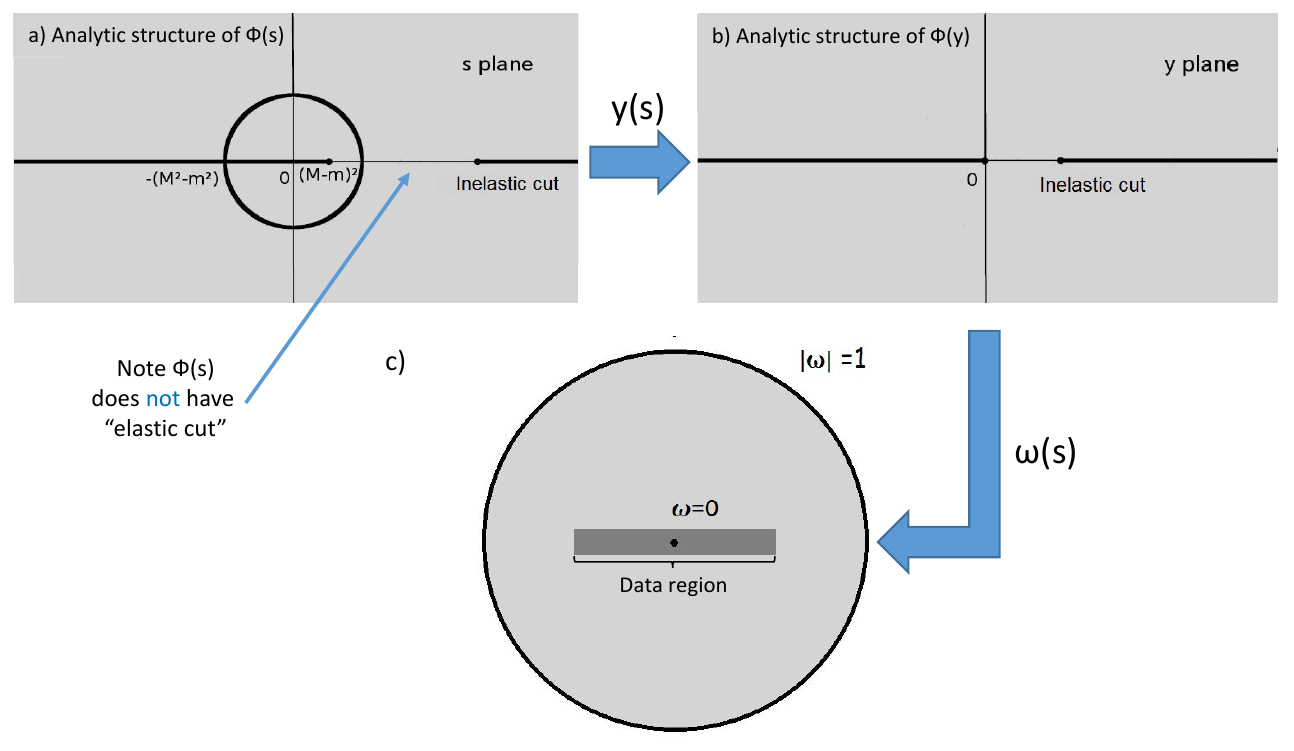}
 \caption{\rm \label{fig:cuts} 
Analytic structure in different variables
of the $\pi K$ effective range function $\Phi_\ell(s)$:
a) $\Phi_\ell(s)$ in the $s$-plane has the same structure as $f(s)$ (see top panel of Fig.~\ref{fig:anstrucpw}), except for the absence of the elastic cut.
b) In the $y(s)$-plane the circular-cut disappears. c) The  conformal variable 
$\omega(y)$ maps the whole analyticity domain of $\Phi_\ell(y)$ inside the unit circle, 
whereas the cut singularities are confined to $\vert \omega \vert=1$.
Note that $\omega$ will be defined so that the data region is roughly centered around $\omega=0$ and not too close to the border.}
\end{center}
\end{figure}

As explained in subsection~\ref{sec:UFDpik}, thanks to the elastic unitarity condition in Eq.~\eqref{eq:elasticpw}, elastic partial-wave amplitudes in the complex $s$-plane can be  recast  as
\begin{equation}
f_\ell(s)=\frac{q^{2\ell}}{\Phi_\ell(s)-iq^{2\ell}\sigma(s)}.
\end{equation}
Note we have introduced the {\it effective range function} $\Phi_\ell(s)$ which {\it in the elastic region in the real axis}, satisfies:
\begin{equation}
    \Phi_\ell(s)=\frac{2q^{2\ell+1}}{\sqrt{s}}\cot\delta_\ell(s).
\end{equation}
Consequently, $\Phi_\ell(s)$ is real in the real axis between $\pi K$ threshold and the first inelastic threshold.
Hence, $\Phi_\ell(s)$ does not have an elastic cut, but still has left-hand, circular, and inelastic cuts, as shown in Fig.~\ref{fig:cuts}. 
Since $\Phi_\ell(s)$ has no singularities from $\pi K$ threshold to the next inelastic threshold $s_0$, we can expand $\Phi_\ell(s)$ in powers of momentum $q$. This is the so-called ``Effective range expansion'', whose radius of convergence is small because the  circular cut 
 lies rather close to $q=0$, i.e., $s=m_+^2$.

 There is however a better way to use the largest possible domain of analyticity, which is 
to perform an expansion in powers of a conformal variable that maps the whole complex plane into the unit circle. 
For $\pi K$ scattering, due to the circular-cut,
it is convenient to introduce first another change of variable
\begin{equation}
y(s)=\left(\frac{s-\Delta_{K\pi}}{s+\Delta_{K\pi}}\right)^2,
\end{equation}
 which maps the circular cut into the left real axis.
As a result, $\Phi_\ell(y(s))$ only has a right-hand ``inelastic'' cut and a left-hand cut, as shown in Fig.~\ref{fig:cuts}.b. It is now that we define the conformal variable
\begin{equation}
w(y)=\frac{\sqrt{y}-\alpha \sqrt{y_0-y}}{\sqrt{y}+\alpha \sqrt{y_0-y}}, \quad y_0=y(s_0),
\end{equation}
to map the cut $y$-plane into the unit circle in the $\omega$-plane.
This leads to the structure seen in Fig.~\ref{fig:cuts}. The $s_0$ constant, with units of energy squared, will set the maximum energy where this conformal series is real and applicable in the real axis. In principle, we could use the {\it truncated} expression beyond that point, but then, it does no longer have the desired analytic properties, and actually $\cot \delta_\ell$ may become a complex number. In practice, we could use this fact to accommodate some inelasticity contributions, in a purely phenomenological way.

Except for the tiny $P^{3/2}$ and $D^{3/2}$-waves, the parameter $\alpha$ is chosen
so that the center of the conformal 
expansion $\omega=0$ corresponds 
to the intermediate point between the $\pi K$ 
threshold and the energy of the last data point to be fitted
with the conformal formula.
In this way, we ensure
that the fitted data
lies well inside the $\omega=1$ circle, far from the border and
roughly centered around $\omega=0$, as shown in Fig.~\ref{fig:cuts}. 
Actually, for the $S^{1/2}$ and $P^{1/2}$ waves, the data fitted with
the elastic formalism lie at
$\vert  \omega\vert<0.45$. In contrast, for the $S^{3/2}$-wave the data lie at $\vert  \omega\vert<0.6$.
The $P^{3/2}$ and $D^{3/2}$-waves are an exception, because their data starts at 1 GeV,
far from the $\pi K$ threshold. Thus we have chosen their $\alpha$ parameters so that the center of the conformal expansion corresponds to the intermediate
point where their data exist. With this choice, the data fitted with this conformal expansion lie at $\vert  \omega\vert<0.6 $.

It is important to realize that, after these two changes of variable,
the singularities now lie at the boundary $\vert \omega\vert =1$. Thus, the effective range function has an analytic expansion $\Phi_\ell(s)=\sum_n{B_n w(s)^n}$  
convergent in the whole $\vert \omega\vert <1$ circle. 
Therefore, in terms of $s$, the domain of analyticity of the conformal mapping extends to 
the whole complex plane outside the circular cut, 
with a left-hand cut from $(M-m)^2$ to $-\infty$ and a right-hand cut above  the first inelastic threshold. This is the power of conformal mappings. 

Finally, as explained in the main text, it is customary to abuse the notation and write:
\begin{equation}
    \cot\delta_\ell(s)=\frac{\sqrt{s}}{2q^{2\ell+1}}\Phi_\ell(s)
\end{equation}
as a function in the complex $s$ plane. With this definition, in the elastic region of the real axis, we can write
\begin{equation}
\cot\delta_\ell(s)=\frac{\sqrt{s}}{2q^{2\ell+1}}\Phi_\ell(s)=\frac{\sqrt{s}}{2q^{2\ell+1}}\sum_n{B_n w(s)^n}.
\end{equation}

This is still not the general form we presented in Eq.~\eqref{eq:generalconformal} in subsection~\ref{sec:UFDpik}, that we repeat here 
\begin{equation}
\cot\delta_\ell(s)=\frac{\sqrt{s}}{2q^{2\ell+1}}F(s)\sum_n{B_n \omega(s)^n}.
\end{equation}
So far we have seen the $F(s)=1$ case. But,
let us recall that, due to chiral symmetry, scalar partial waves have a so-called
Adler zero below threshold, which is easily implemented in the partial waves 
by writing a pole factor in front of the $\Phi_\ell(s)$ expansion, as follows:
\begin{equation}
\Phi_\ell(s)=\frac{1}{s-s_{Adler}}\sum_n{B_n w(s)^n}.
\end{equation}
In other words, choosing $F(s)=1/(s-s_{Adler})$ above.
In addition, when there is a narrow well-established resonance and the phase crosses $\pi/2$ at $m_r$
it is also convenient to extract a factor out of the conformal expansion as:
\begin{equation}
\Phi_\ell(s)=(s-m_r^2)\sum_n{B_n w(s)^n},
\end{equation}
to accelerate the convergence of the fit. This is nothing but choosing $F(s)=(s-m_r^2)$ above. 

These are the expressions, truncated to the minimum number of terms needed to get an acceptable $\chi^2/dof$, that we have used for our parameterizations of $\pi K$ scattering in the elastic region. However, note that the expression above also provides an analytic extension of $\Phi_\ell$ to the complex plane, which, by means of Eq.~\eqref{eq:fintermsofphi}, yields a fairly good approximation to the partial wave $f_\ell$, as long as one is not too far from the elastic region. Actually, we have seen that we find poles for the \kap and $K^*(892)$ resonances using our conformal parameterizations and, in particular, the CFD conformal parameterization yields reasonable values for both  poles and residues. But, of course, these values are model-dependent, since their extraction relies on a particular parameterization. To obtain model-independent results for resonances, one has to rely on dispersion relations, as we do in this report.

\section{Alternative $P$-wave and form factors}
\label{app:palt}

As explained in section~\ref{sec:Data}, there are $P$-wave measurements from other processes than scattering. In particular, Ref.~\cite{Ananthanarayan:2005us} details how  it could be possible to determine the elastic phase shift of the $P$-wave around the resonance region from the results of the  FOCUS  collaboration~\cite{Link:2002ev}. The main motivation is twofold. First, there exists some tension between the $K^*(892)$ parameters coming from production experiments and those coming from heavier decays as listed in the RPP~\cite{Zyla:2020zbs}, particularly for the width of the resonance. Second, should the width be smaller, it would affect our dispersive results, and it could also lead to different results for those works which use the $\pi K$ vector form factor in their analyses~\cite{Moussallam:2007qc,Boito:2008fq,Boito:2010me,Bernard:2011ae,Bernard:2013jxa,Escribano:2014joa,Gonzalez-Solis:2019owk,Rendon:2019awg}. On top of all the above, our $P$-wave UFD does not agree very well with the final CFD result, which is surprising considering how small the data uncertainties are. It seems that there might be a non-trivial systematic deviation within the original experimental data, which could help explain this tension~\cite{Bernard:2013jxa}.

For all these reasons we will implement an alternative fit for the $P$-wave elastic phase shifts, based on the ``pseudo-data'' produced by V. Bernard in~\cite{Bernard:2013jxa} (Fig. 2 therein) as a result of re-sampling the $P$-wave fit by the FOCUS collaboration~\cite{Link:2002ev}. The rest of the partial waves will remain untouched and are given in section~\ref{sec:UFD}. Once we
obtain an Unconstrained Fit to these new Data (UFD)  we then procure an alternative Constrained Fit to Data (CFD), following the approach detailed in section~\ref{sec:DRTests}. We list in Table~\ref{tab:Palt} the alternative UFD and CFD parameters for this $P$-wave.

\begin{table}[!ht] 
\caption{$P^{1/2}$-wave alternative solution parameters. The first four parameters correspond to the elastic parameterization \cref{eq:cot121}, whereas the rest of the parameters correspond to the inelastic formula \cref{eq:pinel}.}
\centering 
\begin{tabular}{c c c } 
\hline\hline  
\rule[-0.05cm]{0cm}{.35cm}Parameters & UFD & CFD \\ 
\hline
\rule[-0.05cm]{0cm}{.35cm}$B_0$ & 1.144 $\pm$0.048      & 1.137   $\pm$0.048\\
\rule[-0.05cm]{0cm}{.35cm}$B_1$ & 0.90   $\pm$0.60        & 0.45  $\pm$0.60\\ 
\rule[-0.05cm]{0cm}{.35cm}$B_2$ & 1.72   $\pm$1.01        & 0.78   $\pm$1.01\\ 
\rule[-0.05cm]{0cm}{.35cm}$m_r$ & 0.8955 $\pm$0.0018GeV & 0.8947 $\pm$0.0018GeV\\
\hline 
\rule[-0.05cm]{0cm}{.35cm}$a_1$ & $-$2.17             $\pm$0.18GeV$^{-2}$  & $-$1.66             $\pm$0.18GeV$^{-2}$  \\ 
\rule[-0.05cm]{0cm}{.35cm}$a_2$ & $-$2.10             $\pm$0.28GeV$^{-2}$   &  $-$1.74           $\pm$0.28GeV$^{-2}$  \\ 
\rule[-0.05cm]{0cm}{.35cm}$a_3$ & $-$1.33             $\pm$0.09GeV$^{-4}$   &  $-$1.36            $\pm$0.09GeV$^{-4}$\\ 
\rule[-0.05cm]{0cm}{.35cm}$\sqrt{s_{r1}}$ & 0.896  GeV (fixed)             &  0.896  GeV (fixed)    \\ 
\rule[-0.05cm]{0cm}{.35cm}$\sqrt{s_{r2}}$ & 1.343 $\pm$0.013GeV     &  1.344 $\pm$0.013GeV \\ 
\rule[-0.05cm]{0cm}{.35cm}$\sqrt{s_{r3}}$ & 1.647 $\pm$0.005GeV     &  1.655 $\pm$0.005GeV \\ 
\rule[-0.05cm]{0cm}{.35cm}$e_1$ & 1   (fixed)                                & 1 (fixed) \\ 
\rule[-0.05cm]{0cm}{.35cm}$e_2$ & 0.048           $\pm$0.008          & 0.067           $\pm$0.008 \\ 
\rule[-0.05cm]{0cm}{.35cm}$e_3$ & 0.307           $\pm$0.016          & 0.320          $\pm$0.016\\ 
\rule[-0.05cm]{0cm}{.35cm}$G_1$ & 0.029           $\pm$0.005GeV     & 0.046           $\pm$0.005GeV  \\ 
\rule[-0.05cm]{0cm}{.35cm}$G_2$ & 0.218           $\pm$0.042GeV     & 0.210        $\pm$0.042GeV \\ 
\rule[-0.05cm]{0cm}{.35cm}$G_3$ & 0.303           $\pm$0.018GeV     & 0.290        $\pm$0.018GeV  \\ 
\hline 
\end{tabular} 
\label{tab:Palt} 
\end{table} 

Including this ``pseudo-data'' into the fits is a bit problematic, as it has been produced from a smooth parameterization. This results in strong, unknown correlations between  different bins, both for the central values and uncertainties. As a consequence, no flawless fit to this partial wave can be attained. For simplicity, we have decided to use an ordinary $\chi^2$ function, for which we have produced many $\mathcal{O}(10^3)$ fits to different re-samplings over the ``pseudo-data''. Hence, instead of using the Minuit~\cite{James:1975dr} crude estimate of the uncertainties, we have calculated the variance of its parameters over the many re-samplings. On top of that, we have taken some of these re-samples as data samples so that we can define several $\chi^2$ over this partial wave, which are combined into an averaged $\chi^2$ afterward. This way we obtain a stable and normalized definition of a pseudo-$\chi^2$-like function for this partial wave and thus all partial waves and dispersion relations can now be imposed following our approach. An alternative description could be obtained using bootstrapping techniques, but this would demand a bigger set of fits to the Montecarlo re-sampling of the initial parameterization and we should deal with the values of each bin rather than the parameters themselves. This is incompatible with our simplistic approach, where we focus on one fit with simple parameter uncertainties, and thus it will not be used.

\begin{figure}[!ht]
\begin{center}
\resizebox{0.8\textwidth}{!}{\input{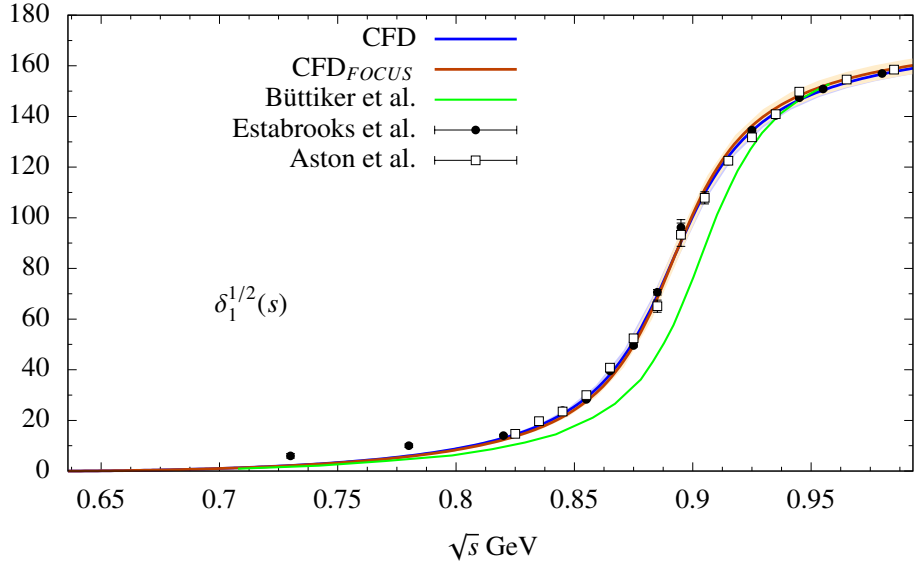}}
\caption{Comparison between the original CFD fit (blue) to the $P$-wave from production data~\cite{Estabrooks:1977xe,Aston:1987ir}, and the alternative CFD (orange) obtained by using the FOCUS collaboration~\cite{Link:2002ev} data.
\label{fig:Pfocuscfd} 
}
\end{center}
\end{figure}

First, notice that the alternative and original inelastic parameters of Tables~\ref{tab:Pinpa} and ~\ref{tab:Palt} are very similar. This result is not surprising as we are using the same data input on this inelastic region. The main difference now is that the alternative UFD describes a narrower $K^*(892)$ resonance, which contributes a bit less to the dispersive inputs.

Additionally, the alternative CFD result is quite compatible with the original fit as seen in Fig.~\ref{fig:Pfocuscfd}, with a small deviation associated with the different width of the $K^*(892)$. Moreover, both CFD results bend towards one another, producing a much similar pole position for the $K^*(892)$ than the UFD fits. On top of that, the result of the rest of the partial waves is barely any different from before.

Taking into account that we do not have access to measured data here, and the null improvement altogether we can not say we favor this alternative solution over the original scattering data. However, we have decided to include its parameters here, in case data from heavier decays were extracted in the future. Once additional data were measured for this partial wave, they could be compared to this alternative solution to see if the new information favors this alternative fit over the older data.

Finally, should one use this fit instead of the original one, the CFD dispersive resonance pole for the $K^*(892)$ would read
\begin{align}
\sqrt{s_p}&=(891\pm 2) -i (23.9\pm 3.0)\,\text{MeV}, \quad \vert g \vert = (5.49\pm0.19), \quad \phi_g=-(0.065\pm0.011), \hspace{1.cm} \text{CFD FOCUS} \nonumber \\
\sqrt{s_p}&=(890\pm 2) -i (25.6\pm 1.2)\,\text{MeV}, \quad  \vert g \vert = (5.69\pm0.12),\quad \phi_g=-(0.076\pm0.008), \hspace{1.cm} \text{CFD Original} 
\end{align}
where we have taken the average between the fixed-$t$ and the two HDR dispersive values. Notice that the mass is almost equal to the original CFD, although the width has moved by around 1.3 $\sigma$ from the original value.

Besides the little changes to the alternative $P$-wave, the rest of the partial waves barely change, if they change at all. The $\kap$ resonance extracted from this analysis is perfectly compatible with our original value, and the new $S$-wave scattering lengths obtained using this $P$-wave are almost identical to the CFD ones, as listed in Tab.~\ref{tab:scattalt}. Notice we have called this new solution CFD$_{\text{FOCUS}}$ for practical purposes.

\begin{table}[!ht] 
\caption{$S$-wave scattering lengths ($m_\pi$ units).}
\vspace{0.3cm}
\centering 
\begin{tabular}{l  c  c c} 
\hline
 & \hspace{0.2cm} \bf{CFD$_{\text{FOCUS}}$} & \hspace{0.2cm} \bf{CFD} & \hspace{0.2cm} Ref. \cite{Buettiker:2003pp}\\
\hline\hline  
\rule[-0.2cm]{-0.1cm}{.55cm} $a^{1/2}_0$ & \hspace{0.2cm} \bf{0.220$\pm$0.011} & \hspace{0.2cm} \bf{0.224$\pm$0.011} & \hspace{0.2cm} 0.224$\pm$0.022\\
\rule[-0.2cm]{-0.1cm}{.55cm} $a^{3/2}_0$ & \hspace{0.2cm} \bf{$-$0.049$\pm$0.007} & \hspace{0.2cm} \bf{$-$0.048$\pm$0.006} & \hspace{0.2cm} $-$0.0448$\pm$0.0077\\
\hline
\end{tabular} 
\label{tab:scattalt} 
\end{table}

In summary, let us recall once again that the idea of implementing an alternative $P$-wave comes from the fact that the production data~\cite{Estabrooks:1977xe,Aston:1987ir} could potentially suffer from a faulty extraction in the region around the $K^*(892)$ mass. Alternatively, in this section, we have provided a dispersive study of a fit compatible with that of the FOCUS collaboration~\cite{Link:2002ev}, which is consistent with new analyses performed in the recent past for form factors, as well as heavier decays and Dalitz plots extractions.

\section{Alternative $g^0_0$ wave}
\label{app:g00alt}

As explained in sections \ref{sec:UFDpipiKK}, \ref{sec:CFDpipiKK} and in \cite{Pelaez:2018qny}  the Brookhaven experiment \cite{Etkin:1982se} used only two resonances, $f_2(1270)$ and $f_2'(1525)$, to describe their data on the $g^0_2(t)$ partial wave. However, we have included a third resonance $f_2(1810)$ following several experiments that claimed its existence and because we have found that it helps obtain a better data fit. 
Of course, since their 
$g_2^0$ wave 
was used to extract their $g_0^0$ data, if one now assumes that the $f_2(1810)$ resonance exists, this should produce non-negligible deviations in the data on the scalar-isoscalar phase above roughly $1.6$ GeV. On top of that, the parameterization used by the Brookhaven collaboration violates Watson's theorem at lower energies. This is evident if one notices that they obtain a different value for their $g^0_2(t)$ phase right above the $K\bar K$ threshold from the $\pi \pi$ phase they should match right below (see Fig.~\ref{fig:g02data}). Hence, the first few data bins by \cite{Etkin:1982se}, shown in Fig.~\ref{fig:g00data}, should also be corrected.

We list in Table \ref{tab:g00phasealt} the parameters of the ``Alternative CFD'' $g^0_0(t)$ solution shown in Fig.~\ref{fig:g00alt}, obtained by extracting the $g^0_0(t)$ phase from~\cite{Etkin:1982se} using our own $g^0_2(t)$ wave, rather than the faulty one used by the experimental collaboration. One could wonder if this modification would produce a significant effect on the dispersive results of the $g^0_0(t)$, and thus propagate it to the rest of the partial waves through the dispersion relations. Fortunately, the affected region lies above $1.6$ GeV, which is already highly suppressed as input to the $g^0_0(t)$ dispersion relation itself, and its modulus is small there, yielding a negligible effect when changing the parameters between the fit to the original  data (Table \ref{tab:g00phase}) and the alternative one. Let us recall here that our dispersion relations can be applied only up to 1.47 GeV, and thus they cannot constrain the region where this new $g^0_0(t)$ deviates substantially from the original one. Moreover, no other dispersion relation is modified by this new solution. 

\begin{table}[!ht]
\caption{Parameters of the alternative $\phi^0_0$.} 
\centering 
\begin{tabular}{c c c c} 
\hline\hline  
\rule[-0.05cm]{0cm}{.35cm}Parameter & UFD & CFD$_\text{B}$ & CFD$_\text{C}$ \\ 
\hline 
\rule[-0.05cm]{0cm}{.35cm}$B_1$ & 23.5  $\pm$1.3  & 22.1 $\pm$1.3 & 23.5  $\pm$1.3\\ 
\rule[-0.05cm]{0cm}{.35cm}$B_2$ & 29.0  $\pm$1.3  & 27.6 $\pm$1.3 & 29.1  $\pm$1.3\\ 
\rule[-0.05cm]{0cm}{.35cm}$B_3$ & 0.01  $\pm$1.60  & 1.7 $\pm$1.6 &   0.6  $\pm$1.6\\ 
\rule[-0.05cm]{0cm}{.35cm}$C_1$ & 12.0890 fixed  & 10.0142 fixed & 9.0032fixed\\ 
\rule[-0.05cm]{0cm}{.35cm}$C_2$ & 13.6  $\pm$2.6  & 11.1 $\pm$2.6 & 10.9  $\pm$2.6\\ 
\rule[-0.05cm]{0cm}{.35cm}$C_3$ &$-$12.9 $\pm$2.3  &$-$16.0 $\pm$5.2 &$-$16.0 $\pm$5.2\\ 
\rule[-0.05cm]{0cm}{.35cm}$C_4$ &$-$13.1 $\pm$2.2  &$-$14.3 $\pm$2.2 &$-$13.9 $\pm$2.2\\ 
\rule[-0.05cm]{0cm}{.35cm}$C_5$ &  4.0 $\pm$2.4  &  4.5 $\pm$2.4 &  4.6 $\pm$2.4\\ 
\hline 
\end{tabular} 
\label{tab:g00phasealt} 
\end{table}

\begin{table}[!ht]
\begin{minipage}[t]{0.49\linewidth}
\caption{Parameters of the UFD$_\text{B}$ and CFD$_\text{B}$ fits to $\vert g^0_0\vert $.} 
\centering 
\begin{tabular}{c c c } 
\hline\hline  
\rule[-0.05cm]{0cm}{.35cm}Parameter & UFD$_\text{B}$ & CFD$_\text{B}$ \\ 
\hline 
\rule[-0.05cm]{0cm}{.35cm}$D_0$ & 0.588 $\pm$0.010  & 0.590 $\pm$0.010\\
\rule[-0.05cm]{0cm}{.35cm}$D_1$ &$-$0.380 $\pm$0.013  &$-$0.339 $\pm$0.013\\ 
\rule[-0.05cm]{0cm}{.35cm}$D_2$ & 0.12 $\pm$0.01  & 0.13 $\pm$0.01\\ 
\rule[-0.05cm]{0cm}{.35cm}$D_3$ &$-$0.09  $\pm$0.01  &$-$0.12  $\pm$0.01\\ 
\rule[-0.05cm]{0cm}{.35cm}$F_1$ &$-$0.04329  fixed  &$-$0.04195  fixed\\
\rule[-0.05cm]{0cm}{.35cm}$F_2$ &$-$0.008  $\pm$0.009  &$-$0.008  $\pm$0.009\\ 
\rule[-0.05cm]{0cm}{.35cm}$F_3$ &$-$0.028  $\pm$0.007  &$-$0.034 $\pm$0.007\\ 
\rule[-0.05cm]{0cm}{.35cm}$F_4$ & 0.026  $\pm$0.007  & 0.038 $\pm$0.007\\ 
\hline 
\end{tabular} 
\label{tab:g00modalt} 
\end{minipage}
\hfill
\begin{minipage}[t]{0.49\linewidth}
\caption{Parameters of the UFD$_\text{C}$ and CFD$_\text{C}$ fits to $\vert g^0_0\vert $.}
\centering 
\begin{tabular}{c c c } 
\hline\hline  
\rule[-0.05cm]{0cm}{.35cm}Parameter & UFD$_\text{C}$ & CFD$_\text{C}$ \\ 
\hline 
\rule[-0.05cm]{0cm}{.35cm}$D_0$ & 0.462 $\pm$0.008  & 0.447 $\pm$0.008\\
\rule[-0.05cm]{0cm}{.35cm}$D_1$ &$-$0.267 $\pm$0.013  &$-$0.237 $\pm$0.013\\ 
\rule[-0.05cm]{0cm}{.35cm}$D_2$ & 0.11 $\pm$0.01  & 0.10 $\pm$0.01\\ 
\rule[-0.05cm]{0cm}{.35cm}$D_3$ &$-$0.078  $\pm$0.009  &$-$0.087 $\pm$0.009\\ 
\rule[-0.05cm]{0cm}{.35cm}$F_1$ &$-$0.04153 fixed   &$-$0.03658 fixed \\
\rule[-0.05cm]{0cm}{.35cm}$F_2$ &$-$0.010  $\pm$0.008  &$-$0.016 $\pm$0.008\\ 
\rule[-0.05cm]{0cm}{.35cm}$F_3$ &$-$0.023  $\pm$0.007  &$-$0.023 $\pm$0.007\\ 
\rule[-0.05cm]{0cm}{.35cm}$F_4$ & 0.021  $\pm$0.006  & 0.027 $\pm$0.006\\ 
\hline 
\end{tabular} 
\label{tab:g00modalt2} 
\end{minipage}
\end{table}

\begin{figure}[!ht]
\begin{center}
\resizebox{0.8\textwidth}{!}{\input{figures/g00modalt.tex}}\\ 
\resizebox{0.8\textwidth}{!}{\input{figures/g00phasealt.tex}}\\
\caption{Alternative \pipikk scattering data on the scalar-isoscalar partial wave $g^0_0$, coming from  \cite{Cohen:1980cq} (Argonne), \cite{Etkin:1981sg} (Brookhaven-I) and \cite{Longacre:1986fh} (Brookhaven-II). As explained in the main text, below $K\bar K$ threshold, due to Watson's Theorem and the fact that no multi-pion states are observed, 
the $\pipikk$ phase shift is precisely that of $\pipi$ scattering.  Thus, in that region, we provide a representative sample of such data coming from scattering experiments which is the data we plot in that region \cite{Grayer:1974cr} (Grayer et al., solution b), \cite{Kaminski:1996da,Kaminski:2001hv} (Kaminski et al.),
or the very precise $K_{\ell4}$ decays from \cite{Batley:2005ax} (NA48/2).}
\label{fig:g00alt}
\end{center}
\end{figure}

 Notice, as shown in Fig.~\ref{fig:g00alt} that the solutions of  the ``Alternative CFD'' $g^0_0(t)$ and the original one shown in Fig.~\ref{fig:g00data} are very similar for the modulus. This is also reflected in the parameters listed in Tables \ref{tab:g00modalt} and \ref{tab:g00modalt2}, which are almost equal to those fitting the original data, shown in Tables~\ref{tab:g00phase} and \ref{tab:g00modufd}. Of course, there is a difference between the phases above $1.6\,$GeV, which could be relevant if one is trying to describe this process up to higher energies.
 
 All in all, we consider this alternative solution as slightly favored over the one in the text, although there we have preferred to stick to the data quoted in the original experimental works, without introducing these further complications. There are several experimental pieces of evidence for the $f_2(1810)$ resonance, which could in principle decay copiously to this channel, and including it clearly improves our fit to the data as explained in~\cite{Pelaez:2018qny}. However, using one solution or the other does not modify in any way our dispersive results nor does it introduce any noticeable systematic effect. We have checked that the scattering lengths and the \kap resonance remain perfectly compatible with the original $g^0_0(t)$ values.

\section{Applicability regions for dispersion relations}
\label{app:Applicability}

In this appendix, we will describe how to calculate the applicability domain of different dispersion relations and how to maximize it, either in the real axis or to make it reach the complex plane in the $\kap$ region. For Hyperbolic Dispersion Relations (HDR) this translates into specific choices of the $a$ parameter defining the hyperbolae.
Our approach will be similar to that in \cite{Hoferichter:2011wk,Ditsche:2012fv} and we will study the applicability range both for the $s$-channel $\pi K \rightarrow \pi K$ and for the $t$-channel $\pi \pi \rightarrow K \bar K$. 

To this end, we will first calculate the double spectral regions, where the imaginary part of the amplitude becomes also imaginary and therefore the Mandelstam analyticity hypothesis does not hold (see \cite{Martin:1970} for a textbook introduction). This is necessary for the partial-wave projections of the dispersion relations themselves and thus applies directly to the ``external'' or ``unprimed'' $s,t,u$ variables.
Second, we have to ensure the convergence of the partial-wave expansion of the imaginary parts inside the dispersive integral, which means that
it is to be used only inside the (large) Lehmann ellipse
\cite{Lehmann:1958,Martin:1966,Martin:19662}. This constraint affects directly the ``internal'' or ``primed'' variables $s',t',u'$ as well as $z_s', \lambda_{s'}$, etc.

\subsection{Double spectral regions and Lehmann ellipses}
\begin{figure}[ht]
\centering
\includegraphics[width=\linewidth]{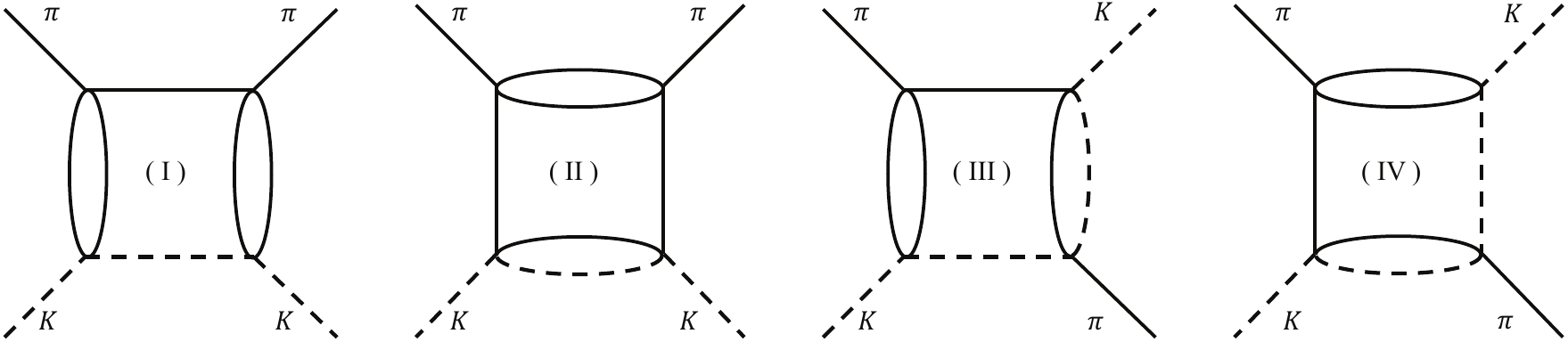}
\caption{Unitarity box diagrams that are used to calculate the double spectral regions of $\pi K$ scattering.  Contrary to Feynmann diagrams, unitarity or Cutkosky diagrams have all internal lines on-shell. Continuous lines denote pions while dashed lines denote kaons. \label{fig:boxdiagrams}}

\end{figure}

Our analysis follows the general scheme of \cite{Hohler:1984ux,Buettiker:2003wj,DescotesGenon:2006uk,Hohler:1984ux,Hoferichter:2011wk,Ditsche:2012fv}.
Namely, we assume that the $F(s,t,u)$ matrix elements, considered as functions of two independent complex variables, obey Mandelstam (or maximal) analyticity \cite{Mandelstam:1958xc}.
This means that the amplitude has only those singularities that are required by bound states (poles in the real axis below threshold) or unitarity (cuts) in each of the $s,t$, and $u$ channels.
Let us very briefly explain the mathematical consequences of this assumption, although for a detailed and pedagogical introduction we refer the reader to \cite{Martin:1970}. 

In the two-body scattering of pions and kaons there are no bound states, at least not at the physical pion mass, so that $F(s,t)$, for a given $t$, can be written as a dispersion relation over the right and left kinematic cuts in the $s$ variable. These cuts are the consequence of intermediate particles becoming real. For instance if real particles can be produced in the $s$-channel for a given energy $s$ then we find a singularity in the form of a discontinuity $D_s(s,t)=[F(s+i\epsilon, t)-F(s-i\epsilon,t)]/2i$, whose value is constrained by $s$-channel unitarity. Note that, due to Schwarz reflection, i.e. $F(s^*,t)=F^*(s,t)$, in the real $s$ axis $D_s(s,t)$ is nothing but the imaginary part of $F$, and therefore real. However, if we now continue $D_s$ in the complex $t$ variable, $D_s$ might turn complex. We can then write a dispersion relation in the $t$ variable, not for the amplitude $F$, but for $D_s$ over the regions where it has a discontinuity due to a kinematic cut in the $t$ channel. 
Such a discontinuity of $D_s$ in the $t$ variable is called the double spectral function
$\rho_{st}=[D_s(s, t+i\epsilon)-D_s(s,t-i\epsilon)]/2i$ and is once again constrained by $t$ channel unitarity. Similarly, we can define $\rho_{su}$ and $\rho_{tu}$.
Therefore, Mandelstam analyticity implies that
$\pi K$ and $\pi\pi\rightarrow K\bar{K}$ scattering 
amplitudes can be written as a sum of double integrals over the regions where the spectral functions $\rho_{st}$, $\rho_{su}$ and $\rho_{tu}$  have support. These areas are called double spectral regions. 
Intuitively, the lowest kinematic discontinuities are found when the lowest possible
number of particles can become on-shell on each diagram. Since the minimum number of legs
in pion and kaon interaction vertices is four,  the ``unitarity'' or Cutkosky 
box diagrams that we show in Fig.~\ref{fig:boxdiagrams} will give us the double spectral regions. 
Since these regions contain the singularities of $F$ in the Mandelstam plane, they have to be avoided when writing dispersion relations and limit the applicability of the dispersive approach. Let us then calculate the boundaries of the double spectral regions, using the unitarity box diagrams in Fig.~\ref{fig:boxdiagrams}.

The equations that describe the boundary of the support of the spectral function $\rho_{s t}$ are:
\begin{align}
&b_I(s,t): (t-16m_\pi^2)\lambda_s-64m_\pi^4 s=0, \label{eq:boundaryst} \\
&b_{II}(s,t):(t-4m_\pi^2)(s-(m_K+3m_\pi)^2)-32m_\pi^3m_+=0,\nonumber
\end{align}
where the subscripts $I$ and $II$ indicate what diagram yields each constraint. For the $s$-channel they apply at 
$s>m_+^2$ and $s>(m_K+3m_\pi)^2 $, respectively.
By means of $s \leftrightarrow u$ crossing, similar equations are obtained for $\rho_{u t}$. The equations that describe the boundary of the support of $\rho_{u s}$ are
\begin{align}
&b_{III}(s,t):
(s-m_-^2)(t+s-m_+^2) 
\Big(\big(( 3 m_\pi-m_K) m_+ + s\big)^2 +t(s-m_+^2)\Big)=0,\\
&b_{IV}(s,t): 
 (s-m_-^2)(t+s-m_+^2) 
\big((m_K^2+2 m_K m_\pi+5 m_\pi^2-s)^2 +t (s-(m_K+3m_\pi)^2)\big)
=0. \nonumber
\end{align}
For the $s$-channel they apply at $s>m_+^2, u>(m_K+3 m_\pi)^2$ and $s>(m_K+3 m_\pi)^2,  u>m_+^2$, respectively.

\begin{figure}[!ht]
\centerline{\includegraphics[width=0.6\linewidth]{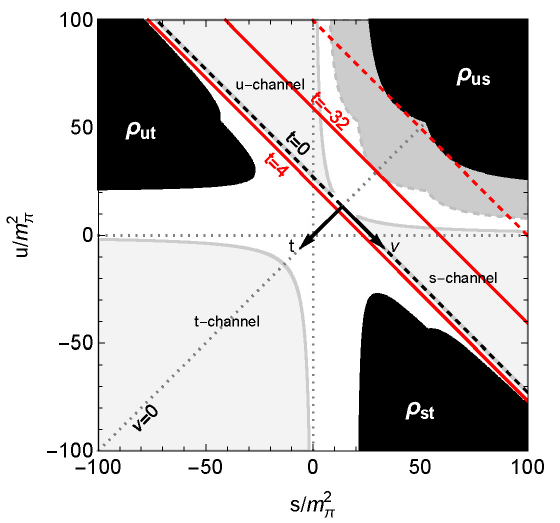}}
\caption{Mandelstam $(s,u)$ plane for the $\pi K$ scattering amplitude. Note we use units of $m_\pi^2$.
The physical regions for $\pik$ scattering ($s$ and $u$-channels)
as well as that for $\pipikk$ ($t$-channel) are shown as light gray areas.
The double spectral regions $\rho_{st},\rho_{ut}$ and $\rho_{us}$ are represented by black areas. The arrows, starting at $(t,\nu)=(0,0)$ show the directions of increasing $t$ and $\nu=s-u$. Also shown are several parallel lines of constant $t$. The dark gray area is the region excluded by the $s$-channel Lehmann ellipse.}
\label{fig:Doublespectral}
\end{figure}

The support of these double spectral regions can be seen in Fig.~\ref{fig:Doublespectral} as black areas in the $(s,u)$ plane. 
Note the $s\leftrightarrow u$ symmetry of the plot.
In order to write a dispersion relation, these areas must be avoided.
This is one kind of constraint on the applicability of dispersion relations.

In addition, there is another constraint because
partial-wave expansions of $\im F(s,t)$ in terms of Legendre polynomials $P_\ell(z)$ converge in the complex $z$ plane within the so-called Lehmann ellipse \cite{Lehmann:1958,Martin:1966,Martin:19662}
\begin{equation}
\frac{(\re z)^2}{A^2}+\frac{(\im z)^2}{B^2}=1,
\end{equation}
of foci $z=\pm1$. Here $A^2-B^2=1$, i.e., the $A$ semi-axis along the real axis is larger than that along the imaginary direction, $B$. The size of this ellipse of convergence is given by the real $z^{max}=A$ where it first touches a singularity of $\im F(s,t)$, i.e., reaches the double spectral region.

 In this report, we have to consider four Lehmann ellipses. The reason is that we have been discussing both $\pik$ and $\pipikk$ scattering, which correspond to the $s$-channel and $t$-channel of the same amplitude $F(s,t,u)$, respectively. Therefore, we are interested in both the $s$-channel partial-wave expansion, using the $z_s$ variable, and the $t$-channel partial waves in terms of the $z_t$ variable. Note that these are ``external'' variables, in the sense that they are not integrated inside dispersion relations. However, we also use each one of these $s$-channel and $t$-channel partial-wave expansions with respect to the $z_{s'}$ and $z_{t'}$ ``internal'' variables, respectively. All these Lehmann ellipses give rise to different constraints on the applicability of the dispersion relations.

Let us now see how all these constraints limit the applicability region of different kinds of dispersion relations.

\subsection{Constraints on fixed-$t$ dispersion relations}

Fixed-$t$ dispersion relations are very effective in the study of both equal mass and different mass particles.  In Fig.~\ref{fig:Doublespectral}
we have plotted several straight lines corresponding to fixed values of $t$.
The first relevant observation is that for forward scattering $t=0$, the double spectral regions are avoided for all $s$. Therefore the forward dispersion relations for $\pi K$ scattering that we have used in Section.~\ref{sec:FDR} are well-defined for any value of $s$. However, it is also obvious that this is an exceptional case. Actually, it can be noticed that the most restrictive boundary for fixed positive $t$ is that of $\rho_{s t}$ (or, by symmetry, $\rho_{u t}$). Thus, if we define $T(s)$ as the solutions of the boundary conditions, by solving Eqs.~\eqref{eq:boundaryst} we find:
\begin{align}
&T_{st}(s)=16m_\pi^2+\frac{64m_\pi^4s}{\lambda_s}, \hspace{0.3cm}\forall s\leq s_0,\label{eq:boundaryfort1}\\
&T_{st}(s)=4m_\pi^2+\frac{32m_\pi^3m_+}{s-(m_K+3m_\pi)^2}, \hspace{0.3cm}\forall s\geq s_0,
\label{eq:boundaryfort2}
\end{align}
where
\begin{equation}
s_0=m_K^2+4m_Km_\pi+5m_\pi^2+2m_\pi\sqrt{5m_K^2+12m_Km_\pi+8m_\pi^2}.
\end{equation}
Using Eq.~\eqref{eq:boundaryfort2}, we find that $t=4 m_\pi^2$ is the first
fixed-$t$ line that touches the double spectral regions, for $t$ positive. We have plotted it as a continuous red line in Fig.~\ref{fig:Doublespectral}, asymptotically tangent to $\rho_{st}$ and $\rho_{ut}$ for $\nu=\infty$ and $\nu=-\infty$, respectively. For negative $t$, if we use Eq.~\eqref{eq:boundaryfort1}, we find that the $\rho_{us}$ region is touched by
the fixed-$t=-16m_\pi m_+\simeq-72.75 m_\pi^2$ line, which is shown as a dotted red line tangent to two points of $\rho_{us}$. Recall that for fixed-$t$ dispersion relations we are integrating $\im F(s',t)$ over some polynomial of $s'$, for all values of $s'$. Thus, if we want $\im F(s',t)$ to be real for all values of $s'$ we find a first constraint:
\begin{equation}
-16m_\pi m_+<t<4m_\pi^2, \label{eq:softfixedtbound}
\end{equation}
required to avoid the double spectral regions.

Let us now remark for completeness that fixed-$s$ or fixed-$u$ dispersion relations are of little use in the physical regions. This is clearly seen in Fig.~\ref{fig:Doublespectral} for $\pi K$ scattering, since for the $s$-channel the highest fixed-$u$ straight line touches the $\rho_{ut}$ double spectral region almost at the $s$ channel threshold, and would therefore be practically useless.
It is actually below threshold once we explain in the next section that we should also avoid the dark grey region.
The situation is similar for the $u$-channel and fixed-$s$. For $\pipikk$ it is not straightforward from the figure, but the applicability region of fixed-$s$ or fixed-$u$ dispersion relations is very limited, well below the $K\bar K$ threshold, as shown for $\pi\pi\rightarrow N \bar N$ in \cite{Steiner:1970fv}.
Moreover, from the figure it is also clear that forward dispersion relations for $\pi\pi\rightarrow K\bar K$, i.e. $\nu=0$, are of no use since that straight line passes right through the $\rho_{us}$ double spectral region. 
Thus, in practice, fixed-variable dispersion relations
for the $\pi K$ amplitude are only useful for fixed-$t$.

However, Eq.~\eqref{eq:softfixedtbound} is not the only constraint. There are others, even more stringent, due to the fact that we have to build $\im F(s',t)$ through the partial-wave expansion in $z_{s'}=1-2s't/\lambda_{s'}$.
The first value where the corresponding Lehmann ellipse touches the double spectral region is  $A_{s'}\equiv z_{s'}^{max}\equiv 1+2s'T_{st}(s')/\lambda_{s'}$, which we have seen that for fixed-$t$ lines 
occurs at $t=4m_\pi^2$.
Once we have obtained the positive end of the ellipse, the negative end is just $-z_{s'}^{max}$. Thus 
the convergence of the internal partial wave series demands $-z_{s'}^{max}\leq z_{s'}\leq z_{s'}^{max}$,
which translates into the following restriction for $t$
\begin{equation}
-\frac{\lambda_{s'}}{s'}-T_{st}(s')\leq t\leq T_{st}(s'). \label{eq:ftbounds}
\end{equation}
The upper applicability bound is given directly by the double-spectral region $T_{st}(s')$ and we have already seen in Fig.~\ref{fig:Doublespectral} that for fixed-$t$ straight lines it lies at $4m_\pi^2$. However the lower bound in Eq.~\eqref{eq:ftbounds} depends also on $\lambda_{s'}/s'$ 
and is given by the dark gray dashed curve that encloses the dark gray region in Fig.~\ref{fig:Doublespectral}, which thus has to be avoided for our fixed-$t$ dispersion relations. This implies $t>-32 m_\pi^2$.
 All in all, the applicability of fixed-$t$ dispersion relations is limited to:
\begin{equation}
-32m_\pi^2<t<4m_\pi^2.
\end{equation}
Both the fixed-$t=-32 m_\pi^2$ and $t=4m_\pi^2$ straight lines are plotted as a red continuous lines in Fig.~\ref{fig:Doublespectral}. It is only between these two lines that fixed-$t$ dispersion relations are well defined.

So far we have not projected the outcome of the dispersion relation in partial waves. What we have seen up to now would be valid to constrain $F(s,t)$. However, we want to constrain partial waves using Roy-Steiner equations to rewrite unphysical cuts in terms of the physical ones within the dispersion relation, as explained in the main text.
This representation has its own constraints on its applicability that we review next.

\subsection{Complex applicability domain of partial-wave relations from fixed-$t$ dispersion relations}
\label{subsub:complexfixedt}

Here we follow closely the excellent account in \cite{DescotesGenon:2006uk}.
Hence, if we want to obtain partial-wave dispersion relations using the Roy-Steiner representation from fixed-$t$ dispersion relations (FTPWDR), we have to project the amplitude obtained from the latter, as follows:
\begin{equation}
f_{\ell}(s)=\frac{1}{32 \pi N} \int_{-1}^{1} d z_{s}\, P_{\ell}(z_s) F\left(s, t\left(z_{s}\right)\right)=\frac{s}{16 \pi N \lambda_s} \int_{-\lambda_s/s}^{0} dt\, P_{\ell}(z_s(t)) F\left(s, t\right),
\label{eq:projec2}
\end{equation}
where $N=1,2$ for non-identical and identical particles, respectively, and $F\left(s, t\left(z_{s}\right)\right)$ will be obtained from fixed-$t$ dispersion relations. Note that now we are interested in the applicability domain within the complex plane of the {\it external} variable $s$. From the definition of the $s$-channel cosine of the scattering angle, $z_s=1+2st/\lambda_s$, which is integrated  over real values in \eqref{eq:projec2}, we see that if $s$ is complex then $t$ is integrated along a complex segment. Hence, in the integrand of the fixed-$t$ dispersion relation we now need $\im F(s',t)$ for real values of $s'$ but complex values of $t$. 

Let us then rewrite the $z_{s'}$ Lehmann-Martin ellipse \cite{Lehmann:1958, Martin:1966, Martin:19662} in terms of the $t$ variable. 
The $t$ ellipse has now foci at real $t=-\lambda_{s'}/s'$ and $t=0$ 
and the new semi-axes are $\tilde A= \lambda_{s'} A/2s'$ and $\tilde A^2-\tilde B^2= (\lambda_{s'}/2s')^2$. Its eccentricity is $\epsilon=\sqrt{1-\tilde B^2/\tilde A^2}=1/A$. 

\begin{figure}[ht]
\centering
\centerline{ \includegraphics[width=0.7\linewidth]{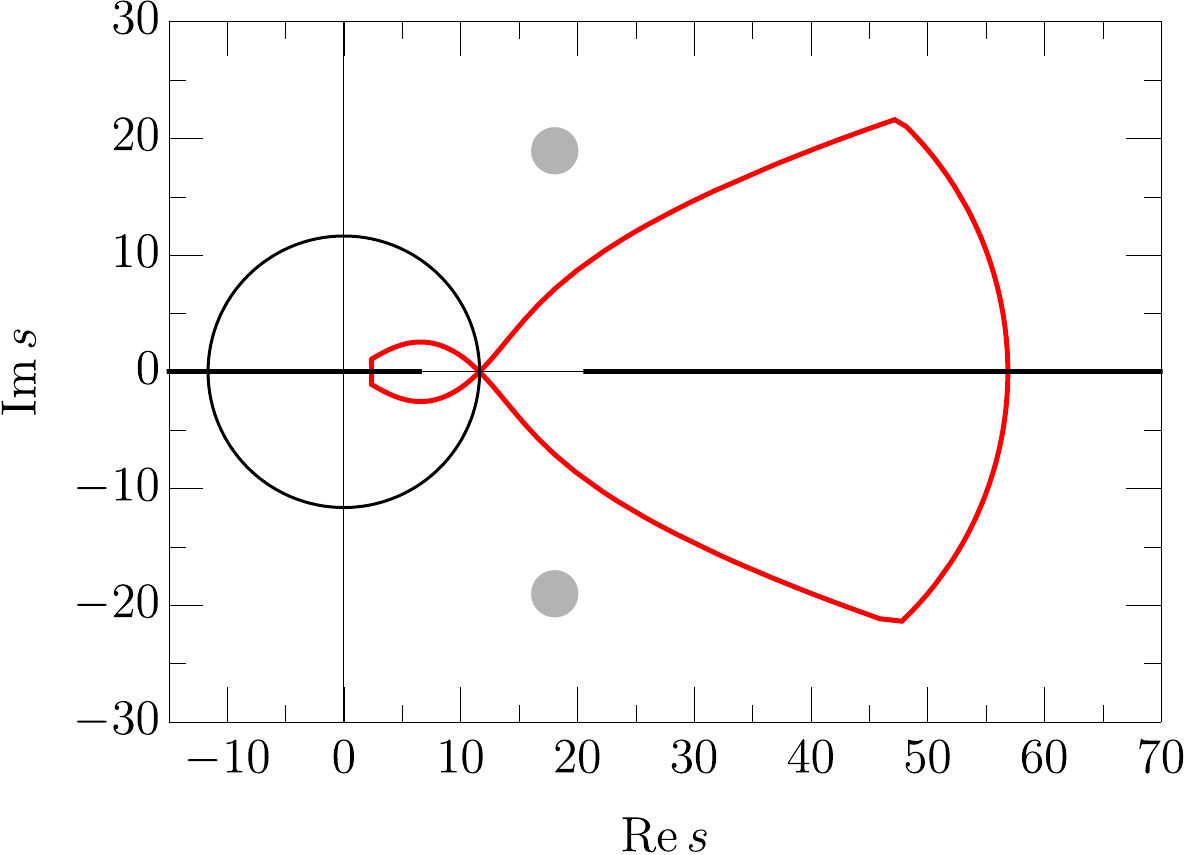}}
\caption{The domain of validity for the Roy-Steiner representation of partial waves obtained from fixed-$t$ dispersion relations (FTPWDR) for \pik scattering, in $m_\pi$ units. The shadowed areas represent the region where the \kap pole lies in the second Riemann sheet.}
\label{fig:rvpik}
\end{figure}

It is now convenient to rewrite the equation of the ellipse in polar coordinates $(T(\theta),\theta)$ with respect to the second foci $t=0$ at the origin of the plane. Namely
\begin{equation}
    T(s',\theta)=\frac{\tilde A (1-\epsilon^2)}{1+\epsilon \cos \theta}=\frac{T_{st}(s')(\lambda_{s'}+s' T_{st}(s'))}{\lambda_{s'} \cos^2 {\frac{\theta}{2}}+s' T_{st}(s')}.
\end{equation}
This equation defines the maximum allowed value for the modulus of $t$ for a given value of the integration variable $s'$. Beware, however, that $t$ lies inside the applicability region only if it falls inside all the ellipses for all possible $s'$ over which we integrate.
Therefore, the boundary of the applicability region is  given by
\begin{equation}
    T(\theta)=\min_{s_{th}\leq s'}T(s',\theta).
\end{equation}
Finally, the allowed values for $s$ in the complex plane are those for which $\vert t\vert\leq T(\theta)$. But $t=T( \theta)\exp(i\theta)=\lambda_s(z_s-1)/2s$, with $0\leq z_s-1\leq 1$ real, so that the largest modulus of $t$ for a given value of $s$ is $t=-\lambda_s/s$. Therefore the boundary in the $s$ complex plane is given by
\begin{equation}
    \lambda_s+s T(\theta)\exp(i\theta)=0,
\end{equation}
which corresponds to the red line shown in Fig.~\ref{fig:rvpik}.
Within that line, we can safely apply FTPWDR.
Note that in the real axis this means that we can use FTPWDR relations up to $s\sim 57m_\pi^2$
, or $\sqrt{s_{max}}\simeq 1.05\,\gev$. Unfortunately, this region does not reach the position of the \kap pole and is the main reason to use partial-wave dispersion relations obtained from hyperbolic dispersion relations (HPWDR), whose applicability region we study next.

\subsection{Constraints on hyperbolic dispersion relations}

Relatively similar constraints appear if, instead of fixed-$t$ dispersion relations we choose other ways of fixing one variable. Actually, Steiner and Hite \cite{Hite:1973pm} showed, for $\pi N$ scattering, that hyperbolae of the form $(s-a)(u-a)=b$ are the most convenient curves to avoid the double spectral regions passing through both direct and crossed channels,  without introducing further cuts and provide reasonably simple integration kernels after the partial-wave projection. The applicability of the $a=0$ case for $\pi K$ scattering was studied in \cite{DescotesGenon:2006uk}. In \cite{Pelaez:2018qny} the general case with $a\neq 0$ was analyzed, paying particular attention to maximizing the $\pi K$ and $\pi\pi\rightarrow K\bar K$ applicability regions in the real axis, in order to constraint data fits. In this appendix, we will review the general case, but now it is also very relevant that, with an appropriate choice of $a$, the $\kap$ pole should lie within the $\pi K$ applicability domain, while still having a rather large applicability region in the real axis. 

In contrast to the fixed-$t$ case, we use hyperbolic dispersion relations (HDR) both for the $s$-channel and $t$-channel, i.e. both for $\pi K$ and $\pi\pi\rightarrow K \bar K$ scattering.  
Let us emphasize again that we have to consider the projection in terms of ``external'' variables of $F(s,t)$ into partial waves for the two channels, according to Eq.~\eqref{eq:projs} and \eqref{eq:gpw}, respectively. In addition, the applicability of both projections will have to be studied concerning
``internal'' integration variables, since in practice $\im F(s',t')$ is built 
from the sum of partial waves.

Thus, on the one hand, for a fixed value of $a$, the family of hyperbolas $(s-a)(u-a)= b$ has to avoid all double spectral regions for all values of $b$ needed to perform the partial-wave projection. On the other hand, for a fixed $a$, we have to calculate the restrictions on $b$ to remain within the corresponding Lehmann ellipse. 

Let us then explain in detail how to calculate these applicability domains.

\subsubsection{Lehmann ellipses for the partial-wave expansion in internal variables}

In contrast to the fixed-$t$ case, now $t'$ also changes in $\im F(s',t')$ 
simultaneously with the integration variable $s'$.
Thus, for a given $a$, neither $s'$ nor $t'$ are fixed, but instead, they determine the value of the parameter 
\begin{equation}
b(s',t',a)=(s'-a)(2\Sigma-s'-t'-a), \label{eq:bfromst}
\end{equation}
that characterizes the hyperbola in the Mandelstam plane. Thus, the applicability constraints will be given in terms of this parameter.

\begin{itemize}
\item{\it Internal $s$-channel partial-wave expansion}

This case is very similar to the fixed-$t$ analysis of $\pi K$ scattering.
Thus, the internal partial-wave expansion for the $s$-channel converges for angles $z_{s'}(s',t')=1+2s't'/\lambda_{s'}$ inside the Lehmann ellipse \cite{Lehmann:1958,Martin:1966,Martin:19662} 
\begin{equation}
\frac{(\re z_{s'})^2}{A^2_s}+\frac{(\im z_{s'})^2}{B^2_s}=1,
\end{equation}
with foci at $z_{s'}=\pm1$. The maximum value of $z_{s'}$, which defines the large semi-axis, touches the double spectral region at $t'=T_{st}(s')$, namely
\begin{equation}
A_s\equiv z^{max}_{s'}=1+\frac{2s'T_{st}(s')}{\lambda_{s'}},
\quad \forall s'\geq m_+^2.
\end{equation}
The  minimum real value of the ellipse is therefore
$z_{s'}^{min}=-z_{s'}^{max}$ and thus $z_{s'}$ is constrained to lie within
\begin{equation}
-z^{max}_{s'}\leq z_{s'}\leq z^{max}_{s'},
\quad  \forall s'\geq m_+^2,
\label{eq:zinorder}
\end{equation}
which translates into the following restriction on $t'$
\begin{equation}
-\frac{\lambda_{s'}}{s'}-T_{st}(s')\leq t'\leq T_{st}(s'),
\quad  \forall s'\geq m_+^2.
\end{equation}
Now, using Eq.~\eqref{eq:bfromst} we obtain a set of bounds for $b$:
\begin{align}
&b^-_s(s',a)\leq b\leq b^+_s(s',a), \quad \forall s'\geq m_+^2>a\nonumber \\
&b^-_s(s',a)\equiv(s'-a)(2\Sigma-s'-T_{st}(s')-a),  \nonumber \\
&b^+_s(s',a)\equiv(s'-a)(2\Sigma-s'+\frac{\lambda_{s'}}{s'}+T_{st}(s')-a)  .
\end{align}

Thus, the final range of values allowed for $b$ to avoid the double spectral regions in the $s$-channel contributions to the HDRs is
\begin{equation}
b^-_s(a)\leq b\leq b^+_s(a), \quad \forall s'\geq m_+^2>a,
\label{eq:bsinterval}
\end{equation}
where
\begin{equation}
b^-_s(a)\equiv\min_{s_{th}\leq s'} b^-_s(s',a),\quad
b^+_s(a)\equiv\max_{s_{th}\leq s'} b^+_s(s',a).
\end{equation}
Note that it is required that $a<m_+^2$, independently of $b$.

\item{\it Internal $t$-channel partial-wave expansion}

For the $t$-channel the partial-wave expansion in the angle  $z_{t'}(s',t')$, 
also converges within a Lehmann ellipse 
\begin{equation}
\frac{(\re z_{t'})^2}{A_{t}^2}+\frac{(\im z_{t'})^2}{B_{t}^2}=1,
\label{eq:ellipseforztp}
\end{equation}
of foci $z_{t'}=\pm1$ and $A_{t}^2-B_{t}^2=1$. However, now we cannot follow a similar argument as for the $s$-channel because  the angle in the $t'$ channel is
\begin{equation}
z^2_{t'}=\frac{\nu'}{4q_\pi(t') q_K(t')}, \quad \nu'=s'-u',
\end{equation}
which becomes pure imaginary in the pseudo-physical region $t_{\pi}\leq t\leq t_K$ and one cannot write an ordering relation as in Eq.~\eqref{eq:zinorder}. Nevertheless, we are interested in the bounds for $b$ and these can be recast in terms of $z_{t'}^2$, since, along the hyperbolae
\begin{equation}
\nu'^2=(t'-2\Sigma+2a)^2-4b(s',t',a),
\end{equation}
so that
\begin{equation}
z^2_{t'}=\frac{(t'-2\Sigma+2a)^2-4b(s',t',a)}{16q_\pi(t')^2 q_K(t')^2},
\end{equation}
where all squares are real, although not necessarily positive.
We can then obtain an ellipse for $z_{t'}^2$ by simply squaring the ellipse for $z_{t'}$ in Eq.~\eqref{eq:ellipseforztp} above. Namely:
\begin{equation}
\frac{(\re z^2_{t'}-\frac{1}{2})^2}{\hat{A}^2_t}+\frac{(\im z^2_{t'})^2}{\hat{B}^2_t}=1,
\label{eq:zt2ellipse}
\end{equation}
where $\hat{A}_t=(A^2_t+B^2_t)/2=A_t^2-1/2$ and $\hat{B}_t=A_tB_t=A_t\sqrt{A_t^2-1}$ are the new semi-axes of the ellipse. The new center lies at $(1/2,0)$ and the new foci are located at $1/2\pm\sqrt{\hat A_t^2-\hat B_t^2}=1/2\pm1/2$. 
Therefore,  for real $z_{t'}^2$, the condition to remain within the ellipse $1/2-\hat A_t^2\leq z_{t'}^2\leq 1/2+\hat A_t^2$ can be recast as
\begin{align}
&1-A^2_t\leq z^2_{t'}\leq A^2_t.
\label{eq:georestriction}
\end{align}

We still have to determine $A_t$. Once again, the most restrictive bound comes from $\rho_{st}$. Given the $s\leftrightarrow u$ symmetry of the $t$ channel, visible in Fig.~\ref{fig:Doublespectral}, and the proportionality between $z_{t'}$ and
$\nu'$, it is convenient to rewrite the double spectral boundaries in
\eqref{eq:boundaryst} in terms of $t, \nu$ to obtain, respectively
\begin{align} 
&\nu_{st}(t)=\frac{1}{t-16m_\pi^2} \Big[ (t-8m_\pi^2)^2+4m_\pi \sqrt{t}\sqrt{(t-16m_\pi^2)m_K^2+16m_\pi^4)}  \Big],\\
&\nu_{st}(t)=\frac{16m_\pi^3m_K+12m_\pi m_+t+t^2}{t-4m_\pi^2}, 
\end{align}
where in the $s$-channel they apply for all $t\geq 4 t_\pi$
and $t\geq t_\pi$, respectively.
Denoting the first value of $\nu$ that touches the boundary by
\begin{equation}
N_{st}(t)\equiv\min \nu_{st}(t),
\label{eq:bound}
\end{equation}
we obtain the maximum value of the angle and therefore the semiaxis as
\begin{equation}
z^{\max}_{t'}(t')=\frac{N_{st}(t')}{4q_\pi(t') q_K(t')}\equiv A_t \hspace{0.3cm}\forall t'\geq t_K.
\end{equation}
Now, using equation \eqref{eq:georestriction} together with \eqref{eq:bound}, we obtain the restriction for $\nu'$

\begin{equation}
16[q_\pi(t') q_K(t')]^2-N_{st}(t')^2\leq \nu'^2\leq N_{st}(t')^2, \hspace{0.3cm}\forall t'\geq t_K.
\end{equation}
Finally, the restriction for $b$ is obtained just by translating $\nu'^2$ into $b$
\begin{equation}
b^-_t(t',a)\leq b\leq b^+_t(t',a),\quad \forall t'\geq t_\pi >a,
\end{equation}
with
\begin{align}
&b^-_t(t',a)=\frac{(t'-2\Sigma+2a)^2-N_{st}(t')^2}{4}, \nonumber \\  
&b^+_t(t',a)=\frac{(t'-2\Sigma+2a)^2-16[q_\pi(t') q_K(t')]^2+N_{st}(t')^2}{4}.
\label{eq:btellipseextremes}
\end{align}

Once again, since we are integrating in the internal variable $t'$, the total bounds are defined as
\begin{align}
&b^-_t(a)=\max_{t'>t_\pi} b^-_t(t',a), \nonumber \\
&b^+_t(a)=\min_{t'>t_\pi} b^+_t(t',a),
\end{align}
and the allowed values of $b$ for a fixed $a$ that do not touch any boundary while expanding in partial waves the $t$-channel contributions inside the HDR are 
\begin{equation}
b^-_t(a)\leq b\leq b^+_t(a),\hspace{0.3cm}\forall t'\geq t_\pi\geq a.
\label{eq:btinterval}
\end{equation}
Note that now we are requiring $t_\pi>a$, independently of $b$.

\end{itemize}

\subsubsection{Lehmann ellipses for the partial-wave projection on external variables. }

In the previous subsection we have studied the constraints due to the fact that
the imaginary parts of the amplitude  {\it inside} dispersion relations are integrated from threshold to infinity either on  the internal variable $s'$ or $t'$.
In practice, these imaginary parts of the amplitude 
are obtained from their partial-wave expansions, which only converge within their respective Lehmann ellipses. Besides, the whole amplitude should not touch the double spectral representation. 

We are now interested in  hyperbolic dispersion relations projected into partial waves (HPWDR), in order to compare with the existing data and to continue them to the complex plane in search for poles associated with resonances. 
We will see next that, in contrast to the FTPWDR, the HPWDR apply to both the $s$-channel and $t$-channel partial waves.
Hence, we will now be integrating HDRs, either with respect to the
external $s$-channel  $\cos z_s$ or the $t$-channel $\cos z_t$, between $-1$ and $1$. In addition, we should require that these partial-wave expansions in external angles should also converge.

\begin{itemize}
\item{$s$-channel partial-wave projection}

We have already seen that the $\rho_{st}$ constraints are the strongest ones. We have also shown that the values of $b$ must lie within
the intervals in Eqs.~\eqref{eq:bsinterval} and \eqref{eq:btinterval}, for all $s'\geq m_+^2$ and $t'\geq t_\pi$. For brevity, we write the two intervals together using the two labels separated by a comma as
$b\in [b^-_{s,t}(a),b^+_{s,t}(a)]$. Now, the external $s,t$ must also fall in the hyperbolae and thus we have to make sure that
$b(s,t,a)=(s-a)(2\Sigma-s-t-a)$
lies within those intervals. 
Since the integration range $-1\leq z_s\leq 1$ translates into
\begin{equation}
-\frac{\lambda_s}{s}\leq t\leq 0,
\end{equation}
then, given a fixed $a$, the parameter $b(s,t)$ due to the $s$-channel projection will vary between
\begin{align}
&b^{min}(s,a)\leq b\leq b^{max}(s,a), \quad s\geq m_+^2>a, \label{eq:bminbmax}\\
&b^{min}(s,a)=(s-a)(2\Sigma-s-a), \nonumber \\
&b^{max}(s,a)=(s-a)\left(2\Sigma-s+\frac{\lambda_s}{s}-a\right)=(s-a)\left(\frac{\Delta^2}{s}-a\right). \nonumber
\end{align}
Therefore, the interval $[b^{min}(s,a),b^{max}(s,a)]$ has to be fully included in the $[b^-_{s,t}(a),b^+_{s,t}(a)]$ intervals. For a given $a$, we define $s^{max}(a)$ as the largest value of $s$ for which this occurs, which is calculated as follows: First, let us define $s^-_{s,t}(a)$ and
$s^+_{s,t}(a)$ as the values of $s$ such that:
\begin{equation}
b_{s,t}^-(a)=b^{min}(s_{s,t}^-(a),a),\quad
b_{s,t}^+(a)=b^{max}(s_{s,t}^+(a),a).
\end{equation}
In view of the definition of $b_{s,t}^{min}$ and $b_{s,t}^{max}$ in Eq.~\eqref{eq:bminbmax}, these are quadratic equations with two solutions each:
\begin{align}
& s_{s,t}^{-(\pm)}=\Sigma\pm\sqrt{(\Sigma-a)^2-b_{s,t}^-(a)},\\
& s_{s,t}^{+(\pm)}=\frac{1}{2a}\left[[\Delta^2+a^2-b_{s,t}^+(a)]^2\pm\sqrt{[\Delta^2+a^2-b_{s,t}^+(a)]^2-4a^2\Delta^2}\right].
\end{align}
Let us consider a given $a<0$, which ensures $a<m_+^2$
(similar arguments follow if $0<a<m_+^2$, but are of less relevance to have large applicability regions).
Then, the applicability of  the HPWDR is reduced to 
those $s$ belonging to the intervals $[s_{s,t}^{-(-)}(a),s_{s,t}^{-(+)}(a)]\cap[s_{s,t}^{+(+)}(a),s_{s,t}^{+(-)}(a)]$. 
Now, since we want to compare with data
we need to choose $a$ so that some physical region $m^2_+\leq s\leq s_{max}(a)$ lies inside those intervals.
The calculation of that $s_{max}$ is done numerically and for $a<0$ it is $s_{max}=min\{s_{s,t}^{-(+)}(a),s_{s,t}^{+(-)}(a)\}$.

For instance, if we choose $a$ to maximize the domain of applicability of the $s$-channel projection, the strongest restriction comes from the $t$-channel Lehmann ellipse 
and we should use
\begin{align}
&a=-13.9 \,m_\pi^2, \quad s_{max}\simeq 50 m_\pi^2\simeq0.98 \,\gev^2 \,,
\quad \sqrt{s_{max}}\simeq 0.989 \,\gev,\nonumber \\
&b^-_t(a)\simeq-592\, m_\pi^4, \quad b^+_t(a)\simeq1070 \,m_\pi^4.
\end{align}

In the left panel of Fig.~\ref{fig:hyperbolae} we show a representative sample of these hyperbolae, showing in blue the intersection with the $s$-channel physical region. The boundary cases correspond to the thicker lines.

However,  in~\ref{subsec:REcomplexdomain} below, we will see that this choice is not so good for our purposes, because the applicability domain in the complex plane does not reach the \kap pole and its uncertainties.

In contrast, 
the authors of \cite{DescotesGenon:2006uk} chose $a=0$, so that $s^{max}\simeq 45m_\pi^2$, i.e. $\sqrt{s^{max}}\simeq0.934\,\gev$, but they showed that the \kap pole lies within their applicability region, as we will also show in subsection~\ref{subsec:REcomplexdomain}.

Finally, in this review, we will mostly use
\begin{align}
&a=-10 \,m_\pi^2, \quad s_{max}\simeq 49 m_\pi^2\simeq 0.954 \,\gev^2 \,,
\quad \sqrt{s_{max}}\simeq 0.976 \,\gev,\nonumber \\
&b^-_t(a)\simeq\,-690 m_\pi^4, \quad b^+_t(a)\simeq 997 \,m_\pi^4,
\end{align}
which ensures that the applicability domain covers the $\kap$ pole region, but still reaches rather high in the real axis, since we only lose 13 MeV of applicability compared to the $a=-13.9 m_\pi^2$ case. Our choice will be explained in detail in \ref{subsec:REcomplexdomain} below.

\begin{figure}
\centerline{\includegraphics[width=0.45
\linewidth]{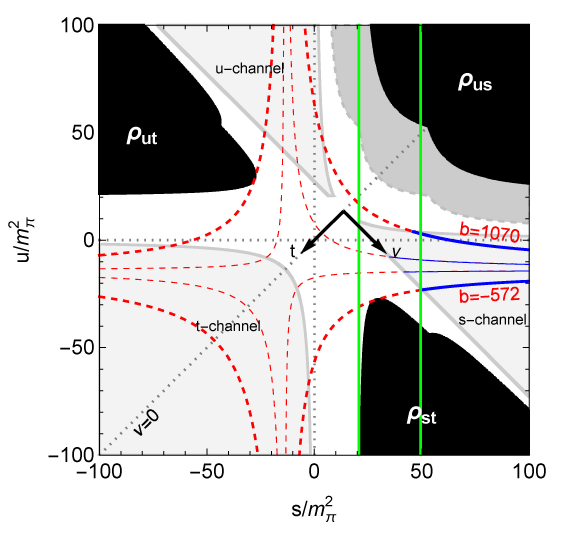}
\includegraphics[width=0.45\linewidth]{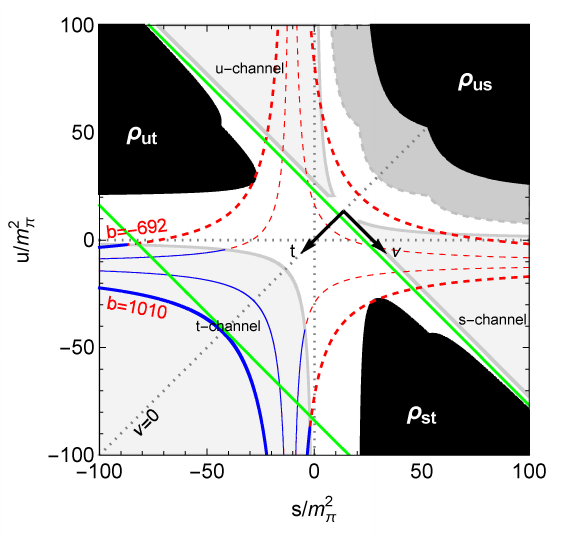}}
 \caption{
 We plot in the Mandelstam $(s,u)$ plane explained in Fig.~\ref{fig:Doublespectral}, a representative sample of the  $(s-a)(u-a)=b$ hyperbolae  used to obtain HPWDR, in $m_\pi$ units. They are represented as dashed red lines when they lie outside the corresponding  physical region and as continuous blue lines when they lie inside. The thicker ones correspond to the limiting $b$ values. On the left panel, we show those that maximize the applicability region  in the real axis for the $\pi K$  analysis ($s$-channel), with $a=-13.9m_\pi^2$. Green straight lines mark the minimum and maximum values of $s$ valid for these dispersion relations. Let us recall that we do not use exactly this value of $a$ when we want to reach the \kap region in the complex plane. For this, it is better to use $a=-10m_\pi^2$ which has only a slightly lower reach, as explained in the main text.  On the right panel
 we show the hyperbolae used for \pipikk ($s$-channel), with $a=-10.9 m_\pi^2$. Green straight lines mark the minimum and maximum  values of $t$ valid for these dispersion relations. }
\label{fig:hyperbolae}
\end{figure}

\item{$t$-channel partial-wave projection}

The relation between $b$ and the cosine of the scattering angle $z_t$ is now quadratic
and is then better to carry out the analysis in terms of $z_t^2$.
Thus, to perform the $t$-channel projection we need to consider  the whole interval
\begin{equation}
0\leq \frac{\nu^2}{16q^2_\pi q^2_K}=\frac{(t-2\Sigma+2a)^2-4b}{16q^2_\pi q^2_K}=\frac{(t-2\Sigma+2a)^2-4b}{(t-t_\pi)(t-t_K)}\leq 1.
\label{proyection}
\end{equation}
For a given $a$, it can be translated into an interval of applicability for $b$, by defining:
\begin{equation}
b^{min}(t,a)=\frac{1}{4}(t-2\Sigma+2a)^2\geq0,\qquad b^{max}(t,a)=a(t-2\Sigma)+a^2+\Delta^2.
\end{equation}
Paying attention to the signs of $t-t_\pi$ and $t-t_K$, then $b$ must lie within:
\begin{align}
&b^{min}(t,a)\leq b\leq b^{max}(t,a), \quad \forall t_\pi\leq t\leq t_K,\\
&b^{max}(t,a)\leq b\leq b^{min}(t,a), \quad \forall t>t_K \quad ({\rm or} \, t<t_\pi).
\end{align}
Thus, we need these ranges above to be fully contained in the
$[b_{s,t}^-(a),b_{s,t}^+(a)]$ that we obtained from the internal variable constraints in Eqs.~\eqref{eq:bsinterval} and \eqref{eq:btinterval}.
In particular, in order to maximize the domain 
of applicability in the real $t$ axis by choosing $a$ we search for the value $t=t_{max}$ where both the maximum and minimum values of $b$ coincide with $b^-_{s,t}(a)$ and $b^+_{s,t}(a)$. Using Eq.~\eqref{proyection} and taking into account that the projection is made between $z^2_t=0$ and $z^2_t=1$ this means
\begin{align}
&z^2_t(t_{max},b^-_{s,t}(a))=1, \nonumber \\
&z^2_t(t_{max},b^+_{s,t}(a))=0.
\end{align}

Once again, the restriction from the $t$-channel is stronger than that from the $s$-channel, and we find that the applicability range is maximized with the following choice:
\begin{align}
&a=-10.9 m_\pi^2, \hspace{0.5cm}  -0.286 \, \gev^2 \,(\sim 15 m_\pi^2) \leq t\leq 2.19\, \gev^2 \,(\sim 112 m_\pi^2), \nonumber \\
&b^-_t(a)=-672 \, m_\pi^4, \hspace{0.5cm} b^+_t(a)=1010 \, m_\pi^4.
\end{align}

Note that this means that the applicability domain in the physical region
extends from $K \bar K$ threshold  $\sqrt{t_K}\simeq 0.992\,\gev$ up to  $\sqrt{t}\simeq \sqrt{2.19}\,\gev\simeq1.47\,\gev$, which is the interval we used in the Roy-Steiner dispersive analysis of $\pi\pi\rightarrow \bar KK$ data \cite{Pelaez:2018qny}.
In the right panel of Fig.~\ref{fig:hyperbolae} we show a representative sample of this family of hyperbolae, whose intersection with the $s$-channel physical region we have highlighted in blue. The boundary cases correspond to the thicker lines.

In contrast, for the choice $a=0$, as in \cite{Buettiker:2003pp},  $\sqrt{t_{max}}\sim 1.3\,\gev$ \footnote{In \cite{Buettiker:2003pp}
the authors wrote $t_{max}=70m_\pi^2$, or $\sqrt{t_{max}}\simeq1.17\,\gev$, but in \cite{DescotesGenon:2006uk} they corrected their double spectral boundary equations. When using the correct ones the actual value is $\sim 1.3\,\gev$.}.
It is then clear that the optimization of $a$ is more relevant in \pipikk than in \pik since an optimal choice of $a$  enlarges the applicability interval to study \pipikk data by an additional 67\% in terms of $t$ for its physical region.

\end{itemize}

\subsection{Complex applicability domain of $s$-channel partial-wave hyperbolic dispersion relations}
\label{subsec:REcomplexdomain}

In the main text we have reviewed the model-independent and precise extraction of the pole parameters associated to strange resonances that are produced in the elastic regime of $\pi K$ scattering, i.e. the scalar \kap and $K^*(892)$ vector resonance.
This can be done because in the elastic regime there is a zero in the first Riemann sheet for each pole in the second and they are easily related through Eqs.~\eqref{ec:firsttosecondsheet}, \eqref{ec:resonancecondition} and \eqref{eq:rescond}. In \ref{subsub:complexfixedt} we just saw that 
the FTPWDR does not reach the region of interest for the \kap. 
We, therefore, need to study the applicability domains of  HPWDR in the complex $s$ plane, which can differ substantially from one another, depending on the choice of $a$. Actually, the domain for the $a=0$ case for $\pi K$ scattering was already shown in  \cite{DescotesGenon:2006uk}. Here  we will give the details  for the general case $a\neq0$, for which we will follow the account given in \cite{Ditsche:2012fv} for $\pi N$ scattering, in order to determine the semi-axes of the Lehmann ellipses, which can be done from real values of $s,t$.

Thus, we are now interested in complex values of the external variable $s$ of the partial waves $f_\ell(s)$. These are obtained as an integration over $0\leq z_s=1+2st/\lambda_s\leq 1$, which means that $t$ is integrated over a complex segment between 0 and $\lambda_s/s$. Therefore, in the dispersive integrals from $s$-channel contributions we will now need $\im F(s',t')$
for real $m_+^2\leq s'$ but complex $t'$. This is relatively similar to the fixed-$t$ case, although now $s'$ and $t'$ are related through the hyperbolae $(s-a)(u-a)=b$.
Consequently, the $b$ parameter will  also be complex.

In addition, in contrast with the case of fixed-$t$ dispersion relations, for HDR we also have $s$-channel contributions where  we have to study the convergence of $\im F(t',s')$ with
$t_\pi\leq t'$ real and  $s'$ related to $t'$ through the hyperbolae equation.

Let us then start with the derivation of the range of validity over the $s$-channel Lehmann-Martin ellipse, which is again defined throughout the ellipse with foci at $t=-\lambda_{s'}/s'$ and $t=0$. However, 
since $s$ and $t$ are now related by a hyperbolae,
it is better to recast our conditions in terms of the parameter
$b=(s-a)(2\Sigma-s-t-a)$. 
With this change of variable, the partial-wave projection  in Eq.~\eqref{eq:projec2} is recast as:
\begin{equation}
f_\ell(s)=\frac{-s}{16\pi N(s-a)\lambda_s}\int^{(s-a)(2\Sigma-s-a)}_{(s-a)(\Delta^2/s-a)} db\,P_\ell(z_s(t(s,b,a))) F\left(s,2\Sigma-s-a-\frac{b}{s-a}\right).
\end{equation}

Thus we have to consider an ellipse on $b$ with foci at $b_s=(s'-a)(2\Sigma-s'-a)$ and $b_s=(s'-a)(2\Sigma-s'+\lambda_{s'}/s'-a)$. The calculation is now more complicated since this ellipse does not have any fixed foci. Nonetheless, we have  seen
in Eqs.~\eqref{eq:bminbmax} above how the maximum and minimum of $b$ can be calculated. Then, denoting the ellipse boundary by $(B_s(s',\theta),\theta)$, calculated in polar coordinates from the origin of coordinates,  $B$ falls inside all possible ellipses if
\begin{equation}
    B_s(\theta)=\min_{s_{th}\leq s'\leq \infty} B_s(s',\theta).
\end{equation}

Next, we have to consider the $t$-channel Lehmann-Martin ellipse. As discussed above, the relation between $b$ and $z_t$ is quadratic,  so it is better to write the ellipse in terms of $z_t^2$. Also, for the $\pi\pi\rightarrow K\bar K$ contributions it is better to use the $t,\nu$ variables. We already did this in Eqs.~\eqref{eq:zt2ellipse} to \eqref{eq:btellipseextremes},
where we found the extremes of the $b_t$ ellipse. Once again we rewrite that ellipse in polar coordinates 
$(B_t(t',\theta),\theta)$. However, since $t'$ is integrated from $\pi\pi$ threshold, we have to consider only the values of $b$ inside all possible ellipses as $t'$ varies. All things considered, the final applicability bound  on $b$ due to the $s$-channel contributions is given by
\begin{equation}
    B_t(\theta)=\min_{4 m_\pi^2\leq t'\leq \infty}B(t',\theta).
\end{equation}

\begin{figure}[ht]
\centering
\centerline{ \includegraphics[width=0.7\linewidth]{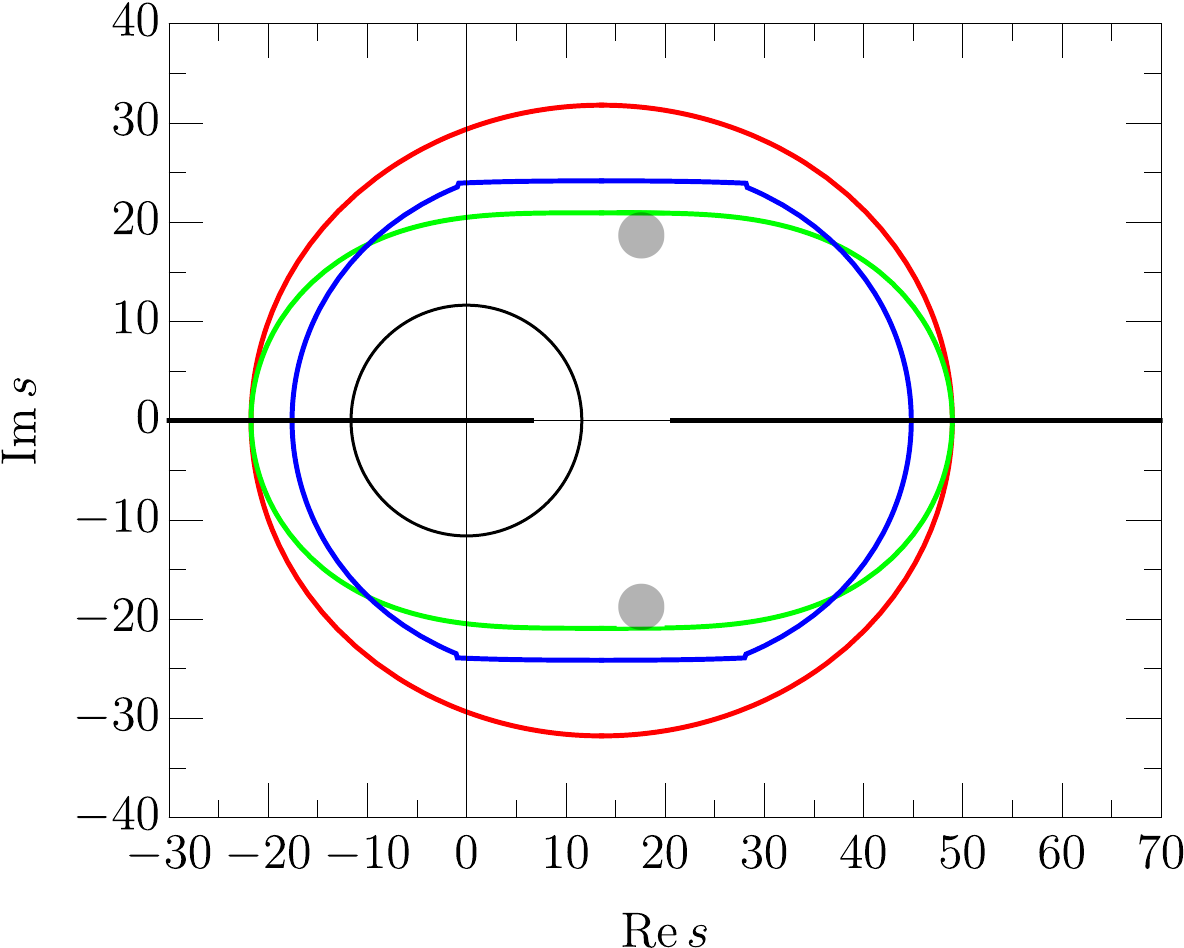}}
\caption{ The domain of validity in the complex $s$-plane of partial-wave hyperbolic dispersion relations for \pik scattering, in $m_\pi$ units. The red curve depicts the allowed region due to the $s$-channel contributions, whereas the green line encloses the very same region for the $t$-channel contributions, both calculated for $a=-10 m_\pi^2$. The blue line represents the applicability area for $a=0$, as used in \cite{DescotesGenon:2006uk}. The shadow areas represent the region where the \kap pole lies at. Although the applicability region can be extended in the real axis by choosing $a=-13.9 m_\pi^2$, then the \kap pole would lie outside the applicability domain. This is why $a=-10 m_\pi^2$ was used in \cite{Pelaez:2020uiw} to obtain its parameters. The black lines represent the physical, left and circular cuts characteristic of $\pi K$ scattering partial waves. }
\label{fig:rvpikhdr}
\end{figure}

Finally, the $s$ variable will remain inside the boundary anytime the $b$ values needed for the $s$-channel partial wave projection are inside their limits. We use again the partial wave projection formula of Eq.~\eqref{eq:projec2}, and perform the corresponding change of variables from $t$ to $b$. At the end of the day, we get that the boundary of applicability in the complex $s$ plane is
given by these two equations
\begin{align}
    (2\Sigma-s-a)(s-a)+B_{s,t}(\theta)\exp(i\theta)=0.
\end{align}

The regions of applicability corresponding to both ellipses are shown in Fig.~\ref{fig:rvpikhdr}, where we have used $a=-10 m_\pi^2$ to maximize the applicability in the real axis while the \kap pole and its uncertainties still lie within the boundaries. There it can be noticed that the boundary due to the $s$-channel contribution (Green line, using $B_t$ in the equation above) is more restrictive than that due to the $s$-channel (Red line, using $B_s$ in the equation above). The blue line represents the
$s$-channel boundary of the $a=0$ case, used in \cite{DescotesGenon:2006uk}, where they first showed that with HPWDR it was possible to reach the \kap pole, although the applicability domain in the real axis is somewhat smaller.

\section{Integral Kernels}
\label{app:Kernels}

In  section \ref{sec:DR}, we detailed the derivation of several Roy-Steiner-like equations, which, after projecting into partial waves, provide us with the necessary tools to perform a rigorous and model-independent  coupled study of both \pipikk and \pik scattering. The integral kernels corresponding to the $t$-channel dispersion relations for $F^+$ and $F^-$, with one and no subtractions, respectively, were already given in \cite{Pelaez:2018qny}. Here we will compile not only those but the whole system of integrands we have used in this review. Namely, we present kernels for fixed-$t$ and hyperbolic dispersion relations in the $s$- and $t$-channels. 

Let us recall that the hyperbolae we use for our dispersion relations are defined through the equation $(s-a)(u-a)=b$, as well as  some previous useful definitions:
\begin{equation}
z_{s}=1+\frac{2st}{\lambda_{s}}, \quad
\lambda_{s}=(s-m_+^2)(s-m_-^2)=s^2-2s\Sigma+\Delta^2=4s\,q_{K\pi}^2(s)\,.\nonumber 
\label{anglet}
\end{equation}
In addition, to  simplify our equations, we will now define
\begin{align}
x(t,s')&=\frac{4q_K(t)q_{\pi}(t)}{2s'+t-2\Sigma},\quad    A(t,s')={\rm artanh}\Big(x(t,s')\Big),\nonumber\\
   B(s,s')&= \frac{s}{\lambda_s}\left[\log\left(s'+s-2\Sigma \right)-\log\left(s'-\frac{\Delta^2}{s}\right)\right],\nonumber\\
   C(s,s')&=1-\frac{2s(s'+s-2\Sigma)}{\lambda_s}.
\end{align}
Beware that $C(s,s')\neq C(s',s)$ and that both expressions will appear below.

Let us recall that the $s$ variable corresponds to the ``output'' partial wave, while $s'$ and $t'$ correspond to the ``internal'' variables of the ``input'' amplitudes that are integrated. At the same time, in what follows, $\ell$ and $\ell'$ stand for the ``output'' and ``input'' angular momenta respectively.

\subsection{Fixed-t kernels}

In this subsection we provide the explicit expressions for the $L^I_{\ell \ell'}(s,s')$ and $L^\pm_{\ell \ell'}(s,t')$ kernels needed in the partial-wave dispersion relations in Eq.~\eqref{eq:sftpwdr}, only up to vector waves, as no tensor wave dispersion relation is implemented in this work. Nevertheless, tensor waves are used as input for the dispersion relations of the other waves.  Then, the dominant $s$-channel kernels we need read
\begin{align}
L^+_{0,0}(s,s')&=\frac{s^2}{s'^2(s'-s)}-\frac{2\Sigma s'-2\Delta^2}{s'\lambda_{s'}}+\frac{\Delta^2+s's+2\Sigma s
}{s'^2 s} +B(s,s'), \nonumber \\
L^+_{0,1}(s,s')&=3\frac{(s'^3+s'^2 s-s' s^2 +s \Delta^2)}{s'^2 \lambda_{s'}} +3C(s',s)B(s,s'), \nonumber \\
L^+_{1,0}(s,s')&=-\frac{2s}{\lambda_s}+C(s,s')B(s,s'), \nonumber \\
L^+_{1,1}(s,s')&=\frac{s\lambda_s}{\lambda_{s'}s'(s'-s)}+\frac{\lambda_s}{s' \lambda_{s'}}+\frac{6 s(\Delta^2-s'(2 s +s'-2\Sigma))}{ \lambda_{s'}\lambda_s} +3\,C(s',s)C(s,s')B(s,s'), \nonumber \\
L^-_{0,0}(s,s')&=\frac{1}{s'-s}-B(s,s'), \nonumber \\
L^-_{0,1}(s,s')&=3\left[\frac{1}{s'-s}-\frac{2s'}{\lambda_{s'}}-\frac{s'\lambda_s}{(s'-s)s\lambda_{s'}}-C(s',s)B(s,s')\right], \nonumber \\
L^-_{1,0}(s,s')&=-C(s,s')B(s,s')-2\frac{s}{\lambda_s}, \nonumber \\
L^-_{1,1}(s,s')&=\frac{s'\lambda_s}{(s'-s)s\lambda_{s'}}-3C(s',s) \left(C(s,s')B(s,s')+2\frac{s C(s',s)}{\lambda_s}\right).
\label{eq:kernelsfixedt}
\end{align}

 In addition, the $t$-channel contribution kernels are way simpler
\begin{align}
L^0_{0,2\ell}(s,t')&=\frac{2(2\ell)+1}{\sqrt{3}}(q_{\pi}(t')q_K(t'))^{2\ell}\frac{s}{\lambda_s}\left[\log\left(1+\frac{\lambda_s}{st'}\right)-\frac{\lambda_s}{st'}\right], \nonumber \\
L^0_{1,2\ell}(s,t')&=\frac{2(2\ell)+1}{\sqrt{3}}(q_{\pi}(t')q_K(t'))^{2\ell}\frac{s}{\lambda_s}\left[\left(1+\frac{2st'}{\lambda_s}\right)\log\left(1+\frac{\lambda_s}{st'}\right)-2\right]. 
\label{eq:tkernelsfixedt}
\end{align}

\subsection{Hyperbolic kernels}

\subsubsection{$t$-channel projection}

In this subsection we provide the explicit expressions for the $G^I_{\ell \ell'}(t,t')$, $G^\pm_{\ell \ell'}(t,s')$, $\hat G^I_{\ell \ell'}(t,t')$ and $\hat G^\pm_{\ell \ell'}(t,s')$ kernels, with $\ell\leq2$, needed in the partial-wave dispersion relations in Eqs.~\eqref{eq:pwhdr} and \eqref{eq:pwhdr1}. 
Note that,
in the input, partial waves with $\ell'>2$ will be safely neglected, except for the $\ell'=4$ partial wave used for the $g^0_2$ equation, which nevertheless gives a very small contribution.

We start by listing the kernels appearing in the $g^1_1(t)$ dispersion relation.
As in the main text of \cite{Pelaez:2018qny}, for the antisymmetric case, those with a hat correspond to one subtraction and those without to the unsubtracted case
\begin{eqnarray}
G^1_{1,3}(t,t')&=&\hat G^1_{1,3}(t,t')=\frac{7}{48}(t'+t-4\Sigma+10a), 
\label{kernels11} \\
G^-_{1,0}(t,s')&=&4\sqrt{2}\left[
\frac{(2s'-2\Sigma+t)A(t,s')-4q_K(t)q_{\pi}(t)}{16(q_K(t)q_{\pi}(t))^3} \right], \nonumber \\
G^-_{1,1}(t,s')&=&12\sqrt{2}\left[P_1(z_{s'})
\frac{(2s'-2\Sigma+t)A(t,s')-4q_K(t)q_{\pi}(t)}{16(q_K(t)q_{\pi}(t))^3}-\frac{2s'}{3(s'-a)\lambda_{s'}} \right], \nonumber \\
G^-_{1,2}(t,s')&=&20\sqrt{2} \left[P_2(z_{s'})
\frac{(2s'-2\Sigma+t)A(t,s')-4q_K(t)q_{\pi}(t)}{16(q_K(t)q_{\pi}(t))^3}-\frac{2s'z_s'}{(s'-a)\lambda_{s'}} \right. \nonumber \\
&+&\left.\frac{s'^2(2s'+t-2\Sigma)^2}{2(s'-a)^2\lambda_{s'}^2} 
-\frac{24s'^2(q_K(t)q_\pi(t))^2}{5(s'-a)^2\lambda_s'^2} \right],\nonumber \\
\hat G^-_{1,0}(t,s')&=&4\sqrt{2}\left[(s'-\Sigma+t/2)
\frac{A(t,s')-4q_K(t)q_{\pi}(t)}{16(q_K(t)q_{\pi}(t))^3}-\frac{1}{3\lambda_{s'}} \right], \nonumber \\
\hat G^-_{1,1}(t,s')&=&12\sqrt{2}\left[(s'-\Sigma+t/2)P_1(z_{s'})
\frac{A(t,s')-4q_K(t)q_{\pi}(t)}{16(q_K(t)q_{\pi}(t))^3}-\frac{1}{3\lambda_{s'}} \right], \nonumber \\
\hat G^-_{1,2}(t,s')&=&20\sqrt{2} \left[(s'-\Sigma+t/2)P_2(z_{s'})
\frac{A(t,s')-4q_K(t)q_{\pi}(t)}{16(q_K(t)q_{\pi}(t))^3}-\frac{1}{3\lambda_{s'}}
-\frac{2s'^2t}{\lambda^2_{s'}(s'-a)} \right], \nonumber
\end{eqnarray}
where $P_l(z_{s'})$ are the Legendre polynomials.
For the symmetric case, we find

\begin{eqnarray}
G^0_{2,4}(t,t')&=&\frac{3}{8} (t+t'-4\Sigma+7a), 
	\label{kernels02}\\
G^+_{2,0}(t,s')&=&\frac{\sqrt{3}(2s'+t-2\Sigma)^2}{32  q_K(t)^5
   q_\pi(t)^5}
\left[(3-x(t,s')^2) A(t,s')-3x(t,s')\right], \nonumber \\
G^+_{2,1}(t,s')&=&\frac{3\sqrt{3}(2s'+t-2\Sigma)^2}{32 q_K(t)^5
   q_\pi(t)^5} 
P_1(z_{s'}) \left[(3-x(t,s')^2) A(t,s')-3x(t,s')\right] , \nonumber \\
G^+_{2,2}(t,s')&=&5\sqrt{3}
\left[\frac{(2s'+t-2\Sigma)^2}{32 q_K(t)^5
   q_\pi(t)^5 }P_2(z_{s'})\Big((3-x(t,s')^2) A(t,s')-3x(t,s')\Big) 
   \frac{16s'^2t}{5(s'-a)^2\lambda_{s'}^2}\right]. \nonumber
\end{eqnarray}
Finally, for the $g_0^0(t)$ dispersion relation the kernels we need are
\begin{eqnarray}
G^0_{0,2}(t,t')&=&\frac{5}{16}(t+t'-4\Sigma+6a), \nonumber	 \\
G^+_{0,0}(t,s')&=&\sqrt{3}\left[\frac{A(t,s')}{q_K(t)q_{\pi}(t)}  +\frac{2(\Sigma-s')}{\lambda_{s'}}\right], \nonumber \\
G^+_{0,1}(t,s')&=&3\sqrt{3}\left[\frac{A(t,s')}{q_K(t)q_{\pi}(t)}P_1(z_{s'})
-\frac{(2s'+2t-2\Sigma)}{\lambda_{s'}}-\frac{2at}{(s'-a)\lambda_{s'}}\right], \nonumber \\
G^+_{0,2}(t,s')&=&5\sqrt{3}\left[\frac{A(t,s')}{q_K(t)q_{\pi}(t)}
P_2(z_{s'})
-\frac{2s-2\Sigma}{\lambda_{s'}}-\frac{6st(\Delta^2+s'(3s'+2t-4\Sigma))}{(s'-a)\lambda_{s'}^2}\right.\nonumber\\
&+&\left.\frac{3s'^2t(2s'+t-2\Sigma)^2}{2(s'-a)^2\lambda_s'^2}
-\frac{8s'^2t(q_K(t)q_{\pi}(t))^2}{(s'-a)^2\lambda_{s'}^2}\right].
\label{eq:kernels00}
\end{eqnarray}

\subsubsection{$s$-channel projection}

Finally, we provide the $K_{\ell,\ell'}^\pm(s,s')$, $\hat K_{\ell,\ell'}^\pm(s,s')$, $K_{\ell,\ell'}^I(s,t')$ and $\hat K_{\ell,\ell'}^I(s,t')$   kernels needed for the dispersive integrals in Eqs.~\eqref{eq:shdrfm} and \eqref{eq:shdrfmi}. Let us recall that in this case we neglect $\ell \geq 2$, since their contribution to the partial waves of interest is very small. The conventions regarding the variables and angular momenta are the ones explained above. 
The $s$-channel kernels of the unsubtracted antisymmetric amplitude thus read
\begin{align}
K^+_{0,0}(s,s')&=\frac{2((s'+\Sigma)\Delta^2-2s'\Sigma)}{s'^2\lambda_{s'}}-\frac{s'^2+2s'(\Sigma-s)-2s\Sigma}{s'^2(s'-s)}+B(s,s'), \\
K^-_{0,0}(s,s')&=\frac{1}{s'-s}-B(s,s'),  \nonumber \\
K^+_{0,1}(s,s')&=3\left[\frac{s(s'+2\Sigma)-\Delta^2}{\lambda_{s'}s}-\frac{s'\lambda_s}{(a-s')\lambda_{s'}s}+C(s',s) B(s,s')\right], \nonumber \\
K^-_{0,1}(s,s')&=-3\left[\frac{(s's+\Delta^2)}{\lambda_{s'}s}-\frac{s'(3s^2-2\Sigma s+\Delta^2)}{(a-s')\lambda_{s'}s}+C(s',s) B(s,s')\right], \nonumber \\
 K^+_{1,0}(s,s')&=-K^-_{1,0}(s,s')=C(s,s')B(s,s') +2\frac{s}{\lambda_s}, \nonumber \\
K^+_{1,1}(s,s')&=3C(s',s)\left[C(s,s')B(s,s') +2\frac{s}{\lambda_s}\right]+\frac{s'(a-s)\lambda_s}{(s'-s)s(a-s')\lambda_{s'}}, \nonumber \\
K^-_{1,1}(s,s')&=-3C(s',s)\left[C(s,s')B(s,s') +2\frac{s}{\lambda_s}\right]+\frac{s'(a-s)\lambda_s}{(s'-s)s(a-s')\lambda_{s'}}, 
\label{eq:kernelsk}
\end{align}
whereas  the once-subtracted kernels for the antisymmetric case read
\begin{align}
\hat K^-_{0,0}(s,s')&=\frac{1}{s'-s}-B(s,s')+\frac{\lambda_s}{2s\lambda_{s'}}-2\frac{(s-\Sigma)}{\lambda_{s'}}, \nonumber \\
\hat K^-_{0,1}(s,s')&=3\left[\frac{1}{s'-s}-2\frac{(s'+s-\Sigma)}{\lambda_{s'}}-\frac{\lambda_s(s'+s)}{2(s'-s)s \lambda_{s'}} -C(s',s)B(s,s')\right] , \nonumber \\
\hat K^-_{1,0}(s,s')&=-C(s,s')B(s,s')-\frac{2s}{\lambda_s}-\frac{\lambda_s}{6s\lambda_{s'}}, \nonumber \\
\hat K^-_{1,1}(s,s')&= -3C(s',s)\left[C(s,s')B(s,s') +\frac{2s}{\lambda_s}\right] +\frac{(s'+s)\lambda_s}{2s(s'-s)\lambda_{s'}}.
\label{eq:kernelsk1}
\end{align}

Next, we provide the kernels for the $t$-channel contribution, which read
\begin{align}
K^0_{0,0}(s,t')&=\frac{1}{\sqrt{3}}\frac{s}{\lambda_s}\left[\log\left(1+\frac{\lambda_s}{st'}\right)-\frac{\lambda_s}{st'}\right], \nonumber \\
K^0_{0,2}(s,t')&=\frac{5}{\sqrt{3}}[q_{\pi}(t')q_K(t')]^2\frac{s}{\lambda_s}\left[\log\left(1+\frac{\lambda_s}{st'}\right)-\frac{\lambda_s}{st'}\right]+\frac{\sqrt{3}\lambda_s(s-a)}{16st'}, \nonumber \\
K^0_{1,0}(s,t')&=\frac{1}{\sqrt{3}}\frac{s}{\lambda_s}\left[P_1(z_s)\log\left(1+\frac{\lambda_s}{st'}\right)-2\right], \nonumber \\
K^0_{1,2}(s,t')&=\frac{5}{\sqrt{3}}\frac{s}{\lambda_s}[q_{\pi}(t')q_K(t')]^2P_1(z_s)\left[\log\left(1+\frac{\lambda_s}{st'}\right)-2\right]+\frac{5\lambda_s(s-a)}{16\sqrt{3}st'}
, \nonumber \\
K^1_{0,1}(s,t')&=\frac{3}{4\sqrt{2}}\frac{s (t'+2s-2\Sigma)}{\lambda_s}\left[\log\left(1+\frac{\lambda_s}{st'}\right)-\frac{\lambda_s}{s (t'+2s-2\Sigma)}\right], \nonumber \\
K^1_{1,1}(s,t')&=\frac{3}{4\sqrt{2}}\frac{s (t'+2s-2\Sigma)}{\lambda_s}\left[P_1(z_s)\log\left(1+\frac{\lambda_s}{st'}\right)-2\right], \nonumber \\
\hat K^1_{0,1}(s,t')&= \frac{3}{4\sqrt{2}}\frac{s (t'+2s-2\Sigma)}{\lambda_s}\left[\log\left(1+\frac{\lambda_s}{st'}\right)-\frac{2\lambda_s}{t' s}+ \frac{\lambda_s^2}{t' s^2 (t'+2s-2\Sigma)}\right], \nonumber \\
\hat K^1_{1,1}(s,t')&= \frac{3}{4\sqrt{2}}\frac{s (t'+2s-2\Sigma)}{\lambda_s}\left[P_1(z_s)\log\left(1+\frac{\lambda_s}{st'}\right)-2\right]-\frac{\lambda_s}{8\sqrt{2}t' s}, 
\label{eq:tkernelsk}
\end{align}
where once again we have used a hat for the once-subtracted kernels of the $F^-$ amplitude.

All these kernels produce smooth integrable inputs in the physical region, they are suppressed at higher energies and match the kinematic behavior of every partial wave. They also produce the left and circular-cut structures required by partial-wave projection.

\section{Muskhelishvili-Omn\`es matching conditions}
\label{app:differenttm}

In sections~\ref{sec:DR} and \ref{sec:DRTests} we detailed our choice of the matching point $t_m$ for the different $\pi \pi \to K \bar K$ partial waves. We decided to modify slightly our previous value in~\cite{Pelaez:2018qny} from $\sqrt{t_m}=1.2$ GeV to $\sqrt{t_m}=1$ GeV for the $g^1_1(t)$ partial wave for practical convenience, and used $\sqrt{t_m}=1.2$ GeV for the other two partial waves studied here, i.e. $g^0_0(t)$ and $g^0_2(t)$.

\begin{figure}[!ht]
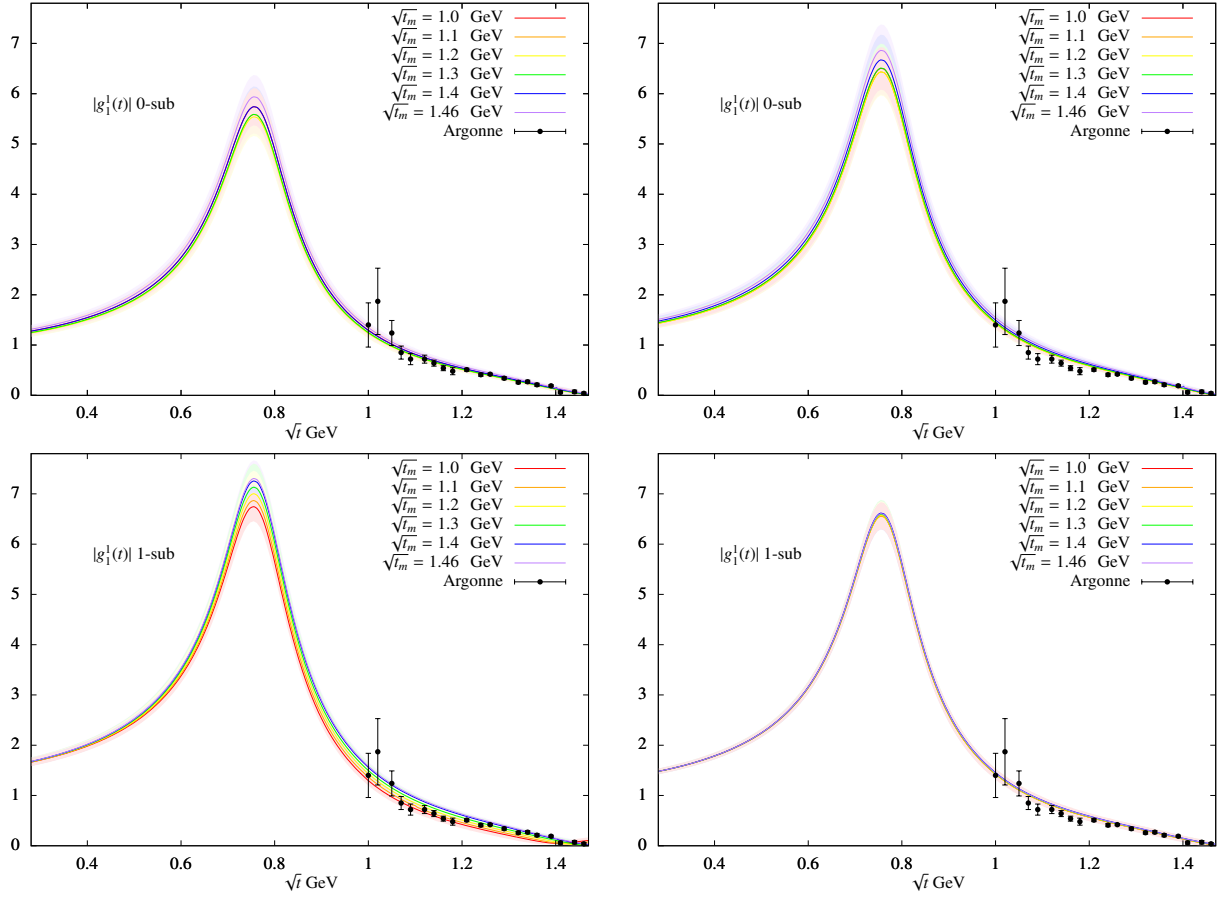

\begin{center}
\resizebox{\textwidth}{!}{\input{figures/g11disptm.tex} \input{figures/g11disptmcfd.tex}}\\
\vspace{0.2cm}
\resizebox{\textwidth}{!}{\input{figures/g11disptmsub.tex} \input{figures/g11disptmsubcfd.tex}}
\caption{Comparison of the 
unsubtracted (top) or once-subtracted (bottom) HPWDR output, when using different matching points $t_m$ for $g^1_1(t)$.
Notice the much better agreement between outputs with different choices of matching points when using as input the CFD (right) rather than the UFD (left), particularly for the subtracted case.
\label{fig:g11tm} 
}
\end{center}
\end{figure}

One may wonder if varying this $t_m$ substantially modifies the dispersive constraint for the crossed-channel partial waves, as it imposes a mathematical matching condition between the dispersion relations of section~\ref{sec:MO}  right below $t_m$ and the fitted partial waves right above. Of course, if the CFD fits were perfect,  this matching condition would not produce any noticeable effect, and thus the dispersive partial wave would  always be compatible within uncertainties regardless of the matching point.

Of the three partial waves, only $g^1_1$ and $g^0_0$ contribute substantially to the $\pi K$ system and $g^0_2$ can be considered ``decoupled'' from the rest. In the following, we will focus on the former two.

\begin{figure}[!ht]
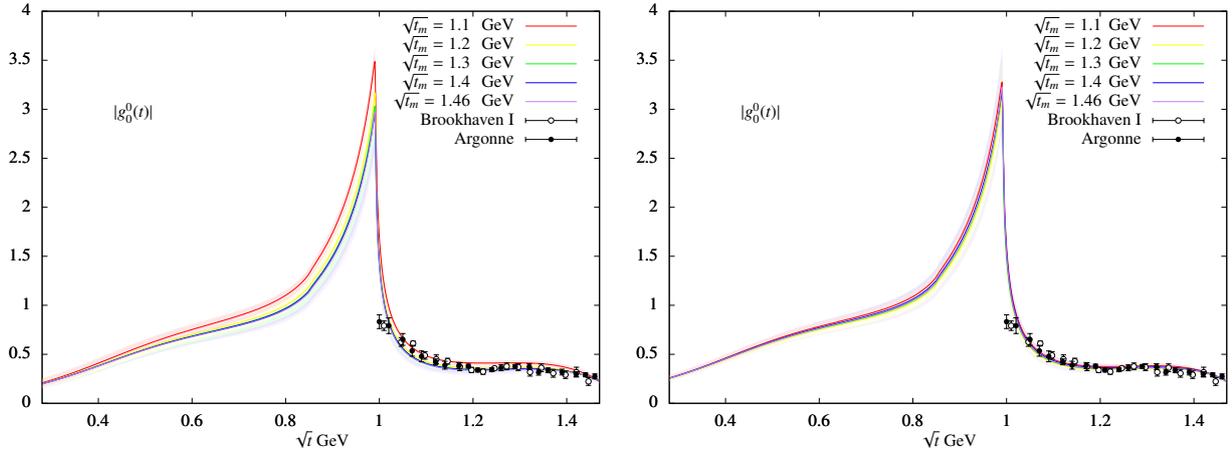

\begin{center}
\resizebox{\textwidth}{!}{\input{figures/g00disptm.tex} \input{figures/g00disptmcfd.tex}}
\caption{Comparison of the HPWDR output, when using different matching points $t_m$ for $g^0_0(t)$.  Notice the much better agreement between the outputs with different choices of matching points when using the CFD (right) instead of the UFD (left) as input.
\label{fig:g00tm} 
}
\end{center}
\end{figure}

Regarding $g^1_1(t)$, different choices of $t_m$ within a rather large region produce perfectly compatible results for its CFD dispersive output, both in the unsubtracted and once-subtracted cases, as shown in Fig.~\ref{fig:g11tm}. This is an improvement upon the UFD case, which nevertheless also yields quite similar results with different choices of $t_m$ within that range.
 However, the higher the matching point $t_m$, the larger the dispersive uncertainty band.  This is the main reason why we have chosen $t_m$ as small as possible, namely, $t_m=1$ GeV$^2$, taking advantage of the fact that the cusp and isospin breaking effects around $K \bar K$ threshold are negligible in this wave.

The situation concerning $g^0_0(t)$ is  more complicated, mostly because using $\sqrt{t_m}\leq1.1$ GeV produces significant deviations from data and thus matching points close to $K \bar K$ threshold should be avoided. As detailed in sections~\ref{sec:DRTests} and~\ref{sec:CFDpipiKK} this could be explained by  isospin breaking effects, which, among other things imply the existence of a double threshold, one for $K^+ K^-$ and another one for $K^0 \bar K^0$. These effects are enhanced in this wave due to the nearby $f_0(980)$ resonance and cannot be well reproduced by our dispersive formalism, which is isospin-symmetric. Nevertheless, as  seen in Fig.~\ref{fig:g00tm}, if one chooses $t_m$ above this energy, different matching points again produce dispersive results fairly compatible within uncertainties in the physical region. Moreover, there is a clear improvement in the pseudo-physical region when using the CFD as input instead of the UFD. In terms of the $\hat{d}^2$  function defined in section~\ref{sec:DRTests}, and excluding the region within 20 MeV of the $K \bar K$ threshold, the CFD $\hat{d}^2$ varies between less than 1 and 1.4 $\sigma$, which we consider perfectly acceptable for our purposes. On top of that, the contribution of this partial wave to the rest of the coupled system is already small, so that varying this matching condition produces negligible effects, well within our original statistical uncertainties.


\bibliography{largebiblio.bib}

\end{document}